\RequirePackage{fix-cm}

\documentclass[pdftex,twocolumn,epjc3]{preprint}   

\usepackage{graphicx}
\usepackage[intlimits,sumlimits,centertags]{amsmath}
\usepackage{amssymb,amsfonts}
\usepackage{gensymb}
\usepackage[bookmarksopen,bookmarksdepth=1]{hyperref}
\usepackage[x11names]{xcolor}
\usepackage{mathpazo} 
\usepackage{microtype}
\usepackage[a4paper, left=1.5cm,right=1.5cm,top=2.5cm,bottom=2cm]{geometry}
\usepackage[normalem]{ulem}
\usepackage{textcomp}

\hypersetup{pdftitle={Probing Particle Physics with IceCube},
pdfsubject={Probing Particle Physics with IceCube},
pdfauthor={Markus Ahlers, Klaus Helbing, and Carlos P\'erez de los Heros},
pdfstartview={FitH},
colorlinks=true,
bookmarksopen=false,
bookmarksnumbered=false,
bookmarksopenlevel=0,
linkcolor=Blue1!50!black,
citecolor=Green1!50!black,
urlcolor=Blue1!50!black
}

\journalname{Eur. Phys. J. C}

\begin{document}

\setlength{\mathindent}{1ex}

\title{Probing Particle Physics with IceCube}
	
\author{Markus Ahlers\thanksref{email01,NBI} \and 
Klaus Helbing\thanksref{email02,WUPP} \and
Carlos P\'erez de los Heros\thanksref{email03,UPP}}

\thankstext{email01}{e-mail: \href{mailto:markus.ahlers@nbi.ku.dk}{markus.ahlers@nbi.ku.dk}}
\thankstext{email02}{e-mail: \href{mailto:helbing@uni-wuppertal.de}{helbing@uni-wuppertal.de}}
\thankstext{email03}{e-mail: \href{mailto:cph@physics.uu.se}{cph@physics.uu.se}}

\institute{Niels Bohr International Academy \& Discovery Centre, Niels Bohr Institute, University of Copenhagen, DK-2100 Copenhagen, Denmark\label{NBI}\and
Department of Physics, University of Wuppertal, D-42119 Wuppertal, Germany\label{WUPP} \and
Department of Physics and Astronomy, Uppsala University, S-75120 Uppsala, Sweden\label{UPP}}

\date{}

\maketitle

\begin{abstract}
 The IceCube observatory located at the South Pole is a cubic-kilometre optical Cherenkov telescope primarily designed for the detection of high-energy astrophysical neutrinos. IceCube became fully operational in 2010, after a seven-year construction phase, and reached a milestone in 2013 by the first observation of cosmic neutrinos in the TeV-PeV energy range. This observation does not only mark an important breakthrough in neutrino astronomy, but it also provides a new probe of particle physics related to neutrino production, mixing, and interaction. In this review we give an overview of the various possibilities how IceCube can address fundamental questions related to the phenomena of neutrino oscillations and interactions, the origin of dark matter, and the existence of exotic relic particles, like monopoles. We will summarize recent results and highlight future avenues.
\end{abstract}

\section{Introduction}
\label{sec:intro}
Not long after the discovery of the neutrino by Cowan and Reines in 1956~\cite{Cowan:1992xc}, the idea emerged that it represented the ideal astronomical messenger~\cite{Reines:1960we}. Neutrinos are only weakly interacting with matter and can cross cosmic distances without being absorbed or scattered. However, this weak interaction is also a challenge for the observation of these particles. Early estimates of the expected flux of high-energy neutrinos associated with the observed flux of extra-galactic cosmic rays indicated that neutrino observatories require gigaton masses as a necessary condition to observe a few neutrino interactions per year~\cite{Waxman:1998yy}. These requirements can only be met by special experimental setups that utilise natural resources. Not only that -- the detector material has to be suitable so that these few interactions can be made visible and separated from large atmospheric backgrounds.

Despite these obstacles, there exist a variety of experimental concepts to detect high-energy neutrinos. One particularly effective method is based on detecting the radiation of optical Cherenkov light produced by relativistic charged particles. This requires the use of optically transparent detector media like water or ice, where the Cherenkov emission can be read out by optical sensors deployed in the medium. This information then allows  to reconstruct the various Cherenkov light patterns produced in neutrino events and infer the neutrino flavour, arrival direction, and energy. The most valuable type of events for neutrino astronomy are charged current interactions of muon-neutrinos with matter near the detector. These events produce muons that can range into the detector and allow the determination of the initial muon-neutrino direction within a precision of better than one degree.

Presently the largest optical Cherenkov telescope is the IceCube Observatory, which uses the deep Glacial Ice at the geographic South Pole as its detector medium. The principal challenge of any neutrino telescope is the large background of atmospheric muons and neutrinos produced in cosmic ray interactions in the atmosphere. High-energy muons produced in the atmosphere have a limited range in ice and bedrock. Nevertheless IceCube, at a depth of 1.5 kilometres, observes about 100 billion atmospheric muon events per year. This large background can be drastically reduced by only looking for {\it up-going} events, {\it i.e.},~events that originate below the horizon. This cut leaves only muons produced by atmospheric neutrinos at a rate of about 100,000 per year. While these large backgrounds are an obstacle for neutrino astronomy they provide a valuable probe for cosmic ray physics in general and for neutrino oscillation and interaction studies in particular.

In this review we want to highlight IceCube's potential as a facility to probe fundamental physics. There exist a variety of methods to test properties of the Standard Model (SM) and its possible extensions. The flux of atmospheric and astrophysical neutrinos observed in IceCube allows to probe fundamental properties in the neutrino sector related to the standard neutrino oscillations (neutrino mass differences, mass ordering, and flavour mixing) and neutrino-matter interactions. It also provides a probe for exotic oscillation effects, {\it e.g.}, related to the presence of sterile neutrinos or non-standard neutrino interactions with matter. The ultra-long baselines associated with the propagation of cosmic neutrinos observed beyond 10~TeV allow for various tests of feeble neutrino oscillation effects that can leave imprints on the oscillation-averaged flavour composition.

One of the fundamental questions in cosmology is the origin of dark matter that today constitutes one quarter of the total energy density of the Universe. Candidate particles for this form of matter include weakly interacting massive particles (WIMPs) that could have been thermally produced in the early Universe. IceCube can probe the existence of these particles by the observation of a flux of neutrinos produced in the annihilation or decay of WIMPs gravitationally clustered in nearby galaxies, the halo of the Milky Way, the Sun, or the Earth. In the case of compact objects, like Sun and Earth, neutrinos are the only SM particles that can escape the dense environments to probe the existence of WIMPs. 

Neutrino telescopes can also probe exotic particles leaving direct or indirect Cherenkov signals during their passage through the detector. One important example are relic magnetic monopoles, topological defects that could have formed during a phase transition in the early Universe. Light exotic particles associated with extensions of the Standard Model can also be produced by the interactions of high-energy neutrinos or cosmic rays. Collisions of neutrinos and cosmic rays with nucleons in the vicinity of the Cherenkov detector can reach center-of-mass energies of the order $\sqrt{s}\simeq 1$~TeV (neutrino energy $E_\nu\simeq10^{15}$~eV) or even $\sqrt{s}\simeq 100$~TeV (cosmic ray energy $E_{\rm CR}\simeq 10^{20}$~eV), respectively, only marginally probed by collider experiments.

The outline of this review is as follows. We will start in sections~\ref{sec:IceCube} and~\ref{sec:analysis} with a description of the IceCube detector, atmospheric backgrounds, standard event reconstructions, and event selections. In section~\ref{sec:std_oscillations} we summarise the phenomenology of three-flavour neutrino oscillation and IceCube's contribution to test the atmospheric neutrino mixing. We will cover standard model neutrino interactions in section~\ref{sec:sm_interactions} and highlight recent measurements of the inelastic neutrino-nucleon cross sections with IceCube. We then move on to discuss IceCube's potential to probe non-standard neutrino oscillation with atmospheric and astrophysical neutrino fluxes in section~\ref{sec:exotic_oscillations}. In section~\ref{sec:dm} we highlight IceCube results on searches for dark matter and section~\ref{sec:monopoles} is devoted to magnetic monopoles while section~\ref{sec:exotics} covers other massive exotic particles and Big Bang relics.

Any review has its limitations, both in scope and timing. We have given priority to present a comprehensive view of the activity of IceCube in areas related to the topic of this review, rather than concentrating on a few recent results. We have also chosen at times to include older results for completeness, or when it was justified as an illustration of the capabilities of the detector on a given topic. The writing of any review develops along its own plot and updated results on some analyses have been made public while this paper was in preparation, and could not be included here. This only reflects on the lively activity of the field.

Throughout this review we will use natural units, $\hbar=c=1$, unless otherwise stated. Electromagnetic expressions will be given in the Heaviside-Lorentz system with $\epsilon_0=\mu_0=1$, $\alpha = e^2/4\pi \simeq 1/137$ and $1{\rm Tesla} \simeq 195{\rm eV}^2$.

\section{The IceCube Neutrino Observatory}
\label{sec:IceCube}

%%%%%%%%%%%%%%%%%%%%%%
\begin{figure*}[t]
  \centering
  \includegraphics[width=0.9\linewidth]{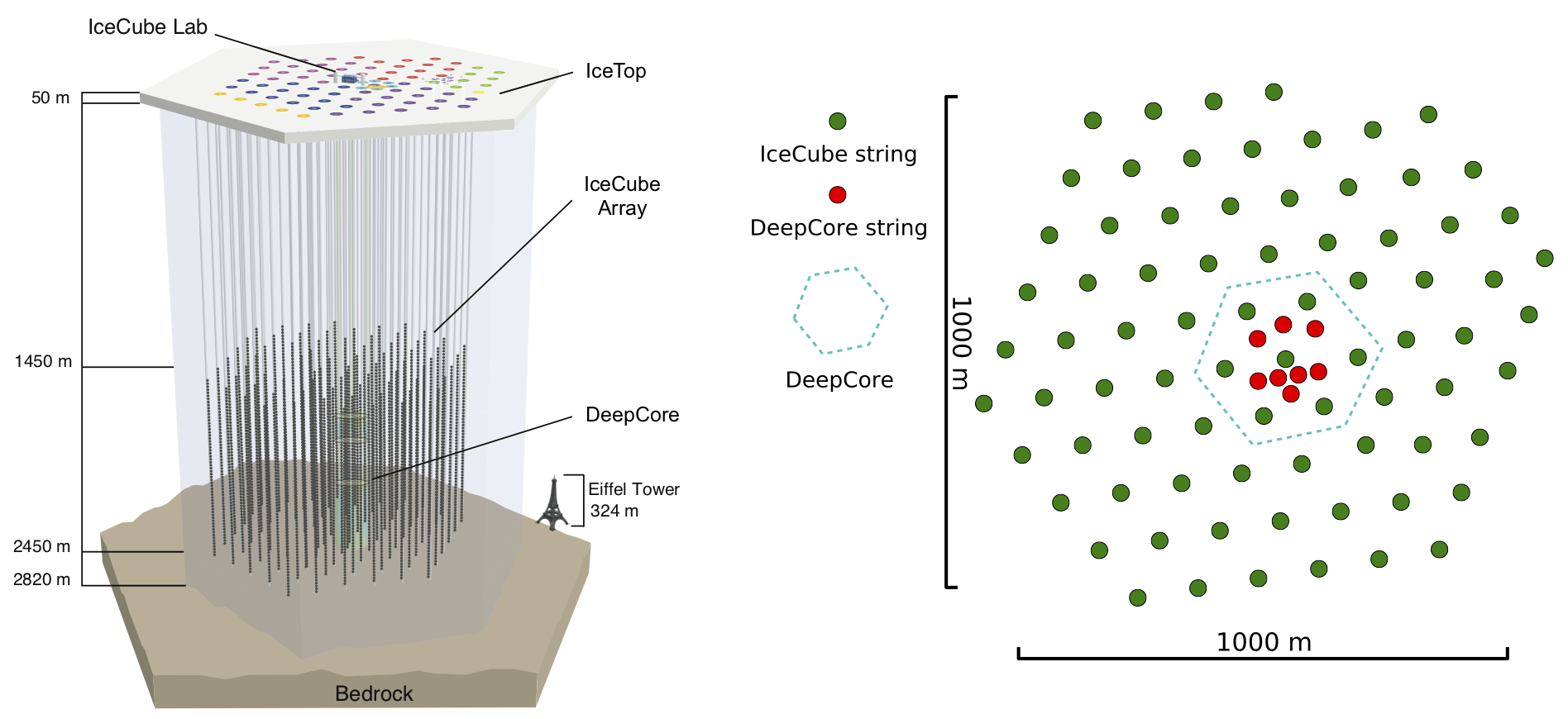}
  \caption[]{Sketch of the IceCube observatory. The right plot shows the surface footprint of IceCube. The green circles represent the standard IceCube strings, separated by 125~m, and the red ones the more densely
 instrumented strings with high quantum efficiency photomultiplier tubes. Strings belonging to the DeepCore sub-array are enclosed by the dashed line.\label{fig:IceCube}}
\end{figure*}
%%%%%%%%%%%%%%%%%%%%%%

The IceCube Neutrino Observatory~\cite{Aartsen:2016nxy} consists of an in-ice array (simply ``IceCube'' hereafter) and a surface air shower array, IceTop~\cite{IceCube:2012nn}. IceCube utilises one cubic kilometre of the deep ultra-clear glacial ice at the South Pole as its detector medium (see left panel of Fig.~\ref{fig:IceCube}). This volume is instrumented with 5,160 Digital Optical Modules (DOMs) that register the Cherenkov light emitted by relativistic charged particles passing through the detector. The DOMs are distributed on 86 read-out and support cables (``strings'') and are deployed between 1.5~km and 2.5~km below the surface. Most strings follow a triangular grid with a width of 125~m, evenly spaced over the volume (see green markers in right panel of Fig.~\ref{fig:IceCube}).

Eight strings are placed in the centre of the array and are instrumented with a denser DOM spacing and typical inter-string separation of 55~m (red markers in right panel of Fig.~\ref{fig:IceCube}). They are equipped with photomultiplier tubes with higher quantum efficiency. These strings, along with the first layer of the surrounding standard strings, form the DeepCore low-energy sub-array~\cite{Collaboration:2011ym}. Its footprint is depicted by a blue dashed line in Fig.~\ref{fig:IceCube}. While the original IceCube array has a neutrino energy threshold of about 100~GeV, the addition of the denser infill lowers the energy threshold to about 10~GeV.  The DOMs are operated to trigger on single photo-electrons and to digitise in-situ the arrival time of charge (``waveforms'') detected in the photomultiplier. The dark noise rate of the DOMs is about 500 Hz for standard modules and 800 Hz for the high-quantum-efficiency DOMs in the DeepCore sub-array.

Some results highlighted in this review were derived from data collected with the AMANDA array~\cite{Andres:1999hm}, the predecessor of IceCube built between 1995 and 2001 at the same site, and in operation until May 2009. AMANDA was not only a proof of concept and a hardware test-bed for the IceCube technology, but a full fledged detector which obtained prime results in the field.

\subsection{Neutrino Event Signatures}

\begin{figure*}[t]
\centering\includegraphics[height=4.5cm]{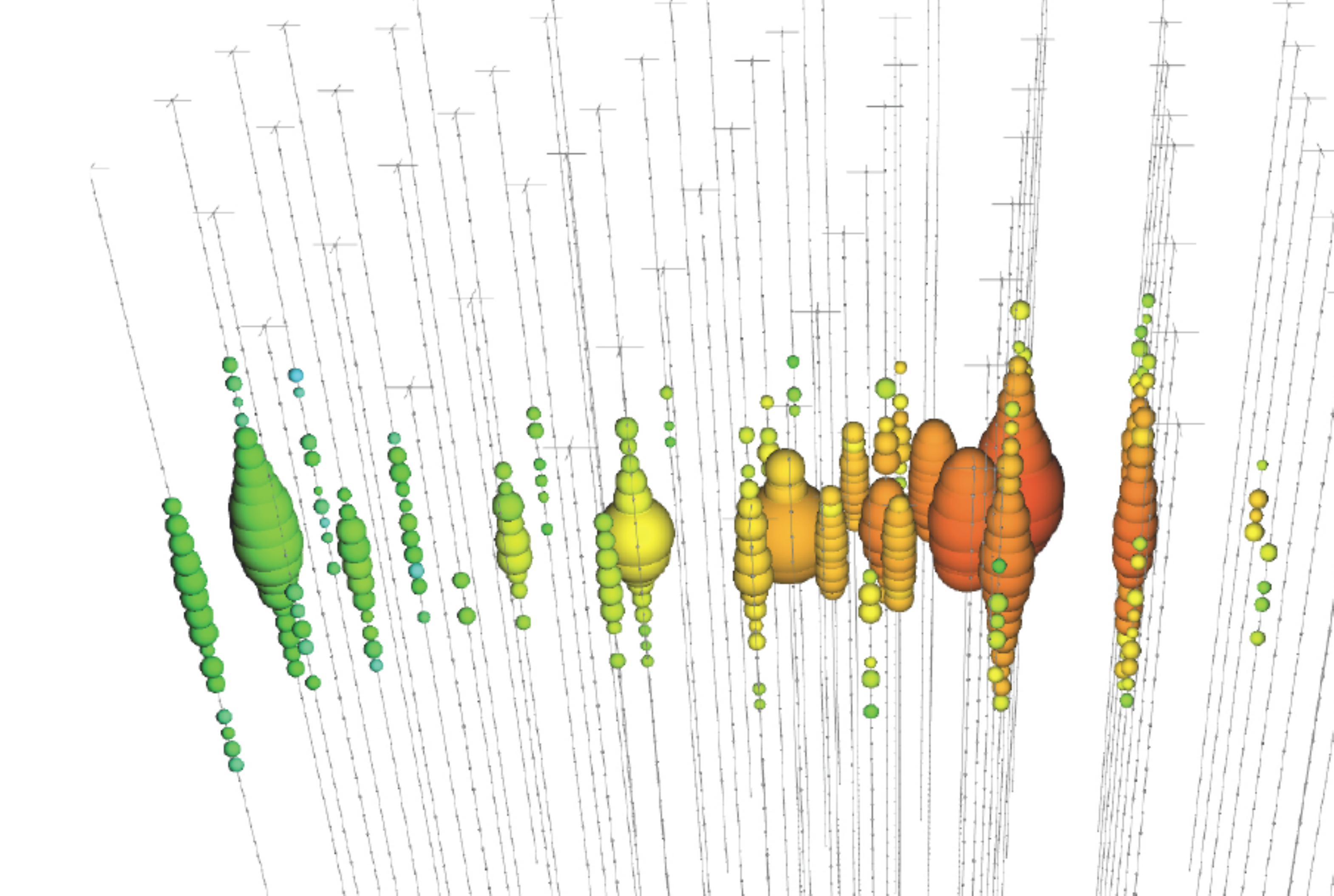}\hspace{1.5cm}\includegraphics[height=4.5cm]{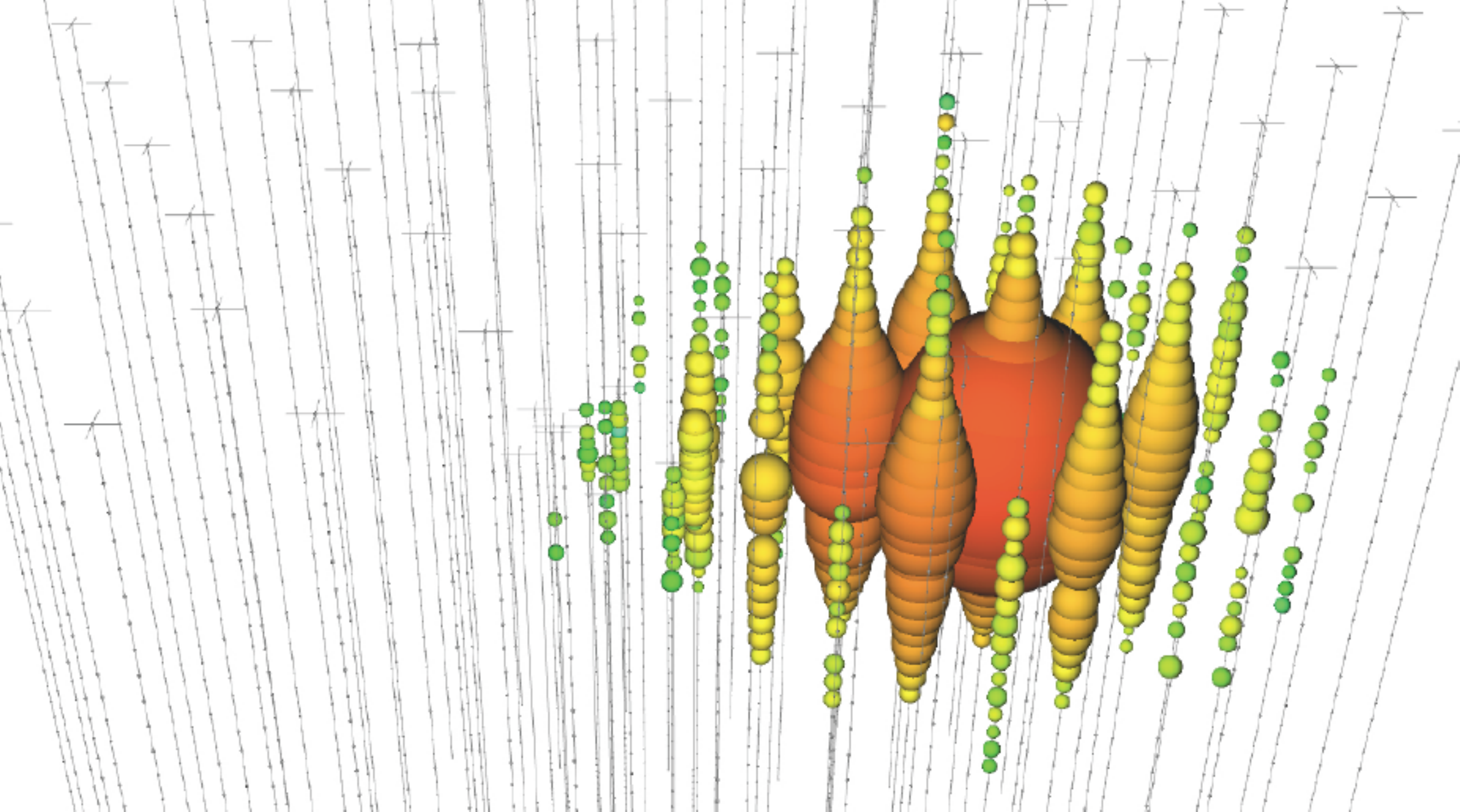}
\caption[]{Two examples of events observed with IceCube. The left plot shows a muon track from a $\nu_{\mu}$ interaction crossing the detector. Each coloured dot represents a hit DOM. The size of the dot is proportional to the amount of light detected and the colour code is related to the relative timing of light detection: read denotes earlier hits, blue, later hits. The right plot shows a $\nu_{e}$ or $\nu_{\tau}$ charged-current (or any flavour neutral-current) interaction inside the detector.}\label{fig:event_examples}
\end{figure*}

As we already highlighted in the introduction, the main event type utilised in high-energy neutrino astronomy are charged current (CC) interactions of muon neutrinos with nucleons ($N$), $\nu_\mu + N \rightarrow \mu^- + X$. These interactions produce high-energy muons that lose energy by ionisation, bremsstrahlung, pair production and photo-nuclear interactions in the ice~\cite{Chirkin:2004hz}. The combined Cherenkov light from the primary muon and secondary relativistic charged particles leaves a track-like pattern as the muon passes through the detector. An example is shown in the left panel of Fig.~\ref{fig:event_examples}. In this figure, the arrival time of Cherenkov light in individual DOMs is indicated by colour (earlier in red and later in blue) and the size of each DOM is proportional to the total Cherenkov light it detected.~\footnote{Note that in this particular example, also the Cherenkov light emission from the hadronic cascade $X$ is visible in the detector.}. Since the average scattering angle between the incoming neutrino and the outgoing muon decreases with energy, $\Psi_{\nu \rightarrow \mu}\sim 0.7{\text{\textdegree}}(E_{\nu}/{\rm TeV})^{-0.7}$~\cite{Learned:2000sw}, an angular resolution below 1{\textdegree} can be achieved for neutrinos with energies above a few TeV, only limited by the detector's intrinsic angular resolution. This changes at low energies, where muon tracks are short and their angular resolution deteriorates rapidly. For neutrino energies of a few tens of GeVs the angular resolution reaches a median of $\sim40${\textdegree}.

 All deep-inelastic interactions of neutrinos, both neutral current (NC), $\nu_{\alpha} + N \rightarrow \nu_{\alpha} + X$ and charged current, $\nu_{\alpha} + N \rightarrow \ell^-_{\alpha} + X$, create hadronic cascades $X$ that are visible by the Cherenkov emission of secondary charged particles. However, these secondaries can not produce elongated tracks in the detector due to their rapid scattering or decay in the medium. Because of the large separation of the strings in IceCube and the scattering of light in the ice, the Cherenkov light distribution from particle cascades in the detector is rather spherical, see right panel of Fig.~\ref{fig:event_examples}. For cascades or tracks fully contained in the detector, the energy resolution is significantly better since the full energy is deposited in the detector and it is proportional to the detected light. The ability to distinguish these two light patterns in any energy range is crucial, since cascades or tracks can contribute to background or signal depending on the analysis performed. 

The electrons produced in charged current interactions of electron neutrinos, $\nu_{e} + N \rightarrow e^- + X$, will contribute to an electromagnetic cascade that overlaps with the hadronic cascade $X$ at the vertex. At energies of $E_\nu\simeq6.3$~PeV, electron anti-neutrinos can interact resonantly with electrons in the ice via a $W$-resonance (``Glashow'' resonance)~\cite{Glashow:1960zz}. The $W$-boson decays either into hadronic states with a branching ratio (BR) of $\simeq67$\%, or into leptonic states (${\rm BR}\simeq11$\% for each flavour). This type of event can be visible by the appearance of isolated muon tracks starting in the detector or by spectral features in the event distribution~\cite{Bhattacharya:2011qu}.

Also the case of charged current interactions of tau neutrinos, $\nu_{\tau} + N \rightarrow \tau + X$, is special. Again, the hadronic cascade $X$ is visible in Cherenkov light. The tau has a lifetime (at rest) of $0.29$~ps and decays to leptons as $\tau^-\to\mu^-+\overline{\nu}_\mu+\nu_\tau$ (BR~$\simeq18\%$) and $\tau^-\to e^-+\overline{\nu}_e+\nu_\tau$ (BR~$\simeq18\%$) or to hadrons (mainly pions and kaons, BR~$\simeq64\%$) as $\tau^-\to\nu_\tau+{\rm mesons}$. With tau energies below $100$~TeV these charged current events will also contribute to track and cascade events. However, the delayed decay of taus at higher energies can become visible in IceCube, in particular above around a PeV when the decay length becomes of the order of 50~m. This allows for a variety of characteristic event signatures, depending on the tau energy and decay channel~\cite{Abbasi:2012cu,Aartsen:2015dlt}. 

%%%%%%%%%%%%%%
%%%%%%%%%%%%%%%%

\section{Event Selection and Reconstruction}
\label{sec:analysis}
In this review we present results from analyses which use different techniques tailored to the characteristics of the signals searched for. It is therefore impossible to give a description of a generic analysis strategy which would cover all aspects of every approach. There are, however, certain levels of data treatment and analysis techniques that are common for all analyses in IceCube, and which we cover in this section.

\subsection{Event Selection}

Several triggers are active in IceCube in order to preselect potentially interesting physics events~\cite{Kelley:2014gra}. They are based on finding causally connected spatial hit distributions in the array, typically requiring a few neighbour or next-to-neighbour DOMs to fire within a predefined time window. Most of the triggers aim at finding relativistic particles crossing the detector and use time windows of the order of a few microseconds. In order to extend the reach of the detector to exotic particles, {\it e.g.}, monopoles catalysing nucleon-decay, which can induce events lasting up to milliseconds, a dedicated trigger sensitive to non-relativistic particles with velocities down to $\beta^{-4}$ has also been implemented.

When a trigger condition is fulfilled the full detector is read out. IceCube triggers at a rate of 2.5~kHz, collecting about 1~TB/day of raw data. To reduce this amount of data to a more manageable level, a series of software filters are applied to the triggered events: fast reconstructions~\cite{Aartsen:2013bfa} are performed on the data and a first event selection carried out, reducing the data stream to about 100~GB/day. These reconstructions are based on the position and time of the hits in the detector, but do not include information about the optical properties of the ice, in order to speed up the computation. The filtered data is transmitted via satellite to several IceCube institutions in the North for further processing.

Offline processing aims at selecting events according to type (tracks or cascades), energy, or specific arrival directions using sophisticated likelihood-based reconstructions~\cite{Ahrens:2003fg,Aartsen:2013vja}. These reconstructions maximise the likelihood function built from the probability of obtaining the actual temporal and spatial information in each DOM (``hit'') given a set of track parameters (vertex, time, energy, and direction). For low-energy events, where the event signature is contained within the volume of the detector, a joint fit of muon track and an hadronic cascade at the interaction vertex is performed. For those events the total energy can be reconstructed with rather good accuracy, depending on further details of the analysis. Typically, more than one reconstruction is performed for each event. This allows, for example, to estimate the probability of each event to be either a track or a cascade. Each analysis will then use complex classification methods based on machine-learning techniques to further separate a possible signal from the background. Variables that describe the quality of the reconstructions, the time development and the spatial distribution of hit DOMs in the detector are usually used in the event selection.

\subsection{Effective Area and Volume}

After the analysis-dependent event reconstruction and selection, the observed event distribution in energy and arrival direction can be compared to the sum of background and signal events. For a given neutrino flux, $\phi_\nu$, the total number of signal events, $\mu_{\rm s}$, expected at the detector can be expressed as
\begin{equation}
 \mu_{\rm s} =  T\sum_\alpha\int{\rm d}\Omega\int {\rm d} E_\nu A^{\rm eff}_{\nu_\alpha}(E_\nu,\Omega) \phi_{\nu_\alpha}(E_\nu,\Omega)\,,
\label{eq:nevents1}
\end{equation}
where $T$ is the exposure time and $A^{\rm eff}_{\nu_\alpha}$ the detector effective area for neutrino flavour $\alpha$. The effective area encodes the trigger and analysis efficiencies and depends on the observation angle and neutrino energy. 

In practice, the figure of merit of a neutrino telescope is the effective volume, $V^{\rm eff}$, the equivalent volume of a detector with 100\% detection efficiency of neutrino events. This quantity is related to the signal events as
\begin{equation}
 \mu_{\rm s} =  \sum_\alpha\int{\rm d}\Omega\int {\rm d} E_\nu V_{\nu_\alpha}^{\rm eff}(E_\nu,\Omega)\left[Tn\sigma(E_\nu)\widetilde{\phi}_{\nu_\alpha}(E_\nu,\Omega)\right]\,,
\label{eq:nevents2}
\end{equation}
where $\widetilde{\phi}$ is the neutrino flux after taking into account Earth absorption and regeneration effects, $n$ is the local target density, and $\sigma$ the neutrino cross section for the relevant neutrino signal. The effective volume allows to express the event number by the local density of events, {\it i.e.}, the quantity within $[\cdot]$. This definition has the practical advantage that the effective volume can be simulated from a uniform distribution of neutrino events: if $n_{\rm gen}(E_\nu,\Omega)$ is the number of Monte-Carlo events generated over a large geometrical generation volume $V^{\rm gen}$ by neutrinos with energy $E_\nu$ injected into the direction $\Omega$, then the effective volume is given by
\begin{equation}
V^{\rm eff}_\nu (E_\nu,\Omega) = \frac{n_s}{n_{\rm gen}(E_\nu,\Omega)} V^{\rm gen}(E_\nu,\Omega)\,,
\label{eq:v_eff}
\end{equation}
where $n_s$ is the number of remaining signal events after all the selection cuts of a given analysis.

\subsection{Background Rejection}

There are two backgrounds in any analysis with a neutrino telescope: atmospheric muons and atmospheric neutrinos, both produced in cosmic-ray interactions in the atmosphere. The atmospheric muon background measured by IceCube~\cite{Aartsen:2015nss} is much more copious than the atmospheric neutrino flux, by a factor up to 10$^6$ depending on declination. Note that cosmic ray interactions can produce several coincident forward muons (``muon bundle'') which are part of the atmospheric muon background. Muon bundles can be easily identified as background in some cases, but they can also mimic bright single tracks (like monopoles for example) and are more difficult to separate from the signal in that case. Even if many of the IceCube analyses measure the atmospheric muon background from the data, the {\texttt{CORSIKA}} package~\cite{Heck:1998a} is generally used to generate samples of atmospheric muons that are used to cross-validate certain steps of the analyses. 

\begin{figure}[t]
\centering\includegraphics[width=\linewidth]{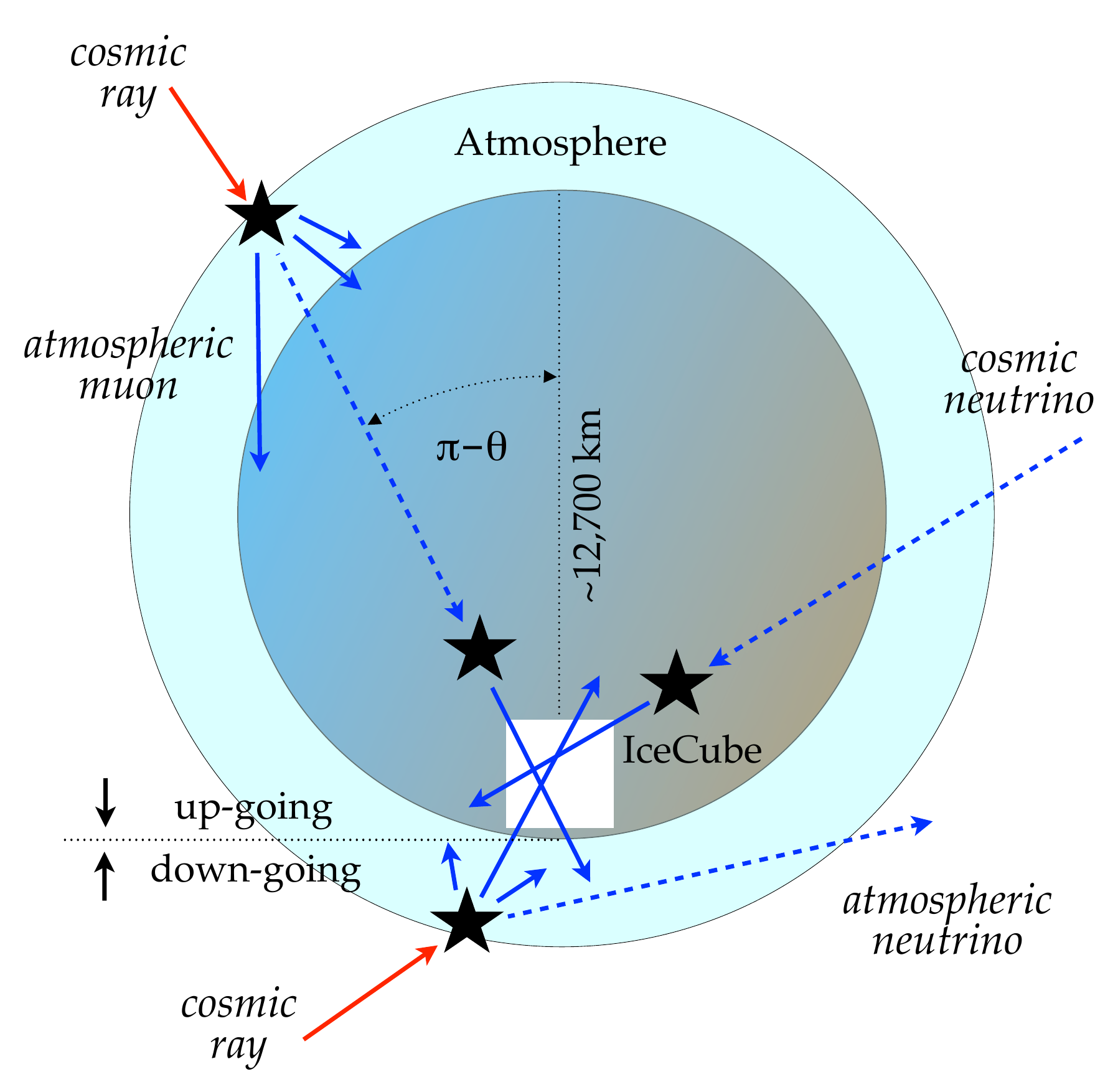}
\caption[]{An illustration of neutrino detection with IceCube located at the South Pole. Cosmic ray interactions in the atmosphere produce a large background of high-energy muon tracks (solid blue arrows) in IceCube. This background can be reduced by looking for up-going tracks produced by muon neutrinos (dashed blue arrows) that cross the Earth and interact close to the detector. The remaining background of up-going tracks produced by atmospheric muon neutrinos can be further reduced by energy cuts.}
\label{fig:sketch_atmo}
\end{figure}

The large background of atmospheric muons can be efficiently reduced by using the Earth as a filter, {\it i.e.}, by selecting {\it up-going} track events, at the expense of reducing the sky coverage of the detector to the Northern Hemisphere (see Fig.~\ref{fig:sketch_atmo}). Still, due to light scattering in the ice and the emission angle of the Cherenkov cone, a fraction of the down-going atmospheric muon tracks can be misreconstructed as up-going through the detector. This typically leads to a mismatch between the predicted atmospheric neutrino rate and the data rate at the final level of many analyses. There are analyses where a certain atmospheric muon contamination can be tolerated and it does not affect the final result. These are searches that look for a difference in the shape of the energy and/or angular spectra of the signal with respect to the background, and are less sensitive to the absolute normalisation of the latter. For others, like monopole searches, misreconstructed atmospheric muons can reduce the sensitivity of the detector. We will describe in more detail how each analysis deals with this background when we touch upon specific analyses in the rest of this review. 

%%%%%%%%%%%%%%%%%%%%%
%%%%%%%%%%%%%%%%%%%%%

The atmospheric neutrino flux constitutes an irreducible background for any search in IceCube, and sets the baseline to define a discovery in many analyses. It is therefore crucial to understand it both quantitatively and qualitatively. The flux of atmospheric neutrinos is dominated by the production and decay of mesons produced by cosmic ray interactions with air molecules~\cite{Gaisser:2002jj}. The behaviour of the neutrino spectra can be understood from the competition of meson ($m$) production and decay in the atmosphere: At high energy, where the meson decay rate is much smaller than the production rate, the meson flux is {\it calorimetric} and simply follows the cosmic ray spectrum, $\Phi_m\propto E^{-\Gamma}$. Below a critical energy $\epsilon_m$, where the decay rate becomes comparable to the production rate, the spectrum becomes harder by one power of energy, $\Phi_m\propto E^{1-\Gamma}$. The corresponding neutrino spectra from the decay of mesons are softer by one power of energy, $\Phi_\nu \propto \Phi_m/E$ due to the energy dependence of the meson decay rate.

%%%%%%%%%%%%%%%%%%%%%
\begin{figure}[t]
  \centering
  \includegraphics[width=\linewidth]{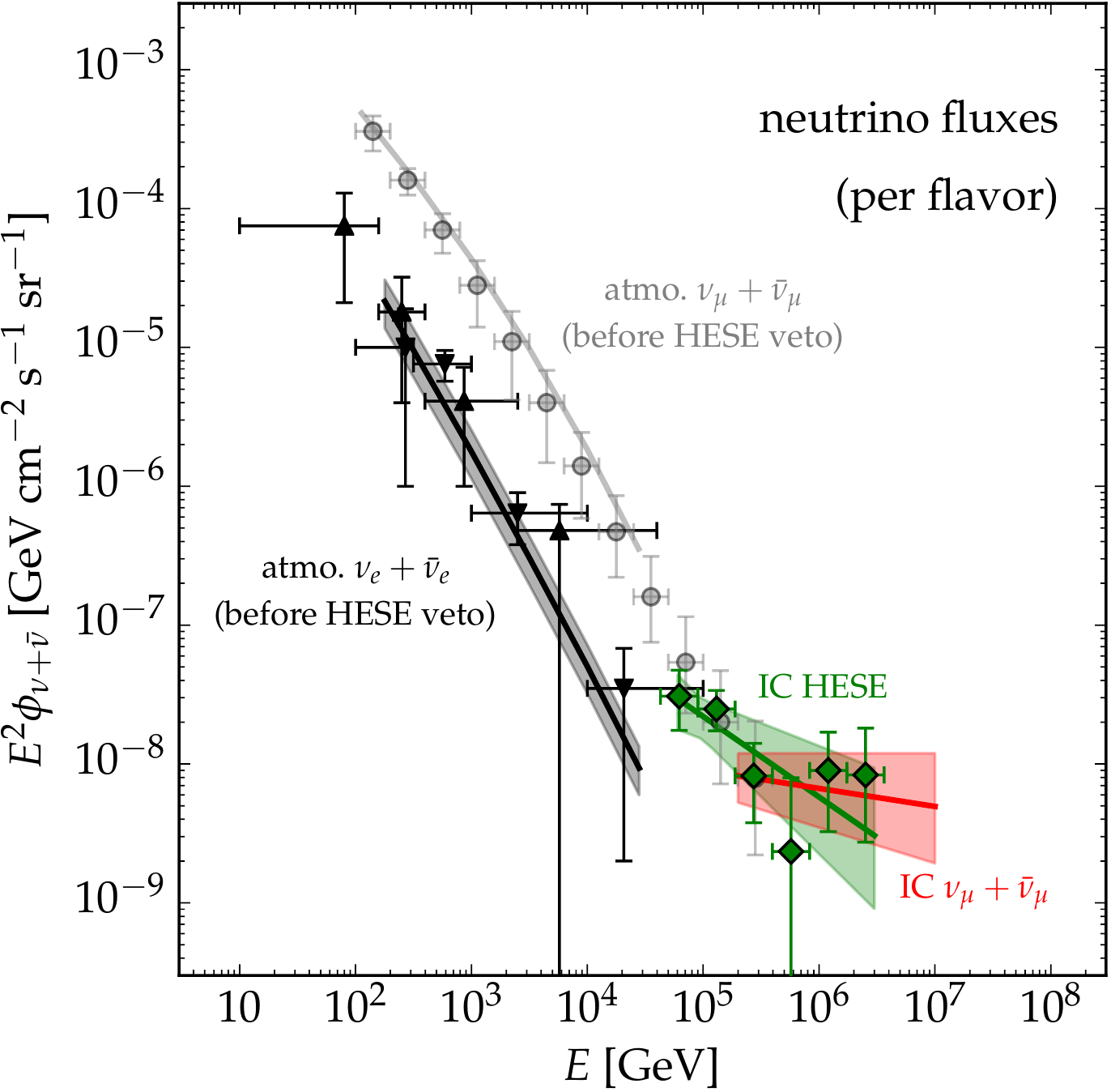}
  \caption[]{Summary of neutrino observations with IceCube (per flavour). The black and grey data shows IceCube's measurement of the atmospheric $\nu_e+\overline\nu_e$~\cite{Aartsen:2012uu,Aartsen:2015xup} and $\nu_\mu +\overline\nu_\mu$~\cite{Abbasi:2010ie} spectra. The green data show the inferred bin-wise spectrum of the four-year high-energy starting event (HESE) analysis~\cite{Aartsen:2013jdh,Aartsen:2014gkd}. The green line and green-shaded area indicate the best-fit and $1\sigma$ uncertainty range of a power-law fit to the HESE data. Note that the HESE analysis vetoes atmospheric neutrinos, and the true background level is much lower as indicated in the plot. The green line and green-shaded area indicate the best-fit and $1\sigma$ uncertainty range of a power-law fit of the up-going muon neutrino analysis~\cite{Aartsen:2016xlq}.\label{fig:fluxsummary}}
\end{figure}
%%%%%%%%%%%%%%%%%%%%%

The neutrino flux arising from pion and kaon decay is reasonably well understood, with an uncertainty in the range 10\%-20\% \cite{Gaisser:2002jj}. Figure~\ref{fig:fluxsummary} shows the atmospheric neutrino fluxes measured by IceCube. The atmospheric muon neutrino spectrum ($\nu_\mu+\overline{\nu}_\mu$) was obtained from one year of IceCube data (April 2008 to May 2009) using up-going muon tracks~\cite{Abbasi:2010ie}. The atmospheric electron neutrino spectra ($\nu_\mu+\overline{\nu}_\mu$) were analysed by looking for contained cascades observed with the low-energy infill array DeepCore between June 2010 and May 2011 in the energy range from 80~GeV to 6~TeV~\cite{Aartsen:2012uu}. This agrees well with a more recent analysis using contained events observed in the full IceCube detector between May 2011 and May 2012 with an extended energy range from 100~GeV to 100~TeV~\cite{Aartsen:2015xup}. All measurements agree well with model prediction of ``conventional'' atmospheric neutrinos produced in pion and kaon decay. IceCube uses the public Monte Carlo software {\texttt{GENIE}}~\cite{Andreopoulos:2009rq} and the internal software {\texttt{NUGEN}} (based on~\cite{Gazizov:2004va}) to generate samples of atmospheric neutrinos for its analyses, following the flux described in~\cite{Honda:2015fha}. 

Kaons with an energy above 1~TeV are also significantly attenuated before decaying and the ``prompt'' component, arising mainly from very short-lived charmed mesons ($D^\pm,$ $D^0,$ $D_s$ and $\Lambda_c$) is expected to dominate the spectrum. The prompt atmospheric neutrino flux, however, is much less understood, because of the uncertainty on the cosmic ray composition and relatively poor knowledge of QCD processes at small Bjorken-$x$~\cite{Enberg:2008te,Bhattacharya:2015jpa,Gauld:2015kvh,Bhattacharya:2016jce,Benzke:2017yjn}. In IceCube analyses the normalisation of the prompt atmospheric neutrino spectrum is usually treated as a nuisance parameter, while the energy distributions follows the model prediction of Ref.~\cite{Enberg:2008te}.

For high enough neutrino energies ($\cal{O}$(10) TeV), the possibility exists of rejecting atmospheric neutrinos by selecting {\it starting events}, where an outer layer of DOMs acts as a virtual veto region for the neutrino interaction vertex. This technique relies on the fact that atmospheric neutrinos are accompanied by muons produced in the same air shower, that would trigger the veto~\cite{Schonert:2008is,Gaisser:2014bja}. The price to pay is a reduced effective volume of the detector for down-going events and a different sensitivity for up-going and down-going events. This approach has been extremely successful, extending the sensitivity of IceCube to the Southern Hemisphere including the Galactic centre. There is not a generic veto region defined for all IceCube analyses, but each analysis finds its optimal definition depending on its physics goal. Events that present more than a predefined number of hits within some time window in the strings included in the definition of the veto volume are rejected. A reduction of the atmospheric muon background by more than 99\%, depending on analysis, can be achieved in this way (see for example~\cite{Gaisser:2014bja,Aartsen:2016xlq}).

This approach has been also the driver behind one of the most exciting recent results in multi-messenger astronomy: the first observation of high-energy astrophysical neutrinos by IceCube. The first evidence of this flux could be identified from a high-energy starting event (HESE) analysis, with only two years of collected data in 2013~\cite{Aartsen:2013bka,Aartsen:2013jdh,Aartsen:2014gkd}. The event sample is dominated by cascade events, with only a rather poor angular resolution of about $10\text{\textdegree}$. The result is consistent with an excess of events above the atmospheric neutrino background observed in up-going muon tracks from the Northern Hemisphere~\cite{Aartsen:2015rwa,Aartsen:2016xlq}. Figure~\ref{fig:fluxsummary} summarises the neutrino spectra inferred from these analyses. Based on different methods for reconstruction and energy measurement, their results agree, pointing at extra-galactic sources whose flux has equilibrated in the three flavours after propagation over cosmic distances~\cite{Aartsen:2015ivb} with $\nu_e:\nu_\mu:\nu_\tau \sim 1:1:1$. While both types of analyses have now reached a significance of more than $5\sigma$ for an astrophysical neutrino flux, the origin of this neutrino emission remains a mystery (see, {\it e.g.}, Ref.~\cite{Ahlers:2015lln}).

\section{Standard Neutrino Oscillations}
\label{sec:std_oscillations}
Over the past decades, experimental evidence for neutrino flavour oscillations has been accumulating in solar ($\nu_e$), atmospheric ($\nu_{e,\mu}$ \& $\overline{\nu}_{e,\mu}$), reactor ($\overline{\nu}_{e}$), and accelerator ($\nu_{\mu}$ \& $\overline{\nu}_{\mu}$) neutrino data (for a review see~\cite{PDG:2018}). These oscillation patterns can be convincingly interpreted as a non-trivial mixing of neutrino flavour and mass states with a small {\it solar} and large {\it atmospheric} mass splitting. Neutrinos $\nu_\alpha$ with flavour $\alpha = e, \mu, \tau$ refer to those neutrinos that couple to leptons $\ell_\alpha$ in weak interactions. Flavour oscillations are based on the effect that these flavour states are a non-trivial superposition of neutrino mass eigenstates $\nu_j$ ($j = 1, 2, 3$) expressed as
\begin{equation}\label{eq:Uneutrino}
  |\nu_\alpha\rangle =
  \sum_j U_{\alpha j}^* |\nu_j\rangle,
\end{equation}
where the $U_{\alpha j}$'s are elements of the unitary neutrino mass-to-flavour mixing matrix, the so-called {\it Pontecorvo-Maki-Nagakawa-Sakata} (PMNS) matrix~\cite{Pontecorvo:1957qd,Maki:1962mu,Pontecorvo:1967fh}. In general, the mixing matrix $U$ has nine degrees of freedom, which can be reduced to six by absorbing three global phases into the flavour states $\nu_\alpha$. The neutrino mixing matrix $U$ is then conveniently parametrised~\cite{PDG:2018} by three Euler rotations $\theta_{12}$, $\theta_{23}$, and $\theta_{13}$, and three $CP$-violating phases $\delta$, $\alpha_1$ and $\alpha_2$,
\begin{multline}\label{stparam}
U =\begin{pmatrix}1&0&0\\0&c_{23}&s_{23}\\0&-s_{23}&c_{23}\end{pmatrix}\begin{pmatrix}c_{13}&0&s_{13}e^{-i\delta}\\0&1&0\\-s_{13}e^{i\delta}&0&c_{13}\end{pmatrix}\begin{pmatrix}c_{12}&s_{12}&0\\-s_{12}&c_{12}&0\\0&0&1\end{pmatrix}\\\cdot{\rm diag}(e^{i\alpha_1/2},e^{i\alpha_2/2},1)\,.
\end{multline}
Here, we have made use of the abbreviations $\sin\theta_{ij}=s_{ij}$ and $\cos\theta_{ij}=c_{ij}$. The phases $\alpha_{1/2}$ are called {\it Majorana} phases, since they have physical consequences only if the neutrinos are Majorana spinors, {\it i.e.}, their own anti-particles. Note, that the phase $\delta$ ({\it Dirac} phase) appears only in combination with non-vanishing mixing $\sin\theta_{13}$.  

Neutrino oscillations can be derived from plane-wave solutions of the Hamiltonian, that coincide with mass eigenstates in vacuum, $\exp(-i(ET - pL))$. To leading order in $m/E$, the neutrino momentum is $p \simeq E-m^2/(2E)$ and a wave packet will travel a distance $L \simeq T$. Therefore, the leading order phase of the neutrino at distance $L$ from its origin is $\exp(-im^2L/(2E))$. From this expression we see that the effect of neutrino oscillations depend on the difference of neutrino masses, $\Delta m_{ij}^2\equiv m_i^2-m_j^2$. After traveling a distance $L$ an initial state $\nu_\alpha$ becomes a superposition of all flavours, with probability of transition to flavour $\beta$ given by $P_{\nu_\alpha \to \nu_\beta} = |\langle\nu_\beta|\nu_\alpha\rangle|^2$. This can be expressed in terms of the PMNS matrix elements as~\cite{PDG:2018}
\begin{multline}\label{pak}
  P_{\nu_\alpha \to \nu_\beta} = \delta_{\alpha \beta} - 4 \sum_{i>j} \Re\, (U_{\alpha i}^*\, U_{\beta i}\,   U_{\alpha j} \, U_{\beta j}^*) \, \sin^2 \Delta_{ij} \\ + 2 \sum_{i>j} \Im\, (U_{\alpha i}^*\, U_{\beta     i}\, U_{\alpha j} \, U_{\beta j}^*) \, \sin 2 \Delta_{ij} \,\,,
\end{multline}
where the oscillation phase $\Delta_{ij}$ can be parametrised as
\begin{align}\label{eq:phase}
  \Delta_{ij} = \frac{\Delta m_{ij}^2 L}{4E_\nu}\simeq 1.27\,\bigg(\frac{\Delta m_{ij}^2}{\rm     eV^2}\bigg)\bigg(\frac{L}{\rm km}\bigg)\bigg(\frac{E_\nu}{\rm GeV}\bigg)^{-1}\,.
\end{align}
Note, that the third term in Eq.~(\ref{pak}) comprises $CP$-violating effects, {\it i.e.}, this term can change sign for the process $P_{\overline\nu_\alpha \to \overline\nu_\beta}$, corresponding to the exchange $U\leftrightarrow U^*$ in Eq.~(\ref{pak}). For the standard parametrisation (\ref{stparam}) the single $CP$-violating contribution can be identified as the Dirac phase $\delta$; oscillation experiments are {\it not} sensitive to Majorana phases.

The first compelling evidence for the phenomenon of atmospheric neutrino oscillations was observed with Super-Kamiokande (SK)~\cite{Fukuda:1998mi}. The simplest and most direct interpretation of the atmospheric data~\cite{Ashie:2005ik} is oscillations of muon neutrinos, most likely converting into tau neutrinos. The survival probability of $\nu_\mu$ can be approximated as an effective two-level system with
\begin{equation}
P_{\nu_\mu \rightarrow \nu_\mu} = 1- \textrm{sin}^2 2\theta_{\rm atm}\,\textrm{sin}^2 \Delta_{\rm atm}
\label{eq:twolevel}
\end{equation}
The angular distribution of contained events in SK shows that for $E_\nu \sim 1~{\rm GeV},$ the deficit comes mainly from $L_{\rm atm} \sim 10^2 - 10^4~{\rm   km}.$ The corresponding oscillation phase must be nearly maximal, $\Delta_{\rm atm} \sim 1,$ which requires a mass splitting $\Delta m_{\rm atm}^2 \sim 10^{-4} - 10^{-2}~{\rm eV}^2$. Moreover, assuming that all up-going $\nu_\mu$'s which would yield multi-GeV events oscillate into a different flavour while none of the down-going ones do, the observed up-down asymmetry leads to a mixing angle very close to maximal, $\sin^2 2\theta_{\rm atm} > 0.92$ at $90\%$C.L. These results were later confirmed by the KEK-to-Kamioka (K2K)~\cite{Ahn:2006zza} and the Main Injector Neutrino Oscillation Search (MINOS)~\cite{Adamson:2008zt} experiments, which observed the disappearance of accelerator $\nu_\mu$'s at a distance of 250~km and 735~km, respectively, as a distortion of the measured energy spectrum. 

Furthermore, solar neutrino data collected by SK~\cite{Hosaka:2005um}, the Sudbury Neutrino Observatory (SNO)~\cite{Aharmim:2009gd} and Borexino~\cite{Arpesella:2008mt} show that solar $\nu_e$'s produced in nuclear processes convert to $\nu_{\mu}$ or $\nu_\tau$. For the interpretation of solar neutrino data it is crucial to account for matter effects that can have a drastic effect on the neutrino flavour evolution. The coherent scattering of electron neutrinos off background electrons with a density $N_e$ introduces a unique\footnote{All neutrino flavours take part in coherent scattering via neutral current interactions, but this corresponds to a flavour-universal potential term, that has no effect on oscillations.} potential term $V_{\rm mat} = \sqrt{2}G_FN_e$, where $G_F$ is the Fermi constant~\cite{Wolfenstein:1977ue}. In the effective two-level system for the survival of electron neutrinos, the effective matter oscillation parameters ($\Delta m_{\rm eff}^2$ \& $\theta_{\rm eff}$) relate to the vacuum values ($\Delta m_\odot^2$ \& $\theta_\odot$) as
\begin{gather}\label{eq:MSW1}
\frac{\Delta m_{\rm eff}^2}{\Delta m^2_\odot} = \left[\left(1-\frac{N_e}{N_{\rm res}}\right)^2\cos^22\theta_\odot +\sin^22\theta_\odot\right]^\frac{1}{2}\,,\\\label{eq:MSW2}
\frac{\tan2\theta_{\rm eff}}{\tan2\theta_\odot} = \left(1-\frac{N_e}{N_{\rm res}}\right)^{-1}\,,
\end{gather}
where the resonance density is given by
\begin{equation}
N_{\rm res} = \frac{\Delta m^2_\odot\cos2\theta_\odot}{\sqrt{2}2EG_F}\,.
\end{equation}
The effective oscillation parameters in the case of electron anti-neutrinos are the same as (\ref{eq:MSW1}) and (\ref{eq:MSW2}) after replacing $N_e\to -N_e$.  

The previous mixing and oscillation parameters are derived under the assumption of a constant electron density $N_e$. If the electron density along the neutrino trajectory is only changing slowly compared to the effective oscillation frequency, the effective mass eigenstates will change adiabatically. Note that the oscillation frequency and oscillation depth in matter exhibits a resonant behaviour~\cite{Mikheev:1986gs,Mikheev:1986wj,Wolfenstein:1977ue}. This {\it Mikheyev-Smirnov-Wolfenstein} (MSW) resonance can have an effect on continuous neutrino spectra, but also on monochromatic neutrinos passing through matter with slowly changing electron densities, like the radial density gradient of the Sun. Once these matter effect is taken into account, the observed intensity of solar electron neutrinos at different energies compared to theoretical predictions can be used to extract the solar mixing parameters. In addition to solar neutrino experiments, the KamLAND Collaboration~\cite{Araki:2004mb} has measured the flux of $\overline \nu_e$ from distant reactors and find that $\overline{\nu}_e$'s disappear over distances of about 180~km. This observation allows a precise determination of the solar mass splitting $\Delta m^2_\odot$ consistent with solar data.

The results obtained by short-baseline reactor neutrino experiments show that the remaining mixing angle $\theta_{13}$ is small. This allows to identify the mixing angle $\theta_{12}$ as the solar mixing angle $\theta_\odot$ and $\theta_{23}$ as the atmospheric mixing angle $\theta_{\rm atm}$. Correspondingly, the mass splitting can be identified as $\Delta m_\odot^2\simeq\Delta m_{21}^2$ and $\Delta m_{\rm atm}^2\simeq|\Delta m_{32}^2|\simeq |\Delta m_{31}^2|$. However, observations by the reactor neutrino experiments Daya-Bay~\cite{An:2012eh} and RENO~\cite{Ahn:2012nd} show that the small reactor neutrino mixing angle $\theta_{13}$ is larger than zero. As pointed out earlier, this is important for the observation of $CP$-violating effects parametrised by the Dirac phase $\delta$ in the PMNS matrix (\ref{stparam}).

%%%%%%%%%%%%%%%%
\begin{table}[t]
  \centering
  \renewcommand\arraystretch{1.2}
  \begin{tabular}{ccc}
    \hline
    & normal ordering & inverted ordering \\
    \hline
   $\Delta m_{21}^2$ [${\rm eV}^2$] & $7.40^{+0.21}_{-0.20}\times10^{-5}$ & $7.40^{+0.21}_{-0.20}\times10^{-5}$\\
   $\Delta m_{31}^2$ [${\rm eV}^2$]  & $ 2.494^{+0.033}_{-0.031}\times10^{-3}$ & ---\\
  $\Delta m_{23}^2$ [${\rm eV}^2$]  & --- & $2.465^{+0.032}_{-0.031}\times10^{-3}$\\
  \hline
  $\theta_{12}$ [${}\text{\textdegree}$] & $33.62^{+0.78}_{-0.76}$ & $33.62^{+0.78}_{-0.76}$\\
  $\theta_{23}$ [${}\text{\textdegree}$] & $47.2^{+1.9}_{-3.9}$ & $48.1^{+1.4}_{-1.9}$\\
  $\theta_{13}$ [${}\text{\textdegree}$] & $8.54^{+0.15}_{-0.15}$ & $8.58^{+0.14}_{-0.14}$\\
  $\delta_{\rm CP}$ [${}\text{\textdegree}$] & $234^{+43}_{-31}$ & $278^{+26}_{-29}$\\
       \hline
  \end{tabular}
  \caption[]{Results of a global analysis~\cite{Esteban:2016qun} of mass splittings, mixing angles, and Dirac phase for normal and inverted mass ordering. We best-fit parameters are shown with $1\sigma$ uncertainty.}\label{tab:oscillation}
\end{table}
%%%%%%%%%%%%%%%%

The global fit to neutrino oscillation data is presently incapable to determine the ordering of neutrino mass states. The fit to the data can be carried out under the assumption of {\it normal} ($m_1<m_2<m_3$) or {\it inverted} ($m_3<m_1<m_2$) mass ordering. A recent combined analysis~\cite{Esteban:2016qun} of solar, atmospheric, reactor, and accelerator neutrino data gives the values for the mass splittings, mixing angles, and  $CP$-violating Dirac phase for normal or inverted mass ordering shown in Table~\ref{tab:oscillation}. Note that, presently, the Dirac phase is inconsistent with $\delta=0$ at the $3\sigma$ level, independent of  mass ordering.

Neutrino oscillation measurements are only sensitive to the relative neutrino mass differences. The absolute neutrino mass scale can be measured by studying the electron spectrum of tritium (${}^3$H) $\beta$-decay. Present upper limits (95\% C.L.) on the (effective) electron anti-neutrino mass are at the level of $m_{\overline{\nu}_e}<2$~eV~\cite{Kraus:2004zw,Aseev:2011dq}. The KATRIN experiment~\cite{KATRIN-design:2005} is expected to reach a sensitivity of $m_{\overline{\nu}_e}<0.2$~eV. Neutrino masses are also constrained by their effect on the expansion history of the Universe and the formation of large-scale structure. Assuming standard cosmology dominated at late times by dark matter and dark energy, the upper limit (95\% C.L.) on the combined neutrino masses is $\sum_im_i<0.23$~eV~\cite{Ade:2015xua}.

The mechanism that provides neutrinos with their small masses is unknown. The existence of right-handed neutrino fields, $\nu_{\rm R}$, would allow to introduce a Dirac mass term of the form $m_{\rm D}\overline{\nu}_L\nu_R + h.c.$, after electroweak symmetry breaking. Such states would be ``neutral'' with respect to the standard model gauge interactions, and therefore sterile~\cite{Pontecorvo:1967fh}. However, the smallness of the neutrino masses would require unnaturally small Yukawa couplings. This can be remedied in {\it seesaw} models (see, {\it e.g.}, Ref.~\cite{Mohapatra:2006gs}). Being electrically neutral, neutrinos can be Majorana spinors, {\it i.e.}, spinors that are identical to their charge-conjugate state, $\psi^{\rm c} \equiv \mathcal{C}\overline{\psi}^T$, where $\mathcal{C}$ is the charge-conjugation matrix. In this case, we can introduce Majorana mass terms of the form $m_L\overline{\nu_{L}}{\nu}^{\rm c}_{L}/2 + h.c.$ and the analogous term for $\nu_R$. In {\it seesaw} models the individual size of the mass terms are such that $m_L\simeq0$ and $m_D \ll m_R$. After diagonalization of the neutrino mass matrix, the masses of active neutrinos are then proportional to $m_i\simeq m_D^2/m_R$. This would explain the smallness of the effective neutrino masses via a heavy sector of particles beyond the Standard Model.

%%%%%%%%%%%%%%%
\begin{figure}[t]
\centering\includegraphics[width=\linewidth]{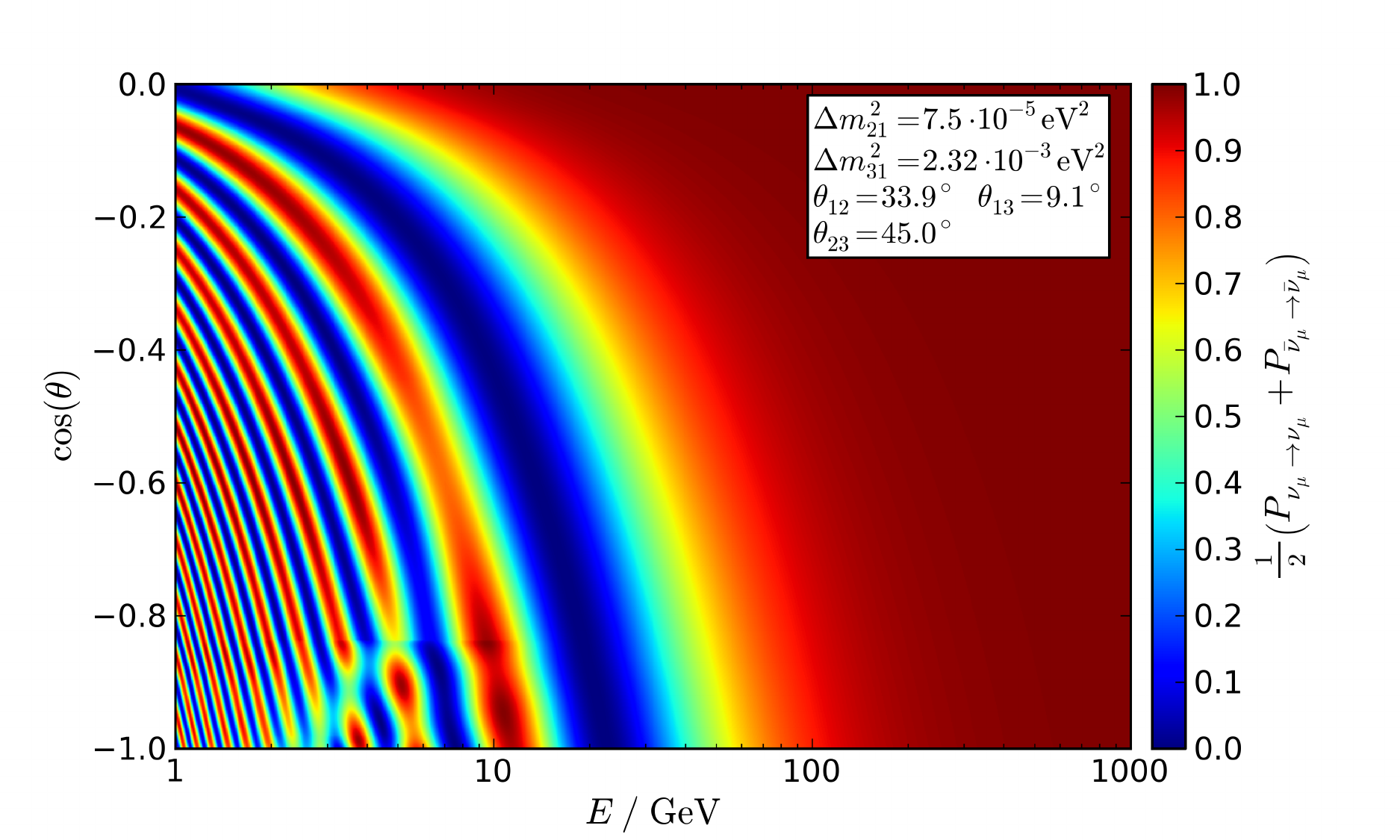}
\caption[]{The survival probability of muon neutrinos (averaged over $\nu_\mu$ and $\overline{\nu}_\mu$) as a function of zenith angle and energy. Figure from Ref.~\cite{Aartsen:2013nla}}
\label{fig:oscillogram}
\end{figure}
%%%%%%%%%%%%%%% 

%%%%%%%%%%%%%%%
\begin{figure*}[t]
\centering\includegraphics[height=0.34\linewidth]{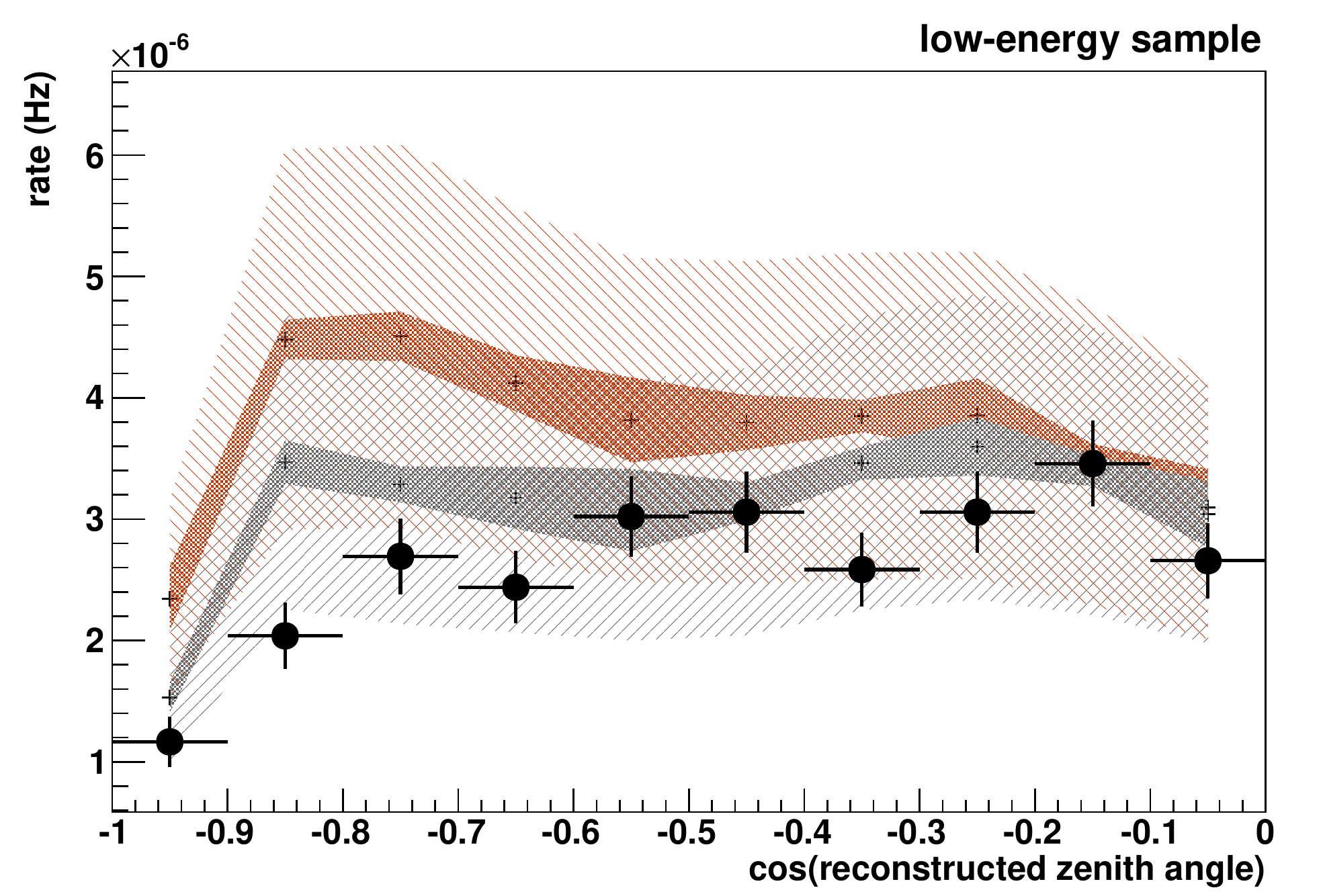}
\hspace{0.3cm}
\includegraphics[height=0.35\linewidth]{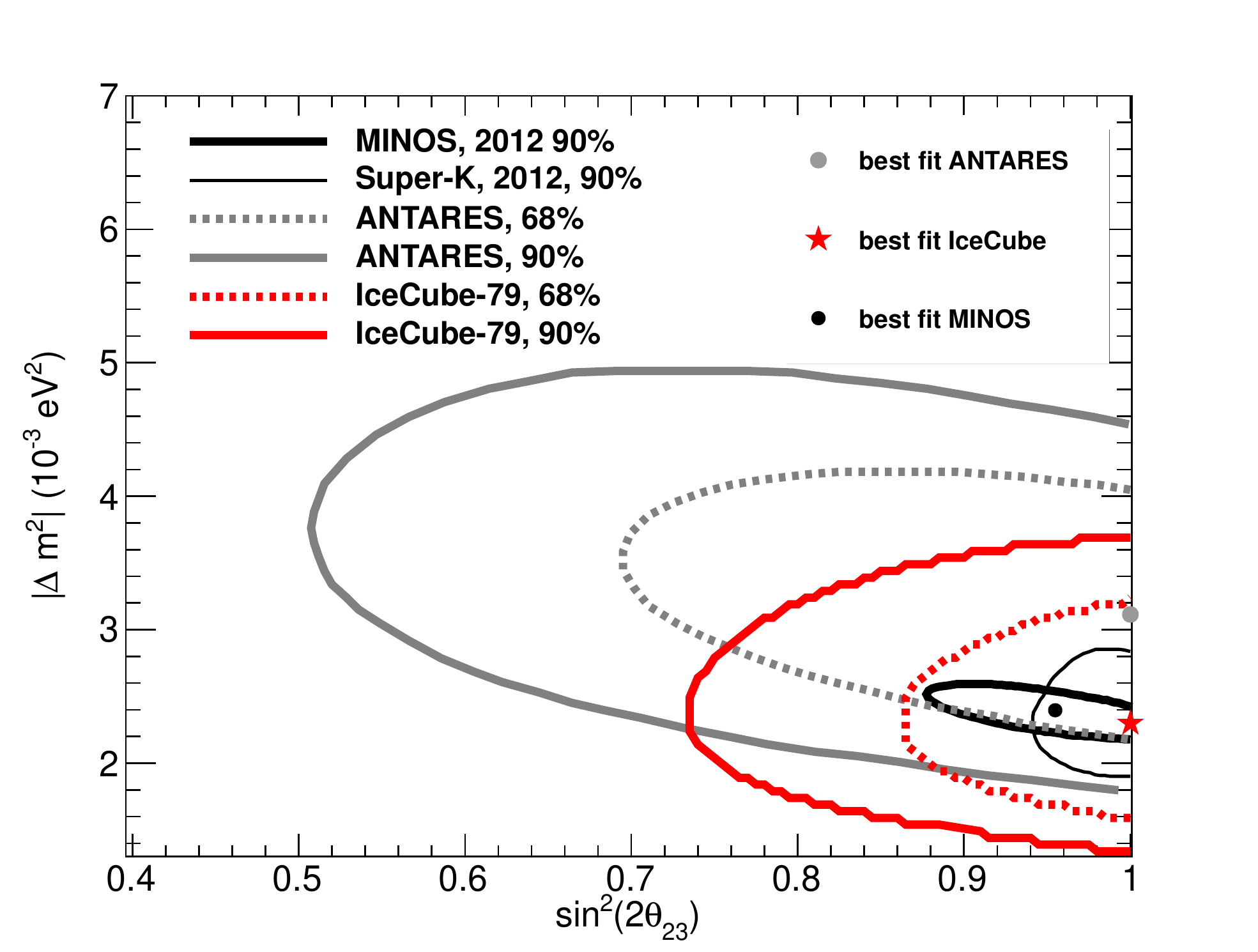}
\caption[]{{\bf Left Panel:} Angular distribution of contained events in DeepCore ({\it i.e.}, with energies between approximately 10~GeV and 60~GeV), compared with the expectation from the non-oscillation scenario (red area) and with oscillations (grey area) assuming current best-fit values of $|\Delta m^2_{32}| = 2.39 \times 10^{-3}$~eV$^2$ and $sin^2(2\theta_{23}) = 0.995$, from~\cite{Fogli:2012ua}. Systematic uncertainties are split into the normalisation contribution (dashed areas) and the shape contribution (filled areas) for each assumption shown. {\bf Right Panel:} Significance contours at 68\% and 90\% C.L. for the best-fit values of the IceCube analysis (red curves), compared with results of the ANTARES~\cite{AdrianMartinez:2012ph}, MINOS~\cite{Nichol:2013caa} and Super-Kamiokande~\cite{Abe:2013fuq} experiments. Figures reprinted with permission from ~Ref.~\cite{Aartsen:2013jza} (Copyright 2013 APS)}
\label{fig:angular_prl}
\end{figure*}
%%%%%%%%%%%%%%%

\subsection{Atmospheric Neutrino Oscillations with IceCube}

The atmospheric neutrino ``beam'' that reaches IceCube allows to perform high-statistics studies of neutrino oscillations at higher energies, and therefore is subject to different systematic uncertainties, than those typically available in reactor- or accelerator-based experiments. Atmospheric neutrinos arrive at the detector from all directions, {\it i.e.}, from travelling more than 12,700 km (vertically up-going) to about 10 km (vertically down-going), see Fig.~\ref{fig:sketch_atmo}. The path length from the production point in the atmosphere to the detector is therefore related to the measured zenith angle $\theta_{\rm zen}$. Combined with a measurement of the neutrino energy, this opens the possibility of measuring $\nu_\mu$ disappearance due to oscillations, exploiting the dependence of the disappearance probability with energy and arrival angle. 

Although the three neutrino flavours play a role in the oscillation process, a two-flavour approximation as in Eq.~(\ref{eq:twolevel}) is usually accurate to the percent level with $\Delta_{\rm atm}\simeq\Delta_{23}$ and $\theta_{\rm atm}\simeq\theta_{23}$. The survival probability of muon neutrinos as a function of path length through the Earth and neutrino energy is shown in Fig.~\ref{fig:oscillogram}. It can be seen that, for the largest distance travelled by atmospheric neutrinos (the diameter of the Earth), Eq.~(\ref{eq:twolevel}) shows a maximum $\nu_{\mu}$ disappearance at about 25~GeV.  This is precisely within the energy range of contained events in DeepCore. Simulations show that the neutrino energy response of DeepCore spans from about 6~GeV to about 60~GeV, peaking at 30~GeV. Neutrinos with higher energies will produce muon tracks that are no longer contained in the DeepCore volume. 

%%%%%%%%%%%%%%%
\begin{figure*}[t]\centering
\includegraphics[width=0.48\linewidth]{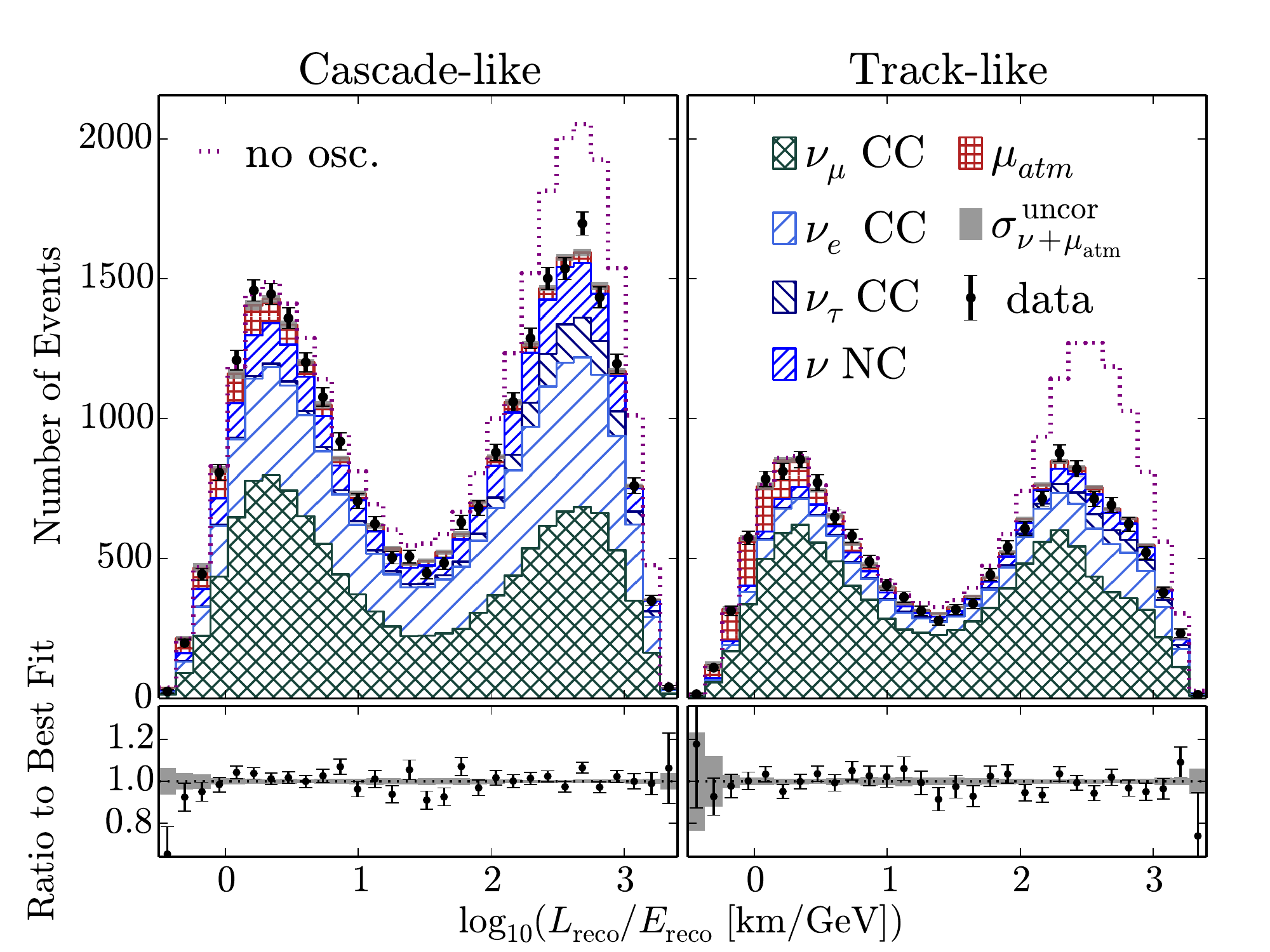}\hspace{0.3cm}
\includegraphics[width=0.48\linewidth]{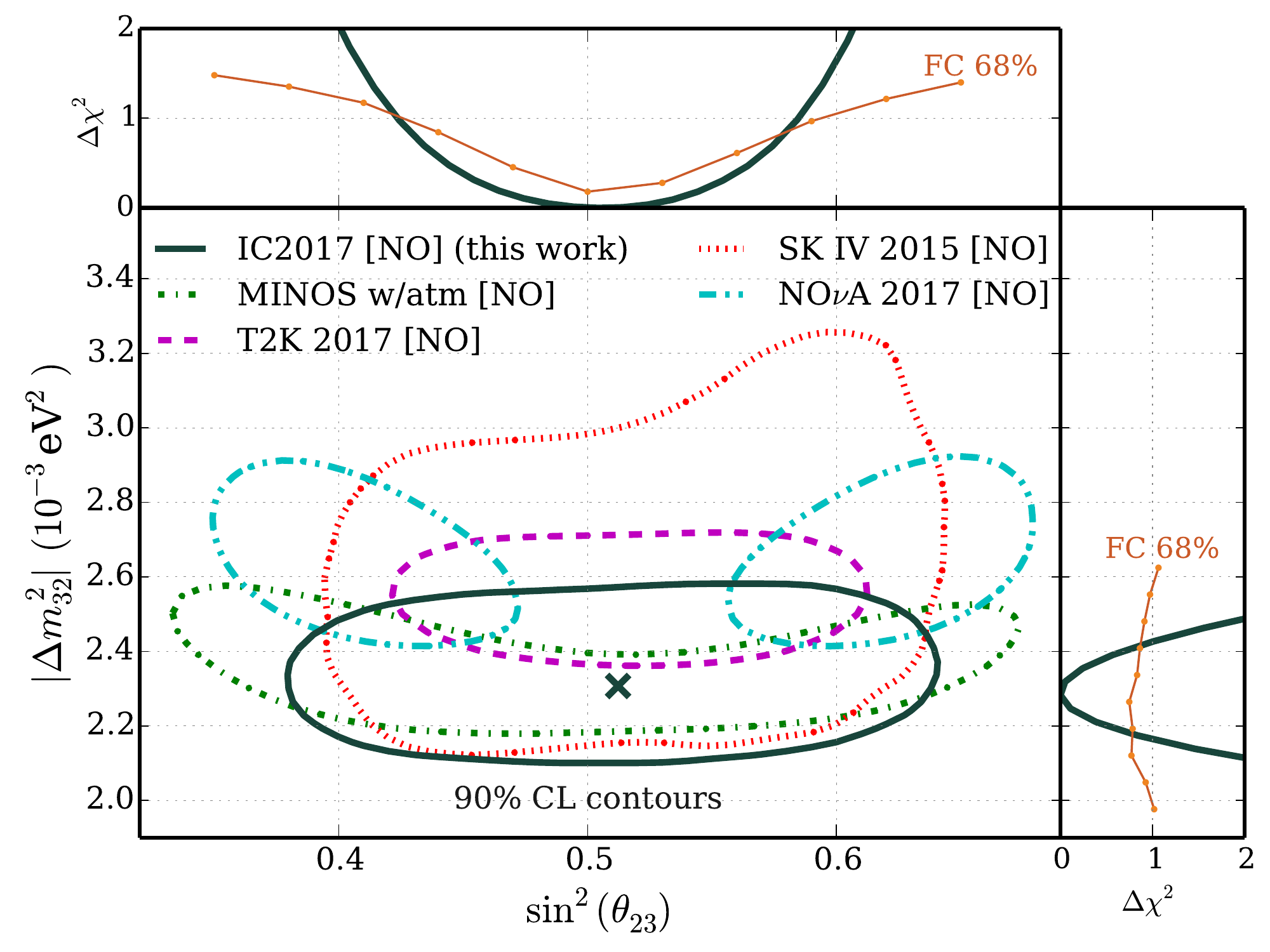}
\caption[]{{\bf Left Panel:} Event count as a function of reconstructed L/E. The expectation with no--oscillations is shown by the dashed line, while the best fit to the data (dots) is shown as a the full line. The hatched histograms show the predicted counts given the best-fit values for each component. $\sigma^{\text{uncor}}_{\nu+\mu_{\text{atm}}}$ represents the uncertainty due to finite Monte Carlo statistics and the data-driven atmospheric muon background estimate. The bottom panel shows the ratio of the data to the best fit hypothesis. {\bf Right Panel:} 90\% confidence contours in the $\sin^2\theta_{23}$--$\Delta m^2_{32}$ plane compared with results of Super-Kamiokande~\cite{Wendell:2014dka}, T2K~\cite{Abe:2017uxa}, MINOS~\cite{Adamson:2013whj} and NOvA~\cite{Adamson:2017qqn}. A normal mass ordering is assumed. Figures from Ref.~\cite{Aartsen:2017nmd}}
\label{fig:osc_LE-fit}
\end{figure*}
%%%%%%%%%%%%%%%

Given this relatively narrow energy response of DeepCore compared with the wide range of path lengths, it is possible to perform a search for $\nu_{\mu}$ disappearance through a measurement of the rate of contained events as a function of arrival direction, even without a precise energy determination. This is the approach taken in Ref.~\cite{Aartsen:2013jza}. Events starting in DeepCore were selected by using the rest of the IceCube strings as a veto. A ``high-energy'' sample of events not contained in DeepCore was used as a reference, since $\nu_{\mu}$ disappearance due to oscillations at higher energies ($\cal{O}$(100)~GeV) is not expected. The atmospheric muon background is reduced to a negligible level by removing tracks that enter the DeepCore fiducial volume from outside, and by only considering up-going events, {\it i.e.}, events that have crossed the Earth ($\cos\theta_{\rm zen} \le 0$), although a contamination of about 10\%-15\% of $\nu_e$ events misidentified as tracks remained, as well as $\nu_{\tau}$ from $\nu_{\mu}$ oscillations. These two effects were included as background. 

After all analysis cuts, a high-purity sample of 719 events contained in DeepCore were detected in a year. The left panel of Fig.~\ref{fig:angular_prl} shows the angular distribution of the remaining events compared with the expected event rate without oscillations (red-shaded area) and with oscillations using current world-average values for $\sin^2\theta_{23}$ and $|\Delta m^2_{32}|$~\cite{Fogli:2012ua} (grey-shaded area). A statistically significant deficit of events with respect to the non-oscillation scenario can be seen near the vertical direction ($-0.6 < \cos\theta_{\rm zen} < -1.0)$, while no discrepancy was observed in the reference high-energy sample (see Fig.~2 in~\cite{Aartsen:2013jza}). The discrepancy between the data and the non-oscillation case can be used to fit the oscillation parameters, without assuming any {\it a priori} value for them. The right panel of Fig.~\ref{fig:angular_prl} shows the result of that fit, with 68\% ($1\sigma$) and 90\% contours around the best-fit values found: $\sin^2(2\theta_{23}) = 1$ and $|\Delta m^2_{32}| =2.3^{+0.6}_{-0.5} \times 10^{-3}$~eV$^2$.

The next step in complexity in an oscillation analysis with IceCube is to add the measurement of the neutrino energy, so the quantities $L$ and $E_\nu$ in equation (\ref{eq:phase}) can be calculated separately. This is the approach followed in Ref.~\cite{Aartsen:2014yll}, where the energy of the neutrinos is obtained by using contained events in DeepCore and the assumption that the resulting muon is minimum ionising. Once the vertex of the neutrino interaction and the muon decay point have been identified, the energy of the muon can be calculated assuming constant energy loss, and it is proportional to the track length. The energy of the hadronic particle cascade at the vertex is obtained by maximising a likelihood function that takes into account the light distribution in adjacent DOMs. The neutrino energy is then the sum of the muon and cascade energies, $E_{\nu}=E_{\rm cascade} + E_{\mu}$. The most recent oscillation analysis from IceCube~\cite{Aartsen:2017nmd} improves on the mentioned techniques in several fronts. It is an all-sky analysis and also incorporates some degree of particle identification by reconstructing the events under two hypotheses: a $\nu_{\mu}$ charged-current interaction which includes a muon track, and a particle-shower only hypothesis at the interaction vertex. This latter hypothesis includes $\nu_{\rm e}$ and $\nu_{\tau}$ charged-current interactions, although these two flavours can not be separately identified. The analysis achieves an energy resolution of about 25\% (30\%) at $\sim$20~GeV for muon-like (cascade-like) events and a median angular resolution of 10{\textdegree} (16\textdegree). Full sensitivity to lower neutrino energies, for example to reach the next oscillation minimum at $\sim$6~GeV, can only be achieved with a denser array, like the proposed PINGU low--energy extension~\cite{Aartsen:2014oha}.

In order to determine the oscillation parameters, the data is binned into a two-dimensional histogram where each bin contains the measured number of events in the corresponding range of reconstructed energy and arrival direction. The expected number of events per bin depend on the mixing angle, $\theta_{23}$, and the mass splitting, $\Delta m^2_{32}$, as shown in Fig.~\ref{fig:oscillogram}. This allows to determine the mixing angle $\theta_{23}$ and the mass splitting $\Delta m^2_{32}$ as the maximum of the binned likelihood. The fit also includes the likelihood of the track and cascade hypotheses. Systematic uncertainties and the effect of the Earth density profile are included as nuisance parameters. In this analysis, a full three-flavour oscillation scheme is used and the rest of the oscillation parameters are kept fixed to $\Delta m^2_{21}=7.53\times10^{-5}$eV$^2$, $\sin^2\theta_{12}=3.04\times10^{-1}$, $\sin^2\theta_{13}=2.17\times10^{-2}$ and $\delta_{\rm CP}=0$. The effect of $\nu_{\mu}$ disappearance due to oscillations is clearly visible in the left panel of Fig.~\ref{fig:osc_LE-fit}, which shows the number of events as a function of the reconstructed $L/E_\nu$, compared with the expected event distribution, shown as a dotted magenta histogram, if oscillations were not present. The results of the best fit to the data are shown in the right panel of Fig.~\ref{fig:osc_LE-fit}. The best-fit values are $\Delta m^2_{32}= 2.31^{+0.11}_{-0.13} \times 10^{-3}$ eV$^2$ and $\sin^2 2\theta_{23}=0.51^{+0.07}_{-0.09}$, assuming a normal mass ordering.

The results of the two analyses mentioned above are compatible within statistics but, more importantly, they agree and are compatible in precision with those from dedicated oscillation experiments. 

%%%%%%%%%%%%%%%% 
\begin{figure}[t]
\centering\includegraphics[width=\linewidth]{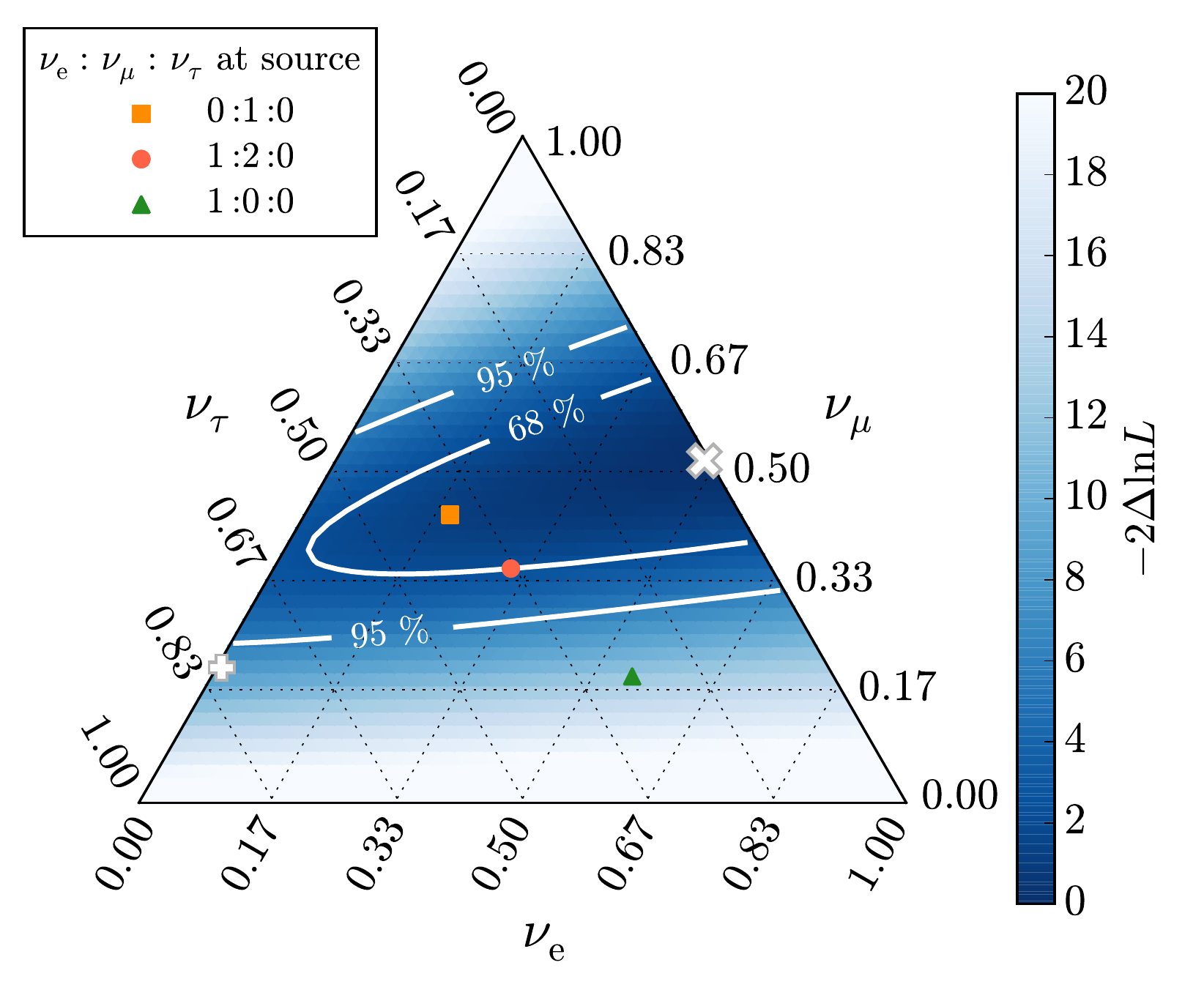}
\caption[]{Observed flavour composition of astrophysical neutrino with IceCube~\cite{Aartsen:2015knd}. The best-fit flavour ratio is indicated by a white ``$\times$'', with $68$\% and $96$\% confidence levels indicated by white lines. The expected oscillation-averaged composition is indicated for three different initial compositions, corresponding to standard pion decay ($1:2:0$), muon-damped pion decay ($0:1:0$), and neutron decay ($1:0:0$). The white ``$+$'' indicate the best-fit from a previous analysis~\cite{Aartsen:2015ivb}. (From Ref.~\cite{Aartsen:2015knd})}\label{fig:triangle}
\end{figure}
%%%%%%%%%%%%%%%% 

\subsection{Flavour of Astrophysical Neutrinos}

The neutrino oscillation phase in equation~(\ref{eq:phase}) depends on the ratio $L/E_\nu$ of distance  travelled, $L$, and neutrino energy, $E_\nu$. For astrophysical neutrinos we have to consider ultra-long oscillation baselines $L$ corresponding to many oscillation periods between source and observer. The initial mixed state of neutrino flavours has to be averaged over $\Delta L$, corresponding to the size of individual neutrino emission zones or the distribution of sources for diffuse emission. In addition, the observation of neutrinos can only decipher energies within an experimental energy resolution $\Delta E_\nu$. The oscillation phase in (\ref{eq:phase}) has therefore an absolute uncertainty that is typically much larger than $\pi$ for astrophysical neutrinos. As a consequence, only the oscillation-averaged flavour ratios can be observed.

The flavour-averaged survival and transition probability of neutrino oscillations in vacuum, can be derived from Eq.~(\ref{pak}) by replacing $\sin^2\Delta_{\ij} \to 1/2$ and  $\sin 2\Delta_{\ij} \to 0$. The resulting expression can be expressed as
\begin{equation}\label{pak_approx}
  P_{\nu_\alpha \to \nu_\beta} 
  \simeq \sum_i|U_{\alpha i}|^2\, |U_{\beta i}|^2\,.
\end{equation}
To a good approximation, neutrinos are produced in astrophysical environments as a mixed state involving $\nu_e$, $\overline{\nu}_e$, $\nu_\mu$, and $\overline{\nu}_\mu$. Due to the similarity of neutrino and anti-neutrino signals in Cherenkov telescopes we consider in the following only flavour ratios of the sum of neutrino and anti-neutrino fluxes $\phi_{\nu+\overline{\nu}}$ with flavour ratios $N_{e}:N_{\mu}:N_{\tau}$. Note, that the mixing angles shown in Table~\ref{tab:oscillation} are very close to the values for ``tri-bi-maximal'' mixing~\cite{Harrison:2002er} corresponding to $\sin^2\theta_{12} \sim 1/3$, $\sin^2\theta_{23} \sim 1/2$ and $\sin^2\theta_{13} \sim 0$. If we use this approximation then the oscillation-averaged spectrum will be close to a flavour ratio
\begin{multline}
N_{e}\!:\!N_{\mu}\!:\!N_{\tau} \simeq \left(\frac{2}{3}+{x_{e}}\right)\!:\!\left(\frac{7}{6}-\frac{x_{e}}{2}\right)\!:\!\left(\frac{7}{6}-\frac{x_e}{2}\right)\,,
\end{multline}
where $x_{e} = N_{e}/N_{\rm tot}$ is the electron neutrino fraction on production. For instance, pion decays $\pi^+ \to \mu^++\nu_\mu$ followed by muon decay $\mu^+ \to e^++\nu_e+\overline{\nu}_\mu$ produces an initial electron fraction of $x_e =1/3$. The resulting flavour ratio is then close to $1:1:1$. It is also feasible that the muon from pion decay loses energy as a result of synchrotron radiation in strong magnetic fields (``muon-damped'' scenario) resulting in  $x_e\simeq 0$ and a flavour ratio of $4:7:7$. Radioactive decay, on the other hand, will produce an initial electron neutrino fraction $x_e\simeq 1$ and a flavour ratio $5:2:2$.

Figure~\ref{fig:triangle} shows a visualisation of the observable neutrino flavour. Each location in the triangle corresponds to a unique flavour composition indicated by the three axis. The coloured markers correspond to the oscillation-averaged flavour ratios from the three scenarios ($x_e=1/3$, $x_e=0$, and $x_e=1$) discussed earlier, where the best-fit oscillation parameters have been used (instead of ``tri-bi-maximal'' mixing). The blue-shaded regions show the relative flavour log-likelihood ratio of a global analysis of IceCube data~\cite{Aartsen:2015knd}. The best-fit is indicated as a white cross. IceCube's observations are consistent with the assumption of standard neutrino oscillations and the production of neutrino in pion decay (full or ``muon-damped''). Neutrino production by radioactive decay is disfavoured at the $2\sigma$ level.

\section{Standard Model Interactions}
\label{sec:sm_interactions}
The measurement of neutrino fluxes requires a precise knowledge of the neutrino interaction probability or, equi\-valently, the cross section with matter. At neutrino energies of less than a few GeV the cross section is dominated by elastic scattering, {\it e.g.}, $\nu_x+p\to\nu_x+p$, and quasi-elastic scattering, {\it e.g.}, $\overline\nu_e+p\to e^++n$. In the energy range of 1-10~GeV, the neutrino-nucleon cross section is dominated by processes involving resonances, {\it e.g.}~$\nu_e+p\to e^-+\Delta^{++}$. At even higher energies neutrino scattering with matter proceeds predominantly via deep inelastic scattering (DIS) off nucleons, {\it e.g.}, $\nu_\mu+p\to \mu^-+X$, where $X$ indicates a secondary particle shower. The neutrino cross sections have been measured up to neutrino energies of a few hundreds of GeV. However, the neutrino energies involved in scattering of atmospheric and astrophysical neutrinos off nucleons far exceed this energy scale and we have to rely on theoretical predictions.

We will discuss in the following the expected cross section of high-energy neutrino-matter interactions. In weak interactions with matter the left-handed neutrino couples via $Z^0$ and $W^\pm$ exchange with the constituents of a proton or neutron. Due to the scale-dependence of the strong coupling constant, the calculation of this process involves both perturbative and non-perturbative aspects due to hard and soft processes, respectively.  

\subsection{Deep Inelastic Scattering}

The gauge coupling of quantum chromodynamics (QCD) increases as the renormalisation scale $\mu$ decreases, a behaviour which leads to the confinement of quarks and gluons at distances smaller that the characteristic size $\Lambda_{\rm QCD}^{-1}\simeq(200{\rm MeV})^{-1}\simeq1$~fm.  In nature (except in high temperature environments ($T\gg\Lambda_{\rm QCD}$) as in the early universe) the only manifestations of coloured representations are composite gauge singlets such as mesons and baryons. These bound states consist of valence quarks, which determine the overall spin, isospin, and flavour of the hadron, and a sea of gluons and quark-anti-quark pairs, which results from QCD radiation and pair-creation. These constituents of baryons and mesons are also called ``partons''.

\begin{figure}[t]\centering
\begin{minipage}[t]{0.5\columnwidth}
\rule{0cm}{0cm}\\[0.4cm]\includegraphics[height=4cm]{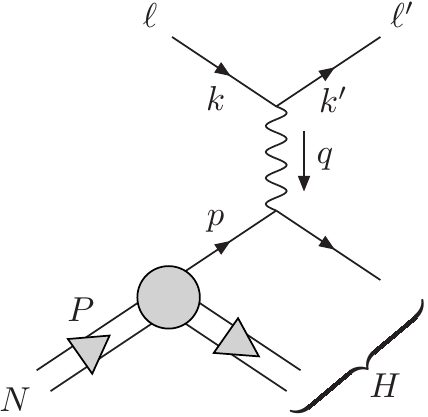}
\end{minipage}\hfill
\begin{minipage}[t]{0.5\linewidth}\centering
\begin{equation*}
\begin{matrix}
s\equiv(P+k)^2&t\equiv q^2\equiv-Q^2\\[0.4cm]
M^2\equiv P^2&W^2\equiv (P+q)^2\\[0.4cm]
\displaystyle x \equiv \frac{Q^2}{2\,q\cdot P}&\displaystyle y \equiv \frac{q\cdot P}{k \cdot P}\\[0.4cm]
{\rm (Bjorken-x)}&{\rm (inelasticity)}\\[0.4cm]
Q\gg M&W\gg M\\[0.1cm]
{\rm (deep)}&{\rm (inelastic)}
\end{matrix}
\end{equation*}
\end{minipage}
\caption[]{The kinematics of deep inelastic scattering. \label{fig:DIS}}
\end{figure}

Due to the strength of the QCD coupling at small scales the neutrino-nucleon interactions cannot be described in a purely perturbative way. However, since the QCD interaction decreases as the renormalisation scale increases (asymptotic freedom) the constituents of a nucleon may be treated as loosely bound objects within sufficiently small distance and time scales ($\Lambda_{\rm QCD}^{-1}$). Hence, in a hard scattering process of a neutrino involving a large momentum transfer to a nucleon the interactions between quarks and gluons may factorise from the sub-process (see Fig.~\ref{fig:DIS}). Due to the renormalisation scale dependence of the couplings this factorisation will also depend on the absolute momentum transfer $Q^2\equiv-q^2$.

Figure \ref{fig:DIS} shows a sketch of a general lepton-nucleon scattering process. A nucleon $N$ with mass $M$ scatters off the lepton $\ell$ by a $t$-channel exchange of a boson. The final state consist of a lepton $\ell'$ and a hadronic state $H$ with centre of mass energy $(P+q)^2=W^2$. This scattering process probes the partons, the constituents of the nucleon with a characteristic size $M^{-1}$ at length scales of the order of $Q^{-1}$. Typically, this probe will be ``deep'' and ``inelastic'', corresponding to $Q\gg M$ and $W\gg M$, respectively. The sub-process between lepton and parton takes place on time scales which are short compared to those of QCD interactions and can be factorised from the soft QCD interactions. The intermediate coloured states, corresponding to the scattered parton and the remaining constituents of the nucleus, will then softly interact and hadronise into the final state $H$.

%%%%%%%%%%%%%%%%%%%%%
%%%%%%%%%%%%%%%%%%%%%

%%%%%%%%%%%%%%%%%%%%%
\begin{figure}[t]\centering
\includegraphics[width=0.8\linewidth,clip=true]{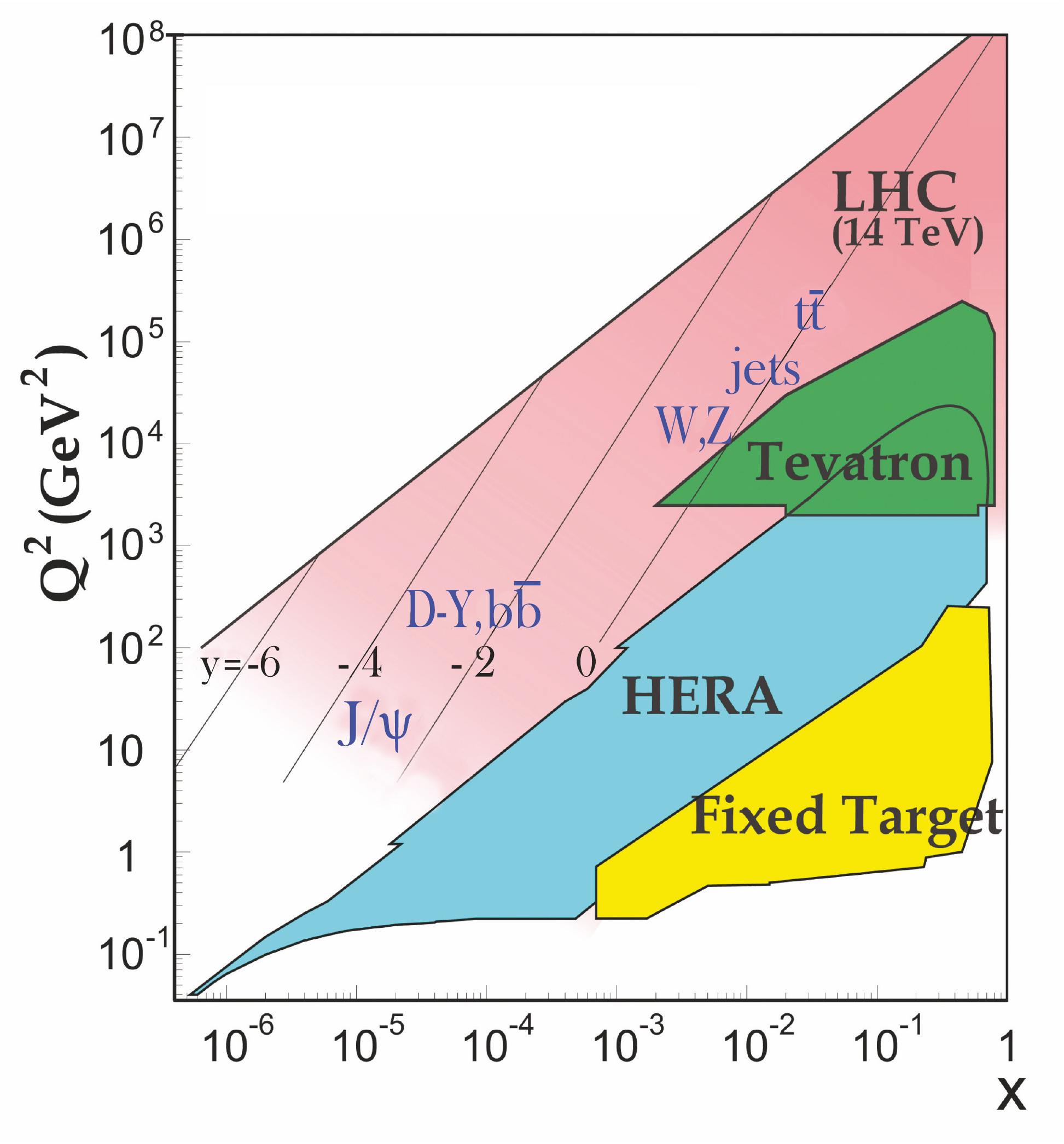}
\caption[]{The kinematic plane investigated by various collider and fixed target experiments in terms of Bjorken-$x$ and momentum transfer $Q^2$. Figure from Ref.~\cite{PDG:2018}.}
\label{KINPLANE}
\end{figure}
%%%%%%%%%%%%%%%%%%%%%

The kinematics of a lepton-nucleon scattering is conveniently described by the Lorentz scalars $x=Q^2/(2q\cdot P)$, also called {\it Bjorken}-$x$, and inelasticity $y=(q\cdot P)/(k\cdot P)$ (see Fig.~\ref{fig:DIS} for definitions). In the kinematic region of deep inelastic scattering (DIS) where $Q\gg M$ and $W\gg M$ we also have $Q^2\simeq2q\cdot p$ and thus $x\simeq(q\cdot p)/(q\cdot P)$. The scalars $x$ and $y$ have simple interpretations in particular reference frames. In a reference frame where the nucleon is strongly boosted along the neutrino 3-momentum $\vec{k}$ the relative transverse momenta of the partons is negligible. The parton momentum $p$ in the boosted frame is approximately aligned with $P$ and the scalar $x$ expresses the momentum fraction carried by the parton. In the rest frame of the nucleus the quantity $y$ is the fractional energy loss of the lepton, $y=(E-E')/E$, where $E$ and $E'$ are the lepton's energy before and after scattering, respectively.

From the previous discussion we obtain the following recipe for the calculation of the total (anti-)neutrino-nucleon cross section $\sigma(\nu (\overline{\nu}) N)$. The differential lepton-parton cross section may be calculated using a perturbative expansion in the weak coupling. The relative contribution of this partonic sub-process with Bjorken-$x$ and momentum transfer $Q^2$ in the nucleon ${N}$ is described by structure functions, which depend on the particular parton distribution functions (PDFs) of quarks ($f_q(x,Q^2)$) and gluons ($f_g(x,Q^2)$) . These functions must be measured in fixed target and accelerator experiments, that only access a limited kinematic region in $x$ and $Q^2$. Figure~\ref{KINPLANE} shows the regions in the kinematical $x$-$Q^2$-plane which have been covered in electron-proton (HERA), anti-proton-proton (Tevatron), and proton-proton (LHC) collisions as well as in fixed target experiments with neutrino, electron, and muon beams (see, {\it e.g.}, Ref.~\cite{Martin:1998sq} and references therein).

\subsection{Charged and Neutral Current Interactions}

\begin{figure}[t]
  \centering
  \includegraphics[width=0.66\linewidth]{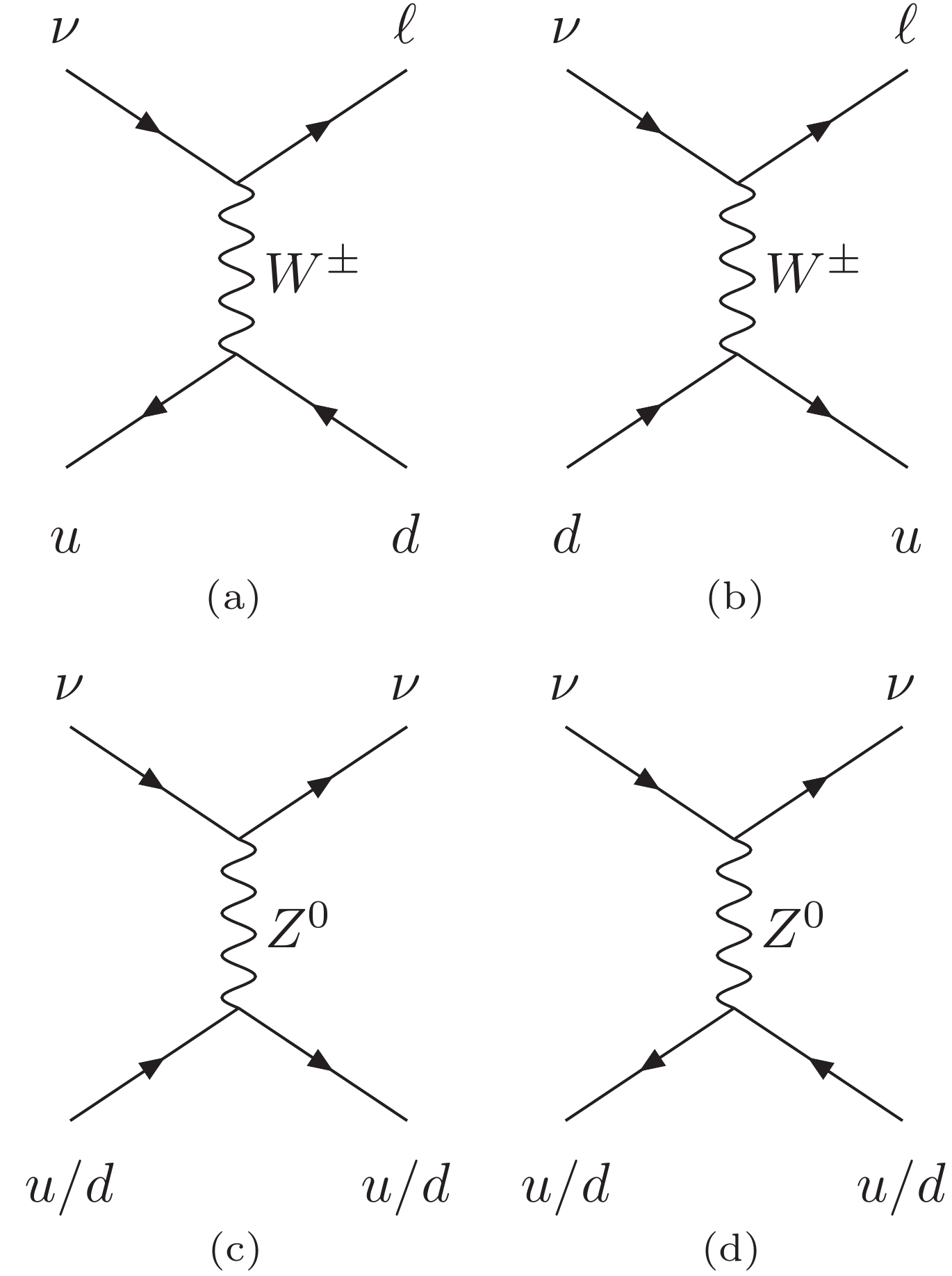}
    \caption[]{The parton level $W$ (a/b) and $Z$ (c/d) boson exchange between neutrinos and light quarks. \label{fig:CCNC}}
\end{figure}

The parton level charged current interactions of neutrinos with nucleons are shown as the top two diagrams (a) and (b) of Fig.~\ref{fig:CCNC}. The leading-order contribution is given by
\begin{multline}\label{CC}
\frac{{\rm d}^2\sigma_{\rm CC}}{{\rm d}Q^2{\rm d}x} = \frac{G_F^2}{\pi}\left(\frac{m_W^2}{Q^2+m_W^2}\right)^2\\\cdot\left(q(x,Q^2) + \overline q(x,Q^2)(1-y^2)\right)\,,
\end{multline}
where $G_F \simeq 1.17\times 10^{-5}{\rm GeV}^{-2}$ is the Fermi coupling constant. The effective parton distribution functions are $q(x,Q^2) = f_d+f_s+f_b$ and $\overline q(x,Q^2) = f_{\overline u}+f_{\overline c}+f_{\overline t}$. For antineutrino scattering we simply have to replace all $f_q$ by $f_{\overline q}$. These structure functions $f_q$ are determined experimentally via deep inelastic lepton-nucleon scattering or hard scattering processes involving nucleons. The corresponding relation of neutron structure function are given by the exchange $u\leftrightarrow d$ and $\overline u\leftrightarrow \overline d$ due to approximate isospin symmetry. In neutrino scattering with matter one usually makes the approximation of an equal mix between protons and neutrons. Hence, for an {\it iso-scalar} target, {\it i.e.}, averaging over isospin, $f_{u/d} \to (f_u+f_d)/2$ and $f_{\overline{u}/\overline{d}} \to (f_{\overline{u}}+f_{\overline{d}})/2$. 

%%%%%%%%%%%%%%%%%%%%%
%%%%%%%%%%%%%%%%%%%%%

Analogously, the parton level neutral current (NC) interactions of the neutrino with nucleons are shown in the bottom two diagrams (c) and (d) of Fig.~\ref{fig:CCNC}. The leading-order double differential neutral current cross section can be expressed as
\begin{multline}\label{NC}
\frac{{\rm d}^2\sigma_{\rm NC}}{{\rm d}Q^2{\rm d}x} = \frac{G_F^2}{\pi}\left(\frac{m_Z^2}{Q^2+m_Z^2}\right)^2\\\cdot\left(q^0(x,Q^2) + \overline q^0(x,Q^2)(1-y^2)\right)\,.
\end{multline}
Here, the structure functions are given by
\begin{multline}
q^0 = (f_u+f_c+f_t)L_u^2 + (f_{\overline u}+f_{\overline c}+f_{\overline t})R_u^2\,,\\+(f_d+f_s+f_b)L_d^2 + (f_{\overline d}+f_{\overline s}+f_{\overline b})R_d^2\,,
\end{multline}
\begin{multline}
\overline q^0 = (f_u+f_c+f_t)R_u^2 + (f_{\overline u}+f_{\overline c}+f_{\overline t})L_u^2\,,\\+(f_d+f_s+f_b)R_d^2 + (f_{\overline d}+f_{\overline s}+f_{\overline b})L_d^2\,.
\end{multline}
The weak couplings after electro-weak symmetry breaking depend on the combination $I_3 - q\sin^2\theta_W$, where $I_3$ is the weak isospin, $q$ the electric charge, and $\theta_W$ the Weinberg angle. More explicitly, the couplings  for left-handed ($I_3=\pm 1/2$) and right-handed ($I_3=0$) quarks are given by
\begin{align}
L_u&=\frac{1}{2}-\frac{2}{3}\sin^2\theta_W\,,&L_d &= -\frac{1}{2}+\frac{1}{3}\sin^2\theta_W\,,\\
R_u&=-\frac{2}{3}\sin^2\theta_W\,,&R_d &= \frac{1}{3}\sin^2\theta_W\,.
\end{align}
As in the case of charged current interactions, the relation of neutron structure function $f_q$ are given by the exchange $u\leftrightarrow d$ and $\overline u\leftrightarrow \overline d$ and for an iso-scalar target one takes $f_{u/d} \to (f_u+f_d)/2$ and $f_{\overline{u}/\overline{d}} \to (f_{\overline{u}}+f_{\overline{d}})/2$.

\subsection{High-Energy Neutrino-Matter Cross Sections}

The expressions for the total charged and neutral current neutrino cross sections are derived from Eqs.~(\ref{CC}) and (\ref{NC}) after integrating over Bjorken-$x$ and momentum transfer $Q^2$ (or equivalently inelasticity $y$). The evolution of PDFs with respect to factorisation scale $\mu$ can be calculated by a perturbative QCD expansion and results in the {\it Dokshitzer-Gribov-Lipatov-Altarelli-Parisi} (DGLAP) equations~\cite{Gribov:1972ri,Lipatov:1974qm,Altarelli:1977zs,Dokshitzer:1977sg}. The solution of the (leading-order) DGLAP equations correspond to a re-summation of powers $(\alpha_s\ln(Q^2/\mu^2))^n$ which appear by QCD radiation in the initial state partons. However, these radiative processes will also generate powers $(\alpha_s\ln(1/x))^n$ and the applicability of the DGLAP formalism is limited to moderate values of Bjor\-ken-$x$ (small $\ln(1/x)$) and large $Q^2$ (small $\alpha_s$). If these logarithmic contributions from a small $x$ become large, a formalism by {\it Balitsky, Fakin, Kuraev, and Lipatov} (BFKL) may be used to re-sum the $\alpha_s\ln(1/x)$ terms~\cite{Kuraev:1977fs,Balitsky:1978ic}. This approach applies for moderate values of $Q^2$, since contributions of $\alpha_s\ln(Q^2/\mu^2)$ have to be kept under control. 

There are unified forms~\cite{Kwiecinski:1997ee} and other improvements of the linear DGLAP and BFKL evolution for the problematic region of small Bjorken-$x$ and large $Q^2$. The extrapolated solutions of the linear DGLAP and BFKL equations predict an unlimited rise of the gluon density at very small $x$. It is expected that, eventually, non-linear effects like gluon recombination $g+g\rightarrow g$ dominate the evolution and screen or even saturate the gluon density~\cite{GolecBiernat:1998js,GolecBiernat:1999qd,Iancu:2003xm}. 

%%%%%%%%%%%%%%%%%%%%%%%%%%%%
\begin{figure}[t]
  \centering
  \includegraphics[width=0.9\linewidth]{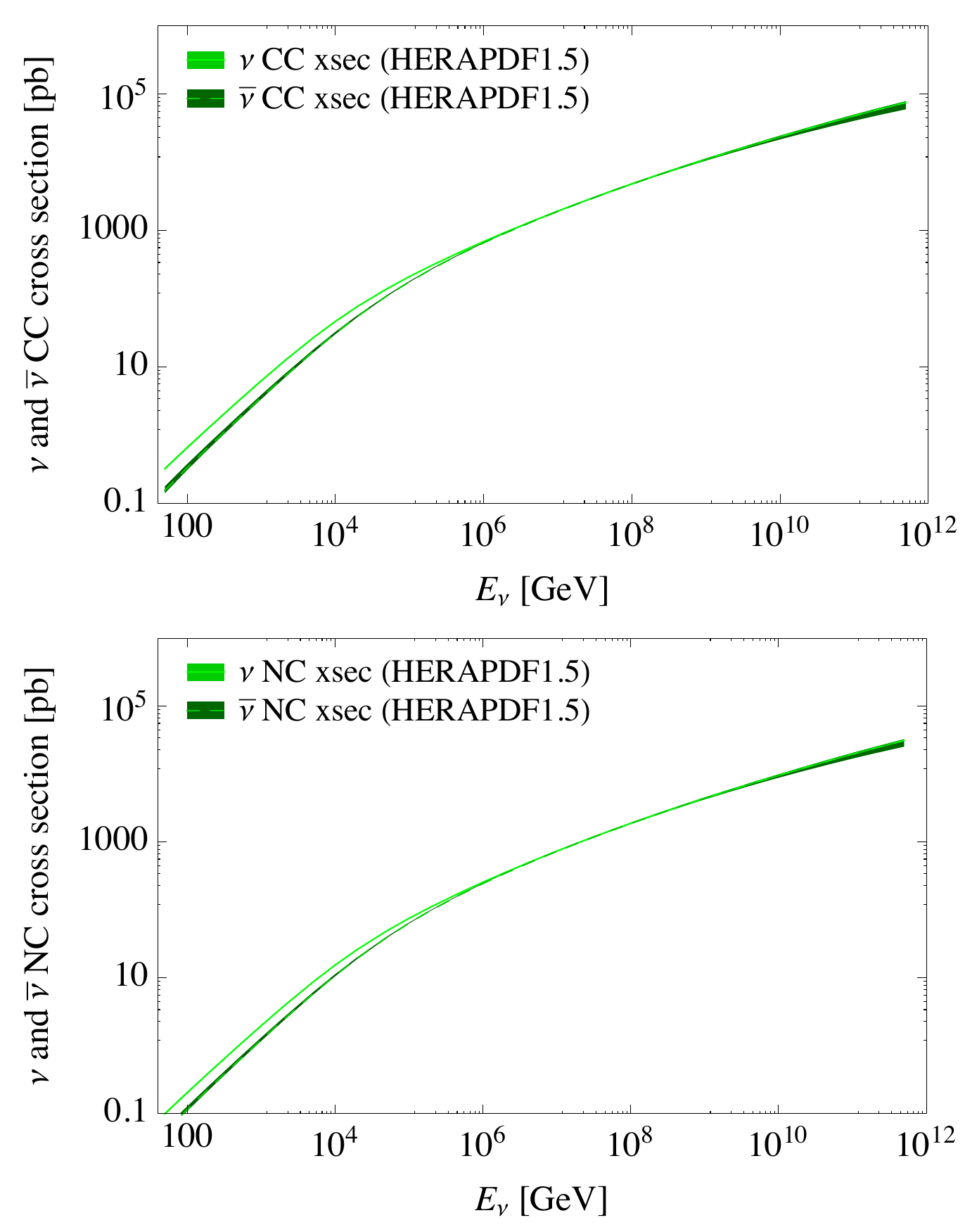}
  \caption[]{ High-energy charged current (top panel) and neutral current (bottom panel) neutrino and anti-neutrino cross sections based on the ZEUS global PDF fits~\cite{CooperSarkar:2011pa}; the width of the lines indicate the uncertainties. Figure from Ref.~\cite{CooperSarkar:2011pa}.}
  \label{fig:nuNcross}
\end{figure}
%%%%%%%%%%%%%%%%%%%%%%%%%%%%

Note, that neutrino-nucleon scattering in charged (\ref{CC}) and neutral (\ref{NC}) current interactions via t-channel exchange of $W$ and $Z$ bosons, respectively, probe the parton content of the nucleus effectively up to momentum transfers of $Q^2 \simeq M^2_{Z/W}$ (see Fig.~\ref{fig:CCNC}). The present range of Bjorken-$x$  probed by experiments only extends down to $x\simeq10^{-4}$ at this $Q$-range, and it is limited to $10^{-6}$ for arbitrary $Q$ values. On the other hand, the Bjorken-x probed by neutrino interactions is, roughly,
\begin{equation}
x\simeq \frac{M^2_{Z/W}}{s-m_N^2}\simeq 10^{-4}\left(\frac{E_\nu}{100 {\rm PeV}}\right)^{-1}\,.
\end{equation}
This shows that high-energy neutrino-nucleon interactions beyond $100$~PeV strongly rely on extrapolations of the structure functions.

Figure~\ref{fig:nuNcross} shows the results of Ref.~\cite{CooperSarkar:2011pa,CooperSarkar:2007cv} which used an update of the PDF fit formalism of the published ZEUS-S global PDF analysis~\cite{Chekanov:2002pv}. The total cross sections in the energy range $10^7 \leq (E_\nu/{\rm GeV}) \leq 10^{12}$ can be approximated to within $\sim10\%$ by the relations~\cite{CooperSarkar:2007cv},
\begin{align}
  \log_{10}\left(\frac{\sigma_{\rm CC}}{{\rm cm}^2}\right) &=
  - 39.59 \left[\log_{10}\left(\frac{E_\nu}{{\rm GeV}}\right)\right]^{-0.0964}\,,\\
  \log_{10}\left(\frac{\sigma_{\rm NC}}{{\rm cm}^2}\right) &= - 40.13   \left[\log_{10}\left(\frac{E_\nu}{{\rm GeV}}\right)\right]^{-0.0983}\,.
  \label{fit_CCNC}
\end{align}
Neutrino electron interactions can often be neglected with respect to neutrino nucleon interactions due to the electron's small mass. There is, however, one exception with $\overline \nu_e+ e^-$ interactions, because of the intermediate-boson resonance formed in the neighbourhood of $E_\nu^{\rm res} = M_W^2/2m_e \simeq 6.3~{\rm PeV}$, generally referred to as the Glashow resonance~\cite{Glashow:1960zz}. The total cross section for the resonant scattering $\overline\nu_e+e^-\to W^-$ is~\cite{PDG:2018}
\begin{equation}
\sigma(s) = B_{\rm in}B_{\rm out}\frac{24\pi}{M_W^2}\frac{\Gamma_W^2s}{(s-M_W^2)^2+(M_W\Gamma_W)^2}\,,
\end{equation}
where $B_{\rm in} = {\rm Br}(W^-\to \overline\nu_e+e^-)$ and $B_{\rm out} = {\rm Br}(W^-\to X)$ are the corresponding branching ratios of $W$ decay and $\Gamma_W\simeq2.1$~GeV the $W$ decay width. The branching ratios into $\overline\nu_\alpha+\ell_\alpha$ are $10.6$\% and into hadronic states $67.4$\%~\cite{PDG:2018}.

\subsection{Neutrino Cross Section Measurement with IceCube}

\begin{figure}[t]
\centering\includegraphics[width=\linewidth,height=0.8\linewidth]{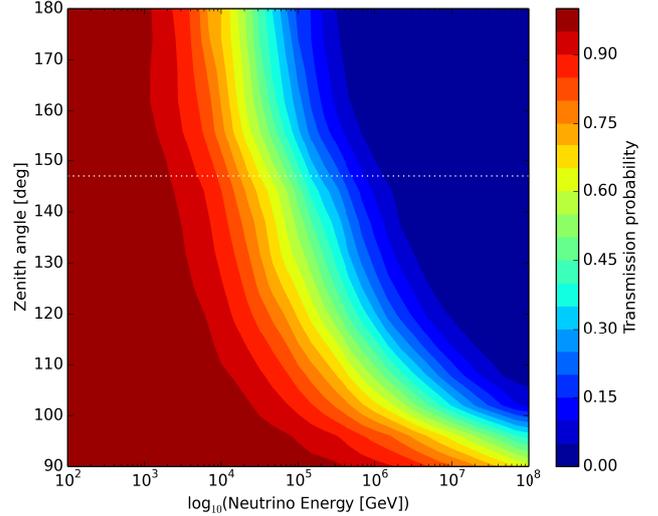}
\caption[]{Transmission probability assuming the Standard Model neutrino-nucleon cross section for neutrinos crossing the Earth, as a function of energy and zenith angle. Neutral-current interactions are included. The horizontal white line shows the location of the core-mantle boundary. Figure from~\cite{Aartsen:2017kpd}.}
\label{fig:nu_transmission}
\end{figure}

Similar to the study of neutrino oscillations, that can be inferred from the low-energy atmospheric neutrino flux that reaches IceCube from different directions, high-energy atmospheric neutrinos can be used to measure the neutrino-nucleon cross section at energies beyond what is currently reached at accelerators. The technique is based on measuring the amount of atmospheric muon-neutrinos as a function of zenith angle $\theta_{\rm zen}$, and compare it with the expected number from the known atmospheric flux assuming the Standard Model neutrino cross sections. Neglecting regeneration effects, the number of events scales as
\begin{equation}
N(\theta_{\rm zen},E_\nu) \propto \sigma_{\nu N}(E_\nu)\exp(-\sigma_{\nu N}(E_\nu)X(\theta_{\rm zen})/m_p)\,,
\end{equation}
where $X(\theta_{\rm zen})$ is the integrated column depth along the line of sight (${\bf n}(\theta_{\rm zen})$) from the location of IceCube (${\bf r}_{\rm IC}$),
\begin{equation}
X(\theta_{\rm zen}) = \int {\rm d}\ell \, \rho_\oplus({\bf r}_{\rm IC}+\ell{\bf n}(\theta_{\rm zen}))\,.
\end{equation}
The neutrino-matter cross section $\sigma_{\nu N}$ increases with neutrino energy, and above 100~TeV the Earth becomes opaque to vertically up-going neutrinos, {\it i.e.}, neutrinos that traverse the whole Earth, see Fig.~\ref{fig:nu_transmission}. Therefore, any deviation from the expected absorption pattern of atmospheric neutrinos can be linked to deviations from the assumed cross section, given all other inputs are known with sufficient precision. 

%%%%%%%%%%%%%%%%%%%%%%%%%%%%
%%%%%%%%%%%%%%%%%%%%%%%%%%%%

IceCube has performed such an analysis~\cite{Aartsen:2017kpd} by a maximum likelihood fit of the neutrino-matter cross section. The data, binned into neutrino energy, $E_\nu$, and neutrino arrival direction, $\cos\theta_{\rm zen}$, was compared to the expected event distribution from atmospheric and astrophysical neutrinos. Deviations from the Standard Model cross section $\sigma_{\rm SM}$ were fitted by the ratio $R=\sigma_{\nu N} / \sigma_{\rm SM}$. The analysis assumes priors on the atmospheric and astrophysical neutrino flux based on the baseline models in Refs.~\cite{Honda:2006qj,Enberg:2008te,Aartsen:2013jdh}.  In practice, the likelihood maximisation uses the product of the flux and the cross section, keeping the observed number of events as a fixed quantity. Thus, trials with higher cross sections must assume lower fluxes (or vice-versa) in order to preserve the total number of events. The procedure is thus sensitive to neutrino absorption in the Earth alone, and not to the total number of observed events. Since the astrophysical flux is still not known to a high precision, the uncertainties in the normalisation and spectral index were included as nuisance parameters in the analysis. Other systematics considered are the Earth density and core radius as obtained from the Preliminary Earth Model~\cite{Dziewonski:1981xy}, the effects of temperature variations in the atmosphere, which impact the neutrino flux during the year, and detector systematics. 

\begin{figure}[t]
\centering\includegraphics[width=\linewidth]{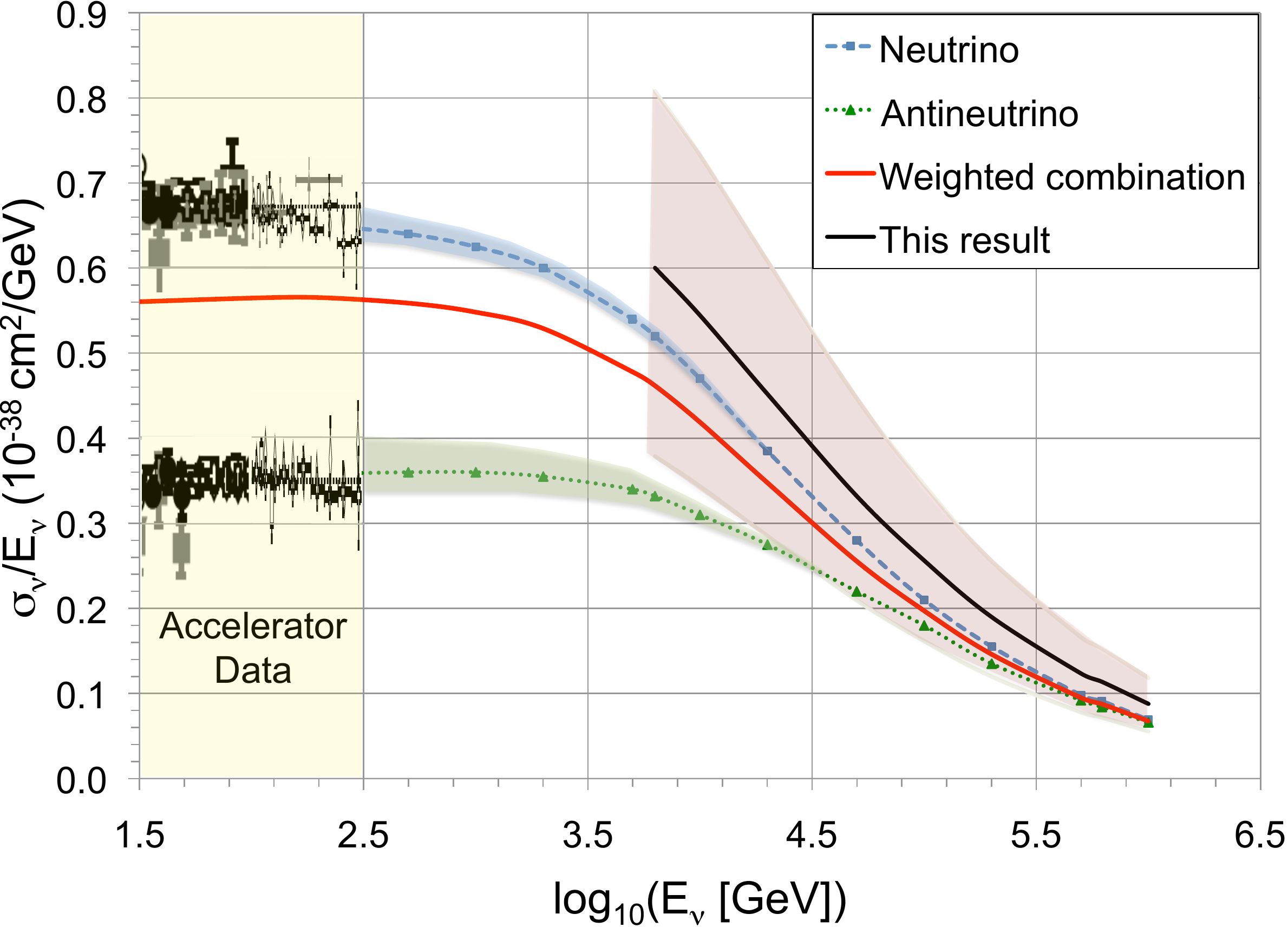}
\caption[]{Charged-current neutrino cross section as a function of energy. Shown are results from previous accelerator measurements (yellow shaded area, from~\cite{PDG:2018}), compared with the result from IceCube for the combined $(\nu+\overline{\nu})+N$ charged-current cross section. The blue and green lines represent the Standard Model expectation for $\nu$ and $\overline{\nu}$ respectively. The dotted red line represents the flux--weighted average of the two cross sections, which is to be compared with the IceCube result, the black line. The light brown shaded area indicates the uncertainty on the IceCube measurement. Figure from~\cite{Aartsen:2017kpd}.}
\label{fig:numu_Xsection}
\end{figure}

The analysis results in a value of $R= 1.30^{+0.21}_{-0.19} ({\rm stat})\\ ^{+0.39}_{-0.43} ({\rm sys})$.
This is compatible with the Standard Model prediction ($R=1$) within uncertainties but, most importantly, it is the first measurement of the neutrino-nucleon cross section at an energy range (few~TeV to about 1~PeV) unexplored so far with accelerator experiments~\cite{PDG:2018}. This is illustrated in Fig.~\ref{fig:numu_Xsection} which shows current accelerator measurements (within the yellow shaded area) and the results of the IceCube analysis as the light brown shaded area. The authors of Ref.~\cite{Bustamante:2017xuy} performed a similar analysis based on six years of high-energy starting event data. Their results are also consistent with perturbative QCD predictions of the neutrino-matter cross section.

\subsection{Probe of Cosmic Ray Interactions with IceCube}

On a slightly different topic, but still related to the products of cosmic ray interactions in the atmosphere, the high rate of atmospheric muons detected by IceCube can be used to perform studies of hadronic interactions at high energies and high momentum transfers. Muons are created from the decays of pions, kaons and other heavy hadrons. For primary energies above about 1~TeV, muons with a high transverse momentum, $p_t \gtrsim 2$~GeV, can be produced alongside the many particles created in the forward direction, the ``core'' of the shower. This will show up in IceCube as two tracks separated by a few hundred meters: one track for the main muon bundle following the core direction, and another track for the high-$p_t$ muon. The muon lateral distribution in cosmic-ray interactions depends on the composition of the primary flux and details of the hadronic interactions~\cite{Apel:2014qqa,Gonzalez:2015rdo}. If the former is sufficiently well known, the measurement of high-$p_t$ muons can be used to probe hadronic processes involving nuclei and to calibrate existing Monte-Carlo codes at energies not accessible with particle accelerators.

%%%%%%%%%%%%%%%%%
%%%%%%%%%%%%%%%% 

The lateral separation, $d_t$, of high $p_t$ muons from the core of the shower is given by $d_t=p_t H/E_{\mu} \cos\theta_{\rm zen}$, where $H$ is the interaction height of the primary with a zenith angle $\theta_{\rm zen}$. The initial muon energy $E_{\mu}$ is close to that at ground level due to minimal energy losses in the atmosphere. That is, turning the argument around, the identification of single, laterally separated muons at a given $d_t$ accompanying a muon bundle in IceCube is a measurement of the transverse momentum of the muon's parent particle, and a handle into the physics of the primary interaction. Given the depth of IceCube, only muons with an energy above $\sim$400 GeV at the surface can reach the depths of the detector. This, along with the inter-string separation of 125~m, sets the level for the minimum $p_t$ accessible in IceCube. However, since the exact interaction height of the primary is unknown and varies with energy, a universal $p_t$ threshold can not be given. For example, a 1~TeV muon produced at 50~km height and detected at 125~m from the shower core has a transverse momentum $p_t$ of 2.5 GeV. 

Our current understanding of lateral muon production in hadronic interactions shows an exponential behaviour at low $p_t$, $\exp(-p_t/T)$, typically below 2~GeV, due to soft, non-perturbative interactions,  and a power-law behaviour at high $p_t$ values, $(1+p_t/p_0)^{-n}$, reflecting the onset of hard processes described by perturbative QCD. The approach traces back to the QCD inspired "modified Hagedorn function"~\cite{Hagedorn:1983wk,Adams:2004zg}. The parameters $T$, $p_0$ and $n$ can be obtained from fits to proton-proton or heavy ion collision data~\cite{Adams:2004zg,Adare:2011vy}.

\begin{figure}[t]
\centering\includegraphics[width=\linewidth]{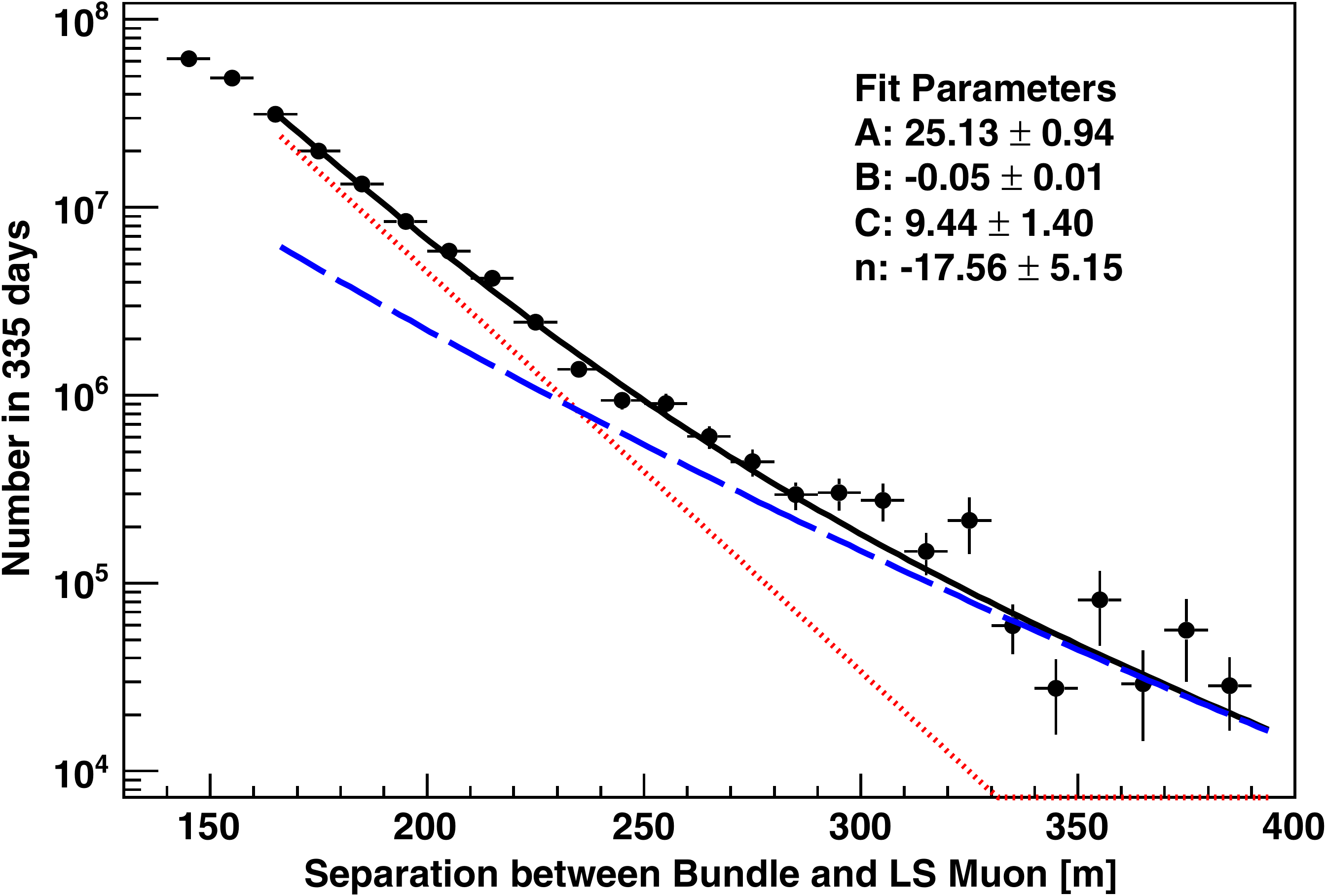}\hspace{0.5cm}
\caption[]{
The lateral muon distribution measured by IceCube, normalised to sea level, along with the best fit parameters to a combined exponential and power law function of the form $\exp(A+Bx)+10^C(1+x/400)^n$.  The exponential part of the fit is plotted as a dotted red line and the power law is shown as a dashed blue line.
Figure reprinted with permission from~\cite{Abbasi:2012kza} (Copyright 2013 APS).}
\label{fig:lateral_and_angular_mu}
\end{figure}

This is also the behaviour seen by IceCube. Fig.~\ref{fig:lateral_and_angular_mu} shows the muon lateral distribution at high momenta obtained from a selection of events reconstructed with a two-track hypothesis in the  59-string detector~\cite{Abbasi:2012kza}, along with a fit to a compound exponential plus power-law function. Due to the size of the 59-string detector and the short live time of the analysis (1 year of data), the statistics for large separations is low and fluctuations in the data appear for track separations beyond 300~m.  Still, the presence of an expected hard component at large lateral distances (high $p_t$) that can be described by perturbative quantum chromodynamics (a power-law behaviour) is clearly visible.

Significant discrepancies also exist between interaction models on the expected $p_t$ distribution of muons~\cite{Soldin:2014ouk}. The high-$p_t$ muon yield per collision depends, both, on the primary composition (protons typically producing higher rates of high-$p_t$ muons than iron interactions) and the hadronic interaction model. Future IceCube analyses with the larger, completed, 86-string detector using IceCube/IceTop coincident events can extend the range of the $\cos\theta_{\rm zen}$ distribution as well as provide an updated comparison of the muon $p_t$ distribution with predictions from existing hadronic models with higher statistics than that shown in Fig.~\ref{fig:lateral_and_angular_mu}. There is definitely complementary information from neutrino telescopes to be added to the efforts in understanding hadronic interactions using air shower arrays and heavy-ion and hadronic accelerator experiments~\cite{Soldin:2015iaa,Dembinski:2017zkb}.

%%%%%%%%%%%%%%%% 
%%%%%%%%%%%%%%%% 

\section{Non-Standard Neutrino Oscillations and Interactions}
\label{sec:exotic_oscillations}
In the previous two sections we have summarised the phenomenology of weak neutrino interactions and standard oscillations based on the mixing between three active neutrino flavour states and the eigenstates of the Hamiltonian (including matter effects). However, the Standard Model of particle physics does not account for neutrino masses and is therefore incomplete. The necessary extensions of the Standard Model that allow for the introduction of neutrino mass terms can also introduce non-standard oscillation effects that are suppressed in a low-energy effective theory. This is one motivation to study non-standard neutrino oscillations. In the following, we will discuss various extensions to the Standard Model that can introduce new neutrino oscillation effects and neutrino interactions. The large energies and very long baselines associated with atmospheric and cosmic neutrinos, respectively, provide a sensitive probe for these effects.

For the following discussion it is convenient to introduce neutrino oscillations via the evolution of the density operator $\rho$ of a mixed state. For a given Hamiltonian $H$, the time evolution of the neutrino density operator is governed by the Liouville equation,
\begin{equation}\label{liouville}
  \dot\rho= -{\rm i}[H_{\rm SM},\rho]\,.
\end{equation}
The solution to the Liouville equation allows to describe neutrino oscillation effects in a basis-independent way. For instance, a neutrino that is produced at times $t=0$ in the flavour state $\nu_\alpha$ can be represented by $\rho(0) =\Pi_\alpha$, where $\Pi_\alpha = |\nu_\alpha\rangle\langle\nu_\alpha|$ is the projection operator onto the flavour state $|\nu_\alpha\rangle$. The transition probability between two flavour states $\nu_\alpha$ and $\nu_\beta$ can then be recovered as the expectation value of the projector $\Pi_\beta$, which is given by trace
\begin{equation}
P_{\alpha\to\beta}(t) = {\rm Tr}(\rho(t)\Pi_\beta)\,.
\end{equation} 
For standard neutrino oscillations the Hamiltonian is composed of two terms, $H_{\rm SM}=H_0 +V_{\rm mat}$, describing the free evolution in vacuum and coherent matter effects, respectively. The free Hamiltonian in vacuum can be written as
\begin{equation}\label{freeH}
  H_{0} \simeq  \sum_i\frac{m_i^2}{2E_\nu}(\Pi_i+\overline{\Pi}_i)\,,
\end{equation}
 where the sum runs over the available flavours and we have introduced the projection operators $\Pi_i \equiv |\nu_i\rangle\langle\nu_i|$ and $\overline{\Pi}_i \equiv |\overline{\nu}_i\rangle\langle\overline{\nu}_i|$ onto neutrino and anti-neutrino mass eigenstates. Matter effects introduced earlier can be cast into the form
\begin{equation}\label{Vmat}
V_{\rm mat} = \sqrt{2}G_FN_e(\Pi_e - \overline{\Pi}_{e})\,,
\end{equation}
where $\Pi_e \equiv |\nu_e\rangle\langle\nu_e|$ and $\overline{\Pi}_e \equiv|\overline{\nu}_e\rangle\langle\overline{\nu}_e|$ are the projections onto electron neutrino and anti-neutrinos states, respectively, and $N_e$ is the number density of electrons in the background.  

In the following, we will discuss non-standard contributions to the Liouville equation that we assume to take on the form
\begin{equation}\label{liouville_mod}
  \dot\rho= -{\rm i}[H_{\rm SM},\rho]-{\rm i}\sum_n[H_n,\rho] + \sum_n\mathcal{D}_n[\rho]\,.
\end{equation}
The second term on the right hand side of Eq.~(\ref{liouville_mod}) accounts for additional terms in the Hamiltonian, that affect the phenomenology of neutrino oscillations. These terms are therefore parametrised by a sequence of Hermitian operators, $H_n = H^\dagger_n$, that preserve the unitarity of the density operator. The third term corresponds to non-unitary effects, related to dissipation or decoherence, that can not only affect the neutrino flavour composition, but also their spectra. In particular, we will focus in the following on cases where the sequence of terms $\mathcal{D}_n$ can be expressed in terms of a set of {\it Lindblad} operators $L_j$ as
\begin{equation}\label{Lindblad} 
\mathcal{D}_n [\rho] = \frac{
1}{2}\sum_j   \left([L_j,\, \rho\, L_j^\dagger] +
    [L_j\, \rho,\, L_j^\dagger]\right)\,.
\end{equation}
These Lindblad operators act on the Hilbert space of the open quantum system, ${\cal H}$, and satisfy $\sum_j L^\dagger_j L_j \in {\cal B} ({\cal H}),$ where ${\cal B} ({\cal H})$ indicates the space of bounded operators acting on ${\cal H}$~\cite{Lindblad:1975ef}. We will discuss in the following various instances of these Lindblad operators for neutrino decoherence, neutrino decay, or hidden neutrino interactions with cosmic backgrounds.

\subsection{Effective Hamiltonians}

Non-trivial mixing terms of the effective Hamiltonian can be generated in various ways, including non-standard interactions with matter and Standard Model extensions that violate the equivalence principle, Lorentz invariance, or CPT symmetry (see Ref.~\cite{GonzalezGarcia:2005xw} for references). We will first study oscillations that are induced by additional terms to the Hamiltonian that can be parametrised in the form~\cite{GonzalezGarcia:2005xw}
\begin{equation}\label{Heff}
H_{\rm eff} =  (E_\nu)^{n} \sum_{\mathfrak{a}} (\delta_\mathfrak{a}\Pi_{\mathfrak{a}} + \overline{\delta}_\mathfrak{a}\overline{\Pi}_{\mathfrak{a}}) \,.
\end{equation}
Here we introduced the projectors $\Pi_{\mathfrak{a}} = |\nu_{\mathfrak{a}}\rangle\langle\nu_{\mathfrak{a}}|$ and $\overline{\Pi}_{\mathfrak{a}} = |\overline{\nu}_{\mathfrak{a}}\rangle\langle\overline{\nu}_{\mathfrak{a}}|$ onto a new set of states $|\nu_\mathfrak{a}\rangle$ that are related to the usual flavor states by a new unitary mixing matrix
\begin{equation}
|\nu_\alpha\rangle = \sum_\mathfrak{a}\widetilde U_{\alpha\mathfrak{a}}^*|\nu_\mathfrak{a}\rangle\,.
\end{equation}
The expansion parameters $\delta_\mathfrak{a}$ have mass dimensions $1-n$ with integer $n$. 

It is convenient to group these contributions into CPT-even terms obeying the relation $\delta_\mathfrak{a} = \overline{\delta}_\mathfrak{a}$ and CPT-odd terms with $\delta_\mathfrak{a} = -\overline{\delta}_\mathfrak{a}$. For a CPT-symmetric process we have $P(\nu_\alpha \to \nu_\beta) = P(\overline{\nu}_\beta\to\overline{\nu}_\alpha)$. In particular, the survival probability between neutrinos and anti-neutrinos is the same. CPT-odd terms break this symmetry. Indeed, we have already encountered such a CPT-odd term as the CP-violating matter effect contributing by the effective potential of Eq.~(\ref{Vmat}). The corresponding expansion in terms the effective Hamiltonian of Eq.~(\ref{Heff}) is given by $\widetilde U = \mathbf{1}$, $\delta_e= -\overline{\delta}_e =\sqrt{2}G_FN_e$, $\delta_{\mu,\tau}=0$, and $n=0$. On the other hand, the free Hamiltonian in Eq.~(\ref{freeH}) is a CPT-even term with $\widetilde U = U_{\rm PMNS}$, $\delta_i= \overline{\delta}_i = m_i^2/2$, and $n=-1$.

%%%%%%%%%%%%%%%%%%
\begin{figure}[t]\centering
  \includegraphics[width=0.9\linewidth]{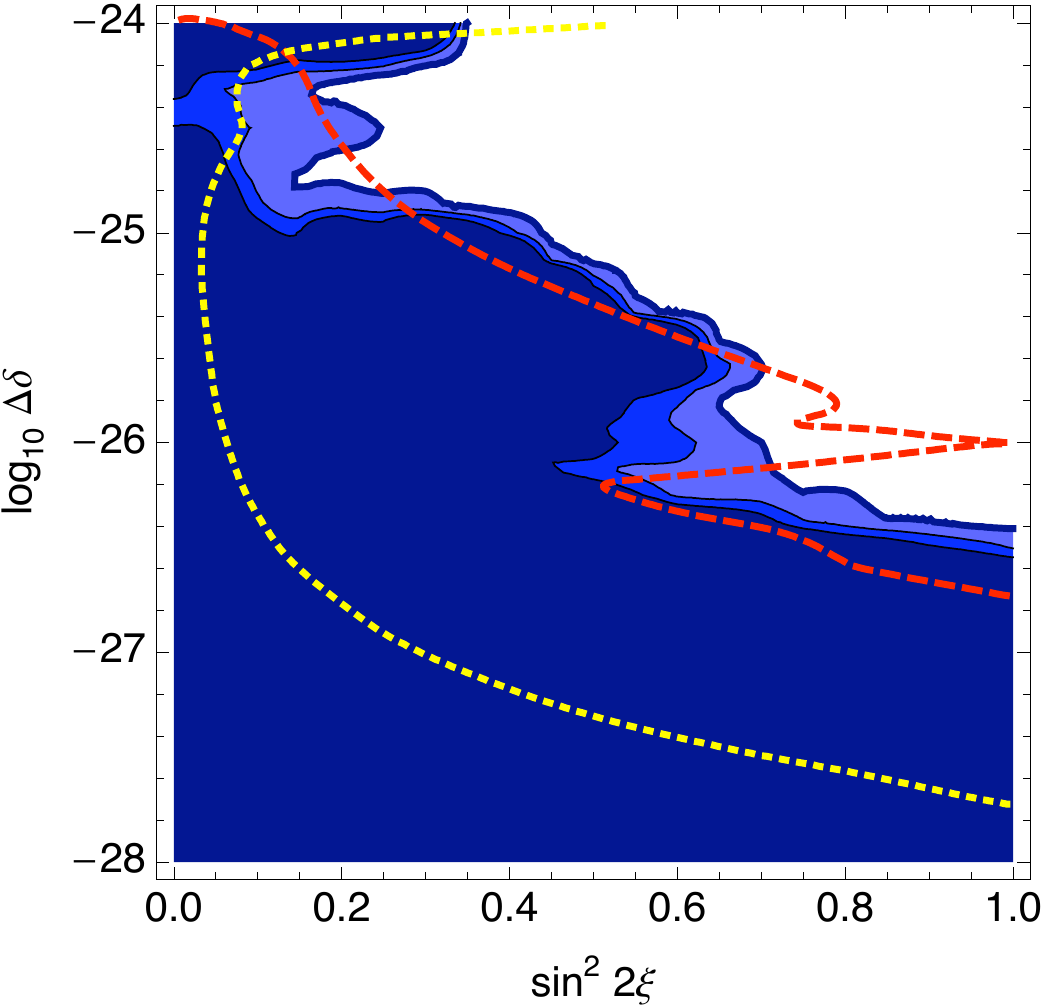}
  \caption[]{Allowed regions at 90\%, 95\%, and 99\% confidence level (from darkest to lightest) for LIV-induced oscillation effects with $n=1$.  Note we plot $\sin^2 2\xi$ to enhance the region of interest. Also shown are the Super-Kamiokande + K2K 90\% contour~\cite{GonzalezGarcia:2004cu} (dashed line), and the projected IceCube 10-year 90\% sensitivity \cite{GonzalezGarcia:2005xw} (dotted line). Reprinted with permission from~\cite{Abbasi:2009nfa}) (Copyright 2013 APS)}.
  \label{fig:chisq}
\end{figure}
%%%%%%%%%%%%%%%%%%

The contribution of effective Hamiltonians can be tested by the oscillation of atmospheric muon neutrinos~\cite{GonzalezGarcia:2005xw}. For simplicity, we will assume that this can be treated effectively as a two-level system, analogous to the case of standard atmospheric neutrino oscillations. The most general form of the unitary matrix $\widetilde U$ is then
\begin{equation}
\widetilde U = \begin{pmatrix}\cos\xi&e^{i\eta}\sin\xi \\- e^{-i\eta}\sin\xi&\cos\xi\end{pmatrix}\,.
\end{equation}
The $\nu_\mu$ survival probability can then be expressed as~\cite{GonzalezGarcia:2005xw}
\begin{equation} \label{eq:effHprob}P_{\nu_\mu \to \nu_\mu} = 1 - \sin^2 2\theta_{\rm eff} \, \sin^2   \frac{\mathcal{R}\Delta m^2_{\rm atm}L}{4E_\nu} \,,
\end{equation}
where we have introduced the scaling parameter
\begin{multline}
\mathcal{R} = \frac{\Delta m^2_{\rm eff}}{\Delta m^2_{\rm atm}} = \big[1 + R^2 + 2 R ( \cos 2\theta_{\rm atm} \cos 2\xi \\+ \sin   2\theta_{\rm atm} \sin 2\xi \cos\eta )\big]^{1/2}\,,
\end{multline}
and
\begin{equation}
 R \equiv 
 \frac{\Delta \delta E_\nu^n}{2}\frac{4E_\nu}{\Delta m_{\rm atm}^2} \,,
\end{equation}
with $\Delta \delta  = \delta_3-\delta_2$. The effective rotation angle $\theta_{\rm eff}$ is given by
\begin{multline}
  \label{eq:Theta}
  \sin^2 2\theta_{\rm eff} = \frac{1}{\mathcal{R}^{2}} \big( \sin^2 2\theta_{\rm atm} \\+   R^2 \sin^2 2\xi + 2 R \sin 2\theta_{\rm atm} \sin 2\xi \cos\eta \big) \,.
\end{multline}
Note that the survival probability in Eq.~(\ref{eq:effHprob}) reduces to the familiar expression of atmospheric oscillation in the limit $R\to0$.

%%%%%%%%%%%%%%%% 
\begin{figure}[t]
\centering
\includegraphics[width=0.8\columnwidth]{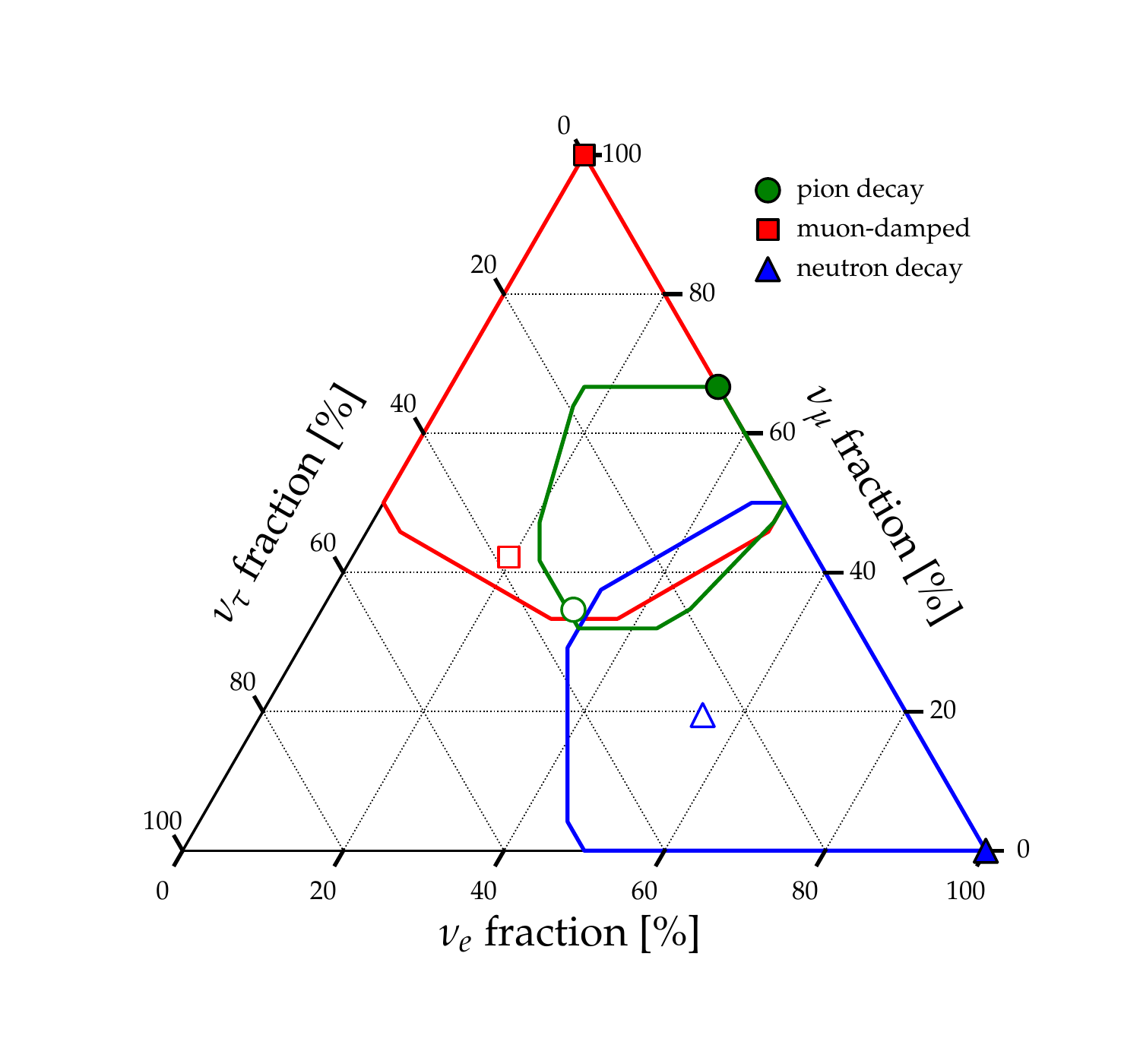}
\caption[]{The range of oscillation-averaged neutrino flavour ratios from astrophysical sources. The filled markers show the initial flavor composition from pion decay (1:2:0), muon-damped pion decay (0:1:0), and neutron decay (1:0:0). The open markers show the expected observed flavour ratio assuming the best-fit oscillation parameters under the assumption of normal ordering of neutrino masses (see Table~\ref{tab:oscillation}). The coloured contours show the range of oscillation-averaged flavor ratios based on the unitarity of the neutrino mixing matrix.}
\label{fig:triangle_unitarity}
\end{figure}
%%%%%%%%%%%%%%%% 

Figure~\ref{fig:chisq} shows the results of an analysis of atmospheric muon neutrino data in the range 100~GeV to 10~TeV taken by AMANDA-II in the years 2000 to 2006~\cite{Abbasi:2009nfa}. The data was binned into two-dimensional histograms in terms of the number of hit optical modules (as a measure of energy) and the zenith angle ($\cos\theta_{\rm zen}$). The predicted effect of non-standard oscillation parameters can be compared to the data via a profile likelihood method. No evidence of non-standard neutrino oscillations  was found and the statistically allowed region of the $[\Delta\delta, \sin^22\xi]$-plane is shown as 90\%, 95\%, 99\% C.L. These results are derived under the assumption that $\eta=\pi/2$ in the unitary mixing matrix. The red dashed line shows the 90\% limit of a combined analysis by Super-Kamiokande and K2K~\cite{GonzalezGarcia:2004cu} which is compatible with the AMANDA-II bound. The projected IceCube 90\% sensitivity after ten years of data taking is given as a yellow dotted line and may improve the limit on $\Delta\delta$ by one order of magnitude~\cite{GonzalezGarcia:2005xw}.

Effective Hamiltonians with positive energy dependence $n>0$ can dominate the standard Hamiltonian $H_0$ at sufficiently high energy. For astrophysical neutrinos this can lead to the situation that the oscillation-averaged flavour transition probability is completely dominated by the new unitary transition matrix leading to
\begin{equation}\label{eq:effectivetrans}
  \widetilde{P}_{\nu_\alpha \to \nu_\beta} 
  \simeq \sum_\mathfrak{a}|\widetilde{U}_{\alpha \mathfrak{a}}|^2\, |\widetilde{U}_{\beta \mathfrak{a}}|^2\,.
\end{equation}
This effect can therefore be tested by looking for statistically significant deviations from the predictions of standard oscillations in Eq.~(\ref{pak_approx}). However, the transition probabilities described by Eq.~(\ref{eq:effectivetrans}) are constrained by lepton unitary. This has been studied in Ref.~\cite{Xu:2014via}, who derived a non-trivial set of unitarity conditions:
\begin{align}
\widetilde{P}_{\nu_\alpha \to \nu_\alpha} &\geq \frac{1}{3}\,,\\
\widetilde{P}_{\nu_\alpha \to \nu_\beta} &\leq \frac{1}{2} \qquad \text{($\alpha\neq\beta$)}\,,\\
\widetilde{P}_{\nu_\alpha \to \nu_\beta} &\leq \frac{1}{24} +  \widetilde{P}_{\nu_\alpha \to \nu_\alpha} \qquad \text{($\alpha\neq\beta$)}\,.
\end{align}
The resulting oscillation-averaged flavour composition from these unitarity bounds are shown as the contours in Fig.~\ref{fig:triangle_unitarity}. We assume three initial flavour composition from pion decay (1:2:0), muon-damped pion decay (0:1:0), and neutron decay (1:0:0).

%%%%%%%%%%%%%%%%
\begin{table}[t]
  \centering
  \renewcommand\arraystretch{1.2}
  \begin{tabular}{cccc}
    \hline
    $n$ & LIV ($\Delta\delta/E_\nu^n$) & QD ($D/E_\nu^n$) & Units \\
    \hline
    1 & $\ 2.8\times10^{-27}\ $ & $\ 1.2\times10^{-27}\ $ & -- \\ 
    2 & $\ 2.7\times10^{-31}\ $ & $\ 1.3\times10^{-31}\ $ & $\ {\rm GeV}^{-1}$     \\
    3 & $\ 1.9\times10^{-35}\ $ & $\ 6.3\times10^{-36}\ $ & $\ {\rm GeV}^{-2}$     \\
    \hline
  \end{tabular}
  \caption[]{\label{tab:vli_qd_limits} (From Ref.~\cite{Abbasi:2009nfa}) 90\% CL     upper limits on Lorentz-invariance violation (LIV)
    and quantum decoherence (QD) effects. LIV upper limits are for the case of
    maximal mixing ($\sin 2\xi = 1$), and quantum decoherence upper limits for     the case of
    a 3-level system with universal decoherence parameters $D$ (see text for     details).} 
\end{table}
%%%%%%%%%%%%%%%%

\subsection{Violation of Lorentz Invariance}

One of the foundations of the Standard Model of particle physics is the principle of Lorentz symmetry: The fundamental laws in nature are thought to be independent of the observer's inertial frame. However, some extensions of the Standard Model, like string theory or quantum gravity, allow for the spontaneous breaking of Lorentz symmetry, that can lead to Lorentz-invariance violating (LIV) effects in the low-energy effective theory. There also exist a deep connection between the appearance of LIV effects with the violation of CPT-invariance\footnote{The invariance of physical laws under simultaneous charge conjugation (C), parity reflection (P), and time reversal (T).} in local quantum field theories~\cite{Jost:1957zz}. Such effects were incorporated in the Standard Model Extension (SME), an effective-field Lorentz-violating extension of the Standard Model, which includes CPT-even and CPT-odd terms~\cite{Colladay:1998fq}. The SME provides a benchmark for experiments to gauge possible Lorentz violating processes in nature, by expressing experimental results in terms of the parameters of the model. The size of LIV effect is expected to be suppressed by Planck scale $M_{\rm P} \simeq 10^{19}$~GeV (or Planck length $\lambda_{\rm P}\simeq10^{-33}$~cm), consistent with the strong experimental limits on the effect~\cite{Kostelecky:2008ts,Liberati:2013xla}.  

%%%%%%%%%%%%%%%% 
\begin{figure*}[t]
\centering
\includegraphics[width=0.6\columnwidth]{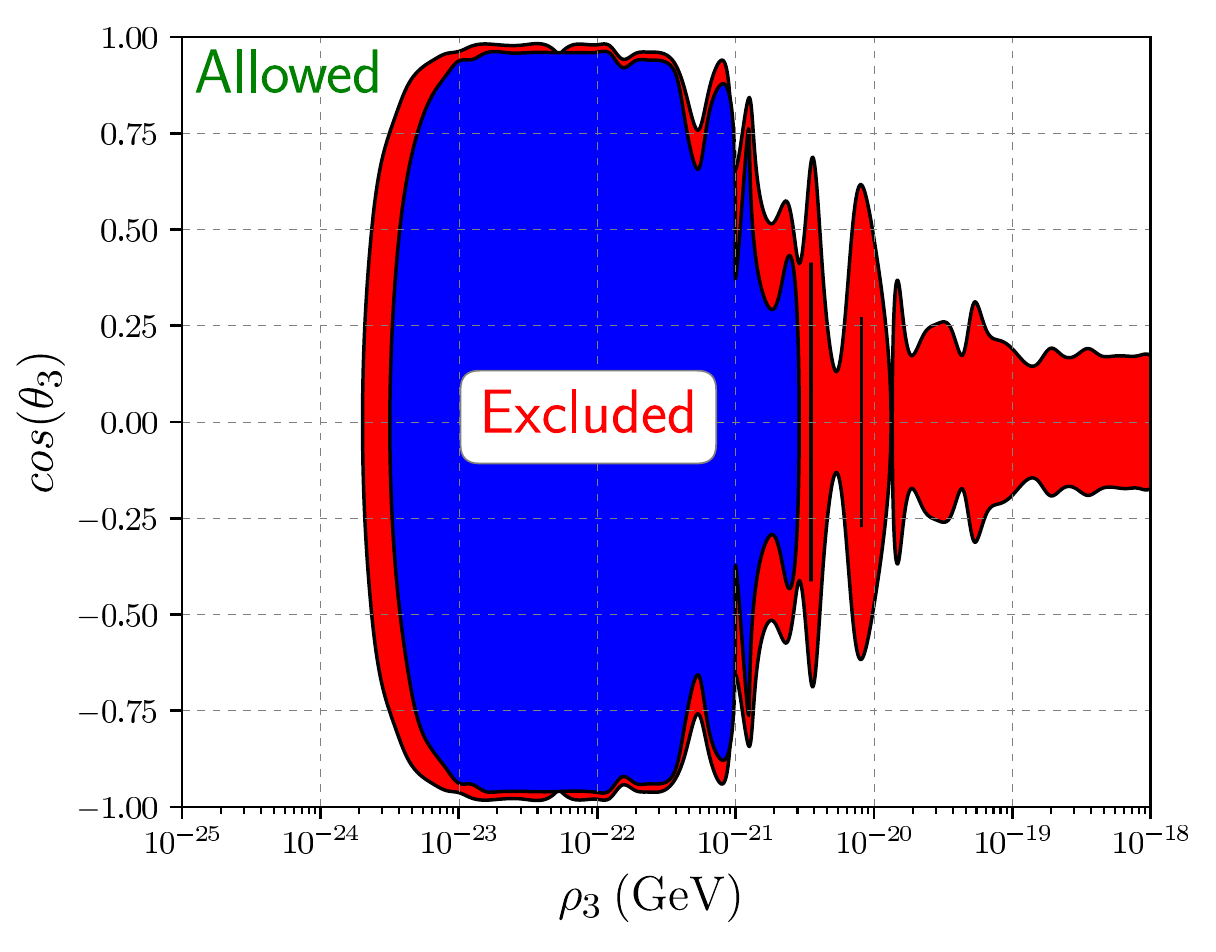}
\includegraphics[width=0.6\columnwidth]{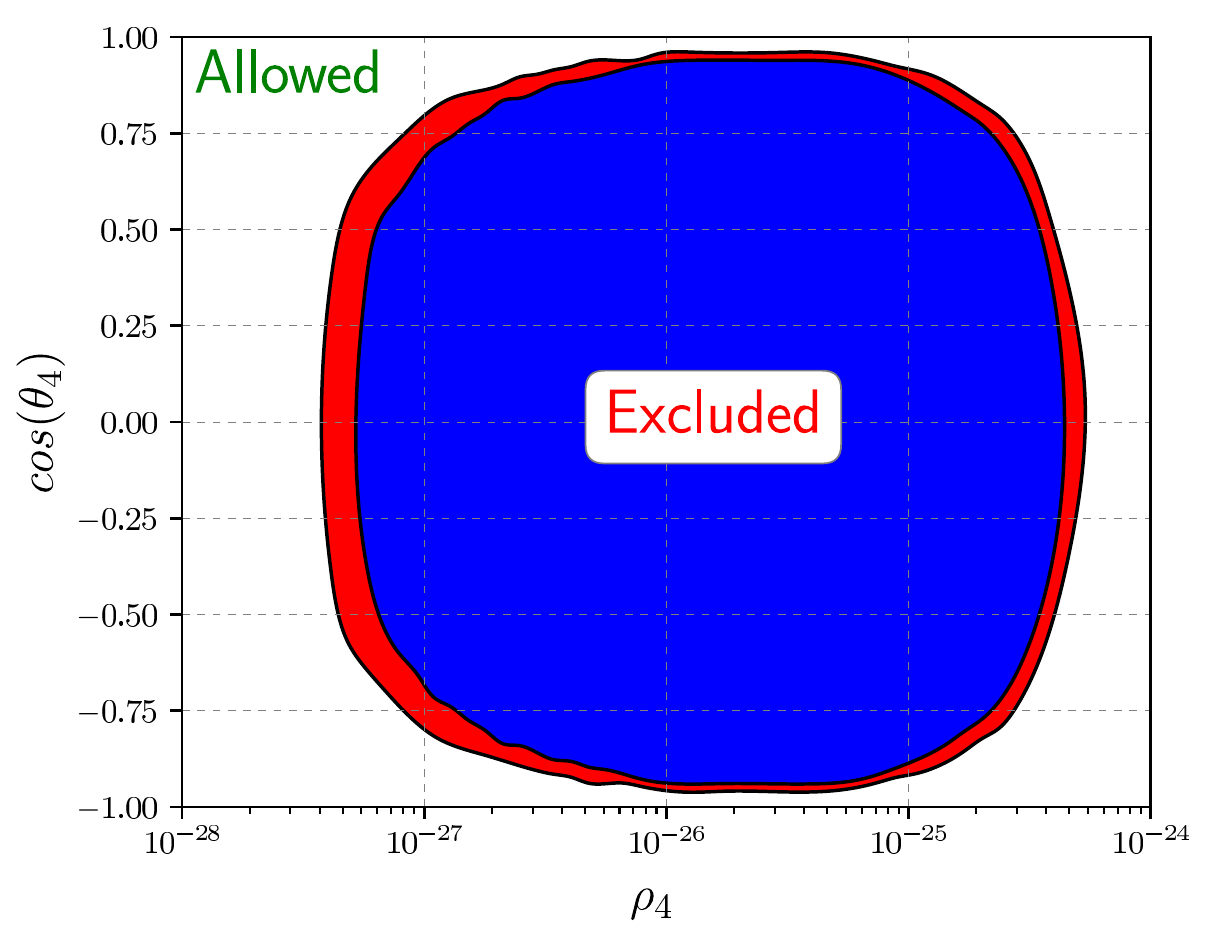}
\includegraphics[width=0.6\columnwidth]{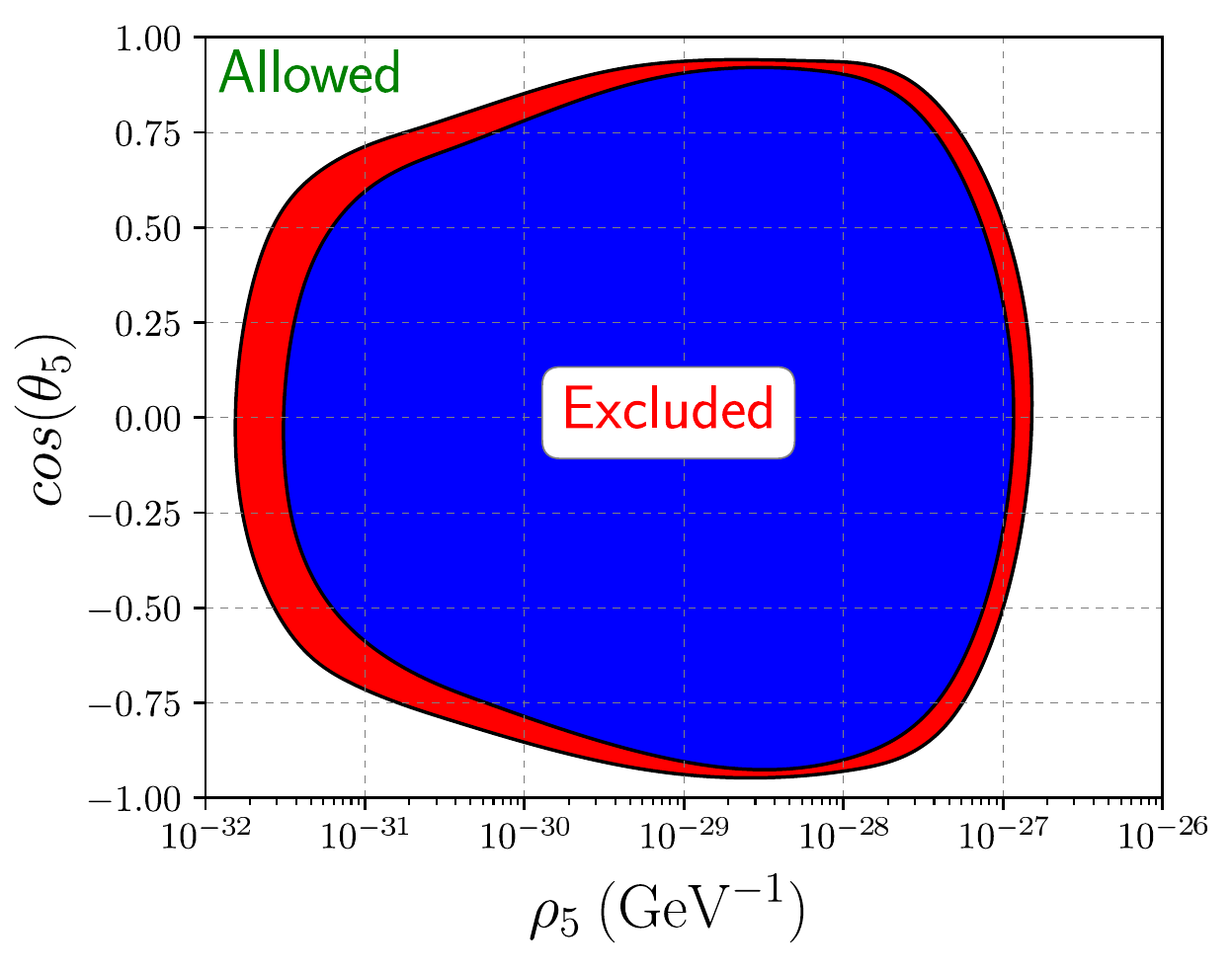}
\includegraphics[width=0.6\columnwidth]{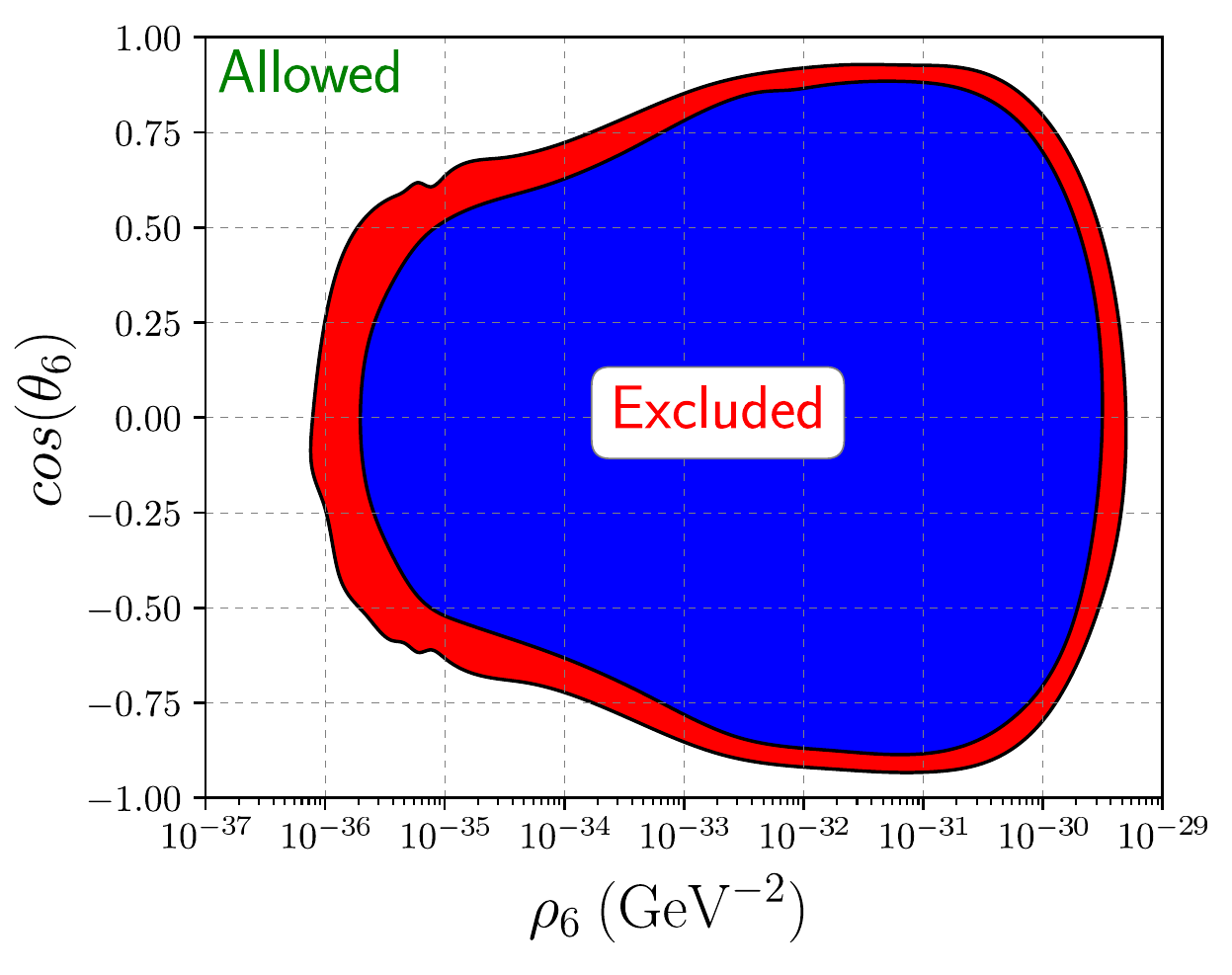}
\includegraphics[width=0.6\columnwidth]{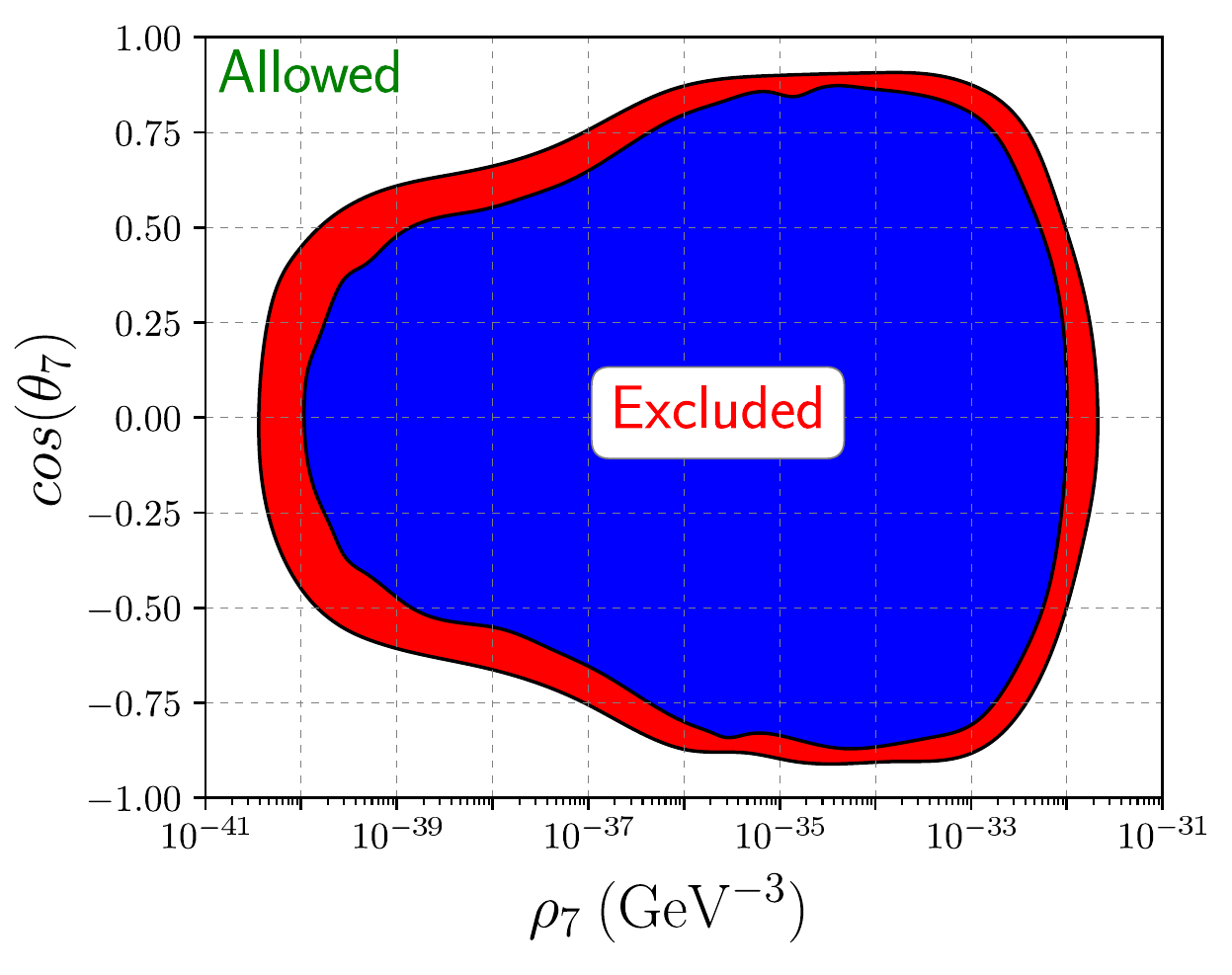}
\includegraphics[width=0.6\columnwidth]{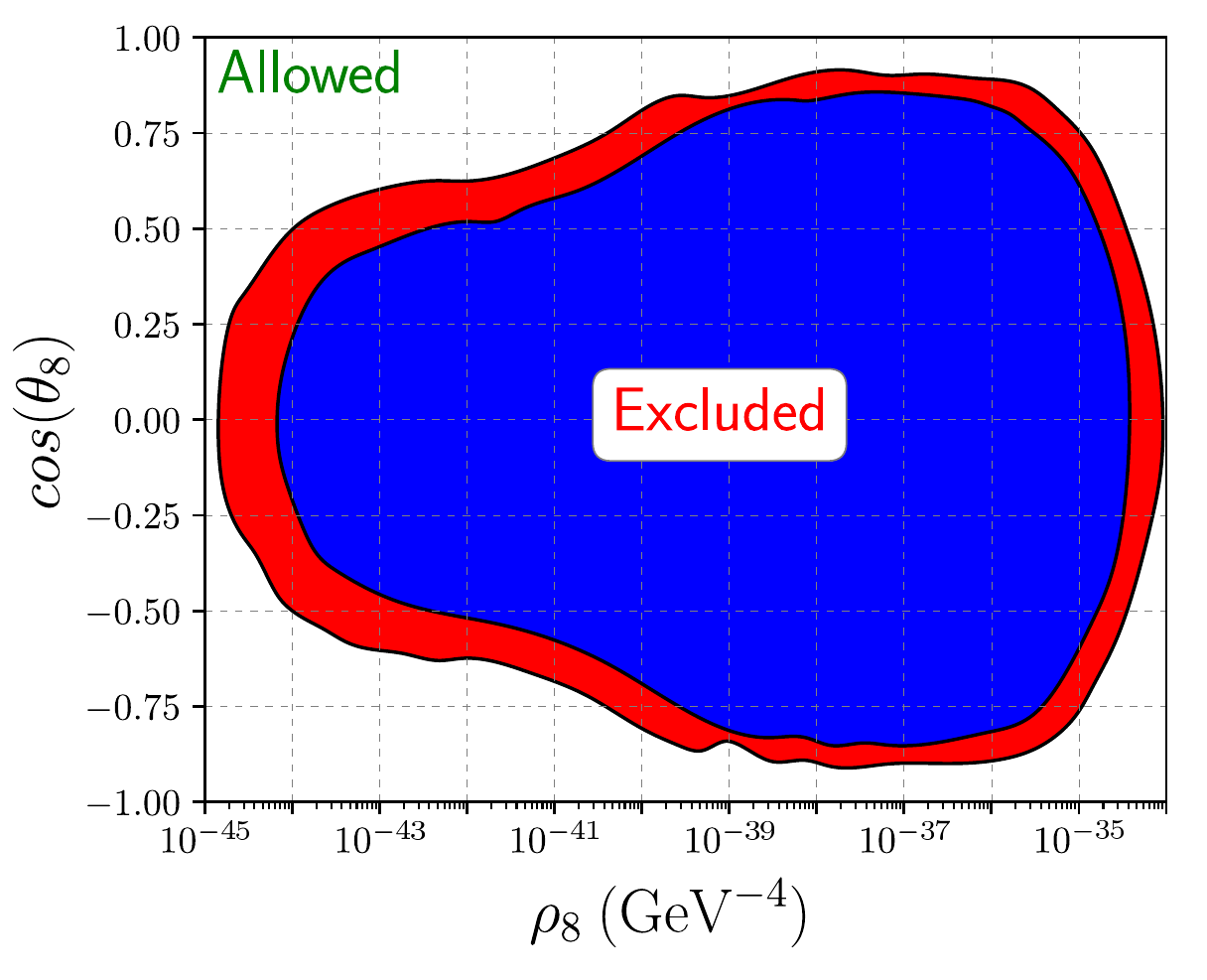}
\caption[]{Exclusion region in the LIV parameter space discussed in the text. The parameter $\rho$ is related to the strength of the Lorentz invariance violation. The parameter $\cos \theta$ is a combination of coefficients that define Lorentz invariance violation in the effective Hamiltonian of Standard Model Extension~\cite{Colladay:1998fq}. Each plot shows results for the dimension $d$ of the operator, from 3 to 8, from left to right, and top to bottom.  The red (blue) regions are excluded at 90\% (99\%) C.L. Figure from~\cite{Aartsen:2017ibm}}
\label{fig:result_g_frequentist_limit}
\label{fig:result_LIVlimit}
\end{figure*}
%%%%%%%%%%%%%%%% 

Oscillations of atmospheric neutrinos with energies above 100~GeV provide a sensitive probe of LIV effects. For instance, LIV in the neutrino sector can lead to small differences in the maximal attainable velocity of neutrino states~\cite{Coleman:1998ti}. Since the ``velocity eigenstates'' are different from the flavour eigenstates, a new oscillation pattern can arise. The effect can be described in the framework of effective Hamiltonians described in the previous section by CPT-even states with $n=1$. For the approximate two-level system with survival probability described by Eq.~(\ref{eq:effHprob}) we can identify effective Hamiltonian parameters as the velocity difference $\Delta \delta = \Delta c$, together with a new mixing angle $\xi$ and a phase $\eta$.  Higher order contributions $n>1$ have been considered for non-renormalisable LIV effects caused by quantum mechanical fluctuations of the space-time metric and topology~\cite{Brustein:2001ik}.  Both the $\Delta\delta \propto E$ ($n=1$) and the $\Delta\delta \propto E^{3}$ ($n=3$) cases have been examined in the context of violations of the equivalence principle (VEP) \cite{Gasperini:1988zf,Halprin:1991gs,Adunas:2000zm}. The 90\% C.L.~upper limits on the corresponding coefficients $\Delta\delta/E_\nu^n$ [GeV${}^{1-n}$] are shown in Tab.~\ref{tab:vli_qd_limits}.

Violations of Lorentz invariance can also manifest themselves by the dependence of neutrino oscillations on a preferred orientation of the neutrino arrival direction. This effect can also be described by effective Hamiltonians similar to Eq.~(\ref{Heff}) where the expansion parameters not only depend on energy but also on direction. The contribution of these terms were studied in the analysis~\cite{Abbasi:2010kx} using data collected by the 40-string configuration of IceCube between April 2008 and May 2009. Due to IceCube's unique position at the geographic South Pole, the field of view of the observatory is constant in time. The atmospheric neutrino data extracted from this period was binned in terms of sidereal time. If neutrino oscillations depend on the relative neutrino momenta in the cosmic rest frame, then the muon neutrino data in this reference frame is expected to show oscillation patterns. The non-observation of this effect allowed to constrain these LIV parameters.

A more recent search for isotropic LIV effects with the complete IceCube detector was the subject of the analysis presented in~\cite{Aartsen:2017ibm}, based on  data collected during the period from May 2010 to May 2012. The LIV effects were parametrised by two parameters $\rho_d$ and $\theta_d$, which are related to the strength of the LIV effect and the a combination of other LIV parameters of the SME~\cite{Colladay:1998fq}, respectively. These parameters accompany the expansion of the Hamiltonian in powers of energy, and the subscript $d$ refers to the power of the corresponding operator in the Hamiltonian. Most experiments are sensitive to effects of dimension $d=3$ and $d=4$, but the energy reach of IceCube makes it possible to extend the search to LIV effects induced by operators of dimension up to $d=8$.  The analysis used binned atmospheric neutrino data into a horizontal ($\cos\theta_{\rm zen} > -0.6$) and a vertical ($\cos\theta_{\rm zen} < -0.6$) sample which allowed to study the LIV effect by the energy-dependent ratio of the samples using a likelihood analysis. No LIV effects were identified, the best fit values of all the $\rho_d$ parameters being compatible with zero. This allowed to place limits in the [$\rho_d$-$\cos\theta_d$] parameter space. The results are shown in Fig.~\ref{fig:result_LIVlimit} for $3\leq d\leq8$.

The violation of Lorentz invariance associated with Planck-scale physics can also affect neutrino spectra over long baselines. The LIV effects can result in modified dispersion relations, {\it e.g.}, $E^2-p^2 = m^2 - \epsilon E^2$, that  introduce non-trivial maximal particle velocities~\cite{Coleman:1998ti}. Whereas at low energies the Lorentz invariance is recovered, $E^2-p^2\simeq m^2$, at high energies we can observe sub- or superluminal maximal particle velocities. These can allow otherwise forbidden neutrino decays, in particular, vacuum pair production, $\nu_\alpha \to \nu_\alpha +e^++e^-$, and vacuum neutrino pair production, $\nu_\alpha \to \nu_\alpha+\overline\nu_\beta+\nu_\beta$. The secondary neutrinos introduce non-trivial flavour compositions as well as spectral bumps and cutoffs. As discussed in Refs.~\cite{Stecker:2014xja,Stecker:2014oxa,Liao:2017yuy}, these small effects can be probed by the IceCube TeV-PeV diffuse flux.

%%%%%%%%%%%%%%%%%%
\begin{figure}[t]\centering
  \includegraphics[width=0.9\linewidth]{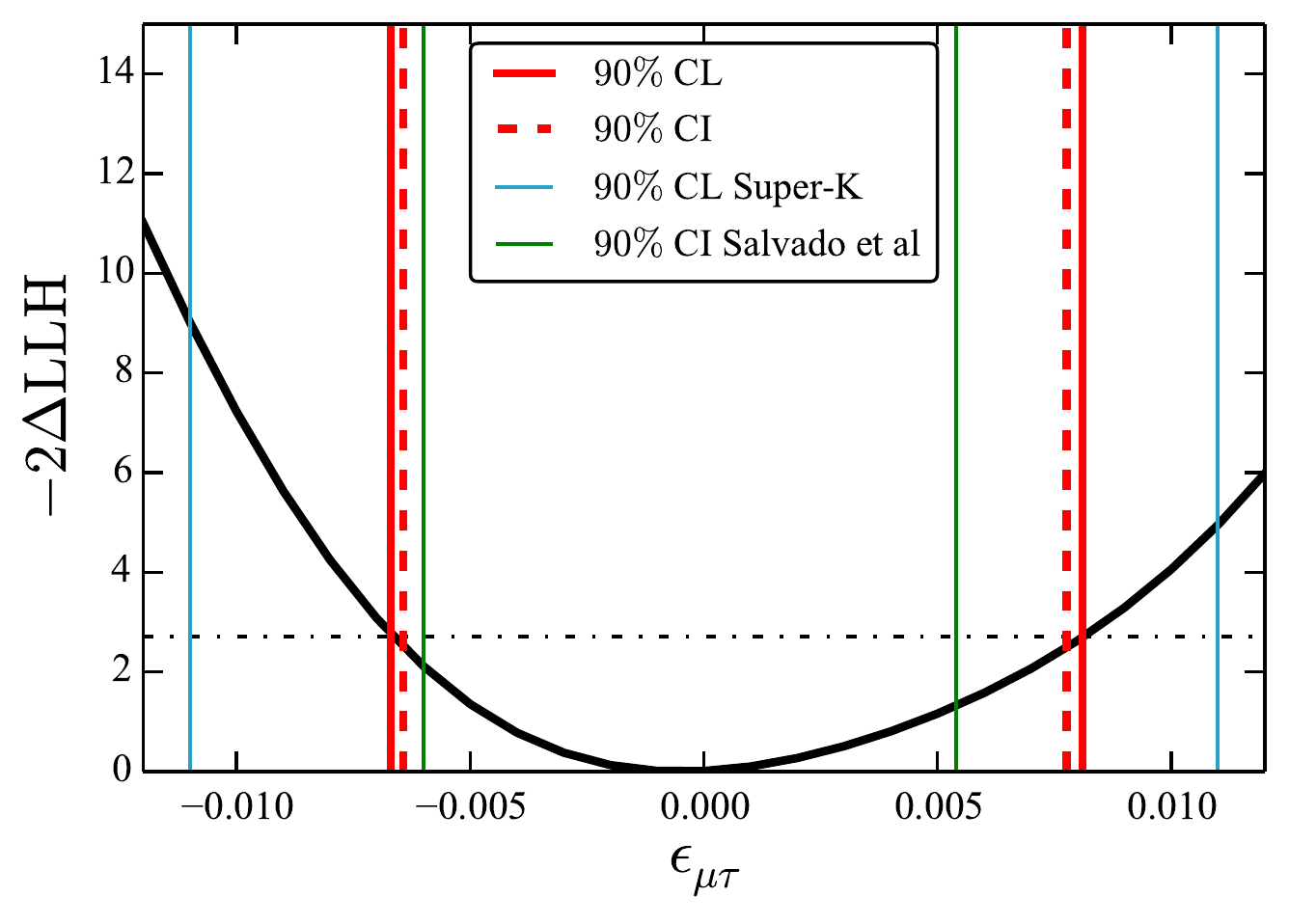}
  \caption[]{Confidence limits on the NSI parameter $\epsilon_{\mu \tau}$ using the event selection from~\cite{Aartsen:2014yll,Aartsen:2017bap} shown as solid vertical red lines.  Similarly, dashed vertical red lines show the 90\% credibility interval using a flat prior on $\epsilon_{\mu\tau}$ and where we have profiled over the nuisance parameters. The light blue vertical lines show the Super-Kamiokande 90$\%$ confidence limit~\cite{Mitsuka:2011ty}. The light green lines show the 90\% credibility region from~\cite{Salvado:2016uqu}. Finally, the horizontal dash-dot line indicates the value of $-2\Delta {\rm LLH}$ that corresponds to a 90$\%$ confidence interval according to Wilks' theorem. Figure from~\cite{Aartsen:2017xtt}.}
\label{fig:NSIresult01}
\end{figure}
%%%%%%%%%%%%%%%%%%

\subsection{Non-Standard Matter Interactions}

We have already discussed coherent scattering of neutrinos and anti-neutrinos in dense matter, that can be accounted for by an effective matter potential in the standard Hamiltonian. Since neutrino oscillations are only sensitive to non-universal matter effects, only the unique charged-current interactions of electron neutrinos and anti-neutrinos with electrons are expected to contribute. However, non-standard interactions (NSI) can change this picture. 

A convenient theoretical framework to study NSI contributions is given by the Hamiltonian,
\begin{equation}
H_{\rm NSI} = \sqrt{2}G_FN_e\sum_{\alpha,\beta}\epsilon_{\alpha\beta}|\nu_\beta\rangle\langle\nu_\alpha|\,.
\end{equation}
The dimensionless coefficients $\epsilon_{\alpha\beta}$ measure the NSI contribution relative to the strength of the standard matter potential. Hermiticity of the Hamiltonian requires that $\epsilon_{\alpha\beta} = \epsilon^*_{\beta\alpha}$. Non-standard interactions with nucleons are not necessarily universal for electrons and valence quarks in matter~\cite{Gonzalez-Garcia:2016gpq}. Therefore, the parameters $\epsilon_{\alpha\beta}$ can be considered as effective parameters of the form
\begin{equation}
\epsilon_{\alpha\beta} = \epsilon^{(e)}_{\alpha\beta} + \frac{\langle N_u\rangle}{N_e}  \epsilon^{(u)}_{\alpha\beta} + \frac{\langle N_d\rangle}{N_e}  \epsilon^{(d)}_{\alpha\beta}\,,
\end{equation}
where $\epsilon_{\alpha\beta}^{(u/d)}$ are the couplings to up- and down-type valence quarks in matter. In charge-neutral matter we must have $N_p = N_e$. The average neutron abundance according to Earth density models is $\langle N_n\rangle \simeq 1.051$~\cite{Dziewonski:1981xy}. Therefore, we have 
\begin{equation}
\langle N_u\rangle \simeq 3.051N_e,\qquad\langle N_d\rangle \simeq 3.102N_e\,.
\end{equation}
In the full three-flavour neutrino system, the hermiticity condition leaves nine degrees of freedom for the $\epsilon_{\alpha\beta}$ coefficients, that reduce to eight after absorption of a global phase.

The case of atmospheric muon neutrino oscillations can again be approximated by oscillations in the $\nu_\mu$-$\nu_\tau$-system. The oscillations induced by NSI interactions has three degrees of freedom that can be parametrized by $\epsilon_{\tau\tau}-\epsilon_{\mu\mu}=\epsilon'$ and $\epsilon_{\mu\tau} = \epsilon\exp(i\alpha)$. We can make use of the fact that the NSI Hamiltonian can be recast as an effective Hamiltonian as in Eq.~(\ref{Heff}). The NSI Hamiltonian in the two level system can be diagonalised by $\eta=\alpha$ and $\tan2\xi = 2\epsilon/\epsilon'$. After removing a global phase, the effective splitting is $\Delta\delta = \sqrt{2}G_FN_e\epsilon'\sqrt{1+4\epsilon^2/\epsilon'^2}$ with $\sigma=-1$.

%%%%%%%%%%%%%%%%%%
\begin{figure}[t]\centering
  \includegraphics[width=0.9\linewidth]{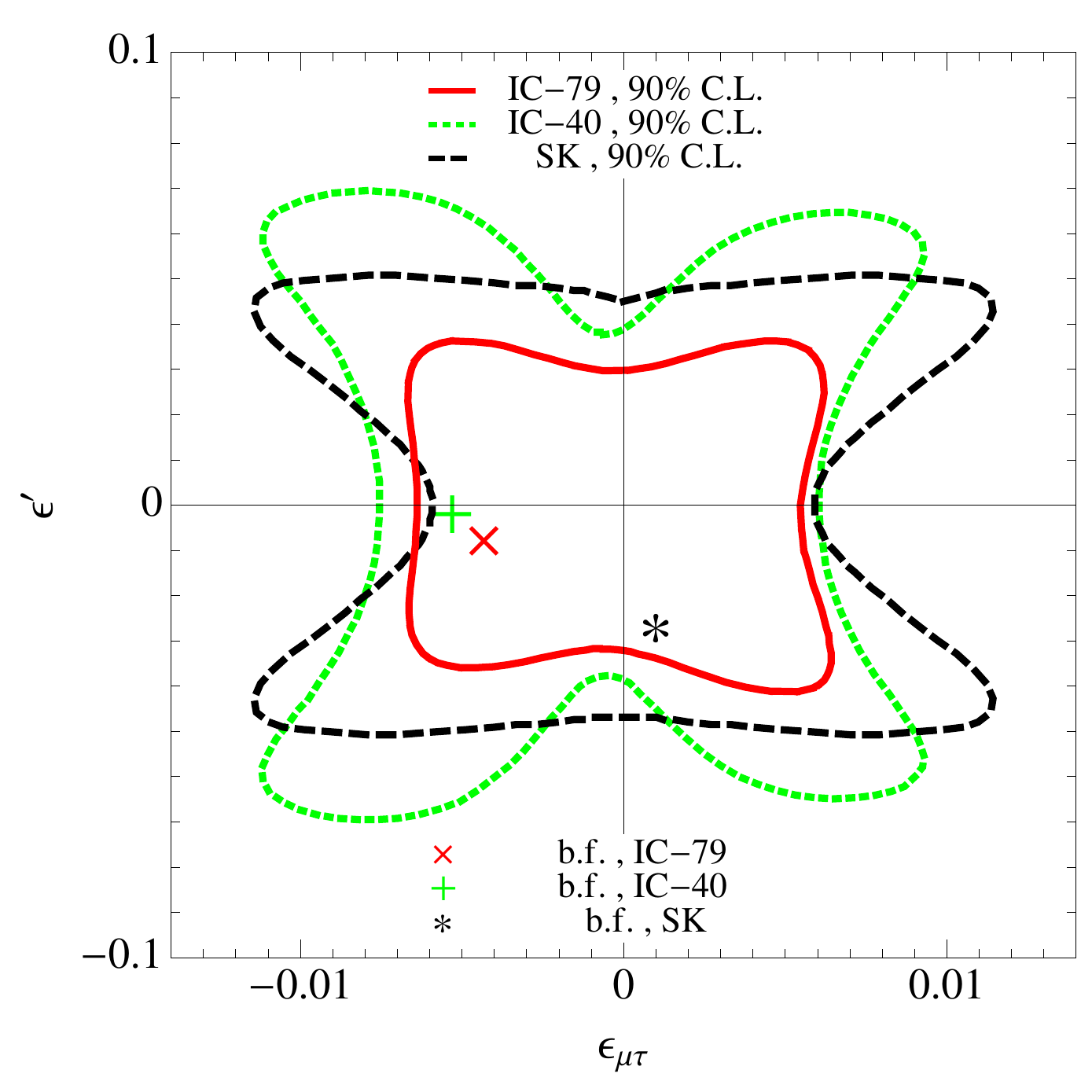}
  \caption[]{Allowed region in the $\epsilon_{\mu \tau}, \epsilon'$ plane  at 90\% C.L. obtained from the combined analysis of low and high energy samples of data
  (IceCube-79 and DeepCore respectively), shown by red solid curve. The black dashed curve shows the allowed region from Super-Kamiokande experiment, taken
from~\cite{Mitsuka:2011ty}, and the green dotted curve is for IceCube-40. The red cross, green + and black star signs show the best-fit values of NSI parameters from IceCube-79,
IceCube-40 and Super-Kamiokande experiments, respectively. Figure from~\cite{Esmaili:2013fva}.}
\label{fig:NSIresult02}
\end{figure}
%%%%%%%%%%%%%%%%%%

The effect of NSI in atmospheric neutrino oscillations with IceCube data were studied in Refs.~\cite{Aartsen:2017xtt} and~\cite{Esmaili:2013fva}. The IceCube analysis of Ref.~\cite{Aartsen:2017xtt} is based on three years of data collected by the low-energy extension DeepCore, that was also used in the standard neutrino oscillation analysis discussed in Section~\ref{sec:std_oscillations}. The data was binned into a two-dimensional histogram in terms of reconstructed neutrino energy and zenith angle, $\cos\theta_{\rm zen}$, and analysed via a profile likelihood method. The analysis focused on NSI interactions with $d$-quarks assuming $\epsilon^{(d)}_{\mu\tau} = |\epsilon^{(d)}_{\mu\tau}|$ and $\epsilon^{(d)}_{\mu\mu}=\epsilon^{(d)}_{\tau\tau}$. The constraints for $\epsilon_{\mu\tau}$ are shown in Fig.~\ref{fig:NSIresult01}. 

A two-dimensional analysis on the parameters $\epsilon_{\mu \tau}$ and $\epsilon'$ was carried out by the authors of Ref.~\cite{Esmaili:2013fva}. The analysis was performed on the atmospheric neutrino measurement done with one year of IceCube-40 data~\cite{Abbasi:2010ie} and with two years of public DeepCore data, which was initially used for the first oscillation analysis by IceCube~\cite{Aartsen:2013jza}. A combined confidence level region of allowed values for $\epsilon_{\mu \tau}$ and $\epsilon'$ is obtained from this analysis, see Fig.~\ref{fig:NSIresult02}. Although the analysis is only based on the arrival direction distribution of muon-neutrinos and not their energy, it is competitive with previous results from Super-Kamiokande, and even produces more restrictive limits. 

\subsection{Neutrino Decoherence}

The Hamiltonian evolution in Eq.~(\ref{liouville}) is a characteristic of physical systems isolated from their surroundings. The time evolution of such a quantum system is given by the continuous group of unitarity transformations, $U_t = {\rm e}^{-i Ht},$ where $t$ is the time. The hermiticity of the Hamiltonian  guarantees the reversibility of the processes, $U^{-1}=U^\dagger$. For open quantum systems, the introduction of dissipative effects lead to modifications of Eq.~(\ref{liouville}) that account for the irreversible nature of the evolution. The transformations responsible for the time evolution of these systems are defined by the operators of the Lindblad quantum dynamical semi-groups~\cite{Lindblad:1975ef}. Since this does not admit an inverse, such a family of transformations has the property of acting only forward in time. The monotonic increase of the von Neumann entropy, $S(\rho) = - {\rm   Tr}\,\, (\rho\, \ln \rho)$, implies the hermiticity of the Lindblad operators, $L_j = L_j^\dagger$~\cite{Benatti:1987dz}. In addition, the conservation of the average value of the energy can be enforced by taking $[H, L_j] = 0$~\cite{Banks:1983by}.

In the three-flavour basis it is convenient to expand the density operator $\rho$, the Hamiltonian $H$ and the Lindblad operators $L_i$ in terms of Hermitian $3\times3$ matrices $F_\mu$ ($\mu=0,\ldots,8$), given by, $F_0 = \mathbf{1}_{3\times3}/\sqrt{6}$ and $F_i =  \lambda_i/2$, where $\lambda_i$ ($i=1,\ldots,8$) are the Gell-Mann matrices~\cite{Gago:2002na}. This set of basis matrices satisfy the orthogonality condition ${\rm   Tr}(F^\dagger_\mu F_\nu) = \delta_{\mu\nu}/2$.
The Liouville equation~(\ref{liouville_mod}) can now be expressed in terms of the expansion coefficients $\rho_i$ and $h_i$ of the density operator and Hamiltonian, respectively, as the matrix equation
\begin{equation}\label{QDdiff}
  \dot\rho_k = \sum_{i,j}h_i\rho_jf_{ijk} - \sum_i D_{ki}\rho_i\,.
\end{equation}
The coefficients $f_{ijk}$ are the $SU(3)$ structure constants defined by $[F_i,F_j] = i\sum_kf_{ijk}F_k$~\cite{Gago:2002na}. The decoherence effects are cast into a matrix 
\begin{equation}\label{defDij}
  D_{ij} = \frac{1}{2}\sum_{k,l,m,n}l^{(n)}_mf_{iml}l^{(n)}_kf_{jkl}\,,
\end{equation}
where $l^{(n)}_i$ are the expansion coefficient of the Lindblad operator $L_n$.

In atmospheric neutrino data we are interested in the survival probability of muon neutrinos (and anti-neutrinos). Typically, it is assumed that the $8\times8$ matrix $D_{ij}$ takes on a diagonal form, {\it i.e.},~$D_{ij} \simeq \delta_{ij}D_{i}$, with positive diagonal elements.
The three-level system including decoherence and oscillations can be further simplified assuming universal decoherence parameters $D_i= D$. The muon neutrino survival probability can then be expressed as~\cite{Gago:2002na},
\begin{multline}
P_{\nu_\mu \to \nu_\mu} = \frac{1}{3}+\frac{2}{3}e^{-DL}-4e^{-DL}\sum_{i<j}|U_{\mu i}U_{\mu j}|^2\sin^2|\Delta_{ij}|\,.
\end{multline}
Note that this reduces to the standard survival probability in the case $D\to0$ and leads to a flavour ratio $1:1:1$ in the case of $DL\to\infty$, independent of initial condition. Table~\ref{tab:vli_qd_limits} shows the limits on the universal decoherence parameter $D/E_\nu^n$ [GeV${}^{n-1}$] for various energy dependencies ($n=1,2,3$) derived from AMANDA-II data of the years 2000 to 2006~\cite{Abbasi:2009nfa}. 

On the other hand, keeping decoherence parameters $D_i$ independent, we can derive the oscillation-averaged flavour survival/appearance probability in the form~\cite{Gago:2002na}
\begin{multline}
P^{\rm avg}_{\nu_\alpha \to \nu_\beta} = \frac{1}{3}+\frac{1}{2}e^{-D_3L}(|U_{\alpha1}|^2-|U_{\alpha2}|^2)(|U_{\beta1}|^2-|U_{\beta2}|^2)\\ +\frac{1}{6}e^{-D_8L}(|U_{\alpha1}|^2+|U_{\alpha2}|^2 -2|U_{\alpha3}|^2)\\\cdot(|U_{\beta1}|^2+|U_{\beta2}|^2 -2|U_{\beta3}|^2)\,.
\end{multline}
Only the parameters $D_3$ and $D_8$ contribute in this case. Again, for $D_3L\to\infty$ and $D_8L\to\infty$ we recover a flavour ratio $1:1:1$, as expected for full decoherence.

\subsection{Neutrino Decay}

The flavour ratio of neutrinos from astrophysical sources can be significantly altered if neutrinos can decay {\it en route} to Earth~\cite{Beacom:2002vi,Shoemaker:2015qul,Bustamante:2016ciw,Rasmussen:2017ert,Denton:2018aml}. Two-body decays of the form $\nu_i\to\nu_j + X$ with a massless light scalar $X$ ({\it e.g.}, a {\it   Majoron}~\cite{Choi:1991aa,Acker:1992eh,Acker:1991ej}) are only weakly limited by solar neutrino data~\cite{Beacom:2002cb} with $\tau_i \gtrsim 10^{-4} {\rm s}({m_i}/{{\rm eV}})$. In the following, we will assume that the decay of neutrino mass eigenstates can be written as a dissipation term in Eq.~(\ref{liouville_mod}) of the form
\begin{multline}\label{eq:Ddec}
\mathcal{D}_{\rm dec}[\rho] = -\frac{1}{2}\sum_i\Gamma_i(E)\big\lbrace\Pi_i,\rho(E)\big\rbrace \\+  \sum_{ij}\int{\rm d}E'\gamma_{i\to j}(E',E)\Pi_{j}{\rm Tr}(\rho(E')\Pi'_{i})\,.
\end{multline}
The first term describes the disappearance of a neutrino mass eigenstate $|\nu_i\rangle$ with total decay width $\Gamma_i(E_i)$. The second term describes the appearance of neutrino mass eigenstates from the transition $|\nu_i\rangle$ and $|\nu_j\rangle$ with differential production rate $\gamma_{i\to j}(E_i,E_j)$. Note, that the corresponding set of Lindblad operators representing the expression in Eq.~(\ref{eq:Ddec}) can be written as
\begin{equation}
L_{ij} = \sqrt{\gamma_{i\to j}(E_i,E_j)}|\nu_j(E_j)\rangle\langle\nu_i(E_i)|\,,
\end{equation}
where the index $i$ runs over active neutrino states and the index $j$ over active and sterile states. In contrast to quantum decoherence discussed in the previous section, this set of Lindblad operators are non-Hermitian and do not commute with the Hamiltonian.

As a further approximation, we will consider neutrino propagation over cosmological distances, {\it i.e.}, all oscillation terms from neutrino mass differences are averaged. As a representative example, we consider transitions that leave the neutrino energy constant, {\it i.e.},~$\gamma_{i\to j}\simeq {\rm Br}_{i\to j}\Gamma_{i}\delta(E_i-E_j)$, and normal ordering of neutrino masses with transitions $3\to2$, $3\to1$, and $2\to1$.
In the mass eigenbasis the evolution of the density matrix has then the form
\begin{equation}
\dot\rho_{ij} = \frac{i\Delta m^2_{ij}}{2E_\nu}\rho_{ij} - \frac{1}{2}(\Gamma_{i}+\Gamma_{j})\rho_{ij} + \delta_{ij}\sum_k \Gamma_k{\rm Br}_{k\to i}\rho_{kk}\,.
\end{equation}
In the oscillation-averaged solution only the diagonal elements $\rho_{ii}$ have non-zero contributions. We can then express the neutrino survival/appearance probability as 
\begin{equation}
P_{\alpha\to\beta} = \sum_{i,j}M_{ji}\left|U_{\beta j }U_{\alpha i}\right|^2\,,
\end{equation}
where the elements of transition matrix, $M_{ji}$, can be written in the form~\cite{Ahlers:2010ty}
\begin{equation}
M_{ji} = \sum_{\bf c} M^{(\bf{c})}_{ji}\,.
\end{equation}
The sum in the previous equation runs over all production chains $c_1\to \ldots \to c_{n_c}$ with $c_1=i$ and $c_{n_c}=j$ and
\begin{equation}
M^{({\bf c})}_{ji} = \left(\prod_{l=1}^{n_c-1} \Gamma_{c_{l+1}c_{l}}\right)\sum_{k=1}^{n_c}e^{-L\Gamma_{c_k}}\prod_{p=1(\neq k)}^{n_c}\frac{1}{\Gamma_{c_p}-\Gamma_{c_k}}\,.
\end{equation}
In this particular scenario, the final neutrino flavour ratio for $L\Gamma_i\gg1$ will thus correspond to the composition of the lowest mass eigenstate. 

Note, that the decay rates $\Gamma_i$ are expected to decrease with energy due to relativistic boosting of the mass eigenstate's lifetime. Therefore, the neutrino flavour composition can experience strong energy dependencies. On the other hand, active neutrino decay into sterile neutrinos can introduce spectral cutoffs due to the energy dependence of the neutrino lifetime. This process is limited by the observation of IceCube's TeV-PeV neutrino flux and could be responsible for a tentative cutoff~\cite{Bustamante:2016ciw}. Neutrino decay has also been considered as a possibility to alleviate a mild tension in the best-fit power-law spectra between cascade- and track-dominated IceCube data~\cite{Denton:2018aml}.

Astrophysical neutrinos propagating over cosmic distances are also susceptible to feeble interactions with cosmic backgrounds. In particular, feeble interactions with the cosmic neutrino background (C$\nu$B) that can be enhanced by resonant interactions, {\it e.g.}, $\overline{\nu}_\alpha+\nu_\alpha\to Z'\to\overline{\nu}_\beta+\nu_\beta$ have been discussed as a source for absorption features~\cite{Blum:2014ewa,Ibe:2014pja,Ng:2014pca,Ioka:2014kca}. The evolution of the neutrino density matrix is identical to that for neutrino decay with dissipation term as in Eq.~(\ref{eq:Ddec}). However, in this case the interaction rates have a non-trivial dependence on redshift via the density evolution of the C$\nu$B.

\subsection{Sterile Neutrinos}

It is conceivable that the three neutrino flavour states, $\nu_e$, $\nu_\mu$, and $\nu_\tau$, are augmented by additional states. The number of light ``active'' neutrinos, {\it i.e.}, neutrinos that take part in the Standard Model weak interactions is limited to the known flavour states from the observed decay width of the $Z$-boson. However, it is feasible that there are light ``sterile'' neutrino states with no Standard Model interactions. These sterile states would imply the existence of additional neutrino mass eigenstates that could impact the three-flavour neutrino oscillation phenomenology, due to an extended PMNS mixing matrix $\widetilde{U}$,
\begin{equation}\label{eq:sterilenu}
  |\nu_\alpha\rangle =
  \sum_{j=1}^{3+n} \widetilde{U}_{\alpha j}^* |\nu_j\rangle\,.
\end{equation}
The term ``sterile'' referring to neutrinos was introduced by {\it Pontecorvo} already in 1968 when discussing the, at that time hypothetical, possibility of vacuum neutrino oscillations~\cite{Pontecorvo:1967fh}. 
   
Many extensions of the Standard Model that relate to the appearance of neutrino mass terms predict the existence of sterile neutrinos. As discussed earlier, the right-handed neutrino field, that can provide a Dirac mass term $m_{\rm D}\overline{\nu}_L\nu_R+h.c.$, does not interact via weak interactions and is therefore sterile. However, in the minimal type I seesaw models (see, {\it e.g.}, Ref.~\cite{Mohapatra:2006gs}) these sterile neutrinos have a large Majorana mass term, $M\overline{\nu}_{R}\mathcal{C}\overline{\nu}_{R}^T/2 + h.c.$, with $m_{\rm D}\ll M$, that give rise to a large effective neutrino mass after diagonalising the mass matrix. These massive sterile states are practically unobservable in low-energy oscillation experiments. On the other hand, for values of $M\ll m_{\rm D}$ (``pseudo-Dirac'' case), the active and sterile state mass states are degenerate after diagonalisation, leading to maximal mixing between the left (active) and right (sterile) states. 

%%%%%%%%%%%%%%%
\begin{figure}[t]
\centering\includegraphics[width=\linewidth]{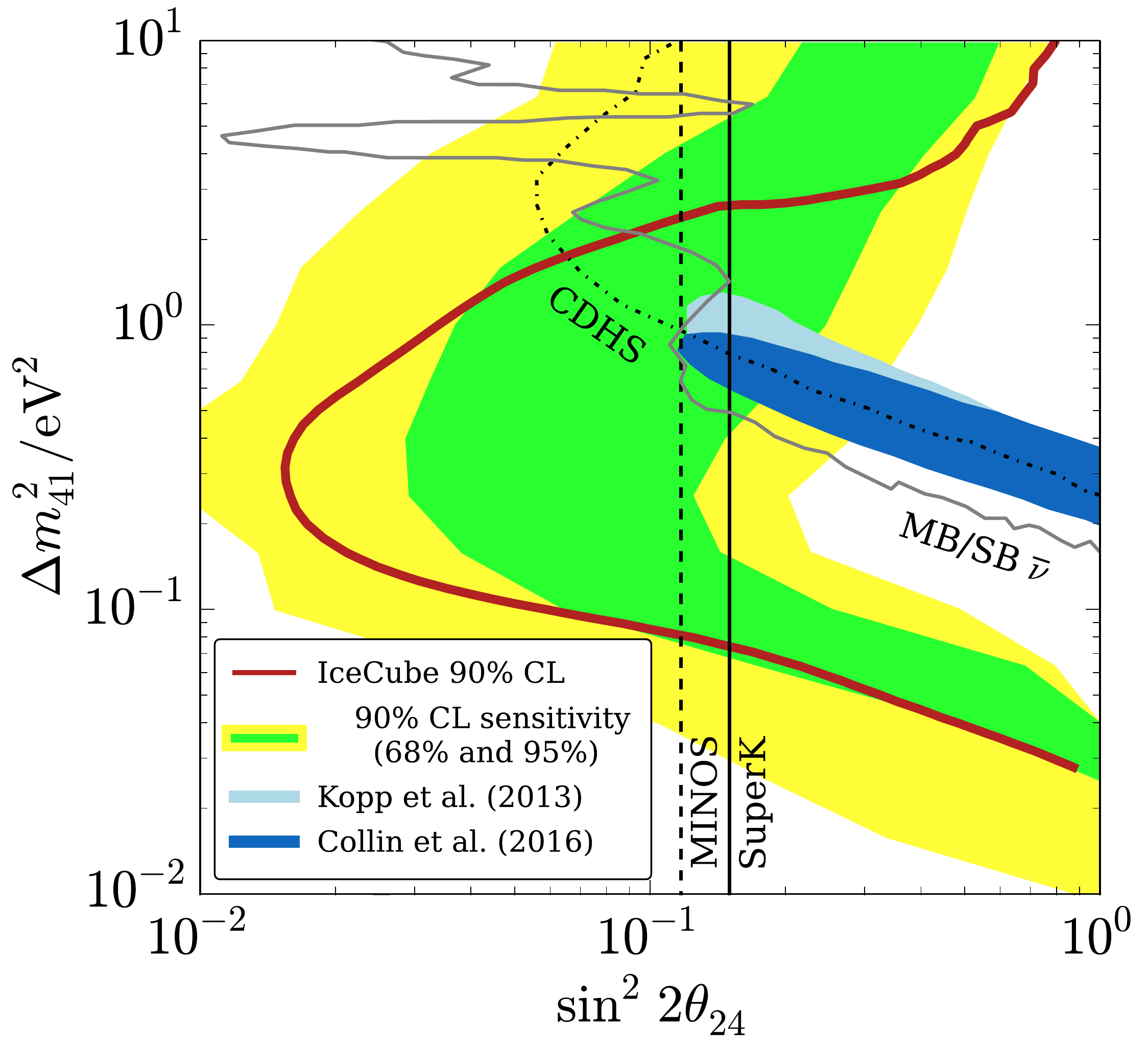}
\caption[]{The 90\% C.L. exclusion contour on the ($\sin^2 2\theta_{24}$, $\Delta m^2_{41})$ plane (orange solid line).
The green and yellow bands contain 68\% and 95\%, respectively, of the 90\% contours in simulated
pseudo-experiments. The contours and bands are overlaid
on 90\% C.L. exclusions from the Super-Kamiokande, MINOS, MiniBooNE and CDHS experiments~\cite{Abe:2014gda,Adamson:2011ku,Cheng:2012yy,Dydak:1983zq}, and
the 99\% C.L. allowed region from global fits to appearance experiments including MiniBooNE and LSND, assuming
$|U_{e4}|^2$=0.0023~\cite{Kopp:2013vaa} and $|U_{e4}|^2$=0.0027~\cite{Conrad:2012qt}, respectively. Figure reprinted with permission from~\cite{Aartsen:2016oqi} (Copyright 2016 APS).}
\label{fig:HE_sterile}
\end{figure}
%%%%%%%%%%%%%%

The minimal sterile neutrino model is a ``3+1'' model where, in addition to the three standard weakly-in\-ter\-ac\-ting neutrino flavours, one additional heavier sterile neutrino state is added. Such a simple extension of the neutrino sector has been advocated to explain certain tensions between experimental results from accelerator~\cite{Athanassopoulos:1996jb,Aguilar:2001ty,Aguilar-Arevalo:2013pmq}, reactor~\cite{Mention:2011rk} and radio-chemical~\cite{Bahcall:1994bq} experiments, and the predictions from the standard three active flavour scenario. In the most general case, the introduction of one sterile neutrino adds six new parameters to the neutrino oscillation phenomenology~\cite{Kopp:2013vaa}: three mixing angles $\theta_{\rm 14}$, $\theta_{\rm 24}$, $\theta_{\rm 34}$, two CP-violating phases, $\delta_{\rm 14}$ and $\delta_{\rm 34}$, and one mass difference, $\Delta m^2_{\rm 41}$, where the indexes '1-3' stand for the known neutrino mass states and '4' for the sterile state. 
  
Although sterile neutrinos can not be detected directly, their existence can leave an imprint on the oscillation pattern of active neutrinos. The sterile neutrino modifies the oscillation pattern of the standard neutrinos since these can now undergo vacuum oscillations into the new state, with a probability that is proportional to the new mixing angles. The period of these oscillations can be small, smaller than the directional resolution of IceCube, and the net effect is then to distort the overall $\nu_{\mu}$ flux normalisation with respect to the three-flavour case. An additional effect arises from the different interactions of flavours with matter when traversing the Earth~\cite{Razzaque:2012tp,Esmaili:2013vza}. The new possibility to oscillate to a state that does not interact results in energy and angular dependent oscillation amplitudes that depend on the mixing angles, but also on the new  $\Delta m^2_{\rm 41}$. More precisely, the comparison between the oscillation pattern in the ${\rm (energy, zenith)}$ phase space predicted by the sterile ``3+1'' model and the pattern seen in experimental data is what IceCube exploits to set limits on the sterile mixing parameters. As for the standard oscillation case described in section~\ref{sec:std_oscillations}, searches for sterile neutrinos in IceCube are based on the detection of the disappearance of muon neutrinos and are more sensitive to $\theta_{\rm 24}$. Therefore the choice of a simplified minimal mixing scenario where only $\theta_{\rm 24}$ is not zero is justified, and is indeed the approach followed in the analyses described below. Nonzero values of the weakly constrained $\theta_{\rm 34}$ within current limits would not significantly affect the results presented here~\cite{Lindner2016}. 

There have been several searches in the past at sub-TeV energies with neutrinos from the Sun, reactors, and accelerator setups (see, {\it e.g.}, Refs.~\cite{Kopp:2013vaa,Gonzalez-Garcia:2015qrr,Collin:2016rao,Capozzi:2016vac} and references therein). The large flux of atmospheric neutrinos that reach IceCube and the wide range of energy response of the detector allow to search for signatures of anomalous $\nu_{\mu}+{\overline{\nu}_{\mu}}$ disappearance caused by oscillations to an sterile neutrino, $\nu_{\rm s}$, at TeV energies, an energy not probed before. Furthermore, the DeepCore detector can extend the search  down to about 6~GeV, improving previous limits from smaller detectors.

%%%%%%%%%%%%%
\begin{figure}[t]
\centering\includegraphics[width=\linewidth]{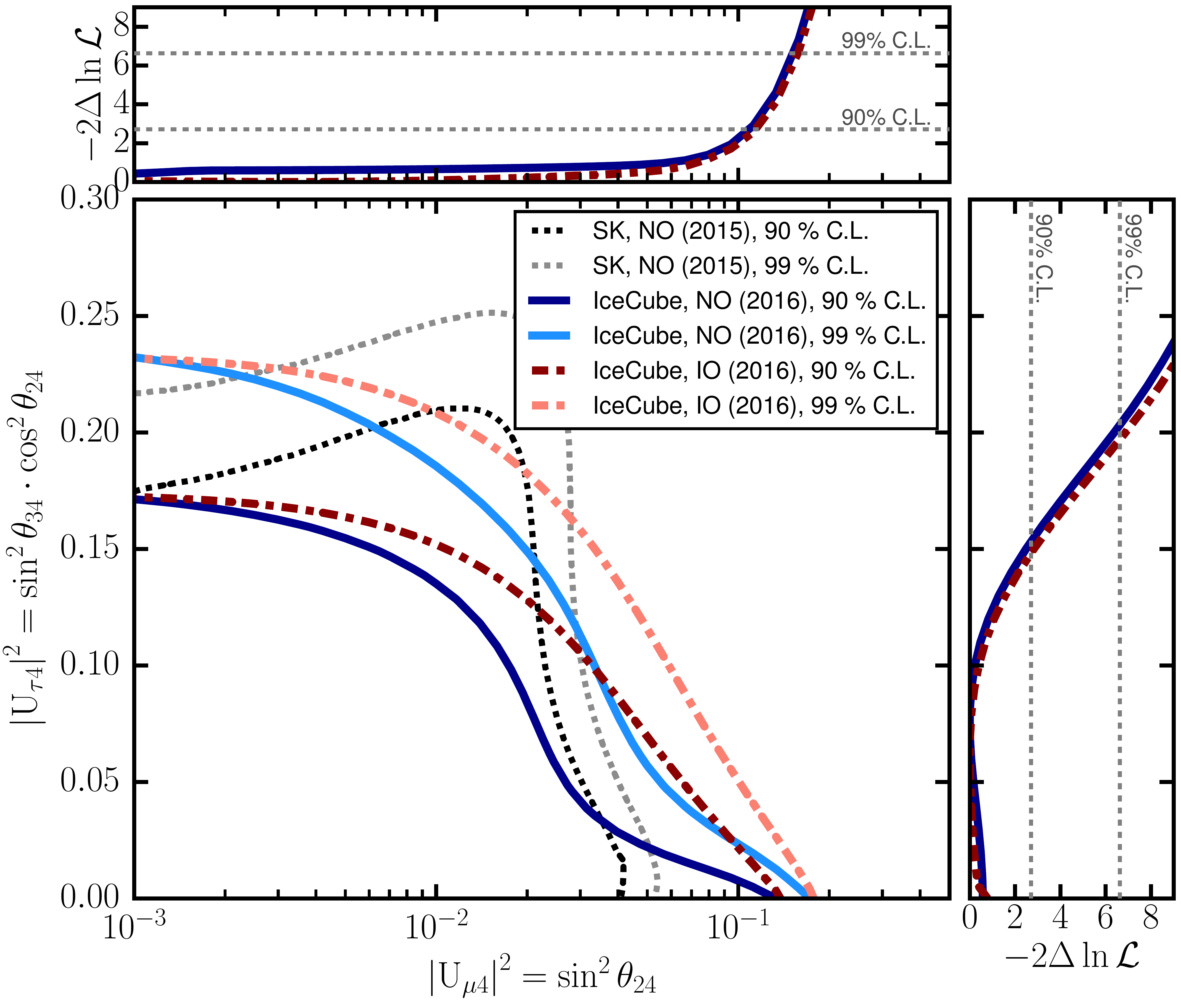}
\caption[]{The exclusion contours on the (sin$^2 \theta_{24}$, $|U_{\tau 4}|^2)$ plane at 90\% and 99\% C.L. (dark and light blue solid lines, respectively). 
The dash lines show results from the Super-Kamiokande~\cite{Abe:2014gda}. Figure reprinted with permission from~\cite{Aartsen:2017bap} (Copyright 2017 APS).}
\label{fig:LE_sterile}
\end{figure}
%%%%%%%%%%%%

The analysis techniques are very similar to the search for standard oscillations and are based on comparing the equivalent oscillogram from Fig.~\ref{fig:oscillogram} for oscillations with an sterile component, to the measured data. Although the low-energy tracks provide a quite short lever arm to reconstruct their direction  with precision, and the angular resolution of the analysis varies between 6$\text{\textdegree}$ and 12$\text{\textdegree}$, depending on energy. This is enough, though, to perform the analysis (see Fig.~5 in Ref.~\cite{Aartsen:2017bap}) since the differences in the oscillation pattern in the presence of a sterile neutrino can still be distinguished from the non-sterile case with such angular resolution. At higher energies the analyses necessarily follow a slightly different approach since the tracks originate outside the detector and the energy deposited in the hadronic shower at the interaction vertex is not accessible. Such analyses use measured muon energies instead of neutrino energies, like the one presented in Ref.~\cite{Aartsen:2016oqi}. The muon energy is reconstructed from the light emission profile from stochastic energy losses of the muon along its trajectory~\cite{Aartsen:2013vja}, achieving an energy resolution of $\sigma_{\rm log_{10}}(E_{\mu}/{\rm GeV})\sim 0.5$. Since at the energies of this analysis the muon track can be well reconstructed, the angular resolution reaches values between 0.2$\text{\textdegree}$ and 0.8$\text{\textdegree}$, depending on incoming angle. 

As with searches for standard oscillations, the aim is to reach a pure sample of atmospheric neutrinos to be able to compare the measured number of events as a function of angle and energy with the model prediction. None of the analyses detected a deviation from the expected standard oscillation scenario and, therefore, limits on the oscillation parameters which depend on the additional neutrino can be set. The low-energy analysis uses a 8$\times$8 binned grid in the (energy, $\cos\theta_{\rm zen}$) plane while, due to its higher energy resolution, the high-energy analysis uses a 10$\times$20 binned grid. A log-likelihood approach is used to find the best fit to the data given the model parameters. Confidence levels are calculated from the difference between the profile log-likelihood and the log-likelihood at the best fit point. The results are illustrated in Figs.~\ref{fig:HE_sterile} and~\ref{fig:LE_sterile}. Figure~\ref{fig:HE_sterile} shows the result of the high-energy analysis expressed in the $(\Delta m_{\rm 14}^2, \sin^2 2\theta_{\rm 24})$ plane. The figure shows the  90\%  confidence level contour (red line) compared with  90\% exclusions from previous disappearance searches. The exclusion is compatible with the sensitivity (green and yellow areas) calculated under the assumption of the no-sterile neutrino hypothesis. The result therefore disfavours much of the parameter space of the ``3+1'' model. On the other hand, Fig.~\ref{fig:LE_sterile} shows the results of an analysis using low-energy events (6~GeV to 60~GeV) contained in DeepCore.  The results are shown in the $(|U_{\tau 4}|^2, |U_{\mu 4}|^2)$ plane, where $|U_{\tau 4}|$ is defined as $\sin^2\theta_{34}\,\cos^2\theta_{34}$ and $|U_{\mu 4}|$ is just $\sin^2 \theta_{24}$. The best-fit value of these elements of the mixing matrix are $|U_{\mu 4}|^2=0.0$ and $|U_{\tau 4}|^2=0.08$, while the 90\% confidence level limits on their values are still compatible with 0, $|U_{\mu 4}|^2 \leq 0.11$ and $|U_{\tau 4}|^2 \leq 0.15$.  Combinations of the parameters shown in the axes above the contours are disfavoured. The figure includes a recent exclusion contour from Super-Kamiokande~\cite{Abe:2014gda} as a comparison.

As in the previous cases, the presence of sterile neutrinos can also have an effect on neutrino spectra from astrophysical sources. In the case of a pseudo-Dirac scenario, where the mass splitting between active and sterile neutrinos is very small, there exist the possibility that the corresponding neutrino oscillation effects are still visible, {\it i.e.},~$L\Delta m^2/E\simeq 1$~\cite{Esmaili:2012ac,Rasmussen:2017ert}.

\section{Indirect Dark Matter Detection}
\label{sec:dm}
There is a large corpus of evidence that supports the existence of a non-baryonic, non-luminous component of matter in the cosmos. A way to understand the rotation curves of galaxies, the peculiar velocities of galaxies in clusters, and the formation of first galaxies growing out of small density perturbations imprinted in the cosmic microwave background, is to introduce a ``dark matter'' component in the energy budget of the Universe~\cite{Bertone:2004pz,Lukovic:2014vma}. Attractive candidates for dark matter consists of stable relic particles whose present density is determined by the thermal history of the early universe~\cite{Jungman:1995df,Feng:2000zu,Feng:2010gw,Bergstrom:2012fi}. The present abundance of dark matter can be naturally explained by physics beyond the Standard Model providing stable, weakly-interacting massive particle (WIMP) in the few GeV-TeV mass range. For thermally produced WIMPs, the upper mass limit arises from theoretical arguments in order to preserve unitarity~\cite{Griest:1989wd}, although higher masses can be accommodated in models where the dark matter candidates are not produced thermally~\cite{Chung:1998ua}. 

There is a vast ongoing experimental effort to try to identify the nature of dark matter through different strategies: production at colliders~\cite{Kahlhoefer:2017dnp} or through the detection of nuclear recoils in a 
selected target in ``direct detection'' experiments~\cite{Undagoitia:2015gya}. A complementary, ``indirect'', approach is based on searching for the products of the annihilation of dark matter particles gravitationally trapped in the halo of galaxies or accumulated in heavy celestial objects like the Sun or Earth~\cite{Press:1985ug,Krauss:1985aaa,Srednicki:1986vj,Gaisser:1986ha,Ritz:1987mh,Gould:1987ww,Gould:1987ir,Bergstrom:1998xh}. In this latter case, neutrinos are the only possible messengers, since other particles produced in the annihilations will be absorbed. These search techniques are competitive since they can set limits on the same physical quantities (the dark matter-nucleon cross section for example). But they are also complementary since they are subject to different backgrounds (the gamma-ray sky is very different from the proton or neutrino sky), different astrophysical inputs (dark matter density and velocity distribution) and different systematics (nucleon and nuclear form factors of different targets).

The strength of the expected neutrino flux emitted from a celestial object depends, among other factors, on the capture rate of WIMPs, which is proportional to the WIMP-nucleon cross section, and the annihilation rate,  which is proportional to the velocity-averaged WIMP-WIMP annihilation cross section, $\langle\sigma_{\rm A} v\rangle$. The evolution of the WIMP number density $n_{\rm DM}$ in compact celestial objects follows the balance equation
\begin{equation}
\dot{n}_{\rm DM} = Q_{\rm C} - 2\langle\sigma_{\rm A} v\rangle(n_{\rm DM}^2/2)\,,
\end{equation}
where $Q_{\rm C}$ is the capture rate per unit volume and the numerical factors account for the fact that annihilations remove two WIMPs per interaction but there are only $1/2$ of the WIMPS available to form pairs. If the capture rate remains constant over a long time, the WIMP density reaches an equilibrium solution
\begin{equation}\label{eq:equilibriumDM}
n_{\rm DM, eq}=\sqrt{\frac{Q_{\rm C}}{\langle\sigma_{\rm A} v\rangle}}\,.
\end{equation}

The WIMP capture rate from interaction with baryonic matter can have spin-dependent, $\sigma_{SD}$, and spin-independent, $\sigma_{SI}$, contributions~\cite{Engel:1992bf}. Since the Sun is primarily a proton target (75\% of H and 24\% of He in mass)~\cite{Grevesse:1998bj} the capture of WIMPs from the halo can be considered to be driven mainly via the spin-dependent scattering. Heavier elements constitute less than 2\% of the mass of the Sun, but can still play a role when considering spin-independent capture, since the spin-independent cross section is proportional to the square of the atomic mass number. Note that these heavy elements can also take part in the spin-dependent capture process if WIMPs present momentum-dependent interactions with normal matter~\cite{Catena:2015uha}. The situation for the Earth is rather different, since the most abundant isotopes of the main components of the Earth inner core, mantle and crust, $^{56}$Fe, $^{28}$Si and $^{16}$O~\cite{Herndon:1980aaa}, are spin 0 nuclei. Searches for dark matter accumulated in the Earth are then more sensitive to the $\sigma_{SI}$ component of the WIMP-nucleon cross section. Another difference with respect to solar searches is that equilibrium between the capture and annihilation rates can not be assumed, and to be able to extract a limit on $\sigma_{SI}$, an assumption on the value of the WIMP self-annihilation cross section must be made. 

Dark matter searches from our own Galaxy, nearby galaxies or galaxy clusters present some distinct features with respect to searches from the Sun or Earth which are advantageous. Firstly, capture is not an issue since the presence of dark matter over-densities has been an essential part in the process of galaxy formation. What can be measured then is the velocity-averaged WIMP self-annihilation cross section, $\langle \sigma_{\mathrm A} v \rangle$. Secondly, the products of the annihilations are not necessarily absorbed at the production site, and other indirect signatures (photons, anti-protons, {\it etc.}) can also be searched for in $\gamma$-ray and cosmic-ray observatories. These multi-wavelength and/or multi-messenger searches can increase the sensitivity of dark matter searches. Neutrinos remain, however, an attractive signature since they do not suffer from uncertainties in their propagation (as charged particles do) and no background or foreground from astrophysical objects is present (as in the case of $\gamma$-rays).  Note, that some of the sources are extended (the Galactic halo for example) and point-source analysis techniques have to be modified. On the other hand we expect that the flux of secondaries from these distant objects is much lower than that predicted from WIMP annihilations in the Sun and, furthermore, there are new systematics effecting the calculations. For example, the assumed shape of the dark matter halo profile effects significantly the interpretation of the results since the annihilation rate depends on the square of the dark matter number density. We will discuss these issues in section~\ref{sec:galaxies}.

\subsection{Neutrinos from WIMP Annihilation and Decay}

The annihilation of dark matter into Standard Model particles can be probed by $\gamma$-ray, cosmic ray, and neutrino emission. We focus in the following on neutrino emission with a spectral production rate per unit volume given by
\begin{equation}
Q^{\rm ann}_{\nu_\alpha}({\bf r}) = \frac{1}{2}\rho^2_{\rm DM}({\bf r})\frac{\langle\sigma_Av\rangle}{m^2_{X}}\frac{{\rm d}N_{\nu_\alpha}}{{\rm d}E_\nu}\,,
\end{equation}
where $\rho_{\rm DM} = m_{X}n_{\rm DM}$ is the WIMP mass density at a given position ${\bf r}$. The energy distribution ${\rm d}N_{\nu_\alpha}/{\rm d}E_\nu$ of neutrinos is normalised to the total number of neutrinos of flavour $\nu_\alpha$ expected from the annihilation. The factor $1/2$ compensates for the symmetry of WIMP combinations. The differential flux observed from the solid angle $\Omega$ can then be expressed as the line-of-sight integral from Earth's location\footnote{Note that in a galactocentric coordinate system considered for WIMP annihilations in the Galactic halo, the Earth's position is ${\bf r}_{\oplus} \simeq {\bf r}_{\odot} \simeq (0,-8.5{\rm kpc},0)$.
} ${\bf r}_\oplus$,
\begin{equation}\label{eq:DMann}
F_{\nu_\beta}(E_\nu) = \sum_{\alpha}P_{\alpha\to\beta}\left[ \int_0^\infty {\rm d}l\rho_{\rm DM}^2({\bf r}(l,\Omega))\right]\frac{\langle\sigma_A v\rangle}{8\pi m_X^2}\frac{{\rm d} N_{\nu_\alpha}}{{\rm d} E_\nu}\,,
\end{equation}
where ${\bf r}(l,\Omega) \equiv {\bf r}_\oplus+l\widehat{{\bf n}}(\Omega)$ and $\widehat{{\bf n}}(\Omega)$ is a unit vector in the direction $\Omega$. The factor $P_{\alpha\to\beta}(E_\nu)$ accounts for flavour oscillations\footnote{Strictly speaking, the flavour transition probabilities depend on the distance $l$ and should appear under the line-of-sight integral. However, in most situations discussed in the following it can be treated as a constant.}. The factor in parenthesis $[\cdot]$ encodes the cosmological and astrophysical dependence on the emission from the dark matter density distribution. This is sometimes called $J$-factor~\cite{Bergstrom:1997fj}.

The decay of dark matter can be treated analogously. Here, the spectral production rate is given by
\begin{equation}
Q^{\rm dec}_{\nu_\alpha}({\bf r}) = \rho_{\rm DM}({\bf r})\frac{\Gamma_X}{m_{X}}\frac{{\rm d}N_{\nu_\alpha}}{{\rm d}E_\nu}\,,
\end{equation}
where $\Gamma_X$ denotes the dark matter lifetime. As before, the distribution ${\rm d}N_{\nu_\alpha}/{\rm d}E_\nu$ of neutrinos is normalized to the total number of neutrinos of flavour $\nu_\alpha$ and it depends implicitly on the branching fraction of the decay. The differential flux observed from the solid angle $\Omega$ can then be expressed as the line-of-sight integral
\begin{equation}\label{eq:DMdec}
F_{\nu_\beta}(E_\nu) = \sum_{\alpha}P_{\alpha\to\beta}\left[ \int_0^\infty {\rm d}l{\rho_{\rm DM}({\bf r}(l,\Omega))}\right]\frac{\Gamma_X}{4\pi m_X}\frac{{\rm d} N_{\nu_\alpha}}{{\rm d} E_\nu}\,.
\end{equation}
Again, the factor in parenthesis $[\cdot]$ encodes the cosmological and astrophysical dependence and is sometimes called $D$-factor.

IceCube analyses are performed in the most model-independent way possible, but the allowed parameter space of the underlying theoretical models force some assumptions to be made. Since the exact branching ratios of WIMP annihilation into different channels is model-dependent, two annihilation channels which give extreme neutrino emission spectra ${\rm d}N/{\rm d}E$ are chosen, and the analysis is optimised for each of them separately. Annihilation into $b\overline{b}$ is taken as a representative case for a soft neutrino energy spectrum. This is in part due to the $b$-quark interaction with the medium (for dense objects like the Sun) and the hadronisation of the final state quarks leading to neutrinos from meson decays. In contrast, the annihilation into $W^+W^-$ or $\tau^+\tau^-$ follows a hard spectrum. Assuming a 100\% branching ratio to each of these channels brackets the expected neutrino spectrum from any realistic model with branching to more channels. The WIMP mass is left free, and independent searches are performed for a few benchmark masses. The total number of signal events, $\mu_{\rm s}$, expected at the detector from a given model is then given by,
\begin{equation}
 \mu_{\rm s} =  T_{\rm live}\sum_\alpha\int{\rm d}\Omega\int {\rm d} E_\nu A^{\rm eff}_{\nu_\alpha}(\Omega,E_\nu) F_{\nu_\alpha}(E_\nu)\,,
\label{eq:dm_nevents}
\end{equation}
where $T_{\rm live}$ is the exposure time, $A^{\rm eff}_{\nu_\alpha}(\Omega,E_\nu)$ the detector effective area for neutrino flavour $\alpha$, that depends on the detector's response with respect to observation angle and neutrino energy. 

In the absence of a signal, the 90\% confidence level limit on the number of signal events, $\mu_{\rm s}^{\rm 90}$, can then be directly translated into a limit on either the velocity-averaged annihilation cross section $\langle\sigma_Av\rangle$ or the dark matter lifetime $\Gamma_X$. The interplay between the total number of observed events in a given data sample, $n_{\rm obs}$, and the estimated number of background events, $n_{\rm bg}$, is the basis to perform a simple event-counting statistical analysis to constrain $\mu_{\rm s}$, {\it i.e.}, to constrain a given model. This was the approach followed in early IceCube publications, {\it e.g.}, Refs.~\cite{Ackermann:2005fr,Abbasi:2011eq}. 

In order to improve the power of the statistical test, a distribution-shape analysis can be used instead. The angular distribution of the expected signal events is expected to peak towards the source, and is very different from the flatter atmospheric neutrino background. If $\Omega$ is the solid angle of a reconstructed track position, we can weight the event by signal and background probability distributions $S(\Omega)$ and $B(\Omega)$, respectively. The likelihood that the data sample contains $\mu_{\rm s}$ signal events out of $n_\mathrm{obs}$ observed events is then defined as
\begin{equation}
\mathcal{L}(\mu_{\rm s}) = \prod_{i=1}^{n_\mathrm{obs}} \left(\frac{\mu_s}{n_\mathrm{obs}}S(\Omega_{i}) + \left(1-\frac{\mu_s}{n_\mathrm{obs}}\right)B(\Omega_{i})\right)\,.
\label{eq:LikelihoodDefinition}
\end{equation}
From this likelihood, one can determine confidence intervals for $\mu_{\rm s}$ by using the likelihood-ratio test statistic as proposed in~\cite{Feldman:1997qc}. This method produces stronger limits than just event counting, and it is less prone to unsimulated background contamination in the final data sample.

There are systematic uncertainties in the translation of the number of detected events into capture cross section values due to uncertainties in the element composition of the Sun~\cite{Basu:2007fp}, the effect of planets on the capture of WIMPS from the halo~\cite{Sivertsson:2012qj} and the uncertainty on the values of the nuclear form factors needed in the rather complex capture calculations~\cite{Engel:1992bf,Bottino:1999ei,Ellis:2008hf}. These effects can be of relevance when comparing results from different search techniques~\cite{deAustri:2013saa}.

The experimental effort to detect neutrinos as a signature of dark matter annihilations in celestial bodies came of age in the mid 90's with underground detectors like MACRO, Baksan, Kamiokande and Super-Kamiokande. These detectors provided the first limits on the flux of neutrinos from dark matter annihilations in the Earth or the Sun~\cite{Ambrosio:1998qj,Boliev:1995xz,Mori:1993tj,Mori:1992yq}. Baksan and Super-Kamiokande continue to be competitive in the field today~\cite{Boliev:2013ai,Choi:2015ara,Mijakowski:2016cph}. Baikal~\cite{Belolaptikov:1997ry} and AMANDA were the first large-scale neutrino detectors with an open geometry to perform dark matter searches in the late 90's, soon followed by ANTARES~\cite{Collaboration:2011nsa}. Early results of these experiments can be found in Refs.~\cite{Belolaptikov:1997ry,Halzen:1998sc,Ahrens:2002eb,Achterberg:2006jf,Ackermann:2005fr,Adrian-Martinez:2013ayv}.

\subsection{Dark Matter Signals from the Sun}\label{sec:solar}

%%%%%%%%%%%%%%%% 
\begin{figure}[t]
\centering\includegraphics[width=\linewidth]{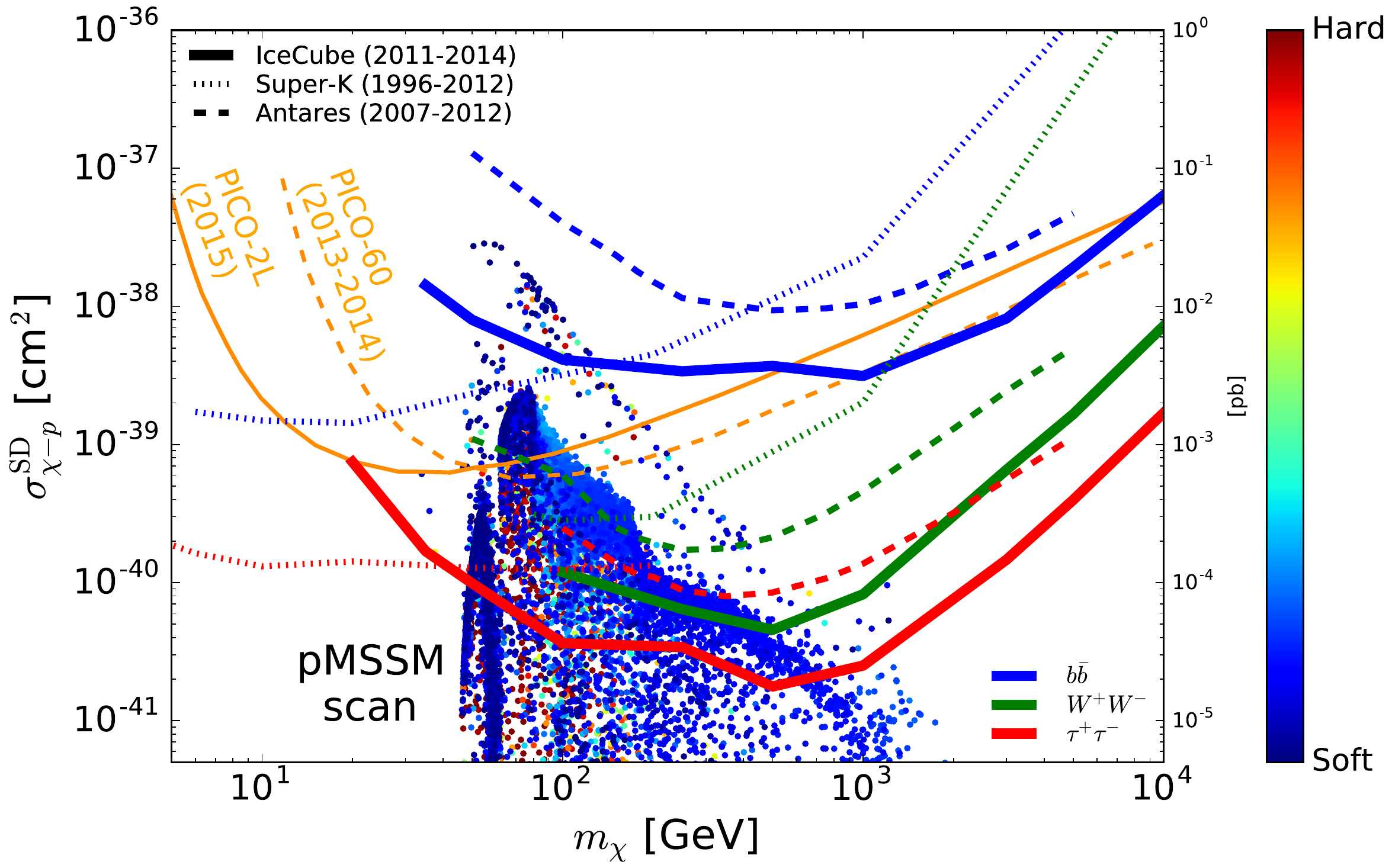}
\caption[]{IceCube limits on the spin-dependent WIMP-proton scattering cross section, $\sigma^{SD}_{\chi-p}$ as a function of WIMP mass, compared to results from other neutrino detectors and direct detection experiments~\cite{Adrian-Martinez:2016gti,Choi:2015ara,Amole:2015pla,Amole:2016pye}. The IceCube limits have been scaled up to the upper edge of the $1\sigma$ systematic uncertainty band. The coloured points correspond to models from a scan of the pMSSM. The model points are shown colour-coded according to the ``hardness'' of the resultant neutrino spectrum. Red points correspond to models that annihilate predominantly into harder channels (such as $\tau^+\tau^-$) and can hence be excluded by the IceCube red line, while blue points correspond to models that favour annihilations into softer channels (such as $b\overline{b}$) and are probed by the blue lines. Similar coding applies for intermediate colours. Figure from~\cite{Aartsen:2016zhm}.}\label{fig:3-yr_solar_limits}
\end{figure}
%%%%%%%%%%%%%%%% 

\begin{figure}[t]
\centering\includegraphics[width=\linewidth]{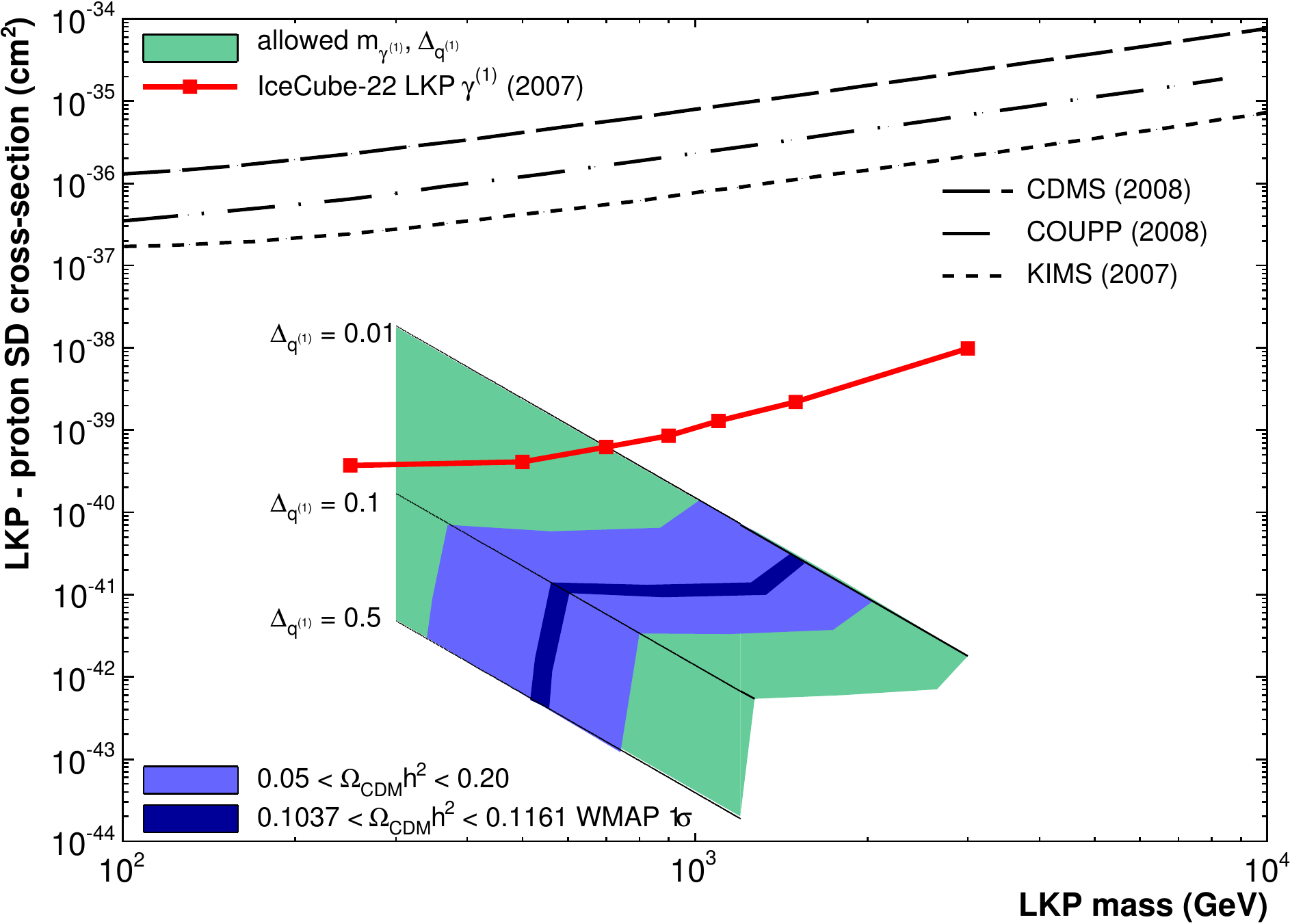}
\caption[]{IceCube limits on the spin-dependent LKP-proton scattering cross section, $\sigma^{SD}_{\chi-p}$ as a function of LKP mass, compared with limits from direct detection experiments~\cite{Ahmed:2008eu,Lee:2007qn,Behnke:2008zza}. The theoretically allowed phase space is indicated by the green shaded~\cite{Arrenberg:2008wy}. The region below $m_{\rm LKP}=300$ GeV is excluded by collider experiments~\cite{Oliver:2002up}, and the upper bound on $m_{\rm LKP}$ arises from arguments to avoid over-closure in the early universe~\cite{Servant:2002aq}. The lighter blue region is allowed when considering a wide range of dark matter relic density $0.05 < \Omega_{CDM}h^2 < 0.20$,  and the darker blue region corresponds to the 1$\sigma$ WMAP relic density, $0.1037 < \Omega_{CDM}h^2 < 0.1161$~\cite{Dunkley:2008ie}. Figure reprinted with permission from~\cite{Abbasi:2009vg}(Copyright 2010 APS).}\label{fig:KK_limits}
\end{figure}

The interplay between the dark matter capture  and annihilation  determines the number of WIMPs accumulated in the Sun. Losses through evaporation due to WIMP-nucleus scattering have been shown to be negligible for WIMP masses above a few GeV~\cite{Krauss:1985aaa,Griest:1986yu,Gould:1987ju} and can therefore be neglected in IceCube analyses. Given the age of the Sun (4.5 Gyr), the estimated dark matter density ($\rho_{\rm local}\sim 0.4\,{\rm GeV/cm}^3$) and a weak-scale interaction between dark matter and baryons, many models predict that dark matter capture and annihilation in the Sun have reached equilibrium with density following Eq.~\ref{eq:equilibriumDM}. In this case, the $J$-factor in Eq.~(\ref{eq:DMann}) becomes proportional to $1/\langle\sigma_{\rm A} v\rangle$ and total neutrino emission only depends on the capture $Q_{\rm C}$ related to the WIMP-nucleon scattering cross section. Under the assumption that the capture rate is fully dominated either by the spin-dependent or spin-independent scattering, conservative limits can be extracted on either the spin-dependent, $\sigma_{\rm SD}$, or spin-independent,  $\sigma_{\rm SI}$,  WIMP-proton scattering cross section. The number of events in the detector observed in a given live time can therefore be translated into a limit on a physical quantity that can be used to compare with other experiments or to test predictions of a specific particle physics model. Different dark matter scenarios can be probed through the predicted neutrino flux in Eq.~(\ref{eq:dm_nevents}). Indeed, IceCube has set limits to the muon flux from annihilations of the lightest Kaluza-Klein mode arising in models of universal extra dimensions, as well as to its scattering cross section with protons~\cite{Abbasi:2009vg} (see Fig.~\ref{fig:KK_limits}). Other, non-standard, scenarios like strongly interacting dark matter or self-interacting dark matter can also be tested since the IceCube event selections are quite generic and model-independent~\cite{Albuquerque:2010bt,Albuquerque:2011ma,Albuquerque:2013xna}.

%%%%%%%%%%%%%%%% 
%%%%%%%%%%%%%%%% 

Traditionally, solar WIMP searches with IceCube have used the muon channel since it gives better pointing and, in the end, dark matter searches from the Sun are really point-source searches. The first analyses used the Earth as a filter of atmospheric muons and ``looked'' at the Sun only in the austral winter, when the Sun is below the horizon at the South Pole~\cite{Abbasi:2009vg,IceCube:2011aj,Abbasi:2009uz}. With the completion of IceCube-79 and DeepCore, it was possible to define effective veto regions to efficiently reject incoming atmospheric muons from above~\cite{Aartsen:2012kia}. Since then the IceCube solar WIMP searches cover also the austral summer, doubling the exposure of the detector per calendar year. DeepCore has also allowed to extend the search for neutrinos from WIMPs with masses as low as 20 GeV/c$^2$, whereas past IceCube searches have only been sensitive above 50 GeV/c$^2$. Additionally, all-flavour analyses are being developed~\cite{Wiebe:2015jaw}, since the addition of $\nu_e$ and $\nu_{\tau}$ events triples the  expected signal. Improved low-energy reconstruction techniques allow to reconstruct electron and tau neutrino interactions with sufficiently good angular resolution to be useful in solar and Earth dark matter searches.

Signal and background differ, though, not only in their angular distribution, but also in their energy spectra. This information can be encapsulated in a likelihood function that includes a number-counting (normalisation) term and an angular and energy (spectral) terms, both for the signal and background p.d.f.'s. This is the approach taken in the latest IceCube solar WIMP analyses~\cite{Aartsen:2016exj,Aartsen:2016zhm}. All-flavour analyses also benefit from this technique since the better energy resolution of cascade events benefits from the use of energy information in the likelihood.  The general form of such an extended likelihood is,  
\begin{equation}
\label{eq:unbinned_like}
\begin{aligned}
\mathcal{L}_{\rm ext}(\mu_s,\mu_{\rm tot}) \,=\, & \frac{\mu_{\rm tot}^{n_{\rm obs}}e^{-\mu_{\rm tot}}}{n_{\rm obs}!} \mathcal{L}(\mu_{\rm s})\,,
\end{aligned}
\end{equation}
where the second term is analogous to Eq.~(\ref{eq:LikelihoodDefinition}) and the prefactor is the Poisson number likelihood for observing $n_{\mathrm{obs}}$ events given a total of $\mu_\mathrm{tot}$ predicted signal and background events. The signal and background probability distribution functions, $S$ and $B$, introduced in Eq.~(\ref{eq:LikelihoodDefinition}) now include the event's reconstructed solid angle $\Omega_i$ and energy $E_i$. For instance, the signal probability distribution is given by
\begin{multline}
\label{eq:generic_likelihood}
S(E_i,\Omega_i,t_i) \equiv   \int {\rm d} \Omega'\int {\rm d} E' \ Q(E_i, \Omega_i | E', \Omega')\\\cdot
\frac{\mathrm{d}^2 P_{\rm S}}{{\mathrm d} E'\,{\mathrm d} \Omega'}(E',\Omega',t_i,{\boldsymbol\xi})\,,
\end{multline}
where ${\rm d}^2 P_S / \mathrm{d} E\,\mathrm{d} \Omega$ is the expected signal distribution of incident neutrino energies and angles at the time of the event's arrival\footnote{The dependence on arrival time is relevant for indirect dark matter searches from the Sun, which can be effectively parametrised by the angular distance $\psi$ between reconstructed arrival direction and the known position of the Sun at the arrival time (corrected for the light-travel time of about 8~min).}, $t_i$. For the signal, this term is a prediction of the model under consideration and it depends on the free parameters of the model, denoted by the vector ${\boldsymbol \xi}$. An equation analogous to Eq.~(\ref{eq:generic_likelihood}) holds for the background distribution $B(E,\Omega)$. For the background, ${\rm d}^2 P_B / \mathrm{d} E\,\mathrm{d} \Omega$ will typically depend only on the energy and angular spectra of the atmospheric neutrino flux, and it is independent of arrival time $t_i$ and ${\boldsymbol \xi}$.

%%%%%%%%%%%%%%%% 
\begin{figure}[t]
\centering\includegraphics[width=\linewidth]{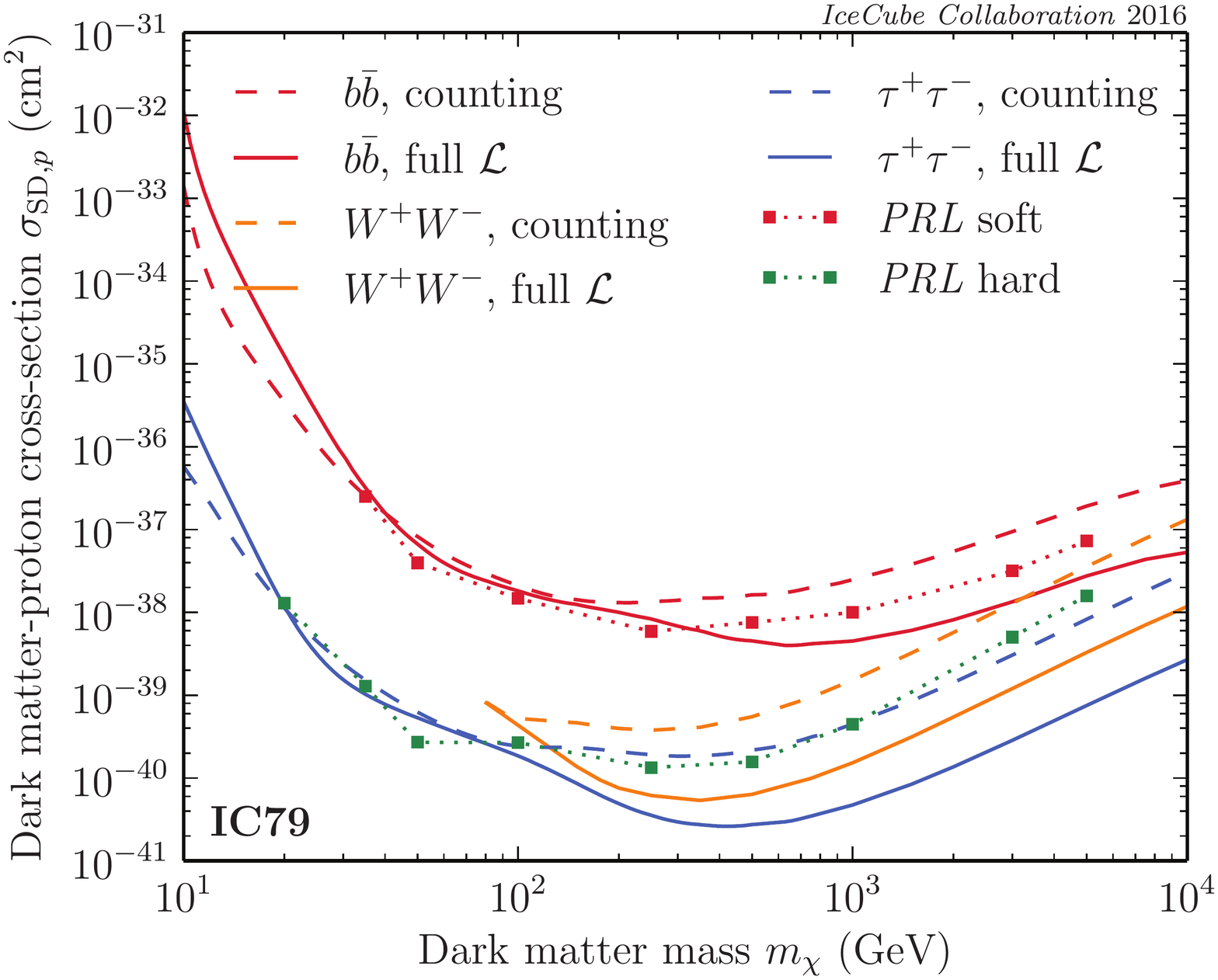}
\caption[]{Improvement on the dark-matter proton cross section due to the use of an event-level likelihood analysis (full lines) compared to a traditional
cut-and-count analysis (dashed lines) and an analysis based on the difference in shape of the space-angle distribution
for signal and background (dotted lines tagged 'PRL' and originally from~\cite{Aartsen:2012kia}).  100\% annihilation to $b\overline{b}$, $W^+W^-$ and $\tau^+\tau^-$ is assumed in each case shown. Figure from~\cite{Aartsen:2016exj}.}
\label{fig:llh_improvement}
\end{figure}
%%%%%%%%%%%%%%%% 

In general, the reconstructed neutrino energy $E$ and solid angle $\Omega$ are different from the true energy $E'$ and true arrival direction $\Omega'$. The relation between these quantities on a statistical basis must be obtained from simulations, and it is specific to the annihilation channel under study. The true energy $E$ can be related to the number of hit DOMs or can be estimated from more elaborate energy reconstructions.  The quantity $Q(E_i, \Omega_i | E', \Omega')$ is the probability density (in effective units of inverse steradian and proxy energy) for reconstructing $E_i$ and $\Omega_i$ for the $i$th event when the true values are $E'$ and $\Omega'$, respectively.

%%%%%%%%%%%%%%%% 
\begin{figure*}[t]
\begin{minipage}[t]{0.45\linewidth}
\centering\includegraphics[width=\linewidth]{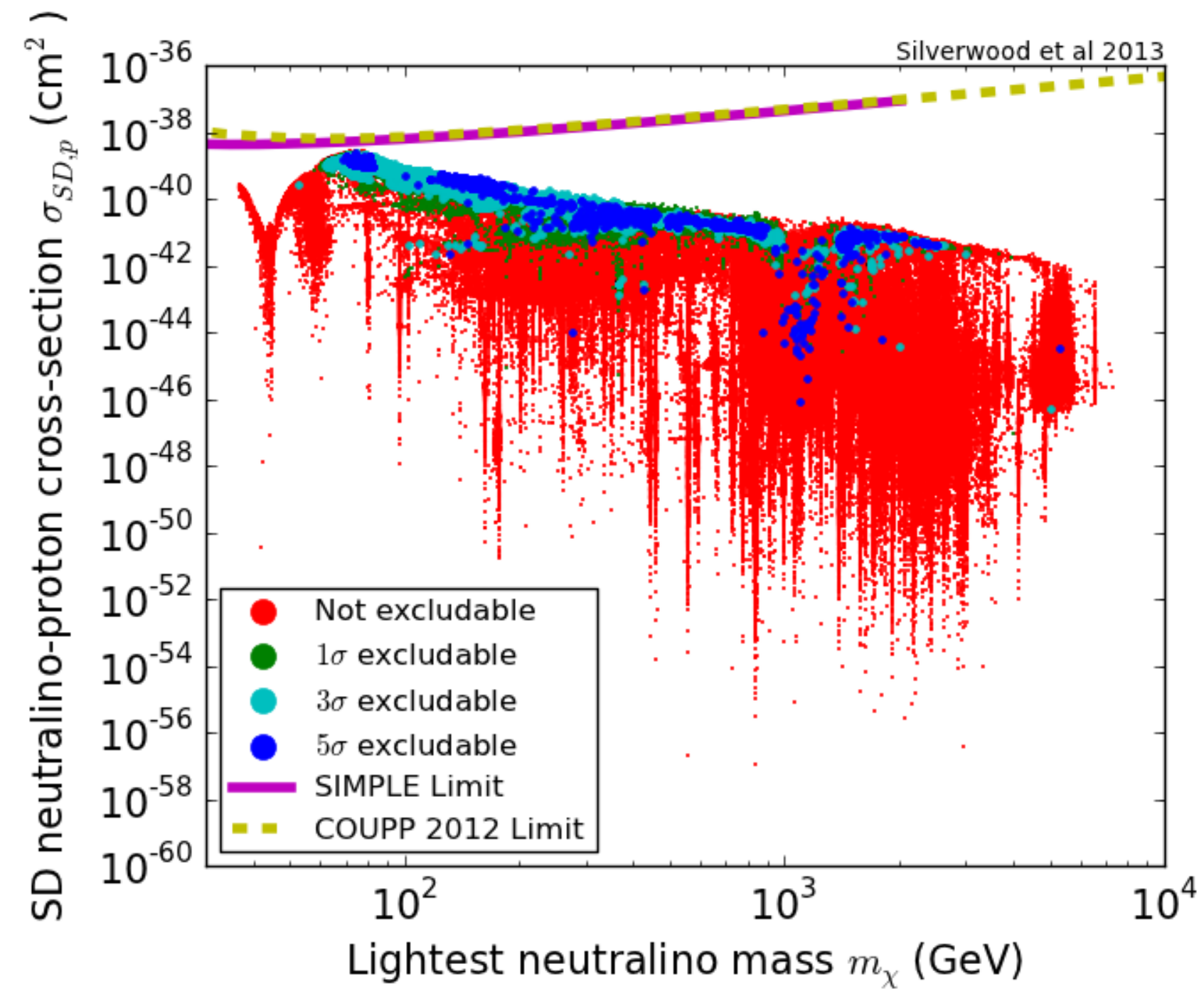}
\caption[]{Spin-dependent WIMP-proton scattering cross section, $\sigma^{SD}_{\chi-p}$ as a function of WIMP mass
 for points derived from explorations of the MSSM-25 parameter space. The corresponding 90\% CL limits from SIMPLE~\cite{Felizardo:2011uw} and COUPP~\cite{Behnke:2012ys}
direct detection experiment are shown as magenta and yellow lines respectively. 
Colour coding indicates predicted IceCube-86 model exclusion levels. The areas of cyan and blue points show that IceCube-86 has the ability to exclude models
beyond the reach of current direct detection experiments. Figure from~\cite{Silverwood:2012tp} (Copyright SISSA Medialab Srl. Reproduced by permission of IOP Publishing. All rights reserved).}
\label{fig:model_exclusion01}
\end{minipage}\hfill
\begin{minipage}[t]{0.45\linewidth}
\centering\includegraphics[width=\linewidth]{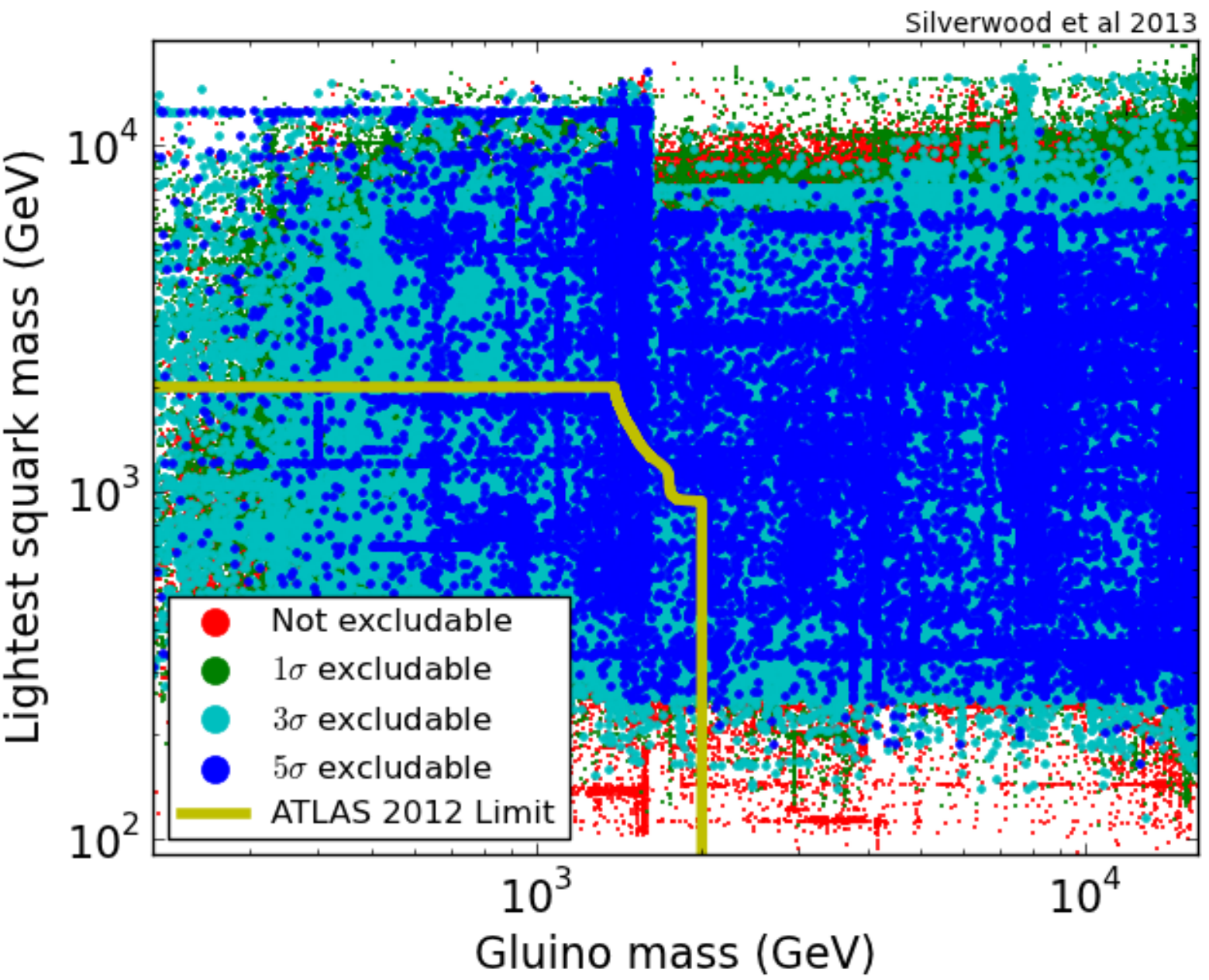}
\caption[]{Lightest squark mass as a function of gluino mass for points derived from explorations of the MSSM-25 parameter space. The points to 
the bottom left of the magenta line are excluded by ATLAS at 95\% CL, based on searches for direct production of coloured sparticles and their decay
into jets and missing transverse energy. Colour coding indicates predicted IceCube-86 model exclusion levels. Figure from~\cite{Silverwood:2012tp} (Copyright SISSA Medialab Srl. Reproduced by permission of IOP Publishing. All rights reserved).}
\label{fig:model_exclusion02}
\end{minipage}
\end{figure*}
%%%%%%%%%%%%%%%% 

Note that systematic uncertainties on the signal and/or background prediction or on the angular or energy resolutions can be easily incorporated in a likelihood approach as nuisance parameters by marginalising over them. The only knowledge needed is the functional form of the nuisance parameters. $P$ is a function of energy and angle, which in a simplified approach can be decomposed in an angle-dependent part (the PSF of the detector) and an energy-dependent part (the energy dispersion of the detector). More generally, the angular response of the detector can depend on energy, and then this decomposition is not valid. For point-source searches, due to the restricted angular region in the sky considered, the PSF and energy dispersion can be taken to be uncorrelated.

Figure~\ref{fig:3-yr_solar_limits} shows the limits on the spin-dependent WIMP-proton cross section using three years of IceCube data, including both angular and energy terms in the likelihood~\cite{Aartsen:2016zhm}. The plot shows the latest IceCube results as full lines for three benchmark annihilation channels (in different colours), compared with the results of Super-Kamiokande~\cite{Choi:2015ara} and ANTARES~\cite{Adrian-Martinez:2016gti}, as well as results from the PICO direct detection experiment~\cite{Amole:2015pla,Amole:2016pye}. The dots correspond to a parameter scan of the phenomenological Minimal Supersymmetric extension of the Standard Model (pMSSM) where the color code represents the leading annihilation channel: channels producing a soft neutrino spectra are marked as blue, and are probed by the soft experimental limits, while harder neutrino spectra are marked in red and are probed by the hard experimental limits. Since large-volume neutrino telescopes are high-energy neutrino detectors, the limits for annihilation channels leading to a soft neutrino spectrum can be  more than two orders of magnitude less restrictive than those resulting in harder spectra. Even in the latter case the limits can decrease rapidly for low WIMP masses.  Direct search experiments do not reach $\sigma_{SD}$  values much below 10$^{-38}$~cm$^2$ at their best point, worsening rapidly away from it, while, together, the limits from IceCube and Super-Kamiokande cover the WIMP mass range between from a few GeV to 100 TeV and reach cross section values of about 10$^{-40}$~cm$^2$. In the case of the $\sigma_{SD}$ cross section, direct-search experiments have the advantage of dedicated spinless targets, and the limits from neutrino telescopes lie about three orders of magnitude above the best limit from LUX at a WIMP mass of about 50 GeV~\cite{Akerib:2013tjd}. 

The improvement due to using a full event-based likelihood in comparison to just an angular shape analysis is illustrated in Fig.~\ref{fig:llh_improvement}, taken from~\cite{Aartsen:2016exj}. The figure shows the limit on the spin-dependent WIMP-proton cross section as a function of WIMP mass obtained with two analyses performed on the same data set taken with IceCube-79. Shown are the limits using an event-count likelihood (dashed lines), the limits obtained using an analysis based on the difference in shape of the space-angle distribution for signal and background (tagged 'PRL' and originally from~\cite{Aartsen:2012kia}) and the limits obtained using a full likelihood like in Eq.~(\ref{eq:generic_likelihood}). Including the event-level energy information has the most impact at high WIMP mass, due to the relatively good energy resolution of IceCube at high neutrino energies. Note that the full likelihood analysis in~\cite{Aartsen:2016exj} used a rather simple energy proxy based on the number of hit DOMs. Better energy reconstruction algorithms being developed within IceCube, particularly at low energies, will further improve the performance of this method~\cite{Aartsen:2014oha}.

There is an additional step in complexity when using neutrino telescope results to probe specific WIMP models, that avoids the need to simulate specific annihilation benchmark channels~\cite{Trotta:2009gr,Scott:2012mq,Silverwood:2012tp}. For a given model, this approach takes into account the full neutrino spectrum from WIMP annihilations including all allowed annihilation channels, {\it i.e.}, the full ${\rm d}^2 P / \mathrm{d} E\,\mathrm{d} \phi(E,\phi,{\boldsymbol \xi})$ is calculated. The expected number of signal events is then obtained through Eq.~(\ref{eq:dm_nevents}) and, through the use of the likelihood function in Eq.~(\ref{eq:unbinned_like}), the  model under consideration can be accepted or rejected at a given confidence level. To explore a wide parameter space, which is typical for supersymmetric extension of the Standard Model, each allowed combination of the free parameters of the model needs to be tested, and the procedure becomes computationally demanding. Even ad-hoc models like the constrained MSSM (cMSSM)~\cite{Kane:1993td} with just 7 parameters pose a computational challenge. But it is a powerful way of assigning a statistically meaningful weight to different areas of the model parameter space. The authors in~\cite{Trotta:2009gr} and~\cite{Silverwood:2012tp} have performed sensitivity studies of IceCube and IceCube-DeepCore under different model assumptions. Figure~\ref{fig:model_exclusion01} shows the results of the model scan in~\cite{Silverwood:2012tp}. The figure shows the spin-dependent neutralino-proton cross section versus neutralino mass. Each point in the plot represents an allowed combination of the cMSSM parameters. The colour code indicates which models can be probed by IceCube and disfavoured at the 1$\sigma$ (green), 2$\sigma$ (light blue) and 5$\sigma$ (dark blue) level. Figure~\ref{fig:model_exclusion02} illustrates the complementarity between  accelerator and neutrino telescope searches for supersymmetry. The plot shows the gluino-squark mass parameter space with colour-coded exclusion levels by IceCube (same colour coding as in Fig.~\ref{fig:model_exclusion01}). The brown line shows the 95\% CL exclusion region from searches for coloured sparticles in jets+missing transverse energy with, at the time, 4.71 fb$^{-1}$ of data at centre-of-mass energies $\sqrt{s}\simeq7$~TeV from ATLAS~\cite{ATLAS:2012sma}. As can be seen, there is a wealth of models not excluded by ATLAS that are under the reach of IceCube (blue dots). Note that more recent results from ATLAS and CMS (see, {\it e.g.}, the summary on the experimental status of Supersymmetry in~\cite{PDG:2018}) do not change the picture of complementarity between the parameter space reach of IceCube and the LHC.

%%%%%%%%%%%%%%%% 
\begin{figure*}[t!]
\begin{minipage}[t]{0.5\linewidth}
\centering\includegraphics[width=0.9\linewidth]{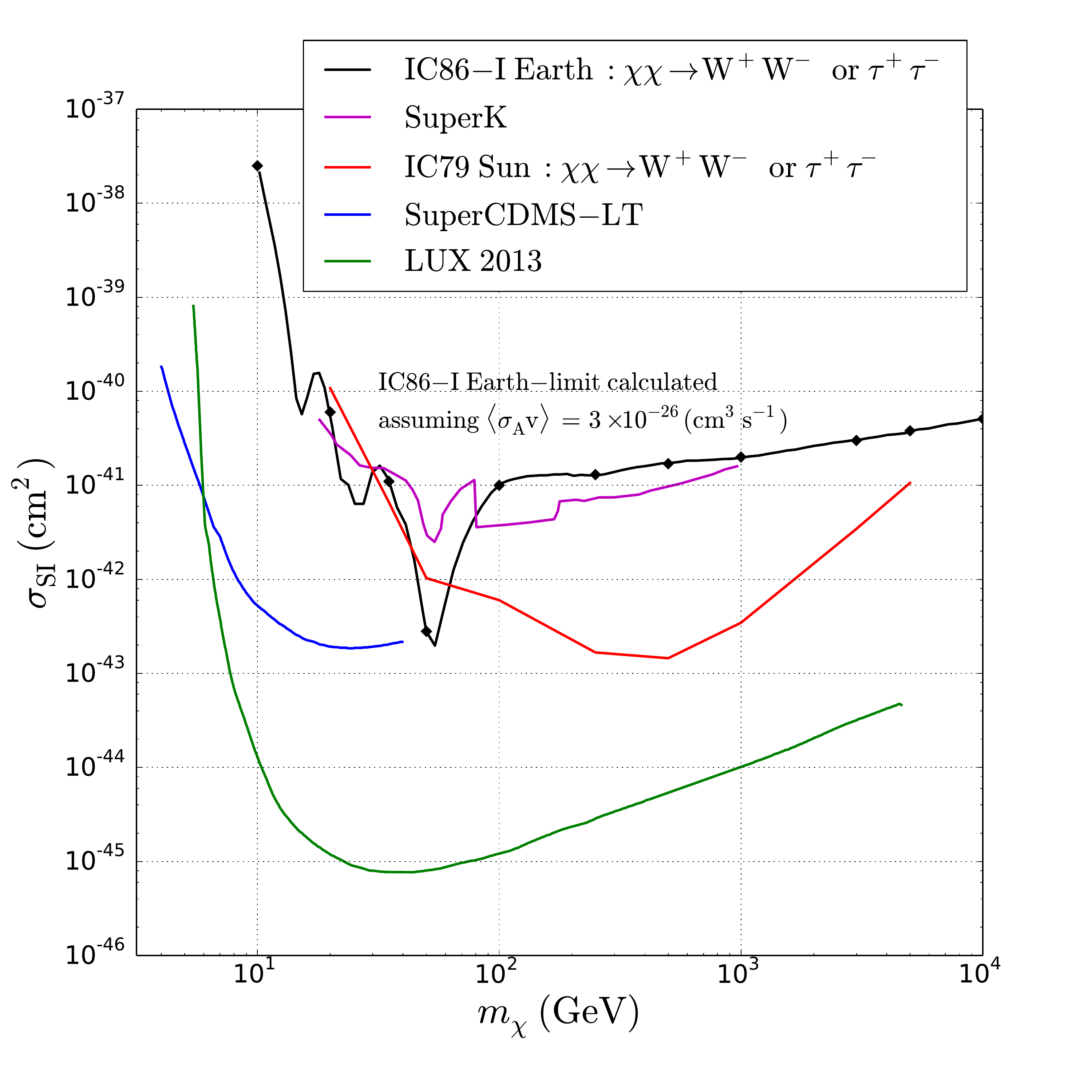}
\caption[]{Upper limits at 90\% confidence level on  $\sigma^{SI}_{\mathrm{WIMP + N}}$ as a function of the WIMP-mass obtained from a dedicated Earth search with IceCube-79~\cite{Aartsen:2016fep}, assuming a WIMP annihilation cross section of $\langle \sigma_{\mathrm A} v \rangle = 3 \times 10^{-26} cm^3 s^{-1}$ and a local dark matter density of 0.3 GeV cm$^{-3}$. Results from Super-Kamiokande~\cite{Desai:2004pq}, SuperCDMS~\cite{Agnese:2013jaa}, LUX~\cite{Akerib:2013tjd} and an IceCube-79 solar search~\cite{Aartsen:2012kia} are shown for comparison. Figure from~\cite{Aartsen:2016fep}}
\label{fig:earth_sigma-mass}
\end{minipage}\hfill
\begin{minipage}[t]{0.45\linewidth}
\centering\includegraphics[width=\linewidth]{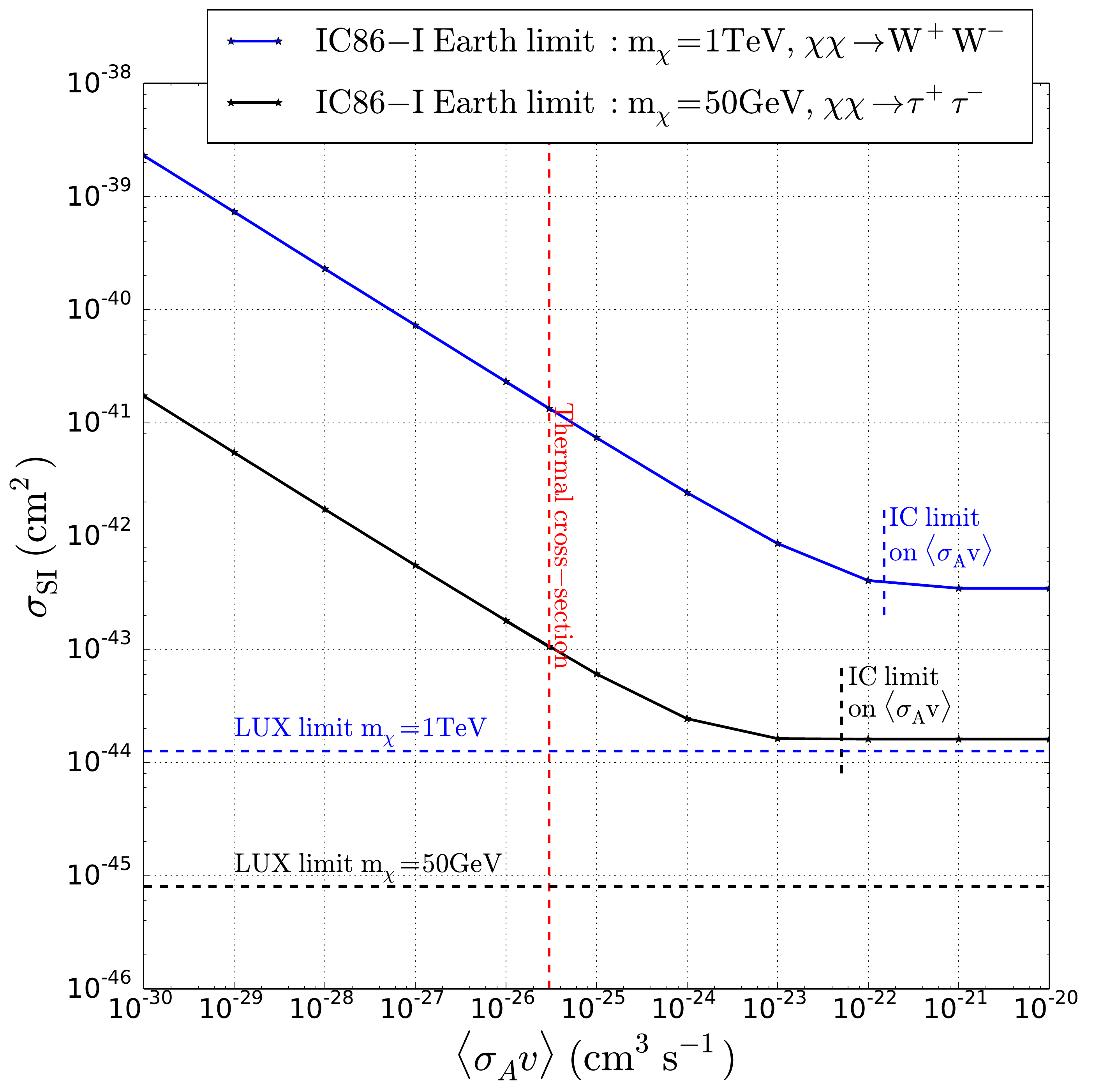}
\caption[]{Upper limits at 90\% confidence level on $\sigma^{SI}_{X + N}$ as a function of the velocity-averaged annihilation cross section $\langle \sigma_{\mathrm A} v \rangle$ for 50 GeV WIMPs annihilating into $\tau^{+}\tau^{-}$ and for 1~TeV WIMPs annihilating into W$^+$W$^-$. Limits from LUX~\cite{Akerib:2013tjd} are shown as dashed lines for comparison. The red vertical line indicates the thermal annihilation cross section for a particle species to constitute the dark matter. Also indicated are IceCube limits on the annihilation cross section for the respective masses and annihilation channels~\cite{Aartsen:2015xej}. Figure from~\cite{Aartsen:2016fep}}
\label{fig:earth_sigma-sv}
\end{minipage}
\end{figure*}
%%%%%%%%%%%%%%%% 

\subsection{Dark Matter Signals from the Earth}\label{sec:earh}

The rationale for searching for dark matter accumulated in the Earth follows a similar line than that of the Sun: WIMPs gravitationally accumulated in the centre of the Earth can annihilate, giving rise to a flux of neutrinos. A signal from dark matter annihilations in the centre of the Earth will produce a unique signature in IceCube as vertically up-going muons, where each string can act as a more or less independent detector. Searches from the vertical direction pose, however, some challenges in IceCube due to its geometry. While in any other point source search an off-source region at the same declination of the source can be defined to measure the background, this is not possible for the vertical direction, and one needs to rely on accurate simulations of the background components (atmospheric neutrinos and muons). Through the use of advanced classification methods to separate signal and background, the amount of misreconstructed atmospheric muons can be reduced to a negligible level, and a likelihood shape analysis can be performed using Eq.~(\ref{eq:LikelihoodDefinition}). The results of such an analysis using 327 days of live time with the IceCube-79 configuration~\cite{Aartsen:2016fep} are shown in Fig.~\ref{fig:earth_sigma-mass}. The shape of the IceCube limits reflects the resonant capture of WIMPs of certain masses that nearly match the mass of the main isotopes of the Earth (the peaks show the Iron, Silicon and Oxygen resonances at 53~GeV/c$^2$, 26~GeV/c$^2$ and 15~GeV/c$^2$, respectively). A self-annihilation cross section, $\langle \sigma_{\mathrm A} v \rangle$, of 3$\times$10$^{-26}$ cm$^3$ s$^{-1}$ has been assumed, a typical value for a particle species to be a thermal relic. The lack of a signal can also be used to set limits on the dependence of $\sigma_{SI}$ on the annihilation cross section, as illustrated in Fig.~\ref{fig:earth_sigma-sv}. Values above the full lines are disfavoured at the 90\% confidence level since the combination of capture and annihilation would have produced a detectable muon flux in IceCube.

\subsection{Dark Matter Signals from Galaxies and Galaxy Clusters}\label{sec:galaxies}

The Milky Way centre and halo, as well as nearby dwarf galaxies and galaxy clusters provide natural large-scale regions of increased dark matter density. Since dark matter played a significant role in the formation of such structures from primordial density fluctuations, the issue of capture is not relevant, and what neutrino telescopes can prove when considering such objects is the WIMP self-annihilation cross section, $\langle \sigma_{\mathrm A} v \rangle$.  In order to predict the rate of annihilation of dark matter particles in galactic halos, the precise size and shape of the halo needs to be known. There is still some controversy on how dark halos evolve and which shape they have. There are different numerical simulations, observational fits, and parametrisations of the dark matter density around visible galaxies, including the Navarro-Frenk-White (NFW) profile~\cite{Navarro:1995iw}, the Kravtsov profile~\cite{Kravtsov:1997dp}, the Moore profile~\cite{Moore:1999gc}, and the Burkert~\cite{Burkert:1995yz}. The common feature of these profiles is a denser spherically symmetric region of dark matter in the centre of the galaxy, with decreasing density as the radial distance to the centre increases. Where they  diverge is in the predicted shape of the central region. Profiles obtained from N-body simulations of galaxy formation and evolution tend to predict a steep power-law type behaviour of the dark matter component in the central region, while profiles based on observational data (stellar velocity fields) tend to favour a constant dark matter density near the core. This is the core-cusp problem~\cite{deBlok:2009sp}, and it is an unresolved issue which affects the signal prediction from dark matter annihilations in neutrino telescopes. Note that the shape of the dark halo can depend on local characteristics of any given galaxy, like the size of the galaxy~\cite{Ricotti:2002qu} or on its evolution history~\cite{Dekek:1986gu,Ogiya:2012jq}

%%%%%%%%%%%%%%%%  
\begin{figure}[t]
\includegraphics[width=\linewidth]{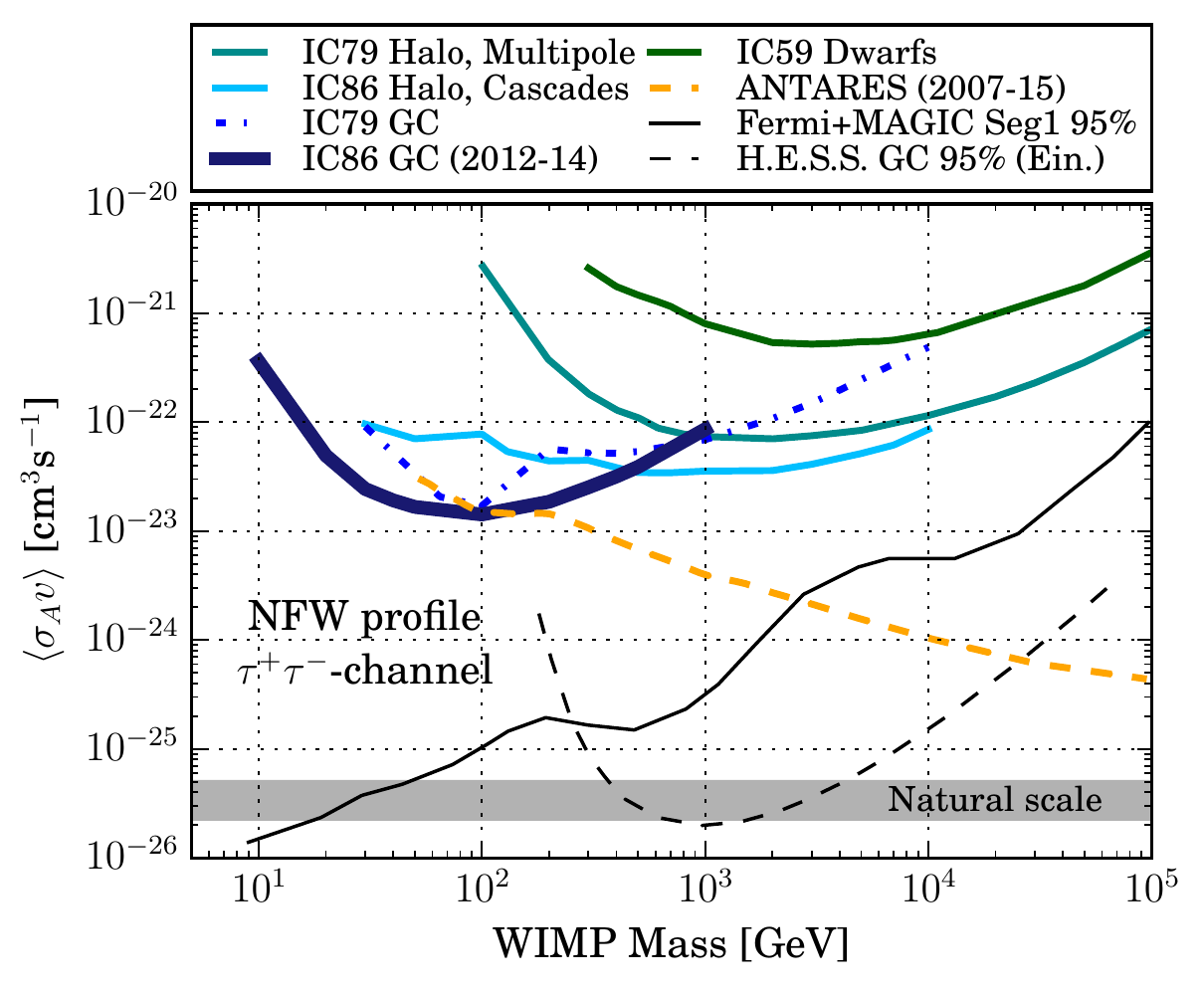}
\caption[]{Comparison of upper limits on $\langle \sigma_\mathrm{A} \mathrm{v} \rangle$ versus WIMP mass, for the annihilation channel $\chi\chi \rightarrow \tau^+\tau^-$. IceCube results obtained with different detector configurations~\cite{Aartsen:2014hva,Aartsen:2013dxa,Aartsen:2015xej,Aartsen:2016pfc} are compared to {ANTARES} \cite{Albert:2016emp} and the latest upper limits from $\gamma$-ray searches from {H.E.S.S.}~\cite{Abdallah:2016ygi} and from a combination of {Fermi-LAT} and {MAGIC} results~\cite{Ahnen:2016qkx}. Figure from~\cite{Aartsen:2017ulx}.}
\label{fig:GCLimits}
\end{figure}
%%%%%%%%%%%%%%%% 

The shape of the dark matter halo is important because the expected annihilation signal depends on the line-of-sight integral from the observation point (the Earth) to the source, and involves an integration over the square of the dark matter density. This is included in the J-factor of Eq.~(\ref{eq:DMann}), which is galaxy-dependent, and absorbs all the assumptions on the shape of the specific halo being considered. In the case of our Galaxy, the expected signal from the Galactic Centre assuming one halo parametrisation or another can differ by as much as four orders of magnitude depending on the halo model used (see, {\it e.g.}, Fig.~2 in Ref.~\cite{Yuksel:2007ac}).

As before, a few benchmark annihilation channels can be chosen to bracket the model expectations, and then $\mathrm{d}N_\nu / \mathrm{d}E$ represents the neutrino flux assuming 100\% annihilation to each of the benchmark scenarios. The analyses use the same likelihood approach as described by Eq.~(\ref{eq:LikelihoodDefinition}). The signal hypothesis (excess of events at small angular distances $\psi$ to the Galactic Centre) can then be tested against the background-only hypothesis (an event distribution isotropic in the sky). There is, however, an additional effect to take into account when dealing with extended sources, like the Galactic halo. Since the signal is allowed to come from anywhere in the halo, the background distribution, which is usually taken from data scrambled over azimuth angle, $B = \widetilde D$, is necessarily contaminated by a potential signal. Thus, the background distribution $B$ depends indirectly on the number of signal events $\mu_{\mathrm S}$ and needs to be corrected. The effective background distribution can be written as
\begin{equation}
B = \left(\widetilde D - \frac{\mu_s}{n_{\rm obs}} \widetilde{S}\right)/\left(1-\frac{\mu_s}{n_{\rm obs}}\right)\,,
\label{eq:backgroundPDF}
\end{equation}
where $\widetilde{S}$ and $\widetilde{D}$ are the probability distribution functions of the azimuth-scrambled arrival directions of signal simulation and data events respectively.

%%%%%%%%%%%%%%%% 
\begin{figure*}[t]\centering
\begin{minipage}[t]{0.45\linewidth}
\centering\includegraphics[width=\linewidth]{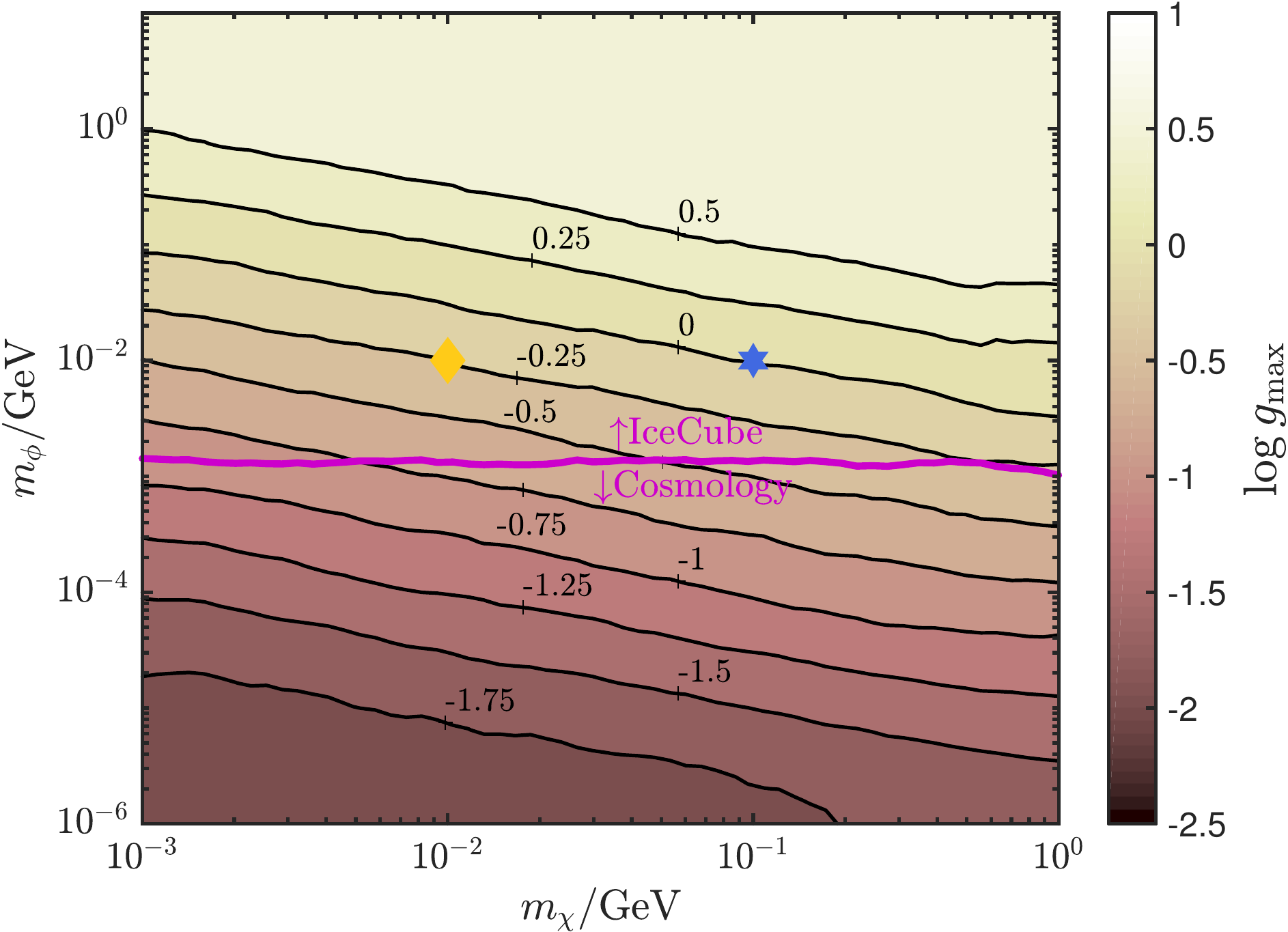}
\end{minipage}\hspace{0.5cm}
\begin{minipage}[t]{0.45\linewidth}
\centering\includegraphics[width=\linewidth]{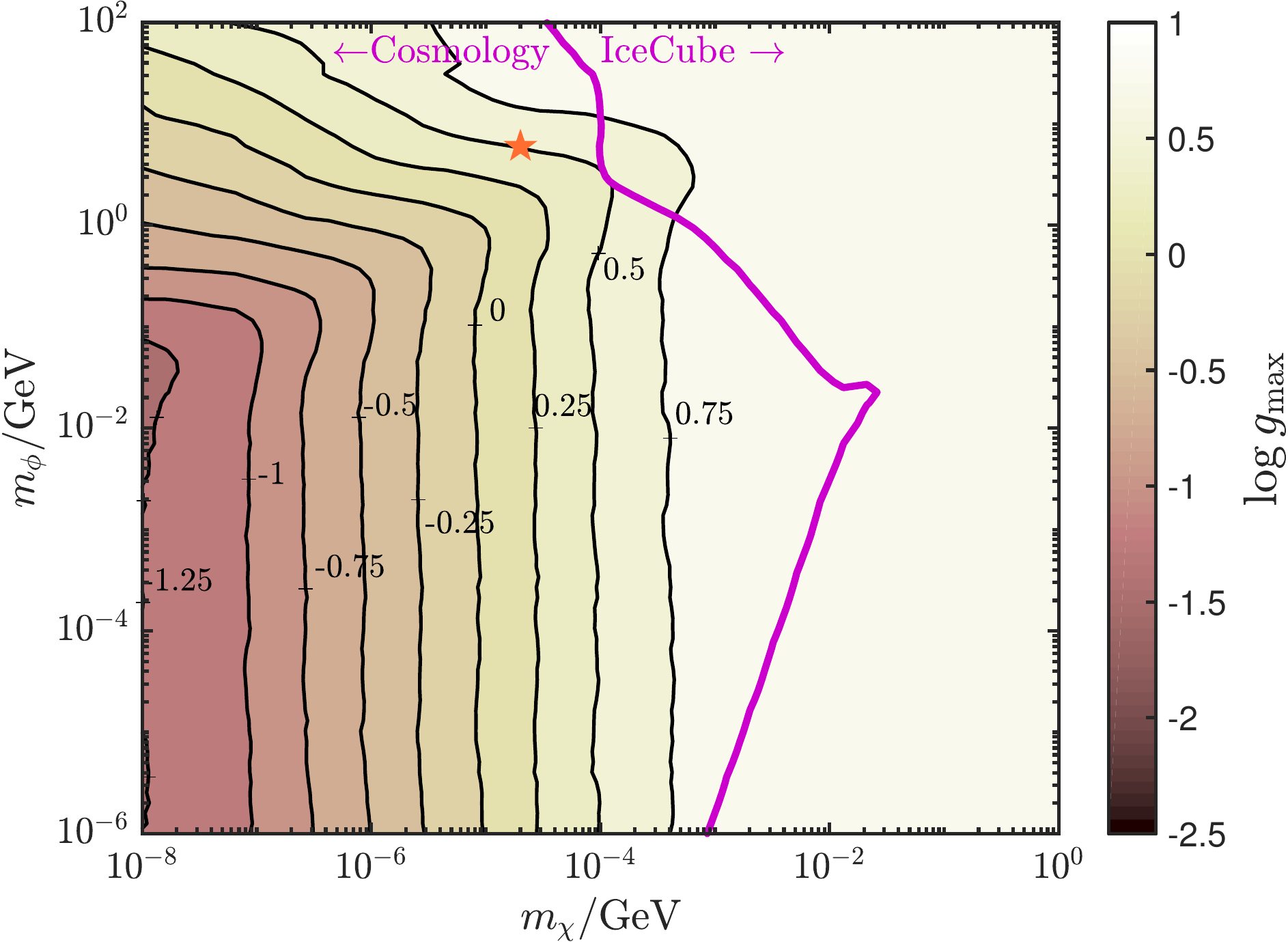}
\end{minipage}
\caption[]{Contours of the maximum allowed value of the coupling of dark matter to neutrinos, $g_{\rm max}$, as a function of the dark matter mass, $m_\chi$, and the mediator mass $m_\phi$. {\bf Left Panel:} fermionic dark matter coupled through a vector mediator. {\bf Right Panel:} scalar dark matter coupled through a fermionic mediator. The magenta line in both plots delimits the region where cosmological observations from large scale structure become more restrictive than the IceCube limits. The plots have been obtained by marginalising over the atmospheric and astrophysical fluxes, allowing the astrophysical spectral index to vary between 2 and 3. The diamond and the stars refer to specific models studied in~\cite{Arguelles:2017atb}.  Reprinted with permission from~\cite{Arguelles:2017atb}. Copyright 2017 by the American Physical Society.}
\label{fig:DM-nu}
\end{figure*}
%%%%%%%%%%%%%%%% 

There is another, complementary, analysis approach for extended sources that naturally incorporates the diffuse character of the signal over a large region in the sky. It is based on a multipole expansion of the sky map of event arrival directions. Dark matter annihilations in the halo would produce a diffuse flux of neutrinos with a characteristic large scale structure following the shape of the halo, while the atmospheric neutrino background presents small anisotropies at smaller scales. The sky map of reconstructed event arrival directions can be constructed as 
\begin{equation}
	f(\theta, \phi) =
		\sum_{i=1}^{n_{\rm obs}}\, 
			\delta^\mathrm{D}(\cos(\theta)-\cos(\theta_i))
			\delta^\mathrm{D}(\phi-\phi_i)
	\, ,
	\label{eq:skymap}
\end{equation}
where $(\theta_i,\phi_i)$ are the reconstructed coordinates (declination and right ascension respectively) of event $i$, $n_{\rm obs}$ is the total number of events in the data sample and $\delta^\mathrm{D}$ is the Dirac-delta-distribution. Such distribution can be mapped onto an expansion in spherical harmonics on the sphere,
\begin{equation}
	f(\Omega) = 
		\sum_{\ell} \sum_{m=-\ell}^{m=\ell} 
			a_\ell^m  Y_\ell^m(\Omega)\,,
	\label{eq:expantion}
\end{equation}
where the expansion coefficients $a_\ell^m$ are given by the sum over events with reconstructed arrival directions $\Omega_i$,
\begin{equation}
a_\ell^m = \sum_{i=1}^{n_{\rm obs}} (Y_\ell^m)^*(\Omega_i)\,.
\end{equation}
Note that this expansion depends on the coordinate system. In particular, in the equatorial coordinate system where the event distribution has strong dependence on declination angle (equivalent to zenith angle) from the detector acceptance, the expansion is dominated by the $m=0$ coefficients. It is possible to design a test statistic of the remaining $m\neq0$ components to separate signal from background (see Eqs.~(8) and (9) in~\cite{Aartsen:2014hva}).

Results from the searches performed by IceCube with different techniques on the Galactic Centre, halo and dwarf spheroid galaxies~\cite{Aartsen:2014hva,Aartsen:2013dxa,Aartsen:2015xej,Aartsen:2016pfc} are shown in Fig.~\ref{fig:GCLimits}, compared with other experiments and theory interpretations. All sources considered showed results compatible with the background-only hypothesis yielding limits on the velocity-averaged annihilation cross section at the level of few 10$^{-23}$ cm$^3$ s$^{-1}$. Recent results from $\gamma$-ray telescopes on dwarf spheroids currently provide the best limits on $\langle \sigma_\mathrm{A} \mathrm{v} \rangle$, due to their accurate pointing and lack of foreground or background from these kind of sources.

The high-energy diffuse astrophysical neutrino flux discovered by IceCube opens a new possibility of probing the galactic dark matter distribution through neutrino-dark matter interactions~\cite{Cherry:2014xra,Davis:2015rza,Arguelles:2017atb}. Indeed dark matter couplings to standard model particles are commonly assumed to exist, and this is the basis of direct detection experiments. Such a coupling can be extended to neutrinos if one assumes the existence of a mediator $\phi$, which can be either bosonic of fermionic in nature, which couples to dark matter with a coupling $g$. For simplicity one can assume that the strength of the $\nu-\phi$ coupling is also given by $g$. Under these assumptions, dark matter-neutrino interactions could distort the isotropy of the astrophysical neutrino flux, resulting in an attenuation of the flux towards the Galactic Centre, where the density of dark matter is higher. Therefore, an analysis of the isotropy of the high-energy astrophysical neutrino data of IceCube can be translated into a limit on the strength of the neutrino-dark matter coupling, $g$. Such an analysis has been performed by the authors in~\cite{Arguelles:2017atb}, where the mass of the dark matter candidate, the mass of the mediator and the strength of the coupling are left as free parameters in a likelihood calculation which aims at evaluating the suppression of astrophysical neutrino events from the direction of the Galactic Centre. The results are shown in Fig.~\ref{fig:DM-nu}. The left panel shows contours of the maximum allowed value of the coupling of fermionic dark matter coupled to neutrinos through a vector mediator, while the right panel shows the case of a scalar dark matter coupled through a fermionic mediator. Interestingly IceCube is sensitive to a region of parameter space complementary to results derived from cosmological arguments alone~\cite{Farzan:2014gza,Boehm:2014vja,Bertoni:2014mva}, as indicated by the magenta line. This line delimits the region where limits from analyses using large-scale structure data become more restrictive than the IceCube limits shown in the plot.

\subsection{TeV-PeV Dark Matter Decay}

The origin of the TeV-PeV diffuse flux observed with IceCube is so far unknown. Whereas most models assume an astrophysical origin of the emission, it is also feasible that the emission is produced via the decay of dark matter, as first proposed in Ref.~\cite{Feldstein:2013kka}. Various studies have argued that heavy dark matter decay can be responsible for various tentative spectral features in the inferred neutrino spectrum of the HESE analysis~\cite{Esmaili:2013gha,Bhattacharya:2014vwa,Boucenna:2015tra,Ko:2015nma,Bhattacharya:2017jaw,Hiroshima:2017hmy,Aisati:2015vma} and also its low energy extension~\cite{Chianese:2016opp,Chianese:2016kpu,Chianese:2017nwe}. Some authors have also discussed the necessary condition for PeV dark matter production in the early Universe, {\it e.g.}, via a secluded dark matter sector~\cite{Dev:2016qbd}, resonantly-enhanced freeze-out~\cite{DiBari:2016guw,Borah:2017xgm}, or freeze-in~\cite{Fong:2014bsa,Roland:2015yoa,Fiorentin:2016avj,Chianese:2016smc}.

%%%%%%%%%%%%%%%% 
\begin{figure*}[t]
\includegraphics[height=6.5cm,viewport=0 20 650 480,clip=true]{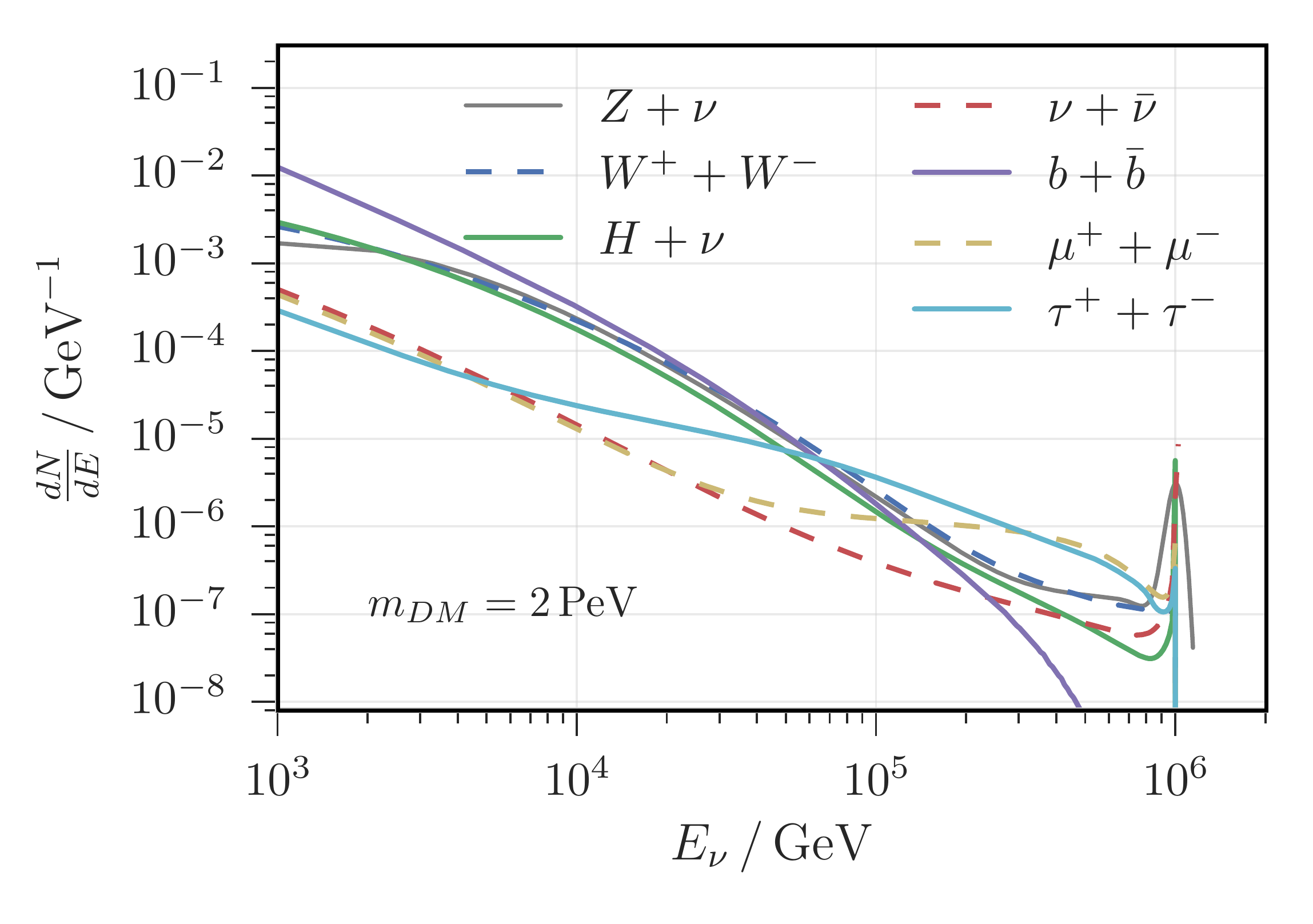}\hspace{0.5cm}\includegraphics[height=6.5cm,viewport=0 -24 750 602,clip=true]{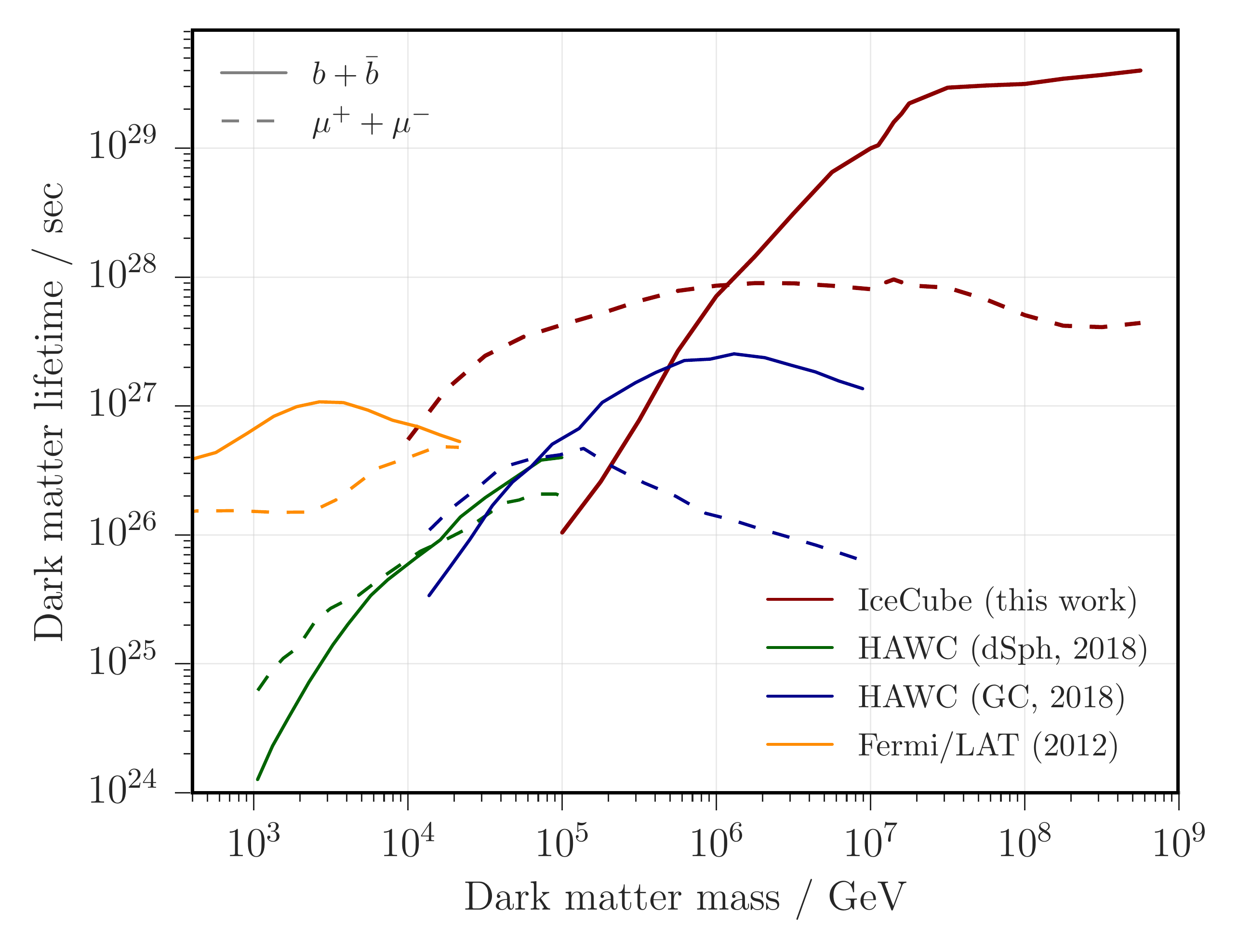}
\caption[]{{\bf Left Panel:} Neutrino yield per decay as a function of neutrino energy (flavour-averaged) for an assumed dark matter mass of 2~PeV. {\bf Right Panel:} IceCube lower limits on dark matter lifetime versus dark matter mass assuming a 100\% branching ratio decay to $b\overline{b}$ (full line), $\mu^+\mu^-$ (dashed line),
compared with limits from HAWC, both for Dwarf Spheroidal Galaxies~\cite{Albert:2017vtb} and the Galactic Centre/Halo~\cite{Abeysekara:2017jxs}, and Fermi/LAT~\cite{Ackermann:2012rg}. Figures from~\cite{Aartsen:2018mxl}.}\label{fig:decay}
\end{figure*}
%%%%%%%%%%%%%%%% 

If dark matter decay is responsible for the high-energy neutrino emission, the arrival directions of TeV-PeV neutrino events observed with IceCube should correlate with the line-of-sight integral of the dark  matter distribution (``$D$-factor''). The contribution of neutrinos from dark matter decay in the Galactic halo can be similar to the isotropic extra-galactic contribution. Half of the neutrino events from dark matter decay in the halo are predicted to fall within $60\text{\textdegree}$ around the Galactic Centre. This introduces a weak large scale anisotropy that can be tested against the observed event distribution~\cite{Bai:2013nga,Esmaili:2014rma,Ahlers:2015moa}. The neutrino emission from galaxies and galaxy clusters could also be identified as (extended) point-source emission in future IceCube searches~\cite{Murase:2015gea}.

It can be expected that the secondary emission from decaying dark matter scenarios will also include other standard model particles that can be constrained by multi-messenger observations. In particle the production of PeV $\gamma$-rays, that have a pair production length of $\mathcal{O}(10)$~kpc in the CMB, would be a smoking gun for a Galactic contributions~\cite{Murase:2015gea}. However, also the the secondary GeV-TeV emission from electro-magnetic cascades initiated by PeV $\gamma$-rays, electrons, and positrons can constrain the heavy dark  matter decay scenario~\cite{Murase:2015gea,Esmaili:2014rma,Esmaili:2015xpa,Cohen:2016uyg,Blanco:2017sbc}.

From an experimental point of view, the search for a neutrino signal from heavy dark matter decay follows closely the strategies used for the search for annihilation signatures. A given analysis can be used for both constraining dark matter lifetime and annihilation, as can be inferred from the similarities between Eqs.~(\ref{eq:DMann}) and~(\ref{eq:DMdec}). A feature of searches for dark matter decay in the $\geqslant$ 100 TeV mass region is that the recently discovered astrophysical neutrino flux becomes a background to the search, and its current uncertainties on normalisation and the presence, or not, of a cut-off effect the interpretation of any search for an additional component due to dark matter decay. IceCube has performed a search for signatures of heavy dark matter decay by assuming that the background is a combination of atmospheric and astrophysical neutrinos, where the normalisation and energy dependence of the astrophysical flux is allowed to flow freely~\cite{Aartsen:2018mxl}. This allows for deviations from a strict power law which could be interpreted as an additional contribution from dark matter decays. The signal consists of both a galactic component and a diffuse component from the contribution of distant galaxies. Typical neutrino spectra from the decay of a 2 PeV dark matter candidate, assuming 100\% decay to each channel,  are shown in the left panel of Fig.~\ref{fig:decay}, featuring the characteristic peak at the dark matter mass from those decays where the neutrino takes most of the initial available energy. Recent results from IceCube on the dark matter lifetime as a function of the dark matter mass are shown in the right panel of Fig.~\ref{fig:decay}, compared with results from HAWC and Fermi/LAT. Values below the lines are disfavoured at 90\% confidence level, with IceCube limits showing the strongest constraint for masses above about 10~TeV.

\section{Magnetic Monopoles}
\label{sec:monopoles}
Maxwell's equations of classical electrodynamics appear to be asymmetric due to the absence of magnetic charges. However, this is merely by choice. We can always redefine electric and magnetic fields by a suitable duality transformation, ${\bf E}' + i{\bf B}' = e^{i\phi}({\bf E} + i{\bf B})$, such that Maxwell's equations in terms of the new fields are completely symmetric. However, this duality transformation requires that for every particle the ratio between magnetic and electric charges are the same. If this is not the case, then it is necessary to include a source term for magnetic charges $q_m$ ({\it i.e.}, magnetic monopoles), that create a  magnetic field of the form ${\bf B} = q_m{\bf e}_r/4\pi r^2$.  

P. Dirac was the first to speculate about the existence of magnetic monopoles, guided\footnote{Previously, he had successfully used this method to propose the existence of electrons with positive charge~\cite{Dirac:1928hu}, {\it i.e.}, positrons.} by ``new mathematical features'' appearing in the quantum-mechanical description of electrodynamics~\cite{Dirac:1931kp}. He showed that the presence of a magnetic monopole with a minimum charge $q_m$ would explain why the electric charge is always quantised, {\it i.e.}, in integer multiples of an elementary charge. His argument was based on the observation that the magnetic potential of the static magnetic monopole field, ${\bf B}={\rm rot}\, {\bf A}$, has to be singular along a semi-infinite line. The wave function of a particle with charge $q_e$ encircling this line will pick up a phase that is equal to $\Delta\phi = \pm q_eq_m$. To be unobservable, this phase should be a multiple of $2\pi$ independent of the type of monopole or electrically charged particle. This leads to the condition\footnote{Note, that in a Gaussian system, where the unit of charge is defined via $e^2 =\hbar c \alpha$, this quantisation condition is expressed as $q_m q_e = n\hbar c/2$. However, to avoid confusion, we use the Heaviside-Lorentz system with the conventional definition $e^2 =4\pi\alpha$ (in natural units).}
\begin{equation}
	 q_m q_e = 2\pi n \qquad\text{($n\in\mathbf{Z}$)}\,.
\label{eqn:dirac-cond}
\end{equation}
At the time of Dirac's seminal work, the smallest electric charge unit was considered to be that of the electron, $q_e = e$, and hence the smallest possible (Dirac) magnetic charge would be
\begin{equation}
g_{\rm D} = 2\pi/e = e/2\alpha \simeq 68.5 e \ .
\label{eqn:dirac-charge}
\end{equation}
This elementary monopole charge derives from basic considerations still valid in modern physics. In the Standard Model the smallest charge is provided by the down-type quarks, increasing the minimal magnetic charge to $3g_{\rm D}$. Note that the large magnetic charge of monopoles predicted by the Dirac quantisation condition~(\ref{eqn:dirac-charge}) leads to distinct electromagnetic features as monopoles pass through matter. We will return to this point shortly. 

In solids, structures have been found which resemble poles. These are sometimes mistakenly called magnetic monopoles, although the poles can only occur in pairs and do not exist as free particles. For distinction, recently the term {\em magnetricity} has been coined for the field that exhibits theses poles. Fundamental magnetic monopoles have not been observed so far.

%%%%%%%%%%%%%%%%
\begin{figure}[t]
\centering\includegraphics[width=0.9\linewidth]{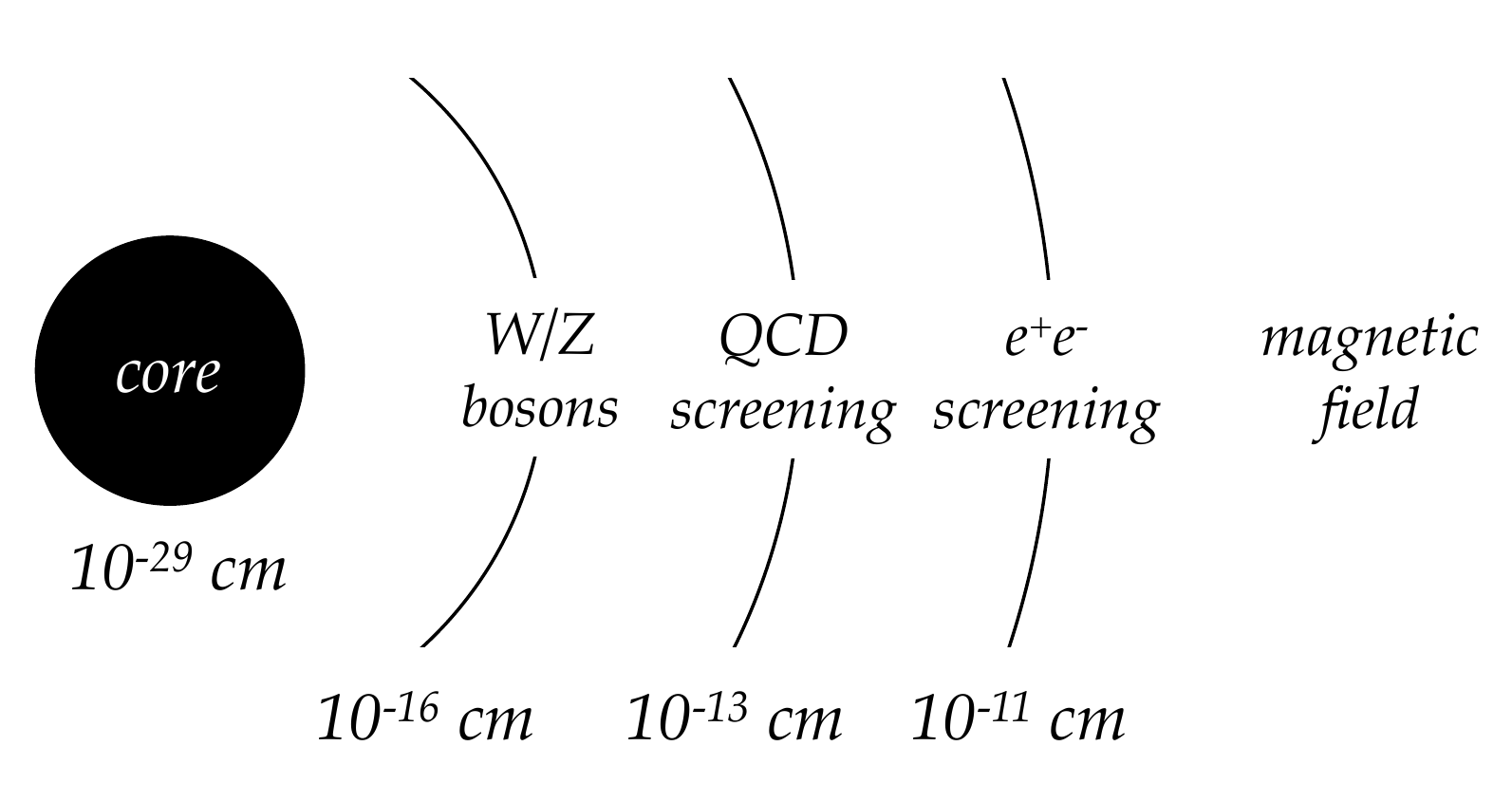}
\caption[]{Structure of a GUT monopole with virtual gauge bosons populating the outer spheres (after Ref.~\cite{Preskill:1984gd}).} 
\label{fig:onion}
\end{figure}
%%%%%%%%%%%%%%%%

Although, as Dirac showed, magnetic monopoles can be consistently described in quantum theory, they do not appear automatically in that framework. As was first found independently by {\it 't Hooft}~\cite{tHooft:1974kcl} and {\it Polyakov}~\cite{Polyakov:1974ek}, this is different in Grand Unified Theories (GUT) which embed the Standard Model interactions into a larger gauge group. These theories are motivated by the observation that the scale-dependent Standard Model gauge couplings seem to unify at very high energies. Generally, a {\it 't Hooft--Polyakov} monopole can arise via spontaneously breaking of the GUT group via the Higgs mechanism. The stability of the monopole is due to the Higgs field configuration which cannot smoothly be transformed to a spatially uniform vacuum configuration. 

An early candidate for a GUT theory studied in this context has been the {\it Georgi--Glashow} model~\cite{Georgi:1974sy}, where matter is unified in SU(5) representations, that spontaneously break to the Standard Model gauge representations below the unification scale $\Lambda_{\rm GUT}\simeq10^{15}$~GeV. This is related to the mass of the monopole as
\begin{equation}
M_{\rm m} \simeq \frac{\Lambda_{\rm GUT}}{\alpha_{\rm GUT}} \simeq 10^{16-17}{\rm GeV}\,,
\end{equation}
where $\alpha_{\rm GUT} = g^2/4\pi$ is the gauge coupling constant at the unification scale. If the original unified group undergoes secondary symmetry breaking at lower energies also monopoles with lower masses may be generated. Depending on details of the GUT model, the monopole mass can range from $10^7$~GeV to $10^{17}$~GeV~\cite{Preskill:1984gd,Wick:2000yc}.

The unification scale is related to the size of the mono\-pole as $r_{\rm M}\simeq \Lambda_{\rm GUT}^{-1}\simeq 10^{-29}$~cm. Larger radii correspond to different energy scales reflecting various transitions in the Standard Model, in particular, the electroweak transition scale $M_Z^{-1}\simeq 10^{-16}$~cm and the confinement scale $\Lambda_{\rm QCD}^{-1}\simeq 10^{-13}$~cm (see Fig.~\ref{fig:onion}). The presence of virtual particles within these ``shells'' influences the monopole's interaction with matter. For instance, within the monopole core, the GUT gauge symmetry is restored, and can mediate baryon-number violating processes. At large distances, only electromagnetic interactions are visible by the magnetic monopole field.

Although the greatest interest has been in the supermassive monopoles that are motivated by GUTs, with the advent of the LHC and the MoEDAL experiment~\cite{Acharya:2014nyr} the possibility of lighter monopoles lately received renewed attention. In the electroweak theory, with $\textrm{SU(2)} \times \textrm{U(1)}$ broken to U(1), there are no topologically nontrivial configurations of the Higgs field, and hence no topologically stable monopole solutions exist. However, there exist specific modifications with a non-minimal coupling of the Higgs field that would allow electroweak mono\-poles with TeV-scale masses, {\it e.g.}, Refs.~\cite{Cho:2013vba,Ellis:2016glu}.

\subsection{Cosmological Bounds}

While inaccessible to collider experiments, very heavy GUT monopoles could have been produced in the early universe if the  temperature exceeds $T_{\rm cr}\sim \Lambda_{\rm GUT}$. In this case, the expected monopole density would be roughly one per correlated volume, which corresponds to the horizon size in a second-order phase transition~\cite{Kibble:1976sj,Kibble:1980mv}. This gives a naive monopole energy density (relative to the critical density) of
\begin{equation}\label{Omono}
  \Omega_{\rm GUT}h^2 \simeq 10^{20}\Big(\frac{T_{\rm cr}}{10^{16}\,{\rm       GeV}}\Big)^3\Big(\frac{M_{\rm m}}{10^{17}\,{\rm GeV}}\Big)\,.
\end{equation}
This prediction is in clear conflict with the observed spatial flatness of the Universe ($\Omega_{\rm tot} \simeq 1$) and known as the ``monopole problem''.

An elegant solution to this problem is an inflationary universe, {\it i.e.}, a universe that underwent an exponential expansion of the scale factor, diluting any initial monopole abundance to an (almost) unobservable level. This inflationary mechanism is a very powerful idea since it simultaneously explains why our Universe has been extremely flat at early times (flatness problem), {\it e.g.}, $\Omega-1\simeq10^{-16}$ at the epoch of big bang nucleosynthesis, and why the Universe appears to be so homogenous over causally disconnected distances (Horizon problem), {\it e.g.}, temperature fluctuations in the CMB of only $10^{-5}$.

The requirement that the contribution of monopoles in Eq.~(\ref{Omono}) does not exceed today's dark matter abundance, $\Omega_{\rm GUT}\lesssim\Omega_{\rm m}$, results in the overclosure bound on the integrated isotropic flux
\begin{equation}
F_{\rm GUT} \lesssim \frac{10^{-15}}{{\rm cm}^{2}\,{\rm sr}\,{\rm s}}\bigg(\frac{\Omega_{\rm       m}h^2}{0.13}\bigg)\bigg(\frac{v}{10^{-3}c}\bigg)\bigg(\frac{10^{17}\,{\rm GeV}}{M_{\rm m}}\bigg)\,.
\end{equation}
If monopoles cluster in our local galaxy this bound can be relaxed by several orders of magnitude. Taking the solar halo density $\rho_{\rm halo} \simeq 10^{-24}\,{\rm g}\,{\rm cm}^{-3}$ we obtain the limit
\begin{equation}
F_{\rm GUT} \lesssim \frac{10^{-11}}{{\rm cm}^{2}\,{\rm sr}\,{\rm s}}\bigg(\frac{v}{10^{-3}c}\bigg)\bigg(\frac{10^{17}\,{\rm GeV}}{M_{\rm m}}\bigg)\,.
\end{equation}

\subsection{Parker Bound}

Magnetic monopoles are accelerated in magnetic fields -- analogously to charged particle acceleration in electric fields. Therefore, relic monopoles that are initially non-relativistic are expected to gain energy while they travel along galactic and intergalactic magnetic fields. The requirement that monopoles have to be rare not to short-circuit these magnetic fields gives the so-called \emph{Parker   bound}~\cite{Parker:1970xv}. The galactic magnetic field with a strength of a few $\mu$G can be generated by a dynamo action on a time scale that is comparable to the Milky Way's rotation period, $\tau\simeq10^8$~yr. A monopole with magnetic charge $q_m$ will gain an energy of $\Delta E_{\rm kin} = \Delta\ell B q_m$ after it travels a distance $\Delta\ell$ along magnetic field lines. The power density of the galactic dynamo $\sim B^2/\tau$ should be larger than the energy drained by the magnetic monopoles.

During its acceleration the monopole encounters different magnetic field orientations coherent over a length scale $\lambda_c$ which are much smaller than the typical size $r$ of the magnetic halo. If the monopoles stay non-relativistic, {\it i.e.}, $M_{\rm m}\ll q_{\rm m} B \lambda_c$, the energy gain is always large compared to the kinetic energy and the particle will be accelerated. The processes can be estimated by a random walk with $N\simeq r/\lambda_c$ encounters with coherent field regions. For weak magnetic fields, {\it i.e.}, $M_{\rm m}\gg q_{\rm m} B \lambda_c$ this process loses its efficiency since the monopole does not follow the potential drop along the field lines. A careful analysis for our own galaxy gives the following bound for masses $M_{\rm m}\lesssim10^{17}$~GeV~\cite{Turner:1982ag}
\begin{equation}\label{parker}
F_{\rm GUT} \lesssim \frac{10^{-15}}{{\rm cm}^{2}\,{\rm sr}\,{\rm s}}\bigg(\frac{B}{3\,\mu{\rm       G}}\bigg)\bigg(\frac{3\times10^7\,{\rm       yr}}{\tau}\bigg)\bigg(\frac{r/\lambda_c}{100}\bigg)^{1/2}\,.
\end{equation}
Very massive monopoles $M_{\rm m}\gtrsim10^{17}$~GeV will not be significantly deflected by the galactic magnetic field, since the acceleration across the galaxy does not change much of the initial virial velocity $v\sim10^{-3}c$. The energy drain of the field by these monopoles depends thus on the initial monopole trajectory with respect to the field lines. To first order, the effect from the motion of these heavy monopoles and their anti-monopoles across the galaxy will cancel. However, the effect will be visible at second order which introduces a mass dependence $\Delta E_{\rm kin}\propto M_{\rm m}$. The Parker bound beyond $M_{\rm m}\gtrsim10^{17}$~GeV is hence weaker than (\ref{parker}) by a factor $(M_{\rm m}/10^{17}~{\rm GeV})$. Applying Parker's arguments to the seed magnetic fields of galaxies or galaxy clusters leads to tighter bounds~\cite{Rephaeli:1982nv,Adams:1993fj}. As these bounds are less secure, and for consistency with other literature on the subject, we compare the experimental results with the original Parker bound (see Fig.~\ref{fig:MM-limits}).

%%%%%%%%%%%%%%%%
\begin{figure}[t]
	\centering
	\includegraphics[width=0.8\linewidth]{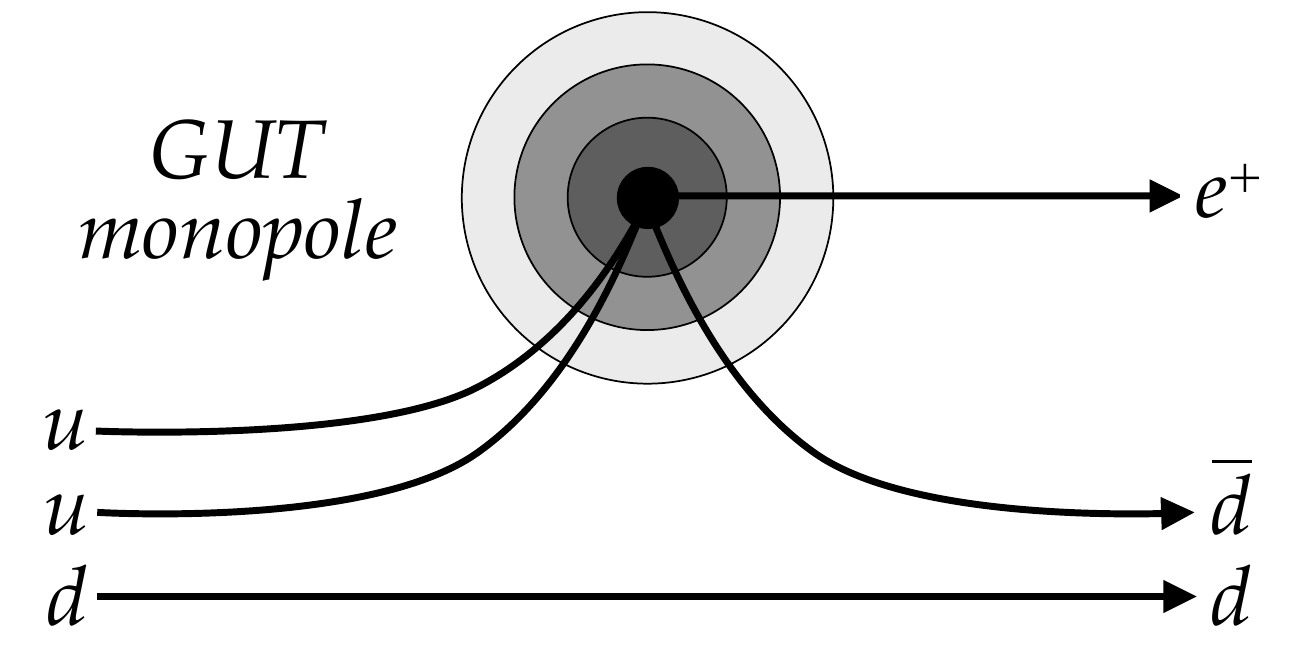}
	\caption[]{Illustration of a proton decay into a positron and a neutral pion catalysed by a GUT monopole (after Ref.~\cite{Patrizii:2015uea}).}
	\label{fig:protondecay}
\end{figure}
%%%%%%%%%%%%%%%%

\subsection{Nucleon Decay Catalysis}\label{sec:MMcatalysis}

%%%%%%%%%%%%%%%%
\begin{figure}[t]
	\centering
	\includegraphics[width=\linewidth]{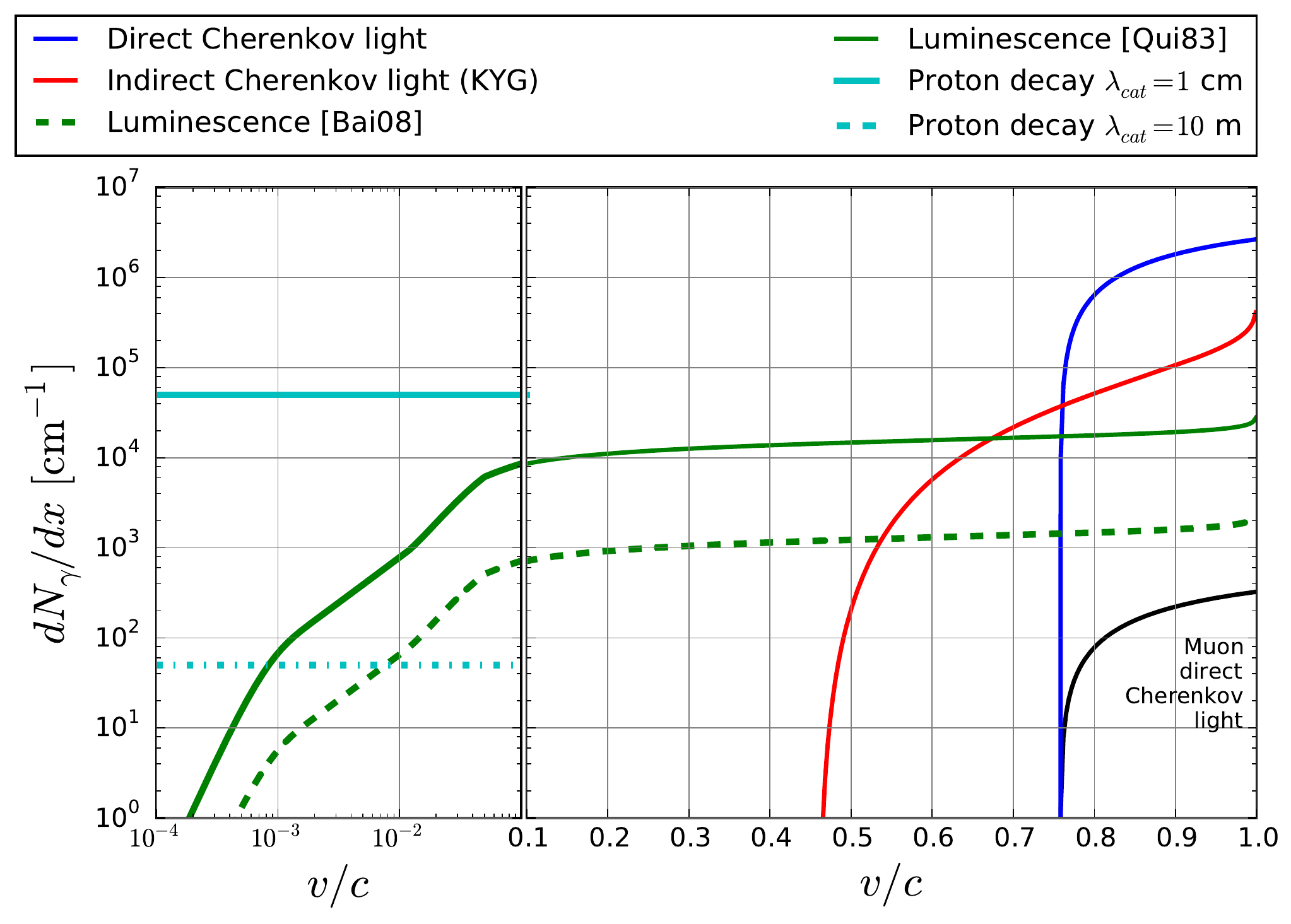}
	\caption[]{Light yield for the different radiation mechanisms of magnetic monopoles~\cite{ObertackePollmann:2016uvi}. For comparison the direct Cherenkov light emitted by a minimum ionising muon is shown.
	}
	\label{fig:lightyield}
\end{figure}
%%%%%%%%%%%%%%%

The central core of a GUT monopole (see Fig.~\ref{fig:onion}) contains the fields of the superheavy gauge bosons that mediate baryon number violation, so one might expect that baryon number conservation could be violated in baryon-monopole scattering and the possibility that a proton or a neutron in contact with a GUT monopole can decay. This feature was pointed out by {\it Callan}~\cite{Rubakov:1982fp} and {\it Rubakov}~\cite{Callan:1983nx}. The cross sections for the catalysis processes such as $p + M \rightarrow e^{+} + \pi^{0} + M$ (Fig.~\ref{fig:protondecay}) are essentially geometric:
\begin{eqnarray}
\sigma_{\mathrm{cat}} = \begin{cases}
\frac{\sigma_0}{\beta} &\text{ for }\beta \geq \beta_0\,,\\
\frac{\sigma_0}{\beta} \cdot F(\beta) &\text{ for }\beta < \beta_0\,.
\end{cases}
\label{eqn:sigma_cat_theo}
\end{eqnarray}
The correction $F(\beta) = \left( \frac{\beta}{\beta_0} \right)^\gamma$ takes into account an additional angular momentum of the monopole-nucleus system. Both parameters $\gamma$ and $\beta_0$ depend on the nucleus~\cite{Arafune:1983uz}. Current estimates for the catalysis cross sections are of the order of $10^{-27}\,\mathrm{cm}^2$ to $10^{-21}\,\mathrm{cm}^2$~\cite{Nath:2006ut}.

\subsection{Monopole Searches with IceCube}

The experimental search for magnetic monopoles has a long history. Searches were pursued at accelerators, in cosmic rays, and for bound monopoles in matter. Detection methods include induction in SQUIDs, the observation of excessive energy loss of the monopole compared to Standard Model particles and particles describing non-helical paths in a uniform magnetic field, or other unusual trajectories like non-relativistic velocities combined with a high stopping power and long ranges. In indirect searches at accelerators virtual monopole processes are assumed to influence the production rates of final states. In a direct search, evidence of the passage of a monopole through material is sought. Here we primarily address direct searches for primordial monopoles with IceCube.

%%%%%%%%%%%%%%%%
\begin{figure*}[t]
\centering
\includegraphics[height=0.3\linewidth]{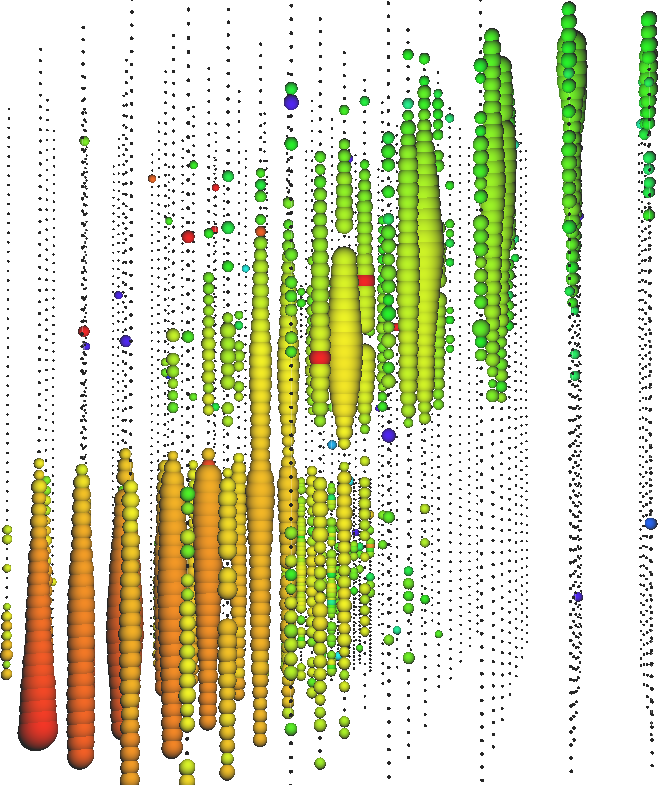}
\hspace{1cm}
\includegraphics[height=0.3\linewidth]{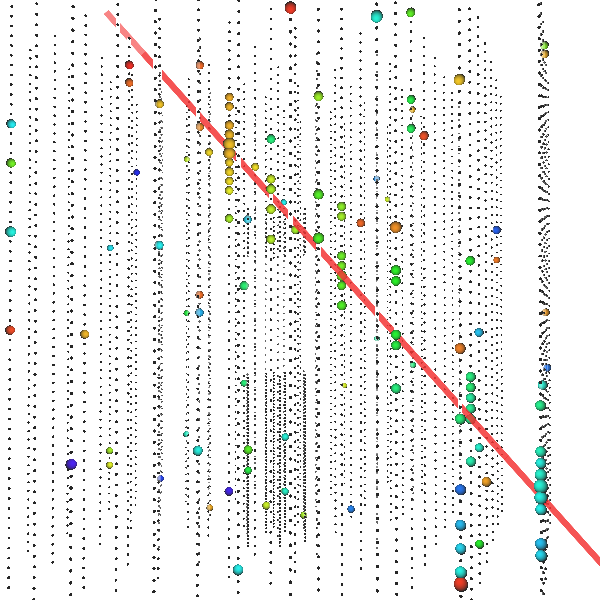}
\hspace{0.5cm}
\includegraphics[height=0.3\linewidth]{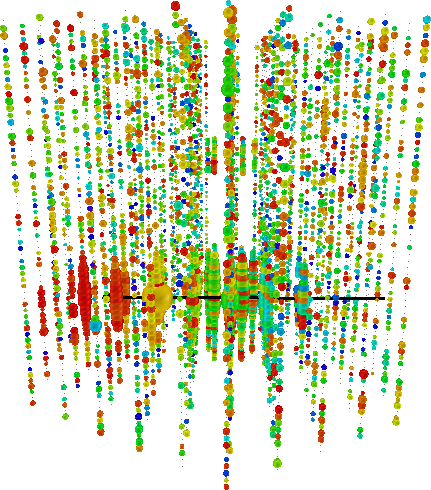}
\caption[]{Simulated event topologies for magnetic monopole of different velocities (see Fig.~\ref{fig:event_examples} for graphical event representation). {\bf Left Panel:} monopole traveling at $0.982 c$ from the bottom to the top of the detector~\cite{Lauber:2017}. The total time of the event is 5000~ns. Less light has been detected in a horizontal plane~\cite{Aartsen:2016nxy} roughly at the half height of the detector due to dust in the ice. {\bf Middle Panel:} monopole at $0.3 c$ moving from the top of the detector to the bottom emitting luminescence light. The simulated track of the particle is indicated in red. Only a few noise hits contribute throughout the detector due to the short time frame of the event~\cite{Lauber:2017}. {\bf Right Panel:} Monopole with $\beta = 10^{-3}$ catalysing nucleon decay with $\lambda_{\mathrm{cat}} = 1\,\mathrm{cm}$ with superimposed background noise~\cite{Aartsen:2014awd}. The black line represents the simulated monopole track.}
\label{fig:MonopolePassage}
\end{figure*}
%%%%%%%%%%%%%%%%
	
Different light production mechanisms induced by monopoles dominate depending on their velocity (see Fig.~\ref{fig:lightyield}): 

\noindent{\it (i)} Direct Cherenkov light is produced at highly relativistic velocities above $ 0.76\, c$ as with any other charged Standard Model particle. Due to the high relative Dirac charge, as shown in Eq.~(\ref{eqn:dirac-charge}), several thousand times more light is radiated with a monopole than from a minimum ionising singly electrically charged particle like a muon~\cite{Aartsen:2015exf}. 
 
\noindent{\it (ii)} Indirect Cherenkov light from secondary knock-off $\delta$-electrons is relevant at mildly relativistic velocities $(\simeq 0.5\, c \textrm{ to } 0.76\, c)$. The high-energy $\delta$-electrons in turn can have velocities above the Cherenkov threshold themselves. The energy transfer of the monopole to the $\delta$-electrons can be inferred from the differential cross section calculated by {\it Kasama}, {\it Yang} and {\it Goldhaber} (KYG)~\cite{Wu:1976ge,Kazama:1976fm}.

\noindent{\it (iii)} Luminescence light from excitation of the ice dominates at low relativistic velocities $(\simeq 0.1\, c \textrm{ to } 0.5\, c)$. The observables of luminescence,
	such as the wavelength spectrum and decay times, are dependent on the properties
	of the medium, in particular, temperature and purity. The signature is relatively dim in comparison to muon signatures. Pending further laboratory measurements in ice and water~\cite{ICRC-Lumi:2017aaa}, the efficiency of luminescence photon production per deposited energy is in the range of  ${\rm d}N_{\gamma}/{\rm d}E=0.2\,\gamma/\mathrm{MeV}$ and $2.4\,\gamma/\mathrm{MeV}$~\cite{Baikal08,Quickenden82}.
	Even for the lower plausible light yield, luminescence is a viable signature due to the high excitation of the medium induced by a monopole~\cite{ObertackePollmann:2016uvi}.

\noindent{\it (iv)} At velocities well below $0.1~c$ luminescence is expected to fall off (see Fig.~\ref{fig:lightyield}). The catalysis of nucleon decays is a plausible scenario for GUT monopoles (see Sec.~\ref{sec:MMcatalysis}) and may be observed if its mean free path is small compared to the detector size. The Cherenkov light from secondaries emitted in nucleon decays along the monopole trajectory can lead to a characteristic slow moving event pattern across the detector~\cite{Aartsen:2014awd}.

For each of these speed ranges, searches for magnetic monopoles at the IceCube experiment are either in pro\-gress (luminescence) or have already set the world's best upper limits on the flux of magnetic monopoles over a wide range of velocities. Examples of magnetic monopole passing through the detector at different velocities are shown in Fig.~\ref{fig:MonopolePassage}. 

\begin{figure*}[t]
\centering
\includegraphics[width=0.8\linewidth]{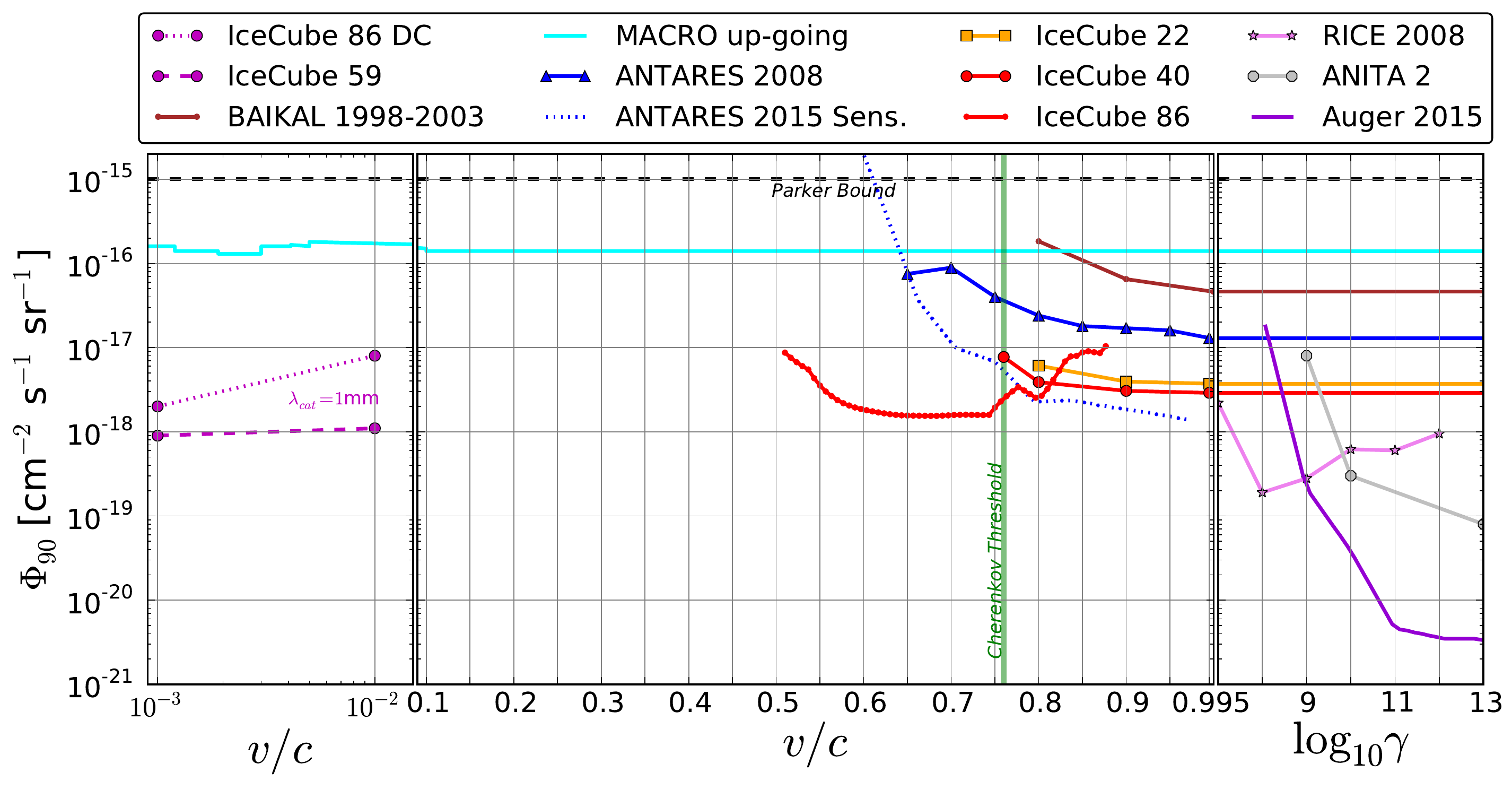}
\caption[]{Upper limits on the flux of magnetic monopoles as a function of the velocity of the monopole at the detector~\cite{ObertackePollmann:2016uvi}. 
Shown are limits from IceCube~\cite{Aartsen:2014awd,Abbasi:2012eda,Aartsen:2015exf}, Baikal~\cite{Baikal08}, ANTARES~\cite{AdrianMartinez:2011xr,Albert:2017fud}, RICE~\cite{Hogan:2008sx}, ANITA~\cite{Detrixhe:2010xi} and Auger~\cite{Aab:2016poe}.}
\label{fig:MM-limits}
\end{figure*}

These unique signatures from monopoles lead to the following general analysis strategies in the different velocity regions:

\noindent{\it (i)} Relativistic monopoles are selected based on their brightness, arrival direction, and velocity~\cite{Aartsen:2015exf}. The high energy astrophysical neutrino flux is an important background to this signal. At similar brightness, monopoles show less stochastic energy loss than Standard Model particles leading to a smoother light yield distribution along the track.

\noindent{\it (ii)} Due to its lower rest mass a Standard Model particle with a velocity below the  speed of light in vacuum, $c$, would not be able to traverse the whole detector. However, the discrimination power of the reconstruction of the velocity is insufficient for the suppression of the vast air shower backgrounds against the identification of mildly relativistic monopoles. Instead variables describing the topology, smoothness, and time distribution of the events are processed in a Boosted Decision Tree (BDT) machine learning~\cite{Aartsen:2015exf}.

\noindent{\it (iii)} Searching for mildly relativistic monopoles using luminescence light can be performed using analysis techniques that combine the non-relativistic reconstructed particle velocity and the continuous but dim light production of a through-going track in the detector~\cite{ObertackePollmann:2016uvi,Lauber:2017}.

\noindent{\it (iv)} 
Catalysed nucleon decay, like $p + M \rightarrow e^{+} + \pi^{0} + M$, transfers almost all of the proton's rest mass to the energy of electromagnetic and hadronic cascades. Because of the high light yield this channel is typically used as a benchmark in analyses. Due to their low speed, the duration of the event is in the order of 10 milliseconds. As obvious from the right event display in Fig.~\ref{fig:MonopolePassage} at such timescales random noise pulses are a significant contribution. Various effects contribute to subtle temporal correlations on long time scales of this noise~\cite{Ackermann:2006gp,Aartsen:2016nxy} complicating an adequate description in simulation. Instead, a background model is established from reshuffling experimental data. For signal identification, time-isolated local coincidences in neighbouring DOMs are searched for along a monopole trajectory hypothesis consistent with a straight particle track of constant speed. A Kalman filter is used to separated noise from monopole signals and a combination of observables are fed to a BDT to further improve the signal purity~\cite{Aartsen:2014awd}.

Figure~\ref{fig:MM-limits} shows a compilation of current flux upper limits of relic monopoles from various experiments. Only in the past decade astrophysical experiments have been able to improve upon the original Parker bound which is shown for comparison. These recent experiments have employed large scale detectors for cosmic rays to achieve the highest sensitivities in the whole $\beta$ range in which GUT magnetic monopoles are expected. Typically it is assumed that the flux at the respective detector site is isotropic implying sufficient kinetic energy in order to cross the Earth or the overburden above the detector due to their large rest masses. This assumption is justified for monopoles of masses in excess of $10^{10}$~GeV. The monopole flux limits commonly assume a single Dirac magnetic charge $q_m = g_D$ (see Eq.~(\ref{eqn:dirac-charge})) with no additional electric charge. In most detectors and velocity ranges, the detection efficiency for larger magnetic charges or for electrically charged monopoles (dyons) is expected to increase.

Operational until 2000, MACRO searched for magnetic monopoles using three types of subdetectors -- liquid scintillation counters, limited streamer tubes and nuclear track detectors. No monopole was found, with an upper flux limit at the 90\% confidence level of $1.4 \times 10^{-16} \rm cm^{-2} s^{-1} sr^{-1}$ for monopoles with velocity between $4 \times 10^{-5} \  c$ to $c$ and magnetic charge with $n \ge 1$~\cite{Ambrosio:2002qu,Ambrosio:2002qq}. Under the assumption that monopoles are gravitationally accumulated in the centre of the Sun, Super-Kamiokande~\cite{Ueno:2012md} could impose stringent limits for non-relativistic velocities in an indirect search for neutrinos from the direction of the Sun. Baikal~\cite{Baikal08} has investigated the direct Cherenkov light from relativistic monopoles. The analysis by ANTARES includes also the mildly relativistic regime employing similar techniques like IceCube~\cite{Albert:2017fud}.  The reduced scattering in water at the ANTARES site compared to ice leads to a better velocity reconstruction which helps with the Standard Model background suppression. This partly compensates the higher noise level in the detector.

Intermediate and low mass monopoles may acquire highly relativistic velocities in intergalactic magnetic fields reaching Lorentz factors of $\gamma \simeq 10^{10}$ for the example of a PeV-mass~\cite{Wick:2000yc}. Ultra-relativistic particles with magnetic charge (or large electric charge) dramatically loose energy in their passage through matter, initiating a large number of bright showers along the track. At high Lorentz boost factors the photo-nuclear effect is the dominant energy loss mechanism generating hadronic showers. While these showers are continuously produced, they may overlap with each other. In the atmosphere of the Earth this leads to a built-up such that the energy deposit increases with slant depth. The Auger experiment has used this feature to distinguish monopoles from the background of electrically singly charged ultrahigh-energy cosmic rays like protons~\cite{Aab:2016poe}. The RICE~\cite{Hogan:2008sx} and ANITA~\cite{Detrixhe:2010xi} experiments have searched for such multiple sub-shower signatures in the Antarctic ice sheet with the Radio-Cherenkov technique, the discriminant against conventional cosmic rays here being primarily the rapid succession of several radio pulses received from each sub-shower.

The extrapolation of IceCube's limit towards highly relativistic velocities by a constant line in Fig.~\ref{fig:MM-limits} is a very conservative approach. It not only neglects the increase in signal detection efficiency with more energy deposited, but it also ignores the onset of the photo-nuclear effect and pair production. These effects would produce showers with light emission orders of magnitude brighter than from the Cherenkov effect considered here, hence visible also from far outside the instrumented detector volume.

While these flux limits reflect the cosmic density, at the electroweak scale monopoles may be created in accelerator collisions, which is studied at the MoEDAL experiment. Also cosmic ray collisions or high energy neutrino interactions in the Earth may produce monopoles. This might be an additional detection opportunity, also for IceCube.

\section{Other Exotic Signals}
\label{sec:exotics}
The detection principle of Cherenkov telescopes is very general in the sense that it applies to any flux of particles that can penetrate the detector shielding and produce light signals inside the detector volume. We have already covered the possibility to observe relic magnetic monopoles in the previous section. In this section we will discuss the detection potential for other exotic candidates like Q-balls and strangelets. We will also address the possibility that long-lived charged particles produced in cosmic ray or neutrino interactions may be discovered via their Cherenkov emission. All these particles have in common that their passage through the IceCube detector produces observable features that can be extracted from backgrounds.

\subsection{Q-Balls}

There exist non-topological solitonic solutions of a field theory, so-called \emph{Q-balls}~\cite{Coleman:1985ki}. Whereas the stability of topological solitons, {\it e.g.}, monopoles is guaranteed by the conservation of a topological charge (winding number) associated with a degeneracy of the vacuum state, a Q-ball can be stable by the conservation of a charge associated with a global symmetry of the theory.  This can happen if its energy configuration is lower than the corresponding multi-particle Fock state. For a single complex scalar field $\phi$ carrying the charge $Q$, this implies that there exists a non-trivial field value $|\phi_0|>0$ where the scalar potential $U(\phi^\dagger\phi)$ obeys the condition $U(\phi_0^\dagger\phi_0) < m_\phi^2\phi_0^\dagger\phi_0$, where $m_\phi$ is the mass of the scalar field (for a review see Ref.~\cite{Lee:1991ax}).

The appearance of these flat scalar potentials is generic in supersymmetric (SUSY) field theories, predicting scalar partners for fermions and gauge bosons, that may carry global charges~\cite{Kusenko:1997zq}. Supersymmetry has to be broken at low energies to provide mass terms for the SUSY partners (for a review see, {\it e.g.}, Ref.~\cite{Martin:1997ns} and references therein). Since a naive spontaneous SUSY breaking by renormalisable terms in the visible sector predicts SUSY masses that are in general too light (below TeV), it is typically assumed that the breaking occurs in a {\it hidden} sector at some unobservable high-energy scale. The mediation of SUSY breaking terms to the visible sector, {\it e.g.}, by gauge interactions or supergravity generates then soft (renormalisable) SUSY breaking masses for SUSY partners of the Standard Model matter and gauge bosons. Generically, SUSY extensions of the Standard Model predict flat directions for some combination of scalar fields. For instance, gauge-mediated SUSY breaking is expected to introduce a scalar potential with $V\sim m_{\rm F}^4(\log(\phi^\dagger\phi/M_m^2))^2$ (for $|\phi|\gg M_m$) with mass scale $m$ and messenger mass $M_m\gg m$. In this case, the mass of a Q-ball with $Q\gg1$ can be estimated as~\cite{Kusenko:1997vp}
\begin{equation}\label{Qmass}
  m_{\rm Q} \simeq \frac{4\pi\sqrt{2}}{3}mQ^{3/4}\,.
\end{equation}
Naturally, the effective mass scale $m$ is expected to lie within the range 
$100~{\rm GeV} < m < 100$~TeV.

If baryogenesis proceeded via the Affleck-Dine mechanism~\cite{Affleck:1984fy}~\footnote{In this scenario, a combination of MSSM scalar fields with non-zero baryon number B develops a large expectation value at the end of inflation and decays.}, stable Q-balls with $10^{12}\lesssim Q_B\lesssim10^{30}$ could have been formed copiously as a dark matter contribution by the fragmentation of the Affleck-Dine condensate~\cite{Kusenko:1997si}. It is an appealing property of this scenario that the baryon and dark-matter abundance, $\Omega_{B}$ and $\Omega_{\rm DM}$, respectively, are related. For $Q_B\simeq10^{26}$ one obtains $\Omega_{\rm DM} \simeq 10\Omega_{\rm B}$ close to the observed values.

Analogously to monopole densities, Q-ball densities are also limited by the dark matter abundance. The maximal contribution of Q-balls to the observed dark matter $\Omega_{\rm Q}h^2\lesssim\Omega_{\rm m}h^2$ results in a bound on the integrated isotropic flux since $4\pi F_{\rm Q} \lesssim v_{\rm   M}\Omega_{\rm m} \rho_{\rm cr}/m_{\rm Q}$ and
\begin{equation}
  F_{\rm Q} \lesssim \frac{5\times10^{-22}}{{\rm cm}^{2}\,{\rm sr}\,{\rm s}}\bigg(\frac{v}{10^{-3}c}\bigg)\bigg(\frac{1\,{\rm TeV}}{m_{\rm soft}}\bigg)\,\bigg(\frac{10^{26}}{Q_B}\bigg)^{3/4}.
\end{equation}
Here the mass term (\ref{Qmass}) from gauge-mediated SUSY breaking is used. If we assume that dark matter Q-balls cluster in our galaxy with $\rho_{\rm halo} \simeq 10^{-24}\,{\rm g}\,{\rm cm}^{-3}$ we obtain the limit
\begin{equation}
  F_{\rm Q} \lesssim \frac{5\times10^{-18}}{{\rm cm}^{2}\,{\rm sr}\,{\rm s}}\bigg(\frac{v}{10^{-3}c}\bigg)\bigg(\frac{1\,{\rm TeV}}{m_{\rm soft}}\bigg)\,\bigg(\frac{10^{26}}{Q_B}\bigg)^{3/4}\,,
\end{equation}
corresponding to a few events per year and square kilometer in a $2\pi$ sky coverage.

The global charge $Q$ associated with the Q-ball can be the same as baryon number (B) or lepton number (L) or some combination of them if these symmetries are connected to a global ${\rm U}(1)$ symmetry. Since the global symmetry is spontaneously broken in the interior of the Q-ball with $\phi\neq0$, the soliton could catalyse nucleon decay traversing the detector volume. This is analogous to the case of monopole-catalysed nucleon decay. Even if the charge is not related to B or L, it is possible that the vacu\-um state associated with the Q-ball interior catalyses nucleon decay. This can happen if the scalar potential is very flat such that $U(\phi_0^\dagger\phi_0) \ll m_\phi^2\phi_0^\dagger\phi_0$. Interactions that lead to nucleon decay induced by new physics at an ultra-violet scale $\Lambda$, for instance, in grand unified theories are typically suppressed by powers of $\langle\phi\rangle/\Lambda$ and can become large in the Q-ball environment~\cite{Dvali:1997qv}.

The cross section of the catalyzed nucleon decay via Q-balls with baryon number $Q_B$ passing through matter can be approximated by its size. In gauge-mediated SUSY breaking, it can be estimated as~\cite{Kusenko:1997vp}
\begin{equation}
\sigma_{\rm cat} \simeq 10^{-20}\bigg(\frac{Q_B}{10^{26}}\bigg)^{1/2}\bigg(\frac{1~{\rm TeV}}{m_{\rm soft}}\bigg)^{2}{\rm cm}^2\,.
\end{equation}
The experimental signature of this process would thus be similarly spectacular as in the case of nucleon-decay catalysed by monopoles. Hence, the upper limits discussed for slow monopoles can be reinterpreted in terms of electrically neutral Q-balls also for neutrino telescopes. An example of such a recalculation is available for the Baikal experiment in~\cite{Bezrukov:1990mr,Belolaptikov:1998mn} leading to a flux upper limit of $F = 3.9 \times 10^{-16} ~{\rm cm^{-2} sr^{-1} s^{-1}}$ at $\beta = 10^{-3}$ for an assumed cross section of $\sigma_{\rm cat} = 1.9 \times 10^{-22} ~{\rm cm^{2}}$. Since the detection efficiency for Q-balls is comparable to that for slow monopoles catalysing nucleon decay with the same cross section it is to be expected that a reinterpretation of the above mentioned IceCube slow monopole limits would also present an improvement of about two orders of magnitude with respect to other limits.

\subsection{Strange Quark Matter}

Using energy and symmetry arguments, it has been speculated that strange quark matter (SQM), a hypothetical form of matter with roughly equal numbers of up ($u$), down ($d$), and strange ($s$) quarks, could be the true ground state of QCD~\cite{Bodmer:1971we,Witten:1984rs,Farhi:1984qu}.  For a plasma of quarks in thermodynamical equilibrium it might be energetically preferable to condense into a phase containing strange quarks instead of ordinary matter with neutrons ($udd$) and protons ($uud$). 

An approximate thermodynamical calculation with massless quarks and neglecting strong interactions shows that the average kinetic energy per quark in ordinary bulk matter could be reduced in bulk SQM by a factor of about $0.89$~\cite{Witten:1984rs}. Therefore, it is feasible that the extra ``penalty'' paid by the presence of more massive strange quarks is over-compensated by the reduction in energy density. However, the meta-stable state of protons, neutrons, and composite nuclei would be very long-lived, since conversion to the SQM ground state would proceed via weak interactions.

Lumbs of SQM, so-called {\it strangelets}, can have a large atomic mass number $A$ and charge $Z$. Classical strangelets have a quark charge $Z\sim 0.1A$ for low mass numbers ($A\ll700$). For total quark charges exceeding $Z\sim\alpha^{-1}\sim 137$ strong field QED corrections lead to screening and $Z\sim 8A^{1/3}$ ($A\gg700$)~\cite{Farhi:1984qu}. It has also been speculated that colour and flavour symmetries at high baryon densities might be broken simultaneously by the condensation of quark Cooper pairs~\cite{Madsen:2001fu}. In this scenario, the ``colour-flavour-locked'' strangelets have charges of $Z\sim0.3A^{2/3}$~\cite{Madsen:2000kb}.

Stable strangelets can absorb ordinary matter in exothermic reactions involving $u\to s+e+\bar\nu_e$ or $u+d\to s+u$~\cite{Farhi:1984qu}. If the strangelet carries a positive charge the Coulomb barrier will usually prevent a strangelet--nucleus system from collapse. However, neutron-rich environments like neutron stars are not protected by this mechanism. In fact, if strange quark matter is stable then all compact stars like white dwarfs or neutron stars are likely to consist of it. Even the capture of a single strangelet would be sufficient to convert a neutron-rich environment very rapidly.

Slowly moving strangelets -- so called {\it nuclearities}~\cite{DeRujula:1984axn} -- lose their energy in matter due to atomic collisions. The excessive energy released will heat the medium and create thermal shocks. The hot expanding plasma will emit Planck radiation from its surface over a wide range of frequencies. IceCube is sensitive to optical photons energies of about $2\div4$~eV (300~nm to 600~nm). 
At nuclearite velocities expected for cold dark matter the fraction of total energy loss emitted in optical photons is of order $10^{-5}c$. Since this leads to signatures similar to slow magnetic monopoles or Q-Balls the detection efficiency is comparable again. Nuclearites with masses in excess of $10^{14}$~GeV and typical velocities of order of $10^{-3}c$ reach underground detectors. 
Correspondingly, the sensitivity to the flux of nuclearites is roughly of the same order as that to slow monopole fluxes catalysing nucleon decays. Results for the MACRO experiments~\cite{Ambrosio:1999gj} and ongoing studies for the ANTARES detector~\cite{Pavalas:2015nab} addressing fluxes of order $10^{-16}-10^{-15}~{\rm cm^{-2} sr^{-1} s^{-1}}$ underline the high potential for a corresponding reinterpretation of IceCube analyses.

\subsection{Long-Lived Charged Massive Particles}

Many extensions of the Standard Model predict the existence of long-lived charged massive particles (CHAMPs). These particles occur naturally in scenarios where the decay of charged particles is limited by (approximate) discrete symmetries and involves final states that have only very weak couplings. Analogously to muons, these CHAMPs have a reduced electro-magnetic energy-loss in matter due to the suppression of bremsstrahlung by the rest mass. Still, they may be detected by their Cherenkov emission and, due to the long range, even with an enlarged effective detection area. 

In the following we will consider SUSY breaking scenarios where the right-handed stau is the next-to-lightest SUSY particle (NLSP). However, most of the arguments apply equally well to other scenarios of CHAMPs with a decay length larger than other experimental scales (see, {\it  e.g.}, Ref.~\cite{Albuquerque:2008zs}).  If $\mathcal{R}$-parity is conserved the stau NLSP can only decay into final states containing the LSP. Depending on the mass and coupling of this particle the lifetime of the stau NLSP can be very long and, in some cases, it can be considered as practically stable on experimental time-scales. 

In the case of a neutralino LSP, the stau NLSP can be very long-lived if its mass is nearly degenerated with that of the neutralino~\cite{Jittoh:2005pq}.  However, there are also super-weakly interacting candidates for the LSP in extensions of the MSSM, which provide the long life-time of the NLSP more naturally. Possible scenarios include SUSY extensions of gravity with a gravitino $\widetilde{G}$ LSP, SUSY versions of the Peccei-Quinn axion and the corresponding axino LSP~\cite{Brandenburg:2005he}, and the MSSM with right-handed chiral neutrinos and a right-handed sneutrino LSP.

Staus produced in SUSY interactions of EHE neutrinos could traverse Cherenkov telescopes at the level of a few per year, assuming that extragalactic neutrino fluxes are close to the existing bounds~\cite{Albuquerque:2003mi,Bi:2004ys,Ahlers:2006pf,Albuquerque:2006am,Ando:2007ds,Canadas:2008ey} or prompt atmospheric neutrino fluxes close to upper theoretical estimates~\cite{Ando:2007ds}. The leading-order SUSY contribution consists of chargino $\chi$ or neutralino $\chi^0$ exchange between neutrinos and quarks, analogous to the parton-level SM contributions shown in Fig.~\ref{fig:CCNC}. The reactions produce sleptons and squarks, which promptly decay into lighter stau NLSPs.

Cosmic ray interactions in the atmosphere could also produce a long-lived stau signal in neutrino telescopes~\cite{Ahlers:2007js} (see also Ref.~\cite{Ando:2007ds}).  At energies above $10^4$~GeV the decay length of weakly decaying nucleons, charged pions and kaons is much larger than their interaction length in the air. Therefore, as they propagate in the atmosphere the probability that a long-lived hadron ($h$) collides with a nucleon to produce SUSY particles is just $\sigma_h^{\rm SUSY} / \sigma_h^{\rm SM}$, where $\sigma_{h}^{\rm SUSY}$ is the cross section to produce the SUSY particles $X$ (gluinos or squarks) and $\sigma_h^{\rm SM}$ the Standard Model cross section of the hadron with nuclei in the atmosphere (which is also the total cross section to a good approximation). However, since $\sigma_{h}^{\rm SM}$ is above $100$~mb, it is apparent that this probability will be very small and that it would be much larger for a neutrino propagating in matter.

The energy of charged particle tracks observed in neutrino detectors is determined by measuring their specific energy loss, ${\rm d}E/{\rm d}x$.  For muons with energies above an energy  of about 500~GeV, the energy loss rises linearly with energy. However, since the energy loss depends on the particle Lorentz boost, a high-energy stau is practically indistinguishable from a muon with reduced energy, $E_\mu/E_{\widetilde{\tau}} \sim m_\mu/m_{\widetilde{\tau}}$. A smoking-gun signal for pair-produced stau NLSPs are parallel tracks in the Cherenkov detector~\cite{Albuquerque:2003mi}.

The detection efficiency of stau pairs, {\it i.e.}, coincident parallel tracks, depends on the energies and directions of the staus, as in the case of single events, but also on their separation. A large fraction of staus reaching the detector will be accompanied by their stau partner from the same interaction. However, not all of the stau pairs might be seen as separable tracks in a Cherenkov telescope if they emerge from interactions too close to or also too far from the detector. The opening angle $\theta_{\widetilde\tau\widetilde\tau}$ between staus can be estimated by the initial opening angle between the SUSY particles in deep inelastic scattering. The separation of stau tracks in the detector is then given as $x\simeq2\,\Delta\ell\,\tan(\theta_{\widetilde\tau\widetilde\tau}/2)$, where $\Delta\ell$ is the distance of the interaction to the detector center.

Double tracks with low track separation are difficult to distinguish from single muons copiously produced either directly in air showers or from atmospheric neutrinos. A required minimum track separation in IceCube of 150~m was found to be necessary to suppress these muons, due to the geometry of the detector~\cite{Kopper:2015rrp}. A possible background to the stau pair signal produced in neutrino interactions consists of parallel muon pair events from random coincidences produced by upgoing atmospheric neutrinos. However, this is bounded from above by the number of muons arriving within a coincidence time-window requiring them to be almost parallel and is several orders of magnitude below the stau pair event rate with $N_{\mu+\mu} \lesssim \mathcal{O}(10^{-12})N_\mu$~(Ref.~\cite{vanderDrift:2013zga}). Muon pairs from charged current muon-neutrino interactions involving final state hadrons that promptly decay into a second muon are expected to be more likely. This has been estimated in Ref.~\cite{Albuquerque:2006am} for the production and decay of charmed hadrons. 

The rate of stau pairs from neutrino production is largely uncertain and depends on the SUSY mass spectrum. If the SUSY mass spectrum close to observational bounds (see, {\it e.g.}, Ref.~\cite{PDG:2018}), the rate might reach a few events per decade in cubic-kilometer Cherenkov telescopes~\cite{Albuquerque:2003mi,Ahlers:2006pf,Albuquerque:2006am,Ando:2007ds}.

\begin{figure}[t]
	\centering\includegraphics[width=0.9\linewidth]{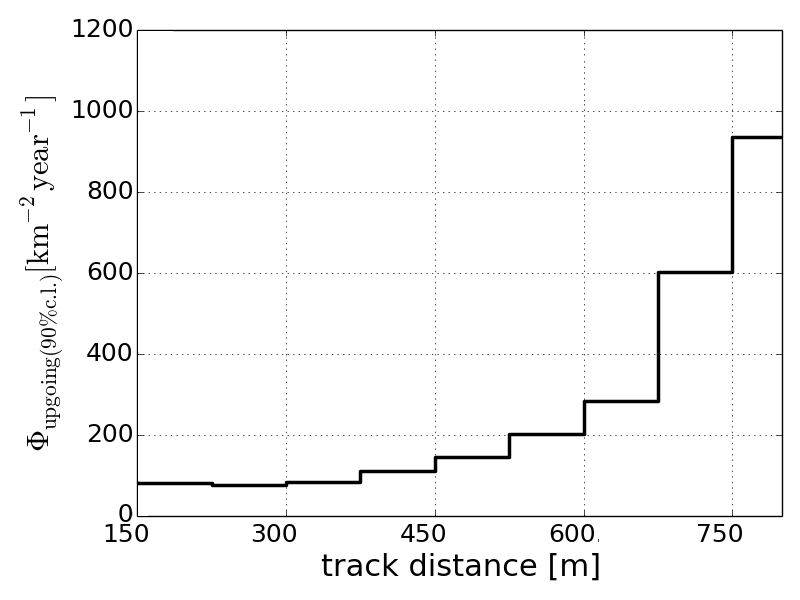}
	\caption[]{Upper limit on the flux of the double tracks at IceCube assuming a uniform arrival direction distribution for up-going tracks~\cite{Kopper:2015rrp}.}
	\label{fig:DoubleTracks}
\end{figure}

Experimentally, the situation over the years has become ever more challenging. The SUSY models typically studied to be in reach of colliders, today have lower limits of their mass scales of the order of TeV from the non-observation at the LHC~\cite{PDG:2018}. This leads to significantly reduced expected opening angles (respectively, track separations) of the double tracks and low production cross sections in these models. However the possibility exists that a model beyond the current Standard Model is realised in nature with properties that escape observation at current collider detectors. Hence, the flux of double tracks has been probed generically with IceCube~\cite{Kopper:2015rrp}. It turns out, a flux of hundreds of double tracks with a separation of above 150 m may pass undetected (Fig.~\ref{fig:DoubleTracks}). This is largely due to the challenge of requiring two very faint tracks to be reconstructed with high accuracy while the detector is optimised for higher light yields and single tracks or cascades. Due to generally higher noise rates, a similar analysis in sea water neutrino telescopes may be even more challenging. Being the first such exploration, the potential for improvements is manifold including the implementation of a dedicated double track trigger, IceCube detector upgrades that allow identification of smaller track separations and the analysis of all years of available data.  

\subsection{Fractional Electric Charges}

The Standard Model intrinsically does not constrain the elementary charge but observationally it appears as a physical constant, {\it i.e.}, all observed colour-singlet particles have integer multiples of elementary charge. As outlined in section~\ref{sec:monopoles} magnetic monopoles would provide a mechanism to explain electrical charge quantisation. In GUT theories the resulting quantisation is driven by the minimum possible magnetic charge rather than directly. Free fractionally charged states hence are often predicted in multiples of $1/2$ ({\it e.g.},~in the Pati-Salam model~\cite{Pati:1974yy}) or $1/3$, but other and smaller fractions are possible. Beyond this, fractional charges may exist in composite objects with large ($10^{12}$ GeV) confinement scales, probably also contributing to dark matter~\cite{Birkel:1998nx}. 

Experimentally, fractionally charged particles in cosmic rays can be observed though the anomalously low energy loss and lower light emission. Discarding one event from the signal region, the lowest upper limits on the flux of such particles have been posed by a study of the MACRO experiment~\cite{Ambrosio:2004ub} assuming a simple charge-squared scaling of the energy loss. Due to its size, it is expected that IceCube can reach considerably improved sensitivity to the signature of fractionally charged particles in future analyses.

\section{Summary}
\label{sec:summary}
IceCube opened a new window to study the non-thermal universe in 2013 through the discovery of a high-energy neutrino flux of astrophysical origin, and a first identification of a high-energy neutrino point source may be possible through joint multi-messenger observations.  While these novel results can be taken as the beginning of neutrino astronomy, the potential exists to use IceCube to probe physics topics beyond astrophysics in particle physics, not least due to its sheer size.

The observation of secondary particles produced in interactions of neutrinos or cosmic rays with matter provides a probe of Standard Model interactions at energies only marginally covered or inaccessible by particle accelerator experiments. The continuous flux of atmospheric neutrinos allows to study standard neutrino flavour oscillations at a precision that is compatible with those of dedicated oscillation experiments. Moreover, the very long oscillation baselines (thousands of kilometres for atmospheric neutrinos or giga-parsec in astrophysical neutrinos) and the very high neutrino energies (up to PeV) can probe feeble deviations from the standard three-flavour oscillation scenario, that are otherwise undetectable. IceCube has indeed provided strict limits on the allowed parameter space for an additional light ``sterile'' neutrino state with no Standard Model interactions, reducing considerably the range of allowed values of the new oscillation variables $\sin^2 2\theta_{24}$ and $\Delta m^2_{41}$. Similar analyses can set limits on the degree of Lorentz invariance violation, an effect that can be factorised in terms of operators proportional to powers of the neutrino energy. The energy reach of IceCube allows to probe higher dimension operators than previous experiments.

Neutrinos are also valuable indirect messengers of dark matter annihilation and decay in the Earth, Sun, Milky Way, local galaxies, or galaxy clusters. In general, neutrino emission does not suffer from large astrophysical fore- and backgrounds like electromagnetic emission and does not suffer from deflections in magnetic fields like cosmic rays. Neither indirect dark matter detection with neutrinos shares the same systematic uncertainties from astrophysical or particle physics inputs with other search techniques. In this way, indirect limits on dark matter properties from neutrino observations are complementary to indirect searches with other messengers or direct searches with accelerator or scattering experiments. Furthermore, neutrinos can be visible from very distant dark matter sources like galaxy clusters and also probe the interior of compact sources like our Sun. Besides probing dark mater capture or self annihilation cross sections, IceCube has current world leading limits on the dark matter lifetime, extending the range of the dark matter masses probed up to two orders of magnitude with respect to results from Cherenkov telescopes. 

Neutrino telescopes are also sensitive to a variety of exotic signatures produced by rare particles, like Big Bang relics, passing through the detector and emitting direct or indirect Cherenkov light, as well as luminescence. Probably the most interesting signal consists of relic monopoles that could be deciphered from atmospheric and astrophysical backgrounds as extremely bright tracks and/or anomalously slow particles. Other heavy exotic particles that could be visible in this are relic Q-balls and strangelets. Cherenkov emission of long-lived supersymmetric particles or fractionally charged particles can also be considered. These are just a few examples of the many possibilities how neutrino observatories can be uses as multi-purpose particle detectors. 

In this review we have summarised the many possibilities how the IceCube Observatory can probe fundamental questions of particle physics. Proposed future extensions of IceCube will enhance the sensitivity of these searches~\cite{Aartsen:2014njl}. A low-energy in-fill, such as PINGU~\cite{TheIceCube-Gen2:2016cap}, would provide highly competitive measurements of the atmospheric neutrino oscillation parameters, the neutrino mass ordering, or the rate of tau neutrino appearance. It would also be more sensitive to indirect signals of low-mass dark matter. On the other hand, high-energy extensions would allow to study the astrophysical flux of neutrinos with better precision and over a wider energy range. This would reduce systematic uncertainties regarding neutrino spectra and flavour composition and help to establish astrophysical neutrinos as a probe of neutrino interactions and oscillations over ultra-long baselines.

\begin{acknowledgements}
We would like to thank our colleagues in the IceCube collaboration for discussions and support. MA acknowledges support by Danmarks Grundforskningsfond (project no.~1041811001) and by VILLUM FONDEN (project no.~18994).\end{acknowledgements}

\bibliographystyle{utphys_mod}
%\bibliography{references}

\begin{thebibliography}{100}

\bibitem{Cowan:1992xc}
C.~L. Cowan, F.~Reines, F.~B. Harrison, H.~W. Kruse and A.~D. McGuire,
\href{http://dx.doi.org/10.1126/science.124.3212.103}{{\em Science} {\bfseries
  124} (1956) 103--104}.
%%CITATION = SCIEA,124,103;%%.

\bibitem{Reines:1960we}
F.~Reines,
\href{http://dx.doi.org/10.1146/annurev.ns.10.120160.000245}{{\em Ann. Rev.
  Nucl. Part. Sci.} {\bfseries 10} (1960) 1--26}.
%%CITATION = ARNUA,10,1;%%.

\bibitem{Waxman:1998yy}
E.~Waxman and J.~N. Bahcall,
  \href{http://dx.doi.org/10.1103/PhysRevD.59.023002}{{\em Phys. Rev.}
  {\bfseries D59} (1999) 023002},
\href{http://arxiv.org/abs/hep-ph/9807282}{{\ttfamily arXiv:hep-ph/9807282}}.
%%CITATION = HEP-PH/9807282;%%.

\bibitem{Aartsen:2016nxy}
M.~G. Aartsen {\em et~al.} (IceCube  Collaboration),
  \href{http://dx.doi.org/10.1088/1748-0221/12/03/P03012}{{\em JINST}
  {\bfseries 12} no.~03, (2017) P03012},
\href{http://arxiv.org/abs/1612.05093}{{\ttfamily arXiv:1612.05093}}.
%%CITATION = ARXIV:1612.05093;%%.

\bibitem{IceCube:2012nn}
R.~Abbasi {\em et~al.} (IceCube  Collaboration),
  \href{http://dx.doi.org/10.1016/j.nima.2012.10.067}{{\em Nucl. Instrum.
  Meth.} {\bfseries A700} (2013) 188--220},
  \href{http://arxiv.org/abs/1207.6326}{{\ttfamily arXiv:1207.6326}}.

\bibitem{Collaboration:2011ym}
R.~Abbasi {\em et~al.} (IceCube  Collaboration),
  \href{http://dx.doi.org/10.1016/j.astropartphys.2012.01.004}{{\em Astropart.
  Phys.} {\bfseries 35} (2012) 615--624},
  \href{http://arxiv.org/abs/1109.6096}{{\ttfamily arXiv:1109.6096}}.

\bibitem{Andres:1999hm}
E.~Andres {\em et~al.},
  \href{http://dx.doi.org/10.1016/S0927-6505(99)00092-4}{{\em Astropart. Phys.}
  {\bfseries 13} (2000) 1--20},
  \href{http://arxiv.org/abs/astro-ph/9906203}{{\ttfamily
  arXiv:astro-ph/9906203}}.

\bibitem{Chirkin:2004hz}
D.~Chirkin and W.~Rhode,
\href{http://arxiv.org/abs/hep-ph/0407075}{{\ttfamily arXiv:hep-ph/0407075}}.
%%CITATION = HEP-PH/0407075;%%.

\bibitem{Learned:2000sw}
J.~G. Learned and K.~Mannheim,
\href{http://dx.doi.org/10.1146/annurev.nucl.50.1.679}{{\em Ann. Rev. Nucl.
  Part. Sci.} {\bfseries 50} (2000) 679--749}.
%%CITATION = ARNUA,50,679;%%.

\bibitem{Glashow:1960zz}
S.~L. Glashow,
\href{http://dx.doi.org/10.1103/PhysRev.118.316}{{\em Phys. Rev.} {\bfseries
  118} (1960) 316--317}.
%%CITATION = PHRVA,118,316;%%.

\bibitem{Bhattacharya:2011qu}
A.~Bhattacharya, R.~Gandhi, W.~Rodejohann and A.~Watanabe,
  \href{http://dx.doi.org/10.1088/1475-7516/2011/10/017}{{\em JCAP} {\bfseries
  1110} (2011) 017},
\href{http://arxiv.org/abs/1108.3163}{{\ttfamily arXiv:1108.3163}}.
%%CITATION = ARXIV:1108.3163;%%.

\bibitem{Abbasi:2012cu}
R.~Abbasi {\em et~al.} (IceCube  Collaboration),
  \href{http://dx.doi.org/10.1103/PhysRevD.86.022005}{{\em Phys. Rev.}
  {\bfseries D86} (2012) 022005},
\href{http://arxiv.org/abs/1202.4564}{{\ttfamily arXiv:1202.4564}}.
%%CITATION = ARXIV:1202.4564;%%.

\bibitem{Aartsen:2015dlt}
M.~G. Aartsen {\em et~al.} (IceCube  Collaboration),
  \href{http://dx.doi.org/10.1103/PhysRevD.93.022001}{{\em Phys. Rev.}
  {\bfseries D93} no.~2, (2016) 022001},
\href{http://arxiv.org/abs/1509.06212}{{\ttfamily arXiv:1509.06212}}.
%%CITATION = ARXIV:1509.06212;%%.

\bibitem{Kelley:2014gra}
J.~L. Kelley (IceCube  Collaboration), {\em AIP Conf. Proc.} {\bfseries 1630}
  (2014) 154--157.

\bibitem{Aartsen:2013bfa}
M.~G. Aartsen {\em et~al.},
  \href{http://dx.doi.org/10.1016/j.nima.2013.10.074}{{\em Nucl. Instrum.
  Meth.} {\bfseries A736} (2014) 143--149},
  \href{http://arxiv.org/abs/1308.5501}{{\ttfamily arXiv:1308.5501}}.

\bibitem{Ahrens:2003fg}
J.~Ahrens {\em et~al.} (AMANDA  Collaboration),
  \href{http://dx.doi.org/10.1016/j.nima.2004.01.065}{{\em Nucl. Instrum.
  Meth.} {\bfseries A524} (2004) 169--194},
  \href{http://arxiv.org/abs/astro-ph/0407044}{{\ttfamily
  arXiv:astro-ph/0407044}}.

\bibitem{Aartsen:2013vja}
M.~G. Aartsen {\em et~al.} (IceCube  Collaboration),
  \href{http://dx.doi.org/10.1088/1748-0221/9/03/P03009}{{\em JINST} {\bfseries
  9} (2014) P03009}, \href{http://arxiv.org/abs/1311.4767}{{\ttfamily
  arXiv:1311.4767}}.

\bibitem{Aartsen:2015nss}
M.~G. Aartsen {\em et~al.} (IceCube  Collaboration),
  \href{http://dx.doi.org/10.1016/j.astropartphys.2016.01.006}{{\em Astropart.
  Phys.} {\bfseries 78} (2016) 1--27},
  \href{http://arxiv.org/abs/1506.07981}{{\ttfamily arXiv:1506.07981}}.

\bibitem{Heck:1998a}
D.~Heck, J.~Knapp, J.~N. Capdevielle, G.~Schatz and T.~Thouw, ``Corsika: A
  monte carlo code to simulate extensive air showers.'' {Forschungszentrum
  Karlsruhe Report FZKA 6019}, 1998.

\bibitem{Gaisser:2002jj}
T.~K. Gaisser and M.~Honda,
  \href{http://dx.doi.org/10.1146/annurev.nucl.52.050102.090645}{{\em Ann. Rev.
  Nucl. Part. Sci.} {\bfseries 52} (2002) 153--199},
\href{http://arxiv.org/abs/hep-ph/0203272}{{\ttfamily arXiv:hep-ph/0203272}}.
%%CITATION = HEP-PH/0203272;%%.

\bibitem{Aartsen:2012uu}
M.~G. Aartsen {\em et~al.} (IceCube  Collaboration),
  \href{http://dx.doi.org/10.1103/PhysRevLett.110.151105}{{\em Phys. Rev.
  Lett.} {\bfseries 110} no.~15, (2013) 151105},
\href{http://arxiv.org/abs/1212.4760}{{\ttfamily arXiv:1212.4760}}.
%%CITATION = ARXIV:1212.4760;%%.

\bibitem{Aartsen:2015xup}
M.~G. Aartsen {\em et~al.} (IceCube  Collaboration),
  \href{http://dx.doi.org/10.1103/PhysRevD.91.122004}{{\em Phys. Rev.}
  {\bfseries D91} (2015) 122004},
\href{http://arxiv.org/abs/1504.03753}{{\ttfamily arXiv:1504.03753}}.
%%CITATION = ARXIV:1504.03753;%%.

\bibitem{Abbasi:2010ie}
R.~Abbasi {\em et~al.} (IceCube  Collaboration),
  \href{http://dx.doi.org/10.1103/PhysRevD.83.012001}{{\em Phys. Rev.}
  {\bfseries D83} (2011) 012001},
  \href{http://arxiv.org/abs/1010.3980}{{\ttfamily arXiv:1010.3980}}.

\bibitem{Aartsen:2013jdh}
M.~G. Aartsen {\em et~al.} (IceCube  Collaboration),
  \href{http://dx.doi.org/10.1126/science.1242856}{{\em Science} {\bfseries
  342} (2013) 1242856}, \href{http://arxiv.org/abs/1311.5238}{{\ttfamily
  arXiv:1311.5238}}.

\bibitem{Aartsen:2014gkd}
M.~G. Aartsen {\em et~al.} (IceCube  Collaboration),
  \href{http://dx.doi.org/10.1103/PhysRevLett.113.101101}{{\em Phys. Rev.
  Lett.} {\bfseries 113} (2014) 101101},
\href{http://arxiv.org/abs/1405.5303}{{\ttfamily arXiv:1405.5303}}.
%%CITATION = ARXIV:1405.5303;%%.

\bibitem{Aartsen:2016xlq}
M.~G. Aartsen {\em et~al.} (IceCube  Collaboration),
  \href{http://dx.doi.org/10.3847/0004-637X/833/1/3}{{\em Astrophys. J.}
  {\bfseries 833} no.~1, (2016) 3},
  \href{http://arxiv.org/abs/1607.08006}{{\ttfamily 1607.08006}}.

\bibitem{Andreopoulos:2009rq}
C.~Andreopoulos {\em et~al.},
  \href{http://dx.doi.org/10.1016/j.nima.2009.12.009}{{\em Nucl. Instrum.
  Meth.} {\bfseries A614} (2010) 87--104},
  \href{http://arxiv.org/abs/0905.2517}{{\ttfamily arXiv:0905.2517}}.

\bibitem{Gazizov:2004va}
A.~Gazizov and M.~P. Kowalski,
  \href{http://dx.doi.org/10.1016/j.cpc.2005.03.113}{{\em Comput. Phys.
  Commun.} {\bfseries 172} (2005) 203--213},
  \href{http://arxiv.org/abs/astro-ph/0406439}{{\ttfamily
  arXiv:astro-ph/0406439}}.

\bibitem{Honda:2015fha}
M.~Honda, M.~Sajjad~Athar, T.~Kajita, K.~Kasahara and S.~Midorikawa,
  \href{http://dx.doi.org/10.1103/PhysRevD.92.023004}{{\em Phys. Rev.}
  {\bfseries D92} no.~2, (2015) 023004},
  \href{http://arxiv.org/abs/1502.03916}{{\ttfamily arXiv:1502.03916}}.

\bibitem{Enberg:2008te}
R.~Enberg, M.~H. Reno and I.~Sarcevic,
  \href{http://dx.doi.org/10.1103/PhysRevD.78.043005}{{\em Phys. Rev.}
  {\bfseries D78} (2008) 043005},
\href{http://arxiv.org/abs/0806.0418}{{\ttfamily arXiv:0806.0418}}.
%%CITATION = ARXIV:0806.0418;%%.

\bibitem{Bhattacharya:2015jpa}
A.~Bhattacharya, R.~Enberg, M.~H. Reno, I.~Sarcevic and A.~Stasto,
  \href{http://dx.doi.org/10.1007/JHEP06(2015)110}{{\em JHEP} {\bfseries 06}
  (2015) 110},
\href{http://arxiv.org/abs/1502.01076}{{\ttfamily arXiv:1502.01076}}.
%%CITATION = ARXIV:1502.01076;%%.

\bibitem{Gauld:2015kvh}
R.~Gauld, J.~Rojo, L.~Rottoli, S.~Sarkar and J.~Talbert,
  \href{http://dx.doi.org/10.1007/JHEP02(2016)130}{{\em JHEP} {\bfseries 02}
  (2016) 130},
\href{http://arxiv.org/abs/1511.06346}{{\ttfamily arXiv:1511.06346}}.
%%CITATION = ARXIV:1511.06346;%%.

\bibitem{Bhattacharya:2016jce}
A.~Bhattacharya, R.~Enberg, Y.~S. Jeong, C.~S. Kim, M.~H. Reno, I.~Sarcevic,
  and A.~Stasto, \href{http://dx.doi.org/10.1007/JHEP11(2016)167}{{\em JHEP}
  {\bfseries 11} (2016) 167},
\href{http://arxiv.org/abs/1607.00193}{{\ttfamily arXiv:1607.00193}}.
%%CITATION = ARXIV:1607.00193;%%.

\bibitem{Benzke:2017yjn}
M.~Benzke, M.~V. Garzelli, B.~Kniehl, G.~Kramer, S.~Moch and G.~Sigl,
  \href{http://dx.doi.org/10.1007/JHEP12(2017)021}{{\em JHEP} {\bfseries 12}
  (2017) 021},
\href{http://arxiv.org/abs/1705.10386}{{\ttfamily arXiv:1705.10386}}.
%%CITATION = ARXIV:1705.10386;%%.

\bibitem{Schonert:2008is}
S.~Schonert, T.~K. Gaisser, E.~Resconi and O.~Schulz,
  \href{http://dx.doi.org/10.1103/PhysRevD.79.043009}{{\em Phys. Rev.}
  {\bfseries D79} (2009) 043009},
  \href{http://arxiv.org/abs/0812.4308}{{\ttfamily arXiv:0812.4308}}.

\bibitem{Gaisser:2014bja}
T.~K. Gaisser, K.~Jero, A.~Karle and J.~van Santen,
  \href{http://dx.doi.org/10.1103/PhysRevD.90.023009}{{\em Phys. Rev.}
  {\bfseries D90} no.~2, (2014) 023009},
\href{http://arxiv.org/abs/1405.0525}{{\ttfamily arXiv:1405.0525}}.
%%CITATION = ARXIV:1405.0525;%%.

\bibitem{Aartsen:2013bka}
M.~G. Aartsen {\em et~al.} (IceCube  Collaboration),
  \href{http://dx.doi.org/10.1103/PhysRevLett.111.021103}{{\em Phys. Rev.
  Lett.} {\bfseries 111} (2013) 021103},
\href{http://arxiv.org/abs/1304.5356}{{\ttfamily arXiv:1304.5356}}.
%%CITATION = ARXIV:1304.5356;%%.

\bibitem{Aartsen:2015rwa}
M.~G. Aartsen {\em et~al.} (IceCube  Collaboration),
  \href{http://dx.doi.org/10.1103/PhysRevLett.115.081102}{{\em Phys. Rev.
  Lett.} {\bfseries 115} no.~8, (2015) 081102},
\href{http://arxiv.org/abs/1507.04005}{{\ttfamily arXiv:1507.04005}}.
%%CITATION = ARXIV:1507.04005;%%.

\bibitem{Aartsen:2015ivb}
M.~G. Aartsen {\em et~al.} (IceCube  Collaboration),
  \href{http://dx.doi.org/10.1103/PhysRevLett.114.171102}{{\em Phys. Rev.
  Lett.} {\bfseries 114} no.~17, (2015) 171102},
\href{http://arxiv.org/abs/1502.03376}{{\ttfamily arXiv:1502.03376}}.
%%CITATION = ARXIV:1502.03376;%%.

\bibitem{Ahlers:2015lln}
M.~Ahlers and F.~Halzen,
\href{http://dx.doi.org/10.1088/0034-4885/78/12/126901}{{\em Rept. Prog. Phys.}
  {\bfseries 78} no.~12, (2015) 126901}.
%%CITATION = RPPHA,78,126901;%%.

\bibitem{PDG:2018}
M.~Tanabashi {\em et~al.} (Particle Data Group  Collaboration), {\em Phys.
  Rev.} {\bfseries D98} (2018) 030001.

\bibitem{Pontecorvo:1957qd}
B.~Pontecorvo, {\em Sov. Phys. JETP} {\bfseries 7} (1958) 172--173.
[Zh. Eksp. Teor. Fiz.34,247(1957)].
%%CITATION = SPHJA,7,172;%%.

\bibitem{Maki:1962mu}
Z.~Maki, M.~Nakagawa and S.~Sakata,
\href{http://dx.doi.org/10.1143/PTP.28.870}{{\em Prog. Theor. Phys.} {\bfseries
  28} (1962) 870--880}.
%%CITATION = PTPKA,28,870;%%.

\bibitem{Pontecorvo:1967fh}
B.~Pontecorvo, {\em Sov. Phys. JETP} {\bfseries 26} (1968) 984--988. [Zh. Eksp.
  Teor. Fiz.53,1717(1967)].

\bibitem{Fukuda:1998mi}
Y.~Fukuda {\em et~al.} (Super-Kamiokande  Collaboration),
  \href{http://dx.doi.org/10.1103/PhysRevLett.81.1562}{{\em Phys. Rev. Lett.}
  {\bfseries 81} (1998) 1562--1567},
\href{http://arxiv.org/abs/hep-ex/9807003}{{\ttfamily arXiv:hep-ex/9807003}}.
%%CITATION = HEP-EX/9807003;%%.

\bibitem{Ashie:2005ik}
Y.~Ashie {\em et~al.} (Super-Kamiokande  Collaboration),
  \href{http://dx.doi.org/10.1103/PhysRevD.71.112005}{{\em Phys. Rev.}
  {\bfseries D71} (2005) 112005},
  \href{http://arxiv.org/abs/hep-ex/0501064}{{\ttfamily arXiv:hep-ex/0501064}}.

\bibitem{Ahn:2006zza}
M.~H. Ahn {\em et~al.} (K2K  Collaboration),
  \href{http://dx.doi.org/10.1103/PhysRevD.74.072003}{{\em Phys. Rev.}
  {\bfseries D74} (2006) 072003},
  \href{http://arxiv.org/abs/hep-ex/0606032}{{\ttfamily arXiv:hep-ex/0606032}}.

\bibitem{Adamson:2008zt}
P.~Adamson {\em et~al.} (MINOS  Collaboration),
  \href{http://dx.doi.org/10.1103/PhysRevLett.101.131802}{{\em Phys. Rev.
  Lett.} {\bfseries 101} (2008) 131802},
  \href{http://arxiv.org/abs/0806.2237}{{\ttfamily arXiv:0806.2237}}.

\bibitem{Hosaka:2005um}
J.~Hosaka {\em et~al.} (Super-Kamiokande  Collaboration),
  \href{http://dx.doi.org/10.1103/PhysRevD.73.112001}{{\em Phys. Rev.}
  {\bfseries D73} (2006) 112001},
  \href{http://arxiv.org/abs/hep-ex/0508053}{{\ttfamily arXiv:hep-ex/0508053}}.

\bibitem{Aharmim:2009gd}
B.~Aharmim {\em et~al.} (SNO  Collaboration),
  \href{http://dx.doi.org/10.1103/PhysRevC.81.055504}{{\em Phys. Rev.}
  {\bfseries C81} (2010) 055504},
  \href{http://arxiv.org/abs/0910.2984}{{\ttfamily arXiv:0910.2984}}.

\bibitem{Arpesella:2008mt}
C.~Arpesella {\em et~al.} (Borexino  Collaboration),
  \href{http://dx.doi.org/10.1103/PhysRevLett.101.091302}{{\em Phys. Rev.
  Lett.} {\bfseries 101} (2008) 091302},
  \href{http://arxiv.org/abs/0805.3843}{{\ttfamily arXiv:0805.3843}}.

\bibitem{Wolfenstein:1977ue}
L.~Wolfenstein,
\href{http://dx.doi.org/10.1103/PhysRevD.17.2369}{{\em Phys. Rev.} {\bfseries
  D17} (1978) 2369--2374}.
%%CITATION = PHRVA,D17,2369;%%.

\bibitem{Mikheev:1986gs}
S.~P. Mikheev and A.~{\relax Yu}. Smirnov, {\em Sov. J. Nucl. Phys.} {\bfseries
  42} (1985) 913--917.
[Yad. Fiz.42,1441(1985)].
%%CITATION = SJNCA,42,913;%%.

\bibitem{Mikheev:1986wj}
S.~P. Mikheev and A.~{\relax Yu}. Smirnov,
\href{http://dx.doi.org/10.1007/BF02508049}{{\em Nuovo Cim.} {\bfseries C9}
  (1986) 17--26}.
%%CITATION = NUCIA,C9,17;%%.

\bibitem{Araki:2004mb}
T.~Araki {\em et~al.} (KamLAND  Collaboration),
  \href{http://dx.doi.org/10.1103/PhysRevLett.94.081801}{{\em Phys. Rev. Lett.}
  {\bfseries 94} (2005) 081801},
  \href{http://arxiv.org/abs/hep-ex/0406035}{{\ttfamily arXiv:hep-ex/0406035}}.

\bibitem{An:2012eh}
F.~P. An {\em et~al.} (Daya Bay  Collaboration),
  \href{http://dx.doi.org/10.1103/PhysRevLett.108.171803}{{\em Phys. Rev.
  Lett.} {\bfseries 108} (2012) 171803},
\href{http://arxiv.org/abs/1203.1669}{{\ttfamily arXiv:1203.1669}}.
%%CITATION = ARXIV:1203.1669;%%.

\bibitem{Ahn:2012nd}
J.~K. Ahn {\em et~al.} (RENO  Collaboration),
  \href{http://dx.doi.org/10.1103/PhysRevLett.108.191802}{{\em Phys. Rev.
  Lett.} {\bfseries 108} (2012) 191802},
\href{http://arxiv.org/abs/1204.0626}{{\ttfamily arXiv:1204.0626}}.
%%CITATION = ARXIV:1204.0626;%%.

\bibitem{Esteban:2016qun}
I.~Esteban, M.~C. Gonz\'alez-Garc\'ia, M.~Maltoni, I.~Martinez-Soler and
  T.~Schwetz, \href{http://dx.doi.org/10.1007/JHEP01(2017)087}{{\em JHEP}
  {\bfseries 01} (2017) 087}, \href{http://arxiv.org/abs/1611.01514}{{\ttfamily
  arXiv:1611.01514}}.
\url{http://www.nu-fit.org}.
%%CITATION = ARXIV:1611.01514;%%.

\bibitem{Kraus:2004zw}
C.~Kraus {\em et~al.}, \href{http://dx.doi.org/10.1140/epjc/s2005-02139-7}{{\em
  Eur. Phys. J.} {\bfseries C40} (2005) 447--468},
\href{http://arxiv.org/abs/hep-ex/0412056}{{\ttfamily arXiv:hep-ex/0412056}}.
%%CITATION = HEP-EX/0412056;%%.

\bibitem{Aseev:2011dq}
V.~N. Aseev {\em et~al.} (Troitsk  Collaboration),
  \href{http://dx.doi.org/10.1103/PhysRevD.84.112003}{{\em Phys. Rev.}
  {\bfseries D84} (2011) 112003},
\href{http://arxiv.org/abs/1108.5034}{{\ttfamily arXiv:1108.5034}}.
%%CITATION = ARXIV:1108.5034;%%.

\bibitem{KATRIN-design:2005}
J.~Angrik {\em et~al.} (KATRIN  Collaboration),
{\em {Forschungszentrum Karlsruhe Report FZKA 7090}} (2005) .
%%CITATION = FZKA-7090;%%.

\bibitem{Ade:2015xua}
P.~A.~R. Ade {\em et~al.} (Planck  Collaboration),
  \href{http://dx.doi.org/10.1051/0004-6361/201525830}{{\em Astron. Astrophys.}
  {\bfseries 594} (2016) A13},
\href{http://arxiv.org/abs/1502.01589}{{\ttfamily arXiv:1502.01589}}.
%%CITATION = ARXIV:1502.01589;%%.

\bibitem{Mohapatra:2006gs}
R.~N. Mohapatra and A.~Y. Smirnov,
  \href{http://dx.doi.org/10.1146/annurev.nucl.56.080805.140534}{{\em Ann. Rev.
  Nucl. Part. Sci.} {\bfseries 56} (2006) 569--628},
\href{http://arxiv.org/abs/hep-ph/0603118}{{\ttfamily arXiv:hep-ph/0603118}}.
%%CITATION = HEP-PH/0603118;%%.

\bibitem{Aartsen:2013nla}
M.~G. Aartsen {\em et~al.} (IceCube  Collaboration), {\em Proceedings, 33rd
  International Cosmic Ray Conference (ICRC2013). Rio de Janeiro, Brazil, July
  2-9, 2013} (2013) ,
\href{http://arxiv.org/abs/1309.7008}{{\ttfamily arXiv:1309.7008}}.
%%CITATION = ARXIV:1309.7008;%%.

\bibitem{Fogli:2012ua}
G.~L. Fogli, E.~Lisi, A.~Marrone, D.~Montanino, A.~Palazzo and A.~M. Rotunno,
  \href{http://dx.doi.org/10.1103/PhysRevD.86.013012}{{\em Phys. Rev.}
  {\bfseries D86} (2012) 013012},
  \href{http://arxiv.org/abs/1205.5254}{{\ttfamily arXiv:1205.5254}}.

\bibitem{AdrianMartinez:2012ph}
S.~Adri\'an-Mart\'inez {\em et~al.} (ANTARES  Collaboration),
  \href{http://dx.doi.org/10.1016/j.physletb.2012.07.002}{{\em Phys. Lett.}
  {\bfseries B714} (2012) 224--230},
  \href{http://arxiv.org/abs/1206.0645}{{\ttfamily arXiv:1206.0645}}.

\bibitem{Nichol:2013caa}
R.~Nichol, (MINOS  Collaboration),
  \href{http://dx.doi.org/10.1016/j.nuclphysbps.2013.03.017}{{\em Nucl. Phys.
  Proc. Suppl.} {\bfseries 235-236} (2013) 105--111}.

\bibitem{Abe:2013fuq}
K.~Abe {\em et~al.} (T2K  Collaboration),
  \href{http://dx.doi.org/10.1103/PhysRevLett.111.211803}{{\em Phys. Rev.
  Lett.} {\bfseries 111} no.~21, (2013) 211803},
  \href{http://arxiv.org/abs/1308.0465}{{\ttfamily arXiv:1308.0465}}.

\bibitem{Aartsen:2013jza}
M.~G. Aartsen {\em et~al.} (IceCube  Collaboration),
  \href{http://dx.doi.org/10.1103/PhysRevLett.111.081801}{{\em Phys. Rev.
  Lett.} {\bfseries 111} no.~8, (2013) 081801},
  \href{http://arxiv.org/abs/1305.3909}{{\ttfamily arXiv:1305.3909}}.

\bibitem{Wendell:2014dka}
R.~Wendell {\em et~al.} (Super-Kamiokande  Collaboration),
  \href{http://dx.doi.org/10.1063/1.4915569}{{\em AIP Conf. Proc. E. Kearns and
  G. Feldman, editors.} {\bfseries 1666} (2015) 100001},
  \href{http://arxiv.org/abs/1412.5234}{{\ttfamily arXiv:1412.5234}}.

\bibitem{Abe:2017uxa}
K.~Abe {\em et~al.} (T2K  Collaboration),
  \href{http://dx.doi.org/10.1103/PhysRevLett.118.151801}{{\em Phys. Rev.
  Lett.} {\bfseries 118} no.~15, (2017) 151801},
  \href{http://arxiv.org/abs/1701.00432}{{\ttfamily arXiv:1701.00432}}.

\bibitem{Adamson:2013whj}
P.~Adamson {\em et~al.} (MINOS  Collaboration),
  \href{http://dx.doi.org/10.1103/PhysRevLett.110.251801}{{\em Phys. Rev.
  Lett.} {\bfseries 110} no.~25, (2013) 251801},
  \href{http://arxiv.org/abs/1304.6335}{{\ttfamily arXiv:1304.6335}}.

\bibitem{Adamson:2017qqn}
P.~Adamson {\em et~al.} (NOvA  Collaboration),
  \href{http://dx.doi.org/10.1103/PhysRevLett.118.151802}{{\em Phys. Rev.
  Lett.} {\bfseries 118} no.~15, (2017) 151802},
  \href{http://arxiv.org/abs/1701.05891}{{\ttfamily arXiv:1701.05891}}.

\bibitem{Aartsen:2017nmd}
M.~G. Aartsen {\em et~al.} (IceCube  Collaboration),
  \href{http://dx.doi.org/10.1103/PhysRevLett.120.071801}{{\em Phys. Rev.
  Lett.} {\bfseries 120} no.~7, (2018) 071801},
  \href{http://arxiv.org/abs/1707.07081}{{\ttfamily arXiv:1707.07081}}.

\bibitem{Aartsen:2014yll}
M.~G. Aartsen {\em et~al.} (IceCube  Collaboration),
  \href{http://dx.doi.org/10.1103/PhysRevD.91.072004}{{\em Phys. Rev.}
  {\bfseries D91} no.~7, (2015) 072004},
\href{http://arxiv.org/abs/1410.7227}{{\ttfamily arXiv:1410.7227}}.
%%CITATION = ARXIV:1410.7227;%%.

\bibitem{Aartsen:2014oha}
M.~G. Aartsen {\em et~al.} (IceCube PINGU  Collaboration),
  \href{http://arxiv.org/abs/1401.2046}{{\ttfamily arXiv:1401.2046}}.

\bibitem{Aartsen:2015knd}
M.~G. Aartsen {\em et~al.} (IceCube  Collaboration),
  \href{http://dx.doi.org/10.1088/0004-637X/809/1/98}{{\em Astrophys. J.}
  {\bfseries 809} no.~1, (2015) 98},
\href{http://arxiv.org/abs/1507.03991}{{\ttfamily arXiv:1507.03991}}.
%%CITATION = ARXIV:1507.03991;%%.

\bibitem{Harrison:2002er}
P.~F. Harrison, D.~H. Perkins and W.~G. Scott,
  \href{http://dx.doi.org/10.1016/S0370-2693(02)01336-9}{{\em Phys. Lett.}
  {\bfseries B530} (2002) 167},
\href{http://arxiv.org/abs/hep-ph/0202074}{{\ttfamily arXiv:hep-ph/0202074}}.
%%CITATION = HEP-PH/0202074;%%.

\bibitem{Martin:1998sq}
A.~D. Martin, R.~G. Roberts, W.~J. Stirling and R.~S. Thorne,
  \href{http://dx.doi.org/10.1007/s100529800904, 10.1007/s100520050220}{{\em
  Eur. Phys. J.} {\bfseries C4} (1998) 463--496},
\href{http://arxiv.org/abs/hep-ph/9803445}{{\ttfamily arXiv:hep-ph/9803445}}.
%%CITATION = HEP-PH/9803445;%%.

\bibitem{Gribov:1972ri}
V.~N. Gribov and L.~N. Lipatov, {\em Sov. J. Nucl. Phys.} {\bfseries 15} (1972)
  438--450.
[Yad. Fiz.15,781(1972)].
%%CITATION = SJNCA,15,438;%%.

\bibitem{Lipatov:1974qm}
L.~N. Lipatov, {\em Sov. J. Nucl. Phys.} {\bfseries 20} (1975) 94--102.
[Yad. Fiz.20,181(1974)].
%%CITATION = SJNCA,20,94;%%.

\bibitem{Altarelli:1977zs}
G.~Altarelli and G.~Parisi,
\href{http://dx.doi.org/10.1016/0550-3213(77)90384-4}{{\em Nucl. Phys.}
  {\bfseries B126} (1977) 298--318}.
%%CITATION = NUPHA,B126,298;%%.

\bibitem{Dokshitzer:1977sg}
Y.~L. Dokshitzer, {\em Sov. Phys. JETP} {\bfseries 46} (1977) 641--653.
[Zh. Eksp. Teor. Fiz.73,1216(1977)].
%%CITATION = SPHJA,46,641;%%.

\bibitem{Kuraev:1977fs}
E.~A. Kuraev, L.~N. Lipatov and V.~S. Fadin, {\em Sov. Phys. JETP} {\bfseries
  45} (1977) 199--204.
[Zh. Eksp. Teor. Fiz.72,377(1977)].
%%CITATION = SPHJA,45,199;%%.

\bibitem{Balitsky:1978ic}
I.~I. Balitsky and L.~N. Lipatov, {\em Sov. J. Nucl. Phys.} {\bfseries 28}
  (1978) 822--829.
[Yad. Fiz.28,1597(1978)].
%%CITATION = SJNCA,28,822;%%.

\bibitem{Kwiecinski:1997ee}
J.~Kwiecinski, A.~D. Martin and A.~M. Stasto,
  \href{http://dx.doi.org/10.1103/PhysRevD.56.3991}{{\em Phys. Rev.} {\bfseries
  D56} (1997) 3991--4006},
\href{http://arxiv.org/abs/hep-ph/9703445}{{\ttfamily arXiv:hep-ph/9703445}}.
%%CITATION = HEP-PH/9703445;%%.

\bibitem{GolecBiernat:1998js}
K.~J. Golec-Biernat and M.~Wusthoff,
  \href{http://dx.doi.org/10.1103/PhysRevD.59.014017}{{\em Phys. Rev.}
  {\bfseries D59} (1998) 014017},
\href{http://arxiv.org/abs/hep-ph/9807513}{{\ttfamily arXiv:hep-ph/9807513}}.
%%CITATION = HEP-PH/9807513;%%.

\bibitem{GolecBiernat:1999qd}
K.~J. Golec-Biernat and M.~Wusthoff,
  \href{http://dx.doi.org/10.1103/PhysRevD.60.114023}{{\em Phys. Rev.}
  {\bfseries D60} (1999) 114023},
\href{http://arxiv.org/abs/hep-ph/9903358}{{\ttfamily arXiv:hep-ph/9903358}}.
%%CITATION = HEP-PH/9903358;%%.

\bibitem{Iancu:2003xm}
E.~Iancu and R.~Venugopalan, in {\em In *Hwa, R.C. (ed.) et al.: Quark gluon
  plasma* 249-3363}.
\newblock World Scientific, 2003.
\newblock
\href{http://arxiv.org/abs/hep-ph/0303204}{{\ttfamily arXiv:hep-ph/0303204}}.
\newblock
%%CITATION = HEP-PH/0303204;%%.

\bibitem{CooperSarkar:2011pa}
A.~Cooper-Sarkar, P.~Mertsch and S.~Sarkar,
  \href{http://dx.doi.org/10.1007/JHEP08(2011)042}{{\em JHEP} {\bfseries 08}
  (2011) 042},
\href{http://arxiv.org/abs/1106.3723}{{\ttfamily arXiv:1106.3723}}.
%%CITATION = ARXIV:1106.3723;%%.

\bibitem{CooperSarkar:2007cv}
A.~Cooper-Sarkar and S.~Sarkar,
  \href{http://dx.doi.org/10.1088/1126-6708/2008/01/075}{{\em JHEP} {\bfseries
  01} (2008) 075},
\href{http://arxiv.org/abs/0710.5303}{{\ttfamily arXiv:0710.5303}}.
%%CITATION = ARXIV:0710.5303;%%.

\bibitem{Chekanov:2002pv}
S.~Chekanov {\em et~al.} (ZEUS  Collaboration),
  \href{http://dx.doi.org/10.1103/PhysRevD.67.012007}{{\em Phys. Rev.}
  {\bfseries D67} (2003) 012007},
\href{http://arxiv.org/abs/hep-ex/0208023}{{\ttfamily arXiv:hep-ex/0208023}}.
%%CITATION = HEP-EX/0208023;%%.

\bibitem{Aartsen:2017kpd}
M.~G. Aartsen {\em et~al.} (IceCube  Collaboration),
  \href{http://dx.doi.org/10.1038/nature24459}{{\em Nature} {\bfseries 551}
  (2017) 596},
\href{http://arxiv.org/abs/1711.08119}{{\ttfamily arXiv:1711.08119}}.
%%CITATION = ARXIV:1711.08119;%%.

\bibitem{Honda:2006qj}
M.~Honda, T.~Kajita, K.~Kasahara, S.~Midorikawa and T.~Sanuki,
  \href{http://dx.doi.org/10.1103/PhysRevD.75.043006}{{\em Phys. Rev.}
  {\bfseries D75} (2007) 043006},
\href{http://arxiv.org/abs/astro-ph/0611418}{{\ttfamily
  arXiv:astro-ph/0611418}}.
%%CITATION = ASTRO-PH/0611418;%%.

\bibitem{Dziewonski:1981xy}
A.~M. Dziewonski and D.~L. Anderson,
  \href{http://dx.doi.org/10.1016/0031-9201(81)90046-7}{{\em Phys. Earth
  Planet. Interiors} {\bfseries 25} (1981) 297--356}.

\bibitem{Bustamante:2017xuy}
M.~Bustamante and A.~Connolly,
\href{http://arxiv.org/abs/1711.11043}{{\ttfamily arXiv:1711.11043}}.
%%CITATION = ARXIV:1711.11043;%%.

\bibitem{Apel:2014qqa}
W.~D. Apel {\em et~al.},
  \href{http://dx.doi.org/10.1016/j.astropartphys.2014.12.001}{{\em Astropart.
  Phys.} {\bfseries 65} (2015) 55--63}.

\bibitem{Gonzalez:2015rdo}
J.~Gonz\'alez (IceCube  Collaboration), {\em Proceedings, 34th International
  Cosmic Ray Conference (ICRC 2015). The Hague, The Netherlands, July 30-August
  6, 2015.} {\bfseries PoS(ICRC2015)} (2016) 338.

\bibitem{Hagedorn:1983wk}
R.~Hagedorn,
\href{http://dx.doi.org/10.1007/BF02740917}{{\em Riv. Nuovo Cim.} {\bfseries
  6N10} (1983) 1--50}.
%%CITATION = RNCIB,6N10,1;%%.

\bibitem{Adams:2004zg}
J.~Adams {\em et~al.} (STAR  Collaboration),
  \href{http://dx.doi.org/10.1103/PhysRevC.70.044901}{{\em Phys. Rev.}
  {\bfseries C70} (2004) 044901},
  \href{http://arxiv.org/abs/nucl-ex/0404020}{{\ttfamily
  arXiv:nucl-ex/0404020}}.

\bibitem{Adare:2011vy}
A.~Adare {\em et~al.} (PHENIX  Collaboration),
  \href{http://dx.doi.org/10.1103/PhysRevC.83.064903}{{\em Phys. Rev.}
  {\bfseries C83} (2011) 064903},
  \href{http://arxiv.org/abs/1102.0753}{{\ttfamily 1102.0753}}.

\bibitem{Abbasi:2012kza}
R.~Abbasi {\em et~al.} (IceCube  Collaboration),
  \href{http://dx.doi.org/10.1103/PhysRevD.87.012005}{{\em Phys. Rev.}
  {\bfseries D87} no.~1, (2013) 012005},
\href{http://arxiv.org/abs/1208.2979}{{\ttfamily arXiv:1208.2979}}.
%%CITATION = ARXIV:1208.2979;%%.

\bibitem{Soldin:2014ouk}
D.~Soldin (IceCube  Collaboration),
  \href{http://dx.doi.org/10.1051/epjconf/20159906001}{{\em EPJ Web Conf.}
  {\bfseries 99} (2015) 06001},
\href{http://arxiv.org/abs/1411.4448}{{\ttfamily arXiv:1411.4448}}.
%%CITATION = ARXIV:1411.4448;%%.

\bibitem{Soldin:2015iaa}
D.~Soldin (IceCube  Collaboration),
{\em Proceedings, 34th International Cosmic Ray Conference (ICRC 2015). The
  Hague, The Netherlands, July 30-August 6, 2015.} {\bfseries PoS(ICRC2015)}
  (2016) 256.
%%CITATION = POSCI,ICRC2015,256;%%.

\bibitem{Dembinski:2017zkb}
H.~Dembinski (IceCube  Collaboration),
\href{http://dx.doi.org/10.1051/epjconf/201614501003}{{\em EPJ Web Conf.}
  {\bfseries 145} (2017) 01003}.
%%CITATION = 00776,145,01003;%%.

\bibitem{Lindblad:1975ef}
G.~Lindblad,
\href{http://dx.doi.org/10.1007/BF01608499}{{\em Commun. Math. Phys.}
  {\bfseries 48} (1976) 119}.
%%CITATION = CMPHA,48,119;%%.

\bibitem{GonzalezGarcia:2005xw}
M.~C. Gonz\'alez-Garc\'ia, F.~Halzen and M.~Maltoni,
  \href{http://dx.doi.org/10.1103/PhysRevD.71.093010}{{\em Phys. Rev.}
  {\bfseries D71} (2005) 093010},
\href{http://arxiv.org/abs/hep-ph/0502223}{{\ttfamily arXiv:hep-ph/0502223}}.
%%CITATION = HEP-PH/0502223;%%.

\bibitem{GonzalezGarcia:2004cu}
M.~C. Gonz\'alez-Garc\'ia, M.~Maltoni and A.~{\relax Yu}. Smirnov,
  \href{http://dx.doi.org/10.1103/PhysRevD.70.093005}{{\em Phys. Rev.}
  {\bfseries D70} (2004) 093005},
\href{http://arxiv.org/abs/hep-ph/0408170}{{\ttfamily arXiv:hep-ph/0408170}}.
%%CITATION = HEP-PH/0408170;%%.

\bibitem{Abbasi:2009nfa}
R.~Abbasi {\em et~al.} (IceCube  Collaboration),
  \href{http://dx.doi.org/10.1103/PhysRevD.79.102005}{{\em Phys. Rev.}
  {\bfseries D79} (2009) 102005},
  \href{http://arxiv.org/abs/0902.0675}{{\ttfamily arXiv:0902.0675}}.

\bibitem{Xu:2014via}
X.-J. Xu, H.-J. He and W.~Rodejohann,
  \href{http://dx.doi.org/10.1088/1475-7516/2014/12/039}{{\em JCAP} {\bfseries
  1412} (2014) 039},
\href{http://arxiv.org/abs/1407.3736}{{\ttfamily arXiv:1407.3736}}.
%%CITATION = ARXIV:1407.3736;%%.

\bibitem{Jost:1957zz}
R.~Jost,
{\em Helv. Phys. Acta} {\bfseries 30} (1957) 409--416.
%%CITATION = HPACA,30,409;%%.

\bibitem{Colladay:1998fq}
D.~Colladay and V.~A. Kostelecky,
  \href{http://dx.doi.org/10.1103/PhysRevD.58.116002}{{\em Phys. Rev.}
  {\bfseries D58} (1998) 116002},
  \href{http://arxiv.org/abs/hep-ph/9809521}{{\ttfamily arXiv:hep-ph/9809521}}.

\bibitem{Kostelecky:2008ts}
V.~A. Kostelecky and N.~Russell,
  \href{http://dx.doi.org/10.1103/RevModPhys.83.11}{{\em Rev. Mod. Phys.}
  {\bfseries 83} (2011) 11--31},
\href{http://arxiv.org/abs/0801.0287}{{\ttfamily arXiv:0801.0287}}.
%%CITATION = ARXIV:0801.0287;%%.

\bibitem{Liberati:2013xla}
S.~Liberati, \href{http://dx.doi.org/10.1088/0264-9381/30/13/133001}{{\em
  Class. Quant. Grav.} {\bfseries 30} (2013) 133001},
\href{http://arxiv.org/abs/1304.5795}{{\ttfamily arXiv:1304.5795}}.
%%CITATION = ARXIV:1304.5795;%%.

\bibitem{Aartsen:2017ibm}
M.~G. Aartsen {\em et~al.} (IceCube  Collaboration),
\href{http://arxiv.org/abs/1709.03434}{{\ttfamily arXiv:1709.03434}}.
%%CITATION = ARXIV:1709.03434;%%.

\bibitem{Coleman:1998ti}
S.~R. Coleman and S.~L. Glashow,
  \href{http://dx.doi.org/10.1103/PhysRevD.59.116008}{{\em Phys. Rev.}
  {\bfseries D59} (1999) 116008},
\href{http://arxiv.org/abs/hep-ph/9812418}{{\ttfamily arXiv:hep-ph/9812418}}.
%%CITATION = HEP-PH/9812418;%%.

\bibitem{Brustein:2001ik}
R.~Brustein, D.~Eichler and S.~Foffa,
  \href{http://dx.doi.org/10.1103/PhysRevD.65.105006}{{\em Phys. Rev.}
  {\bfseries D65} (2002) 105006},
\href{http://arxiv.org/abs/hep-ph/0106309}{{\ttfamily arXiv:hep-ph/0106309}}.
%%CITATION = HEP-PH/0106309;%%.

\bibitem{Gasperini:1988zf}
M.~Gasperini,
\href{http://dx.doi.org/10.1103/PhysRevD.38.2635}{{\em Phys. Rev.} {\bfseries
  D38} (1988) 2635--2637}.
%%CITATION = PHRVA,D38,2635;%%.

\bibitem{Halprin:1991gs}
A.~Halprin and C.~N. Leung,
\href{http://dx.doi.org/10.1103/PhysRevLett.67.1833}{{\em Phys. Rev. Lett.}
  {\bfseries 67} (1991) 1833--1835}.
%%CITATION = PRLTA,67,1833;%%.

\bibitem{Adunas:2000zm}
G.~Z. Adunas, E.~Rodriguez-Milla and D.~V. Ahluwalia,
  \href{http://dx.doi.org/10.1016/S0370-2693(00)00697-3}{{\em Phys. Lett.}
  {\bfseries B485} (2000) 215--223},
\href{http://arxiv.org/abs/gr-qc/0006021}{{\ttfamily arXiv:gr-qc/0006021}}.
%%CITATION = GR-QC/0006021;%%.

\bibitem{Abbasi:2010kx}
R.~Abbasi {\em et~al.} (IceCube  Collaboration),
  \href{http://dx.doi.org/10.1103/PhysRevD.82.112003}{{\em Phys. Rev.}
  {\bfseries D82} (2010) 112003},
\href{http://arxiv.org/abs/1010.4096}{{\ttfamily arXiv:1010.4096}}.
%%CITATION = ARXIV:1010.4096;%%.

\bibitem{Stecker:2014xja}
F.~W. Stecker and S.~T. Scully,
  \href{http://dx.doi.org/10.1103/PhysRevD.90.043012}{{\em Phys. Rev.}
  {\bfseries D90} no.~4, (2014) 043012},
\href{http://arxiv.org/abs/1404.7025}{{\ttfamily arXiv:1404.7025}}.
%%CITATION = ARXIV:1404.7025;%%.

\bibitem{Stecker:2014oxa}
F.~W. Stecker, S.~T. Scully, S.~Liberati and D.~Mattingly,
  \href{http://dx.doi.org/10.1103/PhysRevD.91.045009}{{\em Phys. Rev.}
  {\bfseries D91} no.~4, (2015) 045009},
\href{http://arxiv.org/abs/1411.5889}{{\ttfamily arXiv:1411.5889}}.
%%CITATION = ARXIV:1411.5889;%%.

\bibitem{Liao:2017yuy}
J.~Liao and D.~Marfatia,
  \href{http://dx.doi.org/10.1103/PhysRevD.97.041302}{{\em Phys. Rev.}
  {\bfseries D97} no.~4, (2018) 041302},
\href{http://arxiv.org/abs/1711.09266}{{\ttfamily arXiv:1711.09266}}.
%%CITATION = ARXIV:1711.09266;%%.

\bibitem{Aartsen:2017bap}
M.~G. Aartsen {\em et~al.} (IceCube  Collaboration),
  \href{http://dx.doi.org/10.1103/PhysRevD.95.112002}{{\em Phys. Rev.}
  {\bfseries D95} no.~11, (2017) 112002},
  \href{http://arxiv.org/abs/1702.05160}{{\ttfamily 1702.05160}}.

\bibitem{Mitsuka:2011ty}
G.~Mitsuka {\em et~al.} (Super-Kamiokande  Collaboration),
  \href{http://dx.doi.org/10.1103/PhysRevD.84.113008}{{\em Phys. Rev.}
  {\bfseries D84} (2011) 113008},
\href{http://arxiv.org/abs/1109.1889}{{\ttfamily arXiv:1109.1889}}.
%%CITATION = ARXIV:1109.1889;%%.

\bibitem{Salvado:2016uqu}
J.~Salvado, O.~Mena, S.~Palomares-Ruiz and N.~Rius,
  \href{http://dx.doi.org/10.1007/JHEP01(2017)141}{{\em JHEP} {\bfseries 01}
  (2017) 141},
\href{http://arxiv.org/abs/1609.03450}{{\ttfamily arXiv:1609.03450}}.
%%CITATION = ARXIV:1609.03450;%%.

\bibitem{Aartsen:2017xtt}
M.~Aartsen {\em et~al.} (IceCube  Collaboration),
  \href{http://dx.doi.org/10.1103/PhysRevD.97.072009}{{\em Phys. Rev.}
  {\bfseries D97} no.~7, (2018) 072009},
\href{http://arxiv.org/abs/1709.07079}{{\ttfamily arXiv:1709.07079}}.
%%CITATION = ARXIV:1709.07079;%%.

\bibitem{Gonzalez-Garcia:2016gpq}
M.~C. Gonz\'alez-Garc\'ia, M.~Maltoni, I.~Martinez-Soler and N.~Song,
  \href{http://dx.doi.org/10.1016/j.astropartphys.2016.07.001}{{\em Astropart.
  Phys.} {\bfseries 84} (2016) 15--22},
\href{http://arxiv.org/abs/1605.08055}{{\ttfamily arXiv:1605.08055}}.
%%CITATION = ARXIV:1605.08055;%%.

\bibitem{Esmaili:2013fva}
A.~Esmaili and A.~{\relax Yu}. Smirnov,
  \href{http://dx.doi.org/10.1007/JHEP06(2013)026}{{\em JHEP} {\bfseries 06}
  (2013) 026},
\href{http://arxiv.org/abs/1304.1042}{{\ttfamily arXiv:1304.1042}}.
%%CITATION = ARXIV:1304.1042;%%.

\bibitem{Benatti:1987dz}
F.~Benatti and H.~Narnhofer,
\href{http://dx.doi.org/10.1007/BF00419590}{{\em Lett. Math. Phys.} {\bfseries
  15} (1988) 325}.
%%CITATION = LMPHD,15,325;%%.

\bibitem{Banks:1983by}
T.~Banks, L.~Susskind and M.~E. Peskin,
  \href{http://dx.doi.org/10.1016/0550-3213(84)90184-6}{{\em Nucl. Phys.}
  {\bfseries B244} (1984) 125--134}.

\bibitem{Gago:2002na}
A.~M. Gago, E.~M. Santos, W.~J.~C. Teves and R.~Zukanovich~Funchal,
\href{http://arxiv.org/abs/hep-ph/0208166}{{\ttfamily arXiv:hep-ph/0208166}}.
%%CITATION = HEP-PH/0208166;%%.

\bibitem{Beacom:2002vi}
J.~F. Beacom, N.~F. Bell, D.~Hooper, S.~Pakvasa and T.~J. Weiler,
  \href{http://dx.doi.org/10.1103/PhysRevLett.90.181301}{{\em Phys. Rev. Lett.}
  {\bfseries 90} (2003) 181301},
\href{http://arxiv.org/abs/hep-ph/0211305}{{\ttfamily arXiv:hep-ph/0211305}}.
%%CITATION = HEP-PH/0211305;%%.

\bibitem{Shoemaker:2015qul}
I.~M. Shoemaker and K.~Murase,
  \href{http://dx.doi.org/10.1103/PhysRevD.93.085004}{{\em Phys. Rev.}
  {\bfseries D93} no.~8, (2016) 085004},
\href{http://arxiv.org/abs/1512.07228}{{\ttfamily arXiv:1512.07228}}.
%%CITATION = ARXIV:1512.07228;%%.

\bibitem{Bustamante:2016ciw}
M.~Bustamante, J.~F. Beacom and K.~Murase,
  \href{http://dx.doi.org/10.1103/PhysRevD.95.063013}{{\em Phys. Rev.}
  {\bfseries D95} no.~6, (2017) 063013},
\href{http://arxiv.org/abs/1610.02096}{{\ttfamily arXiv:1610.02096}}.
%%CITATION = ARXIV:1610.02096;%%.

\bibitem{Rasmussen:2017ert}
R.~W. Rasmussen, L.~Lechner, M.~Ackermann, M.~Kowalski and W.~Winter,
  \href{http://dx.doi.org/10.1103/PhysRevD.96.083018}{{\em Phys. Rev.}
  {\bfseries D96} no.~8, (2017) 083018},
\href{http://arxiv.org/abs/1707.07684}{{\ttfamily arXiv:1707.07684}}.
%%CITATION = ARXIV:1707.07684;%%.

\bibitem{Denton:2018aml}
P.~B. Denton and I.~Tamborra,
\href{http://arxiv.org/abs/1805.05950}{{\ttfamily arXiv:1805.05950}}.
%%CITATION = ARXIV:1805.05950;%%.

\bibitem{Choi:1991aa}
K.~Choi and A.~Santamaria,
\href{http://dx.doi.org/10.1016/0370-2693(91)90900-B}{{\em Phys. Lett.}
  {\bfseries B267} (1991) 504--508}.
%%CITATION = PHLTA,B267,504;%%.

\bibitem{Acker:1992eh}
A.~Acker, A.~Joshipura and S.~Pakvasa,
\href{http://dx.doi.org/10.1016/0370-2693(92)91520-J}{{\em Phys. Lett.}
  {\bfseries B285} (1992) 371--375}.
%%CITATION = PHLTA,B285,371;%%.

\bibitem{Acker:1991ej}
A.~Acker, S.~Pakvasa and J.~T. Pantaleone,
\href{http://dx.doi.org/10.1103/PhysRevD.45.1}{{\em Phys. Rev.} {\bfseries D45}
  (1992) 1--4}.
%%CITATION = PHRVA,D45,1;%%.

\bibitem{Beacom:2002cb}
J.~F. Beacom and N.~F. Bell,
  \href{http://dx.doi.org/10.1103/PhysRevD.65.113009}{{\em Phys. Rev.}
  {\bfseries D65} (2002) 113009},
\href{http://arxiv.org/abs/hep-ph/0204111}{{\ttfamily arXiv:hep-ph/0204111}}.
%%CITATION = HEP-PH/0204111;%%.

\bibitem{Ahlers:2010ty}
M.~Ahlers and A.~M. Taylor,
  \href{http://dx.doi.org/10.1103/PhysRevD.82.123005}{{\em Phys. Rev.}
  {\bfseries D82} (2010) 123005},
\href{http://arxiv.org/abs/1010.3019}{{\ttfamily arXiv:1010.3019}}.
%%CITATION = ARXIV:1010.3019;%%.

\bibitem{Blum:2014ewa}
K.~Blum, A.~Hook and K.~Murase,
\href{http://arxiv.org/abs/1408.3799}{{\ttfamily arXiv:1408.3799}}.
%%CITATION = ARXIV:1408.3799;%%.

\bibitem{Ibe:2014pja}
M.~Ibe and K.~Kaneta, \href{http://dx.doi.org/10.1103/PhysRevD.90.053011}{{\em
  Phys. Rev.} {\bfseries D90} no.~5, (2014) 053011},
\href{http://arxiv.org/abs/1407.2848}{{\ttfamily arXiv:1407.2848}}.
%%CITATION = ARXIV:1407.2848;%%.

\bibitem{Ng:2014pca}
K.~C.~Y. Ng and J.~F. Beacom,
  \href{http://dx.doi.org/10.1103/PhysRevD.90.065035,
  10.1103/PhysRevD.90.089904}{{\em Phys. Rev.} {\bfseries D90} no.~6, (2014)
  065035}, \href{http://arxiv.org/abs/1404.2288}{{\ttfamily arXiv:1404.2288}}.
[Erratum: Phys. Rev.D90,no.8,089904(2014)].
%%CITATION = ARXIV:1404.2288;%%.

\bibitem{Ioka:2014kca}
K.~Ioka and K.~Murase, \href{http://dx.doi.org/10.1093/ptep/ptu090}{{\em PTEP}
  {\bfseries 2014} no.~6, (2014) 061E01},
\href{http://arxiv.org/abs/1404.2279}{{\ttfamily arXiv:1404.2279}}.
%%CITATION = ARXIV:1404.2279;%%.

\bibitem{Abe:2014gda}
K.~Abe {\em et~al.} (Super-Kamiokande  Collaboration),
  \href{http://dx.doi.org/10.1103/PhysRevD.91.052019}{{\em Phys. Rev.}
  {\bfseries D91} (2015) 052019},
  \href{http://arxiv.org/abs/1410.2008}{{\ttfamily 1410.2008}}.

\bibitem{Adamson:2011ku}
P.~Adamson {\em et~al.} (MINOS  Collaboration),
  \href{http://dx.doi.org/10.1103/PhysRevLett.107.011802}{{\em Phys. Rev.
  Lett.} {\bfseries 107} (2011) 011802},
  \href{http://arxiv.org/abs/1104.3922}{{\ttfamily arXiv:1104.3922}}.

\bibitem{Cheng:2012yy}
G.~Cheng {\em et~al.} (SciBooNE, MiniBooNE  Collaboration),
  \href{http://dx.doi.org/10.1103/PhysRevD.86.052009}{{\em Phys. Rev.}
  {\bfseries D86} (2012) 052009},
  \href{http://arxiv.org/abs/1208.0322}{{\ttfamily arXiv:1208.0322}}.

\bibitem{Dydak:1983zq}
F.~Dydak {\em et~al.},
  \href{http://dx.doi.org/10.1016/0370-2693(84)90688-9}{{\em Phys. Lett.}
  {\bfseries B134} (1984) 281}.

\bibitem{Kopp:2013vaa}
J.~Kopp, P.~A.~N. Machado, M.~Maltoni and T.~Schwetz,
  \href{http://dx.doi.org/10.1007/JHEP05(2013)050}{{\em JHEP} {\bfseries 05}
  (2013) 050}, \href{http://arxiv.org/abs/1303.3011}{{\ttfamily 1303.3011}}.

\bibitem{Conrad:2012qt}
J.~M. Conrad, C.~M. Ignarra, G.~Karagiorgi, M.~H. Shaevitz and J.~Spitz,
  \href{http://dx.doi.org/10.1155/2013/163897}{{\em Adv. High Energy Phys.}
  {\bfseries 2013} (2013) 163897},
  \href{http://arxiv.org/abs/1207.4765}{{\ttfamily arXiv:1207.4765}}.

\bibitem{Aartsen:2016oqi}
M.~G. Aartsen {\em et~al.} (IceCube  Collaboration),
  \href{http://dx.doi.org/10.1103/PhysRevLett.117.071801}{{\em Phys. Rev.
  Lett.} {\bfseries 117} no.~7, (2016) 071801},
  \href{http://arxiv.org/abs/1605.01990}{{\ttfamily 1605.01990}}.

\bibitem{Athanassopoulos:1996jb}
C.~Athanassopoulos {\em et~al.} (LSND  Collaboration),
  \href{http://dx.doi.org/10.1103/PhysRevLett.77.3082}{{\em Phys. Rev. Lett.}
  {\bfseries 77} (1996) 3082--3085},
\href{http://arxiv.org/abs/nucl-ex/9605003}{{\ttfamily arXiv:nucl-ex/9605003}}.
%%CITATION = NUCL-EX/9605003;%%.

\bibitem{Aguilar:2001ty}
A.~Aguilar-Arevalo {\em et~al.} (LSND  Collaboration),
  \href{http://dx.doi.org/10.1103/PhysRevD.64.112007}{{\em Phys. Rev.}
  {\bfseries D64} (2001) 112007},
\href{http://arxiv.org/abs/hep-ex/0104049}{{\ttfamily arXiv:hep-ex/0104049}}.
%%CITATION = HEP-EX/0104049;%%.

\bibitem{Aguilar-Arevalo:2013pmq}
A.~A. Aguilar-Arevalo {\em et~al.} (MiniBooNE  Collaboration),
  \href{http://dx.doi.org/10.1103/PhysRevLett.110.161801}{{\em Phys. Rev.
  Lett.} {\bfseries 110} (2013) 161801},
\href{http://arxiv.org/abs/1303.2588}{{\ttfamily arXiv:1303.2588}}.
%%CITATION = ARXIV:1303.2588;%%.

\bibitem{Mention:2011rk}
G.~Mention, M.~Fechner, T.~Lasserre, T.~A. Mueller, D.~Lhuillier, M.~Cribier,
  and A.~Letourneau, \href{http://dx.doi.org/10.1103/PhysRevD.83.073006}{{\em
  Phys. Rev.} {\bfseries D83} (2011) 073006},
\href{http://arxiv.org/abs/1101.2755}{{\ttfamily arXiv:1101.2755}}.
%%CITATION = ARXIV:1101.2755;%%.

\bibitem{Bahcall:1994bq}
J.~N. Bahcall, P.~I. Krastev and E.~Lisi,
  \href{http://dx.doi.org/10.1016/0370-2693(95)00111-W}{{\em Phys. Lett.}
  {\bfseries B348} (1995) 121--123},
\href{http://arxiv.org/abs/hep-ph/9411414}{{\ttfamily arXiv:hep-ph/9411414}}.
%%CITATION = HEP-PH/9411414;%%.

\bibitem{Razzaque:2012tp}
S.~Razzaque and A.~{\relax Yu}. Smirnov,
  \href{http://dx.doi.org/10.1103/PhysRevD.85.093010}{{\em Phys. Rev.}
  {\bfseries D85} (2012) 093010},
\href{http://arxiv.org/abs/1203.5406}{{\ttfamily arXiv:1203.5406}}.
%%CITATION = ARXIV:1203.5406;%%.

\bibitem{Esmaili:2013vza}
A.~Esmaili and A.~{\relax Yu}. Smirnov,
  \href{http://dx.doi.org/10.1007/JHEP12(2013)014}{{\em JHEP} {\bfseries 12}
  (2013) 014},
\href{http://arxiv.org/abs/1307.6824}{{\ttfamily arXiv:1307.6824}}.
%%CITATION = ARXIV:1307.6824;%%.

\bibitem{Lindner2016}
M.~Lindner, W.~Rodejohann and X.-J. Xu,
  \href{http://dx.doi.org/10.1007/JHEP01(2016)124}{{\em J. High Energy Phys.}
  {\bfseries 2016} no.~1, (2016) 124}.

\bibitem{Gonzalez-Garcia:2015qrr}
M.~C. Gonz\'alez-Garc\'ia, M.~Maltoni and T.~Schwetz,
  \href{http://dx.doi.org/10.1016/j.nuclphysb.2016.02.033}{{\em Nucl. Phys.}
  {\bfseries B908} (2016) 199--217},
  \href{http://arxiv.org/abs/1512.06856}{{\ttfamily arXiv:1512.06856}}.

\bibitem{Collin:2016rao}
G.~H. Collin, C.~A. Arg\"uelles, J.~M. Conrad and M.~H. Shaevitz,
  \href{http://dx.doi.org/10.1016/j.nuclphysb.2016.02.024}{{\em Nucl. Phys.}
  {\bfseries B908} (2016) 354--365},
\href{http://arxiv.org/abs/1602.00671}{{\ttfamily arXiv:1602.00671}}.
%%CITATION = ARXIV:1602.00671;%%.

\bibitem{Capozzi:2016vac}
F.~Capozzi, C.~Giunti, M.~Laveder and A.~Palazzo,
  \href{http://dx.doi.org/10.1103/PhysRevD.95.033006}{{\em Phys. Rev.}
  {\bfseries D95} no.~3, (2017) 033006},
  \href{http://arxiv.org/abs/1612.07764}{{\ttfamily arXiv:1612.07764}}.

\bibitem{Esmaili:2012ac}
A.~Esmaili and Y.~Farzan,
  \href{http://dx.doi.org/10.1088/1475-7516/2012/12/014}{{\em JCAP} {\bfseries
  1212} (2012) 014},
\href{http://arxiv.org/abs/1208.6012}{{\ttfamily arXiv:1208.6012}}.
%%CITATION = ARXIV:1208.6012;%%.

\bibitem{Bertone:2004pz}
G.~Bertone, D.~Hooper and J.~Silk,
  \href{http://dx.doi.org/10.1016/j.physrep.2004.08.031}{{\em Phys. Rept.}
  {\bfseries 405} (2005) 279--390},
  \href{http://arxiv.org/abs/hep-ph/0404175}{{\ttfamily arXiv:hep-ph/0404175}}.

\bibitem{Lukovic:2014vma}
V.~Lukovic, P.~Cabella and N.~Vittorio,
  \href{http://dx.doi.org/10.1142/S0217751X14430015}{{\em Int. J. Mod. Phys.}
  {\bfseries A29} (2014) 1443001},
  \href{http://arxiv.org/abs/1411.3556}{{\ttfamily arXiv:1411.3556}}.

\bibitem{Jungman:1995df}
G.~Jungman, M.~Kamionkowski and K.~Griest,
  \href{http://dx.doi.org/10.1016/0370-1573(95)00058-5}{{\em Phys. Rept.}
  {\bfseries 267} (1996) 195--373},
  \href{http://arxiv.org/abs/hep-ph/9506380}{{\ttfamily arXiv:hep-ph/9506380}}.

\bibitem{Feng:2000zu}
J.~L. Feng, K.~T. Matchev and F.~Wilczek,
  \href{http://dx.doi.org/10.1103/PhysRevD.63.045024}{{\em Phys. Rev.}
  {\bfseries D63} (2001) 045024},
  \href{http://arxiv.org/abs/astro-ph/0008115}{{\ttfamily
  arXiv:astro-ph/0008115}}.

\bibitem{Feng:2010gw}
J.~L. Feng, \href{http://dx.doi.org/10.1146/annurev-astro-082708-101659}{{\em
  Ann. Rev. Astron. Astrophys.} {\bfseries 48} (2010) 495--545},
  \href{http://arxiv.org/abs/1003.0904}{{\ttfamily arXiv:1003.0904}}.

\bibitem{Bergstrom:2012fi}
L.~Bergstr\"om, \href{http://dx.doi.org/10.1002/andp.201200116}{{\em Annalen
  Phys.} {\bfseries 524} (2012) 479--496},
  \href{http://arxiv.org/abs/1205.4882}{{\ttfamily arXiv:1205.4882}}.

\bibitem{Griest:1989wd}
K.~Griest and M.~Kamionkowski,
\href{http://dx.doi.org/10.1103/PhysRevLett.64.615}{{\em Phys. Rev. Lett.}
  {\bfseries 64} (1990) 615}.
%%CITATION = PRLTA,64,615;%%.

\bibitem{Chung:1998ua}
D.~J.~H. Chung, E.~W. Kolb and A.~Riotto,
  \href{http://dx.doi.org/10.1103/PhysRevLett.81.4048}{{\em Phys. Rev. Lett.}
  {\bfseries 81} (1998) 4048--4051},
  \href{http://arxiv.org/abs/hep-ph/9805473}{{\ttfamily arXiv:hep-ph/9805473}}.

\bibitem{Kahlhoefer:2017dnp}
F.~Kahlhoefer, \href{http://dx.doi.org/10.1142/S0217751X1730006X}{{\em Int. J.
  Mod. Phys.} {\bfseries A32} (2017) 1730006},
  \href{http://arxiv.org/abs/1702.02430}{{\ttfamily arXiv:1702.02430}}.

\bibitem{Undagoitia:2015gya}
T.~Marrod\'an~Undagoitia and L.~Rauch,
  \href{http://dx.doi.org/10.1088/0954-3899/43/1/013001}{{\em J. Phys.}
  {\bfseries G43} no.~1, (2016) 013001},
  \href{http://arxiv.org/abs/1509.08767}{{\ttfamily arXiv:1509.08767}}.

\bibitem{Press:1985ug}
W.~H. Press and D.~N. Spergel, \href{http://dx.doi.org/10.1086/163485}{{\em
  Astrophys. J.} {\bfseries 296} (1985) 679--684}.
[,277(1985)].
%%CITATION = ASJOA,296,679;%%.

\bibitem{Krauss:1985aaa}
L.~M. Krauss, M.~Srednicki and F.~Wilczek,
\href{http://dx.doi.org/10.1103/PhysRevD.33.2079}{{\em Phys. Rev.} {\bfseries
  D33} (1986) 2079--2083}.
%%CITATION = PHRVA,D33,2079;%%.

\bibitem{Srednicki:1986vj}
M.~Srednicki, K.~A. Olive and J.~Silk,
\href{http://dx.doi.org/10.1016/0550-3213(87)90020-4}{{\em Nucl. Phys.}
  {\bfseries B279} (1987) 804--823}.
%%CITATION = NUPHA,B279,804;%%.

\bibitem{Gaisser:1986ha}
T.~K. Gaisser, G.~Steigman and S.~Tilav,
\href{http://dx.doi.org/10.1103/PhysRevD.34.2206}{{\em Phys. Rev.} {\bfseries
  D34} (1986) 2206}.
%%CITATION = PHRVA,D34,2206;%%.

\bibitem{Ritz:1987mh}
S.~Ritz and D.~Seckel,
\href{http://dx.doi.org/10.1016/0550-3213(88)90660-8}{{\em Nucl. Phys.}
  {\bfseries B304} (1988) 877--908}.
%%CITATION = NUPHA,B304,877;%%.

\bibitem{Gould:1987ww}
A.~Gould,
\href{http://dx.doi.org/10.1086/166347}{{\em Astrophys. J.} {\bfseries 328}
  (1988) 919--939}.
%%CITATION = ASJOA,328,919;%%.

\bibitem{Gould:1987ir}
A.~Gould,
\href{http://dx.doi.org/10.1086/165653}{{\em Astrophys. J.} {\bfseries 321}
  (1987) 571}.
%%CITATION = ASJOA,321,571;%%.

\bibitem{Bergstrom:1998xh}
L.~Bergstr\"om, J.~Edsj\"o and P.~Gondolo,
  \href{http://dx.doi.org/10.1103/PhysRevD.58.103519}{{\em Phys. Rev.}
  {\bfseries D58} (1998) 103519},
  \href{http://arxiv.org/abs/hep-ph/9806293}{{\ttfamily arXiv:hep-ph/9806293}}.

\bibitem{Engel:1992bf}
J.~Engel, S.~Pittel and P.~Vogel,
\href{http://dx.doi.org/10.1142/S0218301392000023}{{\em Int. J. Mod. Phys.}
  {\bfseries E1} (1992) 1--37}.
%%CITATION = IMPAE,E1,1;%%.

\bibitem{Grevesse:1998bj}
N.~Grevesse and A.~J. Sauval,
\href{http://dx.doi.org/10.1023/A:1005161325181}{{\em Space Sci. Rev.}
  {\bfseries 85} (1998) 161--174}.
%%CITATION = SPSRA,85,161;%%.

\bibitem{Catena:2015uha}
R.~Catena and B.~Schwabe,
  \href{http://dx.doi.org/10.1088/1475-7516/2015/04/042}{{\em JCAP} {\bfseries
  1504} no.~04, (2015) 042}, \href{http://arxiv.org/abs/1501.03729}{{\ttfamily
  arXiv:1501.03729}}.

\bibitem{Herndon:1980aaa}
J.~M. Herndon, \href{http://dx.doi.org/10.1098/rspa.1980.0106}{{\em Proc. Roy.
  Soc. London. Series A} {\bfseries 372} no.~1748, (1980) 149}.

\bibitem{Bergstrom:1997fj}
L.~Bergstr\"om, P.~Ullio and J.~H. Buckley,
  \href{http://dx.doi.org/10.1016/S0927-6505(98)00015-2}{{\em Astropart. Phys.}
  {\bfseries 9} (1998) 137--162},
  \href{http://arxiv.org/abs/astro-ph/9712318}{{\ttfamily
  arXiv:astro-ph/9712318}}.

\bibitem{Ackermann:2005fr}
M.~Ackermann {\em et~al.} (AMANDA  Collaboration),
  \href{http://dx.doi.org/10.1016/j.astropartphys.2005.09.006}{{\em Astropart.
  Phys.} {\bfseries 24} (2006) 459--466},
  \href{http://arxiv.org/abs/astro-ph/0508518}{{\ttfamily
  arXiv:astro-ph/0508518}}.

\bibitem{Abbasi:2011eq}
R.~Abbasi {\em et~al.} (IceCube  Collaboration),
  \href{http://dx.doi.org/10.1103/PhysRevD.84.022004}{{\em Phys. Rev.}
  {\bfseries D84} (2011) 022004},
\href{http://arxiv.org/abs/1101.3349}{{\ttfamily arXiv:1101.3349}}.
%%CITATION = ARXIV:1101.3349;%%.

\bibitem{Feldman:1997qc}
G.~J. Feldman and R.~D. Cousins,
  \href{http://dx.doi.org/10.1103/PhysRevD.57.3873}{{\em Phys. Rev.} {\bfseries
  D57} (1998) 3873--3889},
  \href{http://arxiv.org/abs/physics/9711021}{{\ttfamily
  arXiv:physics/9711021}}.

\bibitem{Basu:2007fp}
S.~Basu and H.~M. Antia,
  \href{http://dx.doi.org/10.1016/j.physrep.2007.12.002}{{\em Phys. Rept.}
  {\bfseries 457} (2008) 217--283},
  \href{http://arxiv.org/abs/0711.4590}{{\ttfamily arXiv:0711.4590}}.

\bibitem{Sivertsson:2012qj}
S.~Sivertsson and J.~Edsj\"o,
  \href{http://dx.doi.org/10.1103/PhysRevD.85.129905,
  10.1103/PhysRevD.85.123514}{{\em Phys. Rev.} {\bfseries D85} (2012) 123514},
  \href{http://arxiv.org/abs/1201.1895}{{\ttfamily arXiv:1201.1895}}.

\bibitem{Bottino:1999ei}
A.~Bottino, F.~Donato, N.~Fornengo and S.~Scopel,
  \href{http://dx.doi.org/10.1016/S0927-6505(99)00122-X}{{\em Astropart. Phys.}
  {\bfseries 13} (2000) 215--225},
  \href{http://arxiv.org/abs/hep-ph/9909228}{{\ttfamily arXiv:hep-ph/9909228}}.

\bibitem{Ellis:2008hf}
J.~R. Ellis, K.~A. Olive and C.~Savage,
  \href{http://dx.doi.org/10.1103/PhysRevD.77.065026}{{\em Phys. Rev.}
  {\bfseries D77} (2008) 065026},
  \href{http://arxiv.org/abs/0801.3656}{{\ttfamily arXiv:0801.3656}}.

\bibitem{deAustri:2013saa}
R.~Ruiz~de Austri and C.~P\'erez de~los Heros,
  \href{http://dx.doi.org/10.1088/1475-7516/2013/11/049}{{\em JCAP} {\bfseries
  1311} (2013) 049}, \href{http://arxiv.org/abs/1307.6668}{{\ttfamily
  arXiv:1307.6668}}.

\bibitem{Ambrosio:1998qj}
M.~Ambrosio {\em et~al.} (MACRO  Collaboration),
  \href{http://dx.doi.org/10.1103/PhysRevD.60.082002}{{\em Phys. Rev.}
  {\bfseries D60} (1999) 082002},
  \href{http://arxiv.org/abs/hep-ex/9812020}{{\ttfamily arXiv:hep-ex/9812020}}.

\bibitem{Boliev:1995xz}
M.~M. Boliev, E.~V. Bugaev, A.~V. Butkevich, A.~E. Chudakov, S.~P. Mikheev,
  O.~V. Suvorova and V.~N. Zakidyshev,
\href{http://dx.doi.org/10.1016/0920-5632(96)00214-9}{{\em Nucl. Phys. Proc.
  Suppl.} {\bfseries 48} (1996) 83--86}.
%%CITATION = NUPHZ,48,83;%%.

\bibitem{Mori:1993tj}
M.~Mori {\em et~al.} (Kamiokande  Collaboration),
\href{http://dx.doi.org/10.1103/PhysRevD.48.5505}{{\em Phys. Rev.} {\bfseries
  D48} (1993) 5505--5518}.
%%CITATION = PHRVA,D48,5505;%%.

\bibitem{Mori:1992yq}
M.~Mori {\em et~al.},
\href{http://dx.doi.org/10.1016/0370-2693(92)91249-9}{{\em Phys. Lett.}
  {\bfseries B289} (1992) 463--469}.
%%CITATION = PHLTA,B289,463;%%.

\bibitem{Boliev:2013ai}
M.~M. Boliev, S.~V. Demidov, S.~P. Mikheyev and O.~V. Suvorova,
  \href{http://dx.doi.org/10.1088/1475-7516/2013/09/019}{{\em JCAP} {\bfseries
  1309} (2013) 019}, \href{http://arxiv.org/abs/1301.1138}{{\ttfamily
  arXiv:1301.1138}}.

\bibitem{Choi:2015ara}
K.~Choi {\em et~al.} (Super-Kamiokande  Collaboration),
  \href{http://dx.doi.org/10.1103/PhysRevLett.114.141301}{{\em Phys. Rev.
  Lett.} {\bfseries 114} no.~14, (2015) 141301},
  \href{http://arxiv.org/abs/1503.04858}{{\ttfamily arXiv:1503.04858}}.

\bibitem{Mijakowski:2016cph}
P.~Mijakowski, (Super-Kamiokande  Collaboration),
\href{http://dx.doi.org/10.1088/1742-6596/718/4/042040}{{\em J. Phys. Conf.
  Ser.} {\bfseries 718} no.~4, (2016) 042040}.
%%CITATION = 00462,718,042040;%%.

\bibitem{Belolaptikov:1997ry}
I.~A. Belolaptikov {\em et~al.} (BAIKAL  Collaboration),
\href{http://dx.doi.org/10.1016/S0927-6505(97)00022-4}{{\em Astropart. Phys.}
  {\bfseries 7} (1997) 263--282}.
%%CITATION = APHYE,7,263;%%.

\bibitem{Collaboration:2011nsa}
M.~Ageron {\em et~al.} (ANTARES  Collaboration),
  \href{http://dx.doi.org/10.1016/j.nima.2011.06.103}{{\em Nucl. Instrum.
  Meth.} {\bfseries A656} (2011) 11--38},
  \href{http://arxiv.org/abs/1104.1607}{{\ttfamily arXiv:1104.1607}}.

\bibitem{Halzen:1998sc}
F.~Halzen {\em et~al.} (AMANDA  Collaboration),
  \href{http://dx.doi.org/10.1016/S0370-1573(98)00041-6}{{\em Phys. Rept.}
  {\bfseries 307} (1998) 243--252},
  \href{http://arxiv.org/abs/hep-ex/9804007}{{\ttfamily arXiv:hep-ex/9804007}}.

\bibitem{Ahrens:2002eb}
J.~Ahrens {\em et~al.} (AMANDA  Collaboration),
  \href{http://dx.doi.org/10.1103/PhysRevD.66.032006}{{\em Phys. Rev.}
  {\bfseries D66} (2002) 032006},
  \href{http://arxiv.org/abs/astro-ph/0202370}{{\ttfamily
  arXiv:astro-ph/0202370}}.

\bibitem{Achterberg:2006jf}
A.~Achterberg {\em et~al.} (AMANDA  Collaboration), {\em Astropart. Phys.}
  {\bfseries 26} (2006) 129--139.

\bibitem{Adrian-Martinez:2013ayv}
S.~Adri\'an-Mart\'inez {\em et~al.} (ANTARES  Collaboration),
  \href{http://dx.doi.org/10.1088/1475-7516/2013/11/032}{{\em JCAP} {\bfseries
  1311} (2013) 032}, \href{http://arxiv.org/abs/1302.6516}{{\ttfamily
  arXiv:1302.6516}}.

\bibitem{Adrian-Martinez:2016gti}
S.~Adri\'an-Mart\'inez {\em et~al.} (ANTARES  Collaboration),
  \href{http://dx.doi.org/10.1016/j.physletb.2016.05.019}{{\em Phys. Lett.}
  {\bfseries B759} (2016) 69--74},
  \href{http://arxiv.org/abs/1603.02228}{{\ttfamily arXiv:1603.02228}}.

\bibitem{Amole:2015pla}
C.~Amole {\em et~al.} (PICO  Collaboration),
  \href{http://dx.doi.org/10.1103/PhysRevD.93.052014}{{\em Phys. Rev.}
  {\bfseries D93} no.~5, (2016) 052014},
  \href{http://arxiv.org/abs/1510.07754}{{\ttfamily arXiv:1510.07754}}.

\bibitem{Amole:2016pye}
C.~Amole {\em et~al.} (PICO  Collaboration),
  \href{http://dx.doi.org/10.1103/PhysRevD.93.061101}{{\em Phys. Rev.}
  {\bfseries D93} no.~6, (2016) 061101},
  \href{http://arxiv.org/abs/1601.03729}{{\ttfamily arXiv:1601.03729}}.

\bibitem{Aartsen:2016zhm}
M.~G. Aartsen {\em et~al.} (IceCube  Collaboration),
  \href{http://dx.doi.org/10.1140/epjc/s10052-017-4689-9}{{\em Eur. Phys. J.}
  {\bfseries C77} no.~3, (2017) 146},
  \href{http://arxiv.org/abs/1612.05949}{{\ttfamily arXiv:1612.05949}}.

\bibitem{Griest:1986yu}
K.~Griest and D.~Seckel, \href{http://dx.doi.org/10.1016/0550-3213(87)90293-8,
  10.1016/0550-3213(88)90409-9}{{\em Nucl. Phys.} {\bfseries B283} (1987)
  681--705}.
[Erratum: Nucl. Phys.B296,1034(1988)].
%%CITATION = NUPHA,B283,681;%%.

\bibitem{Gould:1987ju}
A.~Gould,
\href{http://dx.doi.org/10.1086/165652}{{\em Astrophys. J.} {\bfseries 321}
  (1987) 560}.
%%CITATION = ASJOA,321,560;%%.

\bibitem{Abbasi:2009vg}
R.~Abbasi {\em et~al.} (IceCube  Collaboration),
  \href{http://dx.doi.org/10.1103/PhysRevD.81.057101}{{\em Phys. Rev.}
  {\bfseries D81} (2010) 057101},
  \href{http://arxiv.org/abs/0910.4480}{{\ttfamily arXiv:0910.4480}}.

\bibitem{Albuquerque:2010bt}
I.~F.~M. Albuquerque and C.~P\'erez de~los Heros,
  \href{http://dx.doi.org/10.1103/PhysRevD.81.063510}{{\em Phys. Rev.}
  {\bfseries D81} (2010) 063510},
  \href{http://arxiv.org/abs/1001.1381}{{\ttfamily arXiv:1001.1381}}.

\bibitem{Albuquerque:2011ma}
I.~F.~M. Albuquerque, L.~J. Beraldo~e Silva and C.~P\'erez de~los Heros,
  \href{http://dx.doi.org/10.1103/PhysRevD.85.123539}{{\em Phys. Rev.}
  {\bfseries D85} (2012) 123539},
  \href{http://arxiv.org/abs/1107.2408}{{\ttfamily arXiv:1107.2408}}.

\bibitem{Albuquerque:2013xna}
I.~F.~M. Albuquerque, C.~P\'erez de~los Heros and D.~S. Robertson,
  \href{http://dx.doi.org/10.1088/1475-7516/2014/02/047}{{\em JCAP} {\bfseries
  1402} (2014) 047}, \href{http://arxiv.org/abs/1312.0797}{{\ttfamily
  arXiv:1312.0797}}.

\bibitem{Ahmed:2008eu}
Z.~Ahmed {\em et~al.} (CDMS  Collaboration),
  \href{http://dx.doi.org/10.1103/PhysRevLett.102.011301}{{\em Phys. Rev.
  Lett.} {\bfseries 102} (2009) 011301},
  \href{http://arxiv.org/abs/0802.3530}{{\ttfamily arXiv:0802.3530}}.

\bibitem{Lee:2007qn}
H.~S. Lee {\em et~al.} (KIMS  Collaboration),
  \href{http://dx.doi.org/10.1103/PhysRevLett.99.091301}{{\em Phys. Rev. Lett.}
  {\bfseries 99} (2007) 091301},
  \href{http://arxiv.org/abs/0704.0423}{{\ttfamily arXiv:0704.0423}}.

\bibitem{Behnke:2008zza}
E.~Behnke {\em et~al.} (COUPP  Collaboration),
  \href{http://dx.doi.org/10.1126/science.1149999}{{\em Science} {\bfseries
  319} (2008) 933--936}, \href{http://arxiv.org/abs/0804.2886}{{\ttfamily
  arXiv:0804.2886}}.

\bibitem{Arrenberg:2008wy}
S.~Arrenberg, L.~Baudis, K.~Kong, K.~T. Matchev and J.~Yoo,
  \href{http://dx.doi.org/10.1103/PhysRevD.78.056002}{{\em Phys. Rev.}
  {\bfseries D78} (2008) 056002},
  \href{http://arxiv.org/abs/0805.4210}{{\ttfamily arXiv:0805.4210}}.

\bibitem{Oliver:2002up}
J.~F. Oliver, J.~Papavassiliou and A.~Santamaria,
  \href{http://dx.doi.org/10.1103/PhysRevD.67.056002}{{\em Phys. Rev.}
  {\bfseries D67} (2003) 056002},
  \href{http://arxiv.org/abs/hep-ph/0212391}{{\ttfamily arXiv:hep-ph/0212391}}.

\bibitem{Servant:2002aq}
G.~Servant and T.~M.~P. Tait,
  \href{http://dx.doi.org/10.1016/S0550-3213(02)01012-X}{{\em Nucl. Phys.}
  {\bfseries B650} (2003) 391--419},
  \href{http://arxiv.org/abs/hep-ph/0206071}{{\ttfamily arXiv:hep-ph/0206071}}.

\bibitem{Dunkley:2008ie}
J.~Dunkley {\em et~al.} (WMAP  Collaboration),
  \href{http://dx.doi.org/10.1088/0067-0049/180/2/306}{{\em Astrophys. J.
  Suppl.} {\bfseries 180} (2009) 306--329},
  \href{http://arxiv.org/abs/0803.0586}{{\ttfamily arXiv:0803.0586}}.

\bibitem{IceCube:2011aj}
R.~Abbasi {\em et~al.} (IceCube  Collaboration),
  \href{http://dx.doi.org/10.1103/PhysRevD.85.042002}{{\em Phys. Rev.}
  {\bfseries D85} (2012) 042002},
  \href{http://arxiv.org/abs/1112.1840}{{\ttfamily arXiv:1112.1840}}.

\bibitem{Abbasi:2009uz}
R.~Abbasi {\em et~al.} (IceCube  Collaboration),
  \href{http://dx.doi.org/10.1103/PhysRevLett.102.201302}{{\em Phys. Rev.
  Lett.} {\bfseries 102} (2009) 201302},
  \href{http://arxiv.org/abs/0902.2460}{{\ttfamily arXiv:0902.2460}}.

\bibitem{Aartsen:2012kia}
M.~G. Aartsen {\em et~al.} (IceCube  Collaboration),
  \href{http://dx.doi.org/10.1103/PhysRevLett.110.131302}{{\em Phys. Rev.
  Lett.} {\bfseries 110} no.~13, (2013) 131302},
\href{http://arxiv.org/abs/1212.4097}{{\ttfamily arXiv:1212.4097}}.
%%CITATION = ARXIV:1212.4097;%%.

\bibitem{Wiebe:2015jaw}
K.~Wiebe and A.~Steuer, (IceCube  Collaboration),
{\em Proceedings, 34th International Cosmic Ray Conference (ICRC 2015). The
  Hague, The Netherlands, July 30-August 6, 2015.} {\bfseries PoS(ICRC2015)}
  (2016) 1224.
%%CITATION = POSCI,ICRC2015,1224;%%.

\bibitem{Aartsen:2016exj}
M.~G. Aartsen {\em et~al.} (IceCube  Collaboration),
  \href{http://dx.doi.org/10.1088/1475-7516/2016/04/022}{{\em JCAP} {\bfseries
  1604} no.~04, (2016) 022},
\href{http://arxiv.org/abs/1601.00653}{{\ttfamily arXiv:1601.00653}}.
%%CITATION = ARXIV:1601.00653;%%.

\bibitem{Felizardo:2011uw}
M.~Felizardo {\em et~al.},
  \href{http://dx.doi.org/10.1103/PhysRevLett.108.201302}{{\em Phys. Rev.
  Lett.} {\bfseries 108} (2012) 201302},
  \href{http://arxiv.org/abs/1106.3014}{{\ttfamily arXiv:1106.3014}}.

\bibitem{Behnke:2012ys}
E.~Behnke {\em et~al.} (COUPP  Collaboration),
  \href{http://dx.doi.org/10.1103/PhysRevD.86.052001,
  10.1103/PhysRevD.90.079902}{{\em Phys. Rev.} {\bfseries D86} no.~5, (2012)
  052001}, \href{http://arxiv.org/abs/1204.3094}{{\ttfamily arXiv:1204.3094}}.
  [Erratum: Phys. Rev.D90,no.7,079902(2014)].

\bibitem{Silverwood:2012tp}
H.~Silverwood, P.~Scott, M.~Danninger, C.~Savage, J.~Edsj\"o, J.~Adams, A.~M.
  Brown and K.~Hultqvist,
  \href{http://dx.doi.org/10.1088/1475-7516/2013/03/027}{{\em JCAP} {\bfseries
  1303} (2013) 027}, \href{http://arxiv.org/abs/1210.0844}{{\ttfamily
  1210.0844}}.

\bibitem{Akerib:2013tjd}
D.~S. Akerib {\em et~al.} (LUX  Collaboration),
  \href{http://dx.doi.org/10.1103/PhysRevLett.112.091303}{{\em Phys. Rev.
  Lett.} {\bfseries 112} (2014) 091303},
  \href{http://arxiv.org/abs/1310.8214}{{\ttfamily arXiv:1310.8214}}.

\bibitem{Trotta:2009gr}
R.~Trotta, R.~Ruiz~de Austri and C.~P\'erez de~los Heros,
  \href{http://dx.doi.org/10.1088/1475-7516/2009/08/034}{{\em JCAP} {\bfseries
  0908} (2009) 034}, \href{http://arxiv.org/abs/0906.0366}{{\ttfamily
  arXiv:0906.0366}}.

\bibitem{Scott:2012mq}
P.~Scott {\em et~al.} (IceCube  Collaboration),
  \href{http://dx.doi.org/10.1088/1475-7516/2012/11/057}{{\em JCAP} {\bfseries
  1211} (2012) 057},
\href{http://arxiv.org/abs/1207.0810}{{\ttfamily arXiv:1207.0810}}.
%%CITATION = ARXIV:1207.0810;%%.

\bibitem{Kane:1993td}
G.~L. Kane, C.~F. Kolda, L.~Roszkowski and J.~D. Wells,
  \href{http://dx.doi.org/10.1103/PhysRevD.49.6173}{{\em Phys. Rev.} {\bfseries
  D49} (1994) 6173--6210},
  \href{http://arxiv.org/abs/hep-ph/9312272}{{\ttfamily arXiv:hep-ph/9312272}}.

\bibitem{ATLAS:2012sma}
ATLAS Collaboration, {\em ATLAS-CONF-2012-033} (2012) .

\bibitem{Aartsen:2016fep}
M.~G. Aartsen {\em et~al.} (IceCube  Collaboration),
  \href{http://dx.doi.org/10.1140/epjc/s10052-016-4582-y}{{\em Eur. Phys. J.}
  {\bfseries C77} no.~2, (2017) 82},
\href{http://arxiv.org/abs/1609.01492}{{\ttfamily arXiv:1609.01492}}.
%%CITATION = ARXIV:1609.01492;%%.

\bibitem{Desai:2004pq}
S.~Desai {\em et~al.} (Super-Kamiokande  Collaboration),
  \href{http://dx.doi.org/10.1103/PhysRevD.70.083523,
  10.1103/PhysRevD.70.109901}{{\em Phys. Rev.} {\bfseries D70} (2004) 083523},
  \href{http://arxiv.org/abs/hep-ex/0404025}{{\ttfamily arXiv:hep-ex/0404025}}.
  [Erratum: Phys. Rev.D70,109901(2004)].

\bibitem{Agnese:2013jaa}
R.~Agnese {\em et~al.} (SuperCDMS  Collaboration),
  \href{http://dx.doi.org/10.1103/PhysRevLett.112.041302}{{\em Phys. Rev.
  Lett.} {\bfseries 112} no.~4, (2014) 041302},
  \href{http://arxiv.org/abs/1309.3259}{{\ttfamily arXiv:1309.3259}}.

\bibitem{Aartsen:2015xej}
M.~G. Aartsen {\em et~al.} (IceCube  Collaboration),
  \href{http://dx.doi.org/10.1140/epjc/s10052-015-3713-1}{{\em Eur. Phys. J.}
  {\bfseries C75} no.~10, (2015) 492},
\href{http://arxiv.org/abs/1505.07259}{{\ttfamily arXiv:1505.07259}}.
%%CITATION = ARXIV:1505.07259;%%.

\bibitem{Navarro:1995iw}
J.~F. Navarro, C.~S. Frenk and S.~D.~M. White,
  \href{http://dx.doi.org/10.1086/177173}{{\em Astrophys. J.} {\bfseries 462}
  (1996) 563--575}, \href{http://arxiv.org/abs/astro-ph/9508025}{{\ttfamily
  arXiv:astro-ph/9508025}}.

\bibitem{Kravtsov:1997dp}
A.~V. Kravtsov, A.~A. Klypin, J.~S. Bullock and J.~R. Primack,
  \href{http://dx.doi.org/10.1086/305884}{{\em Astrophys. J.} {\bfseries 502}
  (1998) 48}, \href{http://arxiv.org/abs/astro-ph/9708176}{{\ttfamily
  arXiv:astro-ph/9708176}}.

\bibitem{Moore:1999gc}
B.~Moore, T.~R. Quinn, F.~Governato, J.~Stadel and G.~Lake,
  \href{http://dx.doi.org/10.1046/j.1365-8711.1999.03039.x}{{\em Mon. Not. Roy.
  Astron. Soc.} {\bfseries 310} (1999) 1147--1152},
  \href{http://arxiv.org/abs/astro-ph/9903164}{{\ttfamily
  arXiv:astro-ph/9903164}}.

\bibitem{Burkert:1995yz}
A.~Burkert, \href{http://dx.doi.org/10.1086/309560}{{\em IAU Symp.} {\bfseries
  171} (1996) 175}, \href{http://arxiv.org/abs/astro-ph/9504041}{{\ttfamily
  arXiv:astro-ph/9504041}}. [Astrophys. J.447,L25(1995)].

\bibitem{deBlok:2009sp}
W.~J.~G. de~Blok, \href{http://dx.doi.org/10.1155/2010/789293}{{\em Adv.
  Astron.} {\bfseries 2010} (2010) 789293},
  \href{http://arxiv.org/abs/0910.3538}{{\ttfamily arXiv:0910.3538}}.

\bibitem{Ricotti:2002qu}
M.~Ricotti, \href{http://dx.doi.org/10.1046/j.1365-8711.2003.06910.x}{{\em Mon.
  Not. Roy. Astron. Soc.} {\bfseries 344} (2003) 1237},
  \href{http://arxiv.org/abs/astro-ph/0212146}{{\ttfamily
  arXiv:astro-ph/0212146}}.

\bibitem{Dekek:1986gu}
A.~Dekel and J.~Silk,
\href{http://dx.doi.org/10.1086/164050}{{\em Astrophys. J.} {\bfseries 303}
  (1986) 39--55}.
%%CITATION = ASJOA,303,39;%%.

\bibitem{Ogiya:2012jq}
G.~Ogiya and M.~Mori, \href{http://dx.doi.org/10.1088/0004-637X/793/1/46}{{\em
  Astrophys. J.} {\bfseries 793} (2014) 46},
  \href{http://arxiv.org/abs/1206.5412}{{\ttfamily arXiv:1206.5412}}.

\bibitem{Aartsen:2014hva}
M.~G. Aartsen {\em et~al.} (IceCube  Collaboration),
  \href{http://dx.doi.org/10.1140/epjc/s10052-014-3224-5}{{\em Eur. Phys. J.}
  {\bfseries C75} no.~99, (2015) 20},
\href{http://arxiv.org/abs/1406.6868}{{\ttfamily arXiv:1406.6868}}.
%%CITATION = ARXIV:1406.6868;%%.

\bibitem{Aartsen:2013dxa}
M.~G. Aartsen {\em et~al.} (IceCube  Collaboration),
  \href{http://dx.doi.org/10.1103/PhysRevD.88.122001}{{\em Phys. Rev.}
  {\bfseries D88} (2013) 122001},
\href{http://arxiv.org/abs/1307.3473}{{\ttfamily arXiv:1307.3473}}.
%%CITATION = ARXIV:1307.3473;%%.

\bibitem{Aartsen:2016pfc}
M.~G. Aartsen {\em et~al.} (IceCube  Collaboration),
  \href{http://dx.doi.org/10.1140/epjc/s10052-016-4375-3}{{\em Eur. Phys. J.}
  {\bfseries C76} no.~10, (2016) 531},
  \href{http://arxiv.org/abs/1606.00209}{{\ttfamily 1606.00209}}.

\bibitem{Albert:2016emp}
A.~Albert {\em et~al.},
  \href{http://dx.doi.org/10.1016/j.physletb.2017.03.063}{{\em Phys. Lett.}
  {\bfseries B769} (2017) 249--254},
  \href{http://arxiv.org/abs/1612.04595}{{\ttfamily arXiv:1612.04595}}.

\bibitem{Abdallah:2016ygi}
H.~Abdallah {\em et~al.} (H.E.S.S.  Collaboration),
  \href{http://dx.doi.org/10.1103/PhysRevLett.117.111301}{{\em Phys. Rev.
  Lett.} {\bfseries 117} no.~11, (2016) 111301},
  \href{http://arxiv.org/abs/1607.08142}{{\ttfamily arXiv:1607.08142}}.

\bibitem{Ahnen:2016qkx}
M.~L. Ahnen {\em et~al.} (Fermi-LAT, MAGIC  Collaboration),
  \href{http://dx.doi.org/10.1088/1475-7516/2016/02/039}{{\em JCAP} {\bfseries
  1602} no.~02, (2016) 039}, \href{http://arxiv.org/abs/1601.06590}{{\ttfamily
  arXiv:1601.06590}}.

\bibitem{Aartsen:2017ulx}
M.~G. Aartsen {\em et~al.} (IceCube  Collaboration),
  \href{http://dx.doi.org/10.1140/epjc/s10052-017-5213-y}{{\em Eur. Phys. J.}
  {\bfseries C77} no.~9, (2017) 627},
  \href{http://arxiv.org/abs/1705.08103}{{\ttfamily arXiv:1705.08103}}.

\bibitem{Yuksel:2007ac}
H.~Yuksel, S.~Horiuchi, J.~F. Beacom and S.~Ando,
  \href{http://dx.doi.org/10.1103/PhysRevD.76.123506}{{\em Phys. Rev.}
  {\bfseries D76} (2007) 123506},
  \href{http://arxiv.org/abs/0707.0196}{{\ttfamily arXiv:0707.0196}}.

\bibitem{Arguelles:2017atb}
C.~A. Arg\"uelles, A.~Kheirandish and A.~C. Vincent,
  \href{http://dx.doi.org/10.1103/PhysRevLett.119.201801}{{\em Phys. Rev.
  Lett.} {\bfseries 119} no.~20, (2017) 201801},
\href{http://arxiv.org/abs/1703.00451}{{\ttfamily arXiv:1703.00451}}.
%%CITATION = ARXIV:1703.00451;%%.

\bibitem{Cherry:2014xra}
J.~F. Cherry, A.~Friedland and I.~M. Shoemaker,
\href{http://arxiv.org/abs/1411.1071}{{\ttfamily arXiv:1411.1071}}.
%%CITATION = ARXIV:1411.1071;%%.

\bibitem{Davis:2015rza}
J.~H. Davis and J.~Silk,
\href{http://arxiv.org/abs/1505.01843}{{\ttfamily arXiv:1505.01843}}.
%%CITATION = ARXIV:1505.01843;%%.

\bibitem{Farzan:2014gza}
Y.~Farzan and S.~Palomares-Ruiz,
  \href{http://dx.doi.org/10.1088/1475-7516/2014/06/014}{{\em JCAP} {\bfseries
  1406} (2014) 014}, \href{http://arxiv.org/abs/1401.7019}{{\ttfamily
  arXiv:1401.7019}}.

\bibitem{Boehm:2014vja}
C.~Boehm, J.~A. Schewtschenko, R.~J. Wilkinson, C.~M. Baugh and S.~Pascoli,
  \href{http://dx.doi.org/10.1093/mnrasl/slu115}{{\em Mon. Not. Roy. Astron.
  Soc.} {\bfseries 445} (2014) L31--L35},
  \href{http://arxiv.org/abs/1404.7012}{{\ttfamily arXiv:1404.7012}}.

\bibitem{Bertoni:2014mva}
B.~Bertoni, S.~Ipek, D.~McKeen and A.~E. Nelson,
  \href{http://dx.doi.org/10.1007/JHEP04(2015)170}{{\em JHEP} {\bfseries 04}
  (2015) 170}, \href{http://arxiv.org/abs/1412.3113}{{\ttfamily
  arXiv:1412.3113}}.

\bibitem{Feldstein:2013kka}
B.~Feldstein, A.~Kusenko, S.~Matsumoto and T.~T. Yanagida,
  \href{http://dx.doi.org/10.1103/PhysRevD.88.015004}{{\em Phys. Rev.}
  {\bfseries D88} no.~1, (2013) 015004},
\href{http://arxiv.org/abs/1303.7320}{{\ttfamily arXiv:1303.7320}}.
%%CITATION = ARXIV:1303.7320;%%.

\bibitem{Esmaili:2013gha}
A.~Esmaili and P.~D. Serpico,
  \href{http://dx.doi.org/10.1088/1475-7516/2013/11/054}{{\em JCAP} {\bfseries
  1311} (2013) 054},
\href{http://arxiv.org/abs/1308.1105}{{\ttfamily arXiv:1308.1105}}.
%%CITATION = ARXIV:1308.1105;%%.

\bibitem{Bhattacharya:2014vwa}
A.~Bhattacharya, M.~H. Reno and I.~Sarcevic,
  \href{http://dx.doi.org/10.1007/JHEP06(2014)110}{{\em JHEP} {\bfseries 06}
  (2014) 110},
\href{http://arxiv.org/abs/1403.1862}{{\ttfamily arXiv:1403.1862}}.
%%CITATION = ARXIV:1403.1862;%%.

\bibitem{Boucenna:2015tra}
S.~M. Boucenna, M.~Chianese, G.~Mangano, G.~Miele, S.~Morisi, O.~Pisanti and
  E.~Vitagliano, \href{http://dx.doi.org/10.1088/1475-7516/2015/12/055}{{\em
  JCAP} {\bfseries 1512} no.~12, (2015) 055},
\href{http://arxiv.org/abs/1507.01000}{{\ttfamily arXiv:1507.01000}}.
%%CITATION = ARXIV:1507.01000;%%.

\bibitem{Ko:2015nma}
P.~Ko and Y.~Tang, \href{http://dx.doi.org/10.1016/j.physletb.2015.10.021}{{\em
  Phys. Lett.} {\bfseries B751} (2015) 81--88},
\href{http://arxiv.org/abs/1508.02500}{{\ttfamily arXiv:1508.02500}}.
%%CITATION = ARXIV:1508.02500;%%.

\bibitem{Bhattacharya:2017jaw}
A.~Bhattacharya, A.~Esmaili, S.~Palomares-Ruiz and I.~Sarcevic,
  \href{http://dx.doi.org/10.1088/1475-7516/2017/07/027}{{\em JCAP} {\bfseries
  1707} no.~07, (2017) 027},
\href{http://arxiv.org/abs/1706.05746}{{\ttfamily arXiv:1706.05746}}.
%%CITATION = ARXIV:1706.05746;%%.

\bibitem{Hiroshima:2017hmy}
N.~Hiroshima, R.~Kitano, K.~Kohri and K.~Murase,
  \href{http://dx.doi.org/10.1103/PhysRevD.97.023006}{{\em Phys. Rev.}
  {\bfseries D97} no.~2, (2018) 023006},
\href{http://arxiv.org/abs/1705.04419}{{\ttfamily arXiv:1705.04419}}.
%%CITATION = ARXIV:1705.04419;%%.

\bibitem{Aisati:2015vma}
C.~El~Aisati, M.~Gustafsson and T.~Hambye,
  \href{http://dx.doi.org/10.1103/PhysRevD.92.123515}{{\em Phys. Rev.}
  {\bfseries D92} no.~12, (2015) 123515},
  \href{http://arxiv.org/abs/1506.02657}{{\ttfamily arXiv:1506.02657}}.

\bibitem{Chianese:2016opp}
M.~Chianese, G.~Miele, S.~Morisi and E.~Vitagliano,
  \href{http://dx.doi.org/10.1016/j.physletb.2016.03.084}{{\em Phys. Lett.}
  {\bfseries B757} (2016) 251--256},
\href{http://arxiv.org/abs/1601.02934}{{\ttfamily arXiv:1601.02934}}.
%%CITATION = ARXIV:1601.02934;%%.

\bibitem{Chianese:2016kpu}
M.~Chianese, G.~Miele and S.~Morisi,
  \href{http://dx.doi.org/10.1088/1475-7516/2017/01/007}{{\em JCAP} {\bfseries
  1701} no.~01, (2017) 007},
\href{http://arxiv.org/abs/1610.04612}{{\ttfamily arXiv:1610.04612}}.
%%CITATION = ARXIV:1610.04612;%%.

\bibitem{Chianese:2017nwe}
M.~Chianese, G.~Miele and S.~Morisi,
  \href{http://dx.doi.org/10.1016/j.physletb.2017.09.016}{{\em Phys. Lett.}
  {\bfseries B773} (2017) 591--595},
\href{http://arxiv.org/abs/1707.05241}{{\ttfamily arXiv:1707.05241}}.
%%CITATION = ARXIV:1707.05241;%%.

\bibitem{Dev:2016qbd}
P.~S.~B. Dev, D.~Kazanas, R.~N. Mohapatra, V.~L. Teplitz and Y.~Zhang,
  \href{http://dx.doi.org/10.1088/1475-7516/2016/08/034}{{\em JCAP} {\bfseries
  1608} no.~08, (2016) 034},
\href{http://arxiv.org/abs/1606.04517}{{\ttfamily arXiv:1606.04517}}.
%%CITATION = ARXIV:1606.04517;%%.

\bibitem{DiBari:2016guw}
P.~Di~Bari, P.~O. Ludl and S.~Palomares-Ruiz,
  \href{http://dx.doi.org/10.1088/1475-7516/2016/11/044}{{\em JCAP} {\bfseries
  1611} no.~11, (2016) 044},
\href{http://arxiv.org/abs/1606.06238}{{\ttfamily arXiv:1606.06238}}.
%%CITATION = ARXIV:1606.06238;%%.

\bibitem{Borah:2017xgm}
D.~Borah, A.~Dasgupta, U.~K. Dey, S.~Patra and G.~Tomar,
  \href{http://dx.doi.org/10.1007/JHEP09(2017)005}{{\em JHEP} {\bfseries 09}
  (2017) 005},
\href{http://arxiv.org/abs/1704.04138}{{\ttfamily arXiv:1704.04138}}.
%%CITATION = ARXIV:1704.04138;%%.

\bibitem{Fong:2014bsa}
C.~S. Fong, H.~Minakata, B.~Panes and R.~Zukanovich~Funchal,
  \href{http://dx.doi.org/10.1007/JHEP02(2015)189}{{\em JHEP} {\bfseries 02}
  (2015) 189},
\href{http://arxiv.org/abs/1411.5318}{{\ttfamily arXiv:1411.5318}}.
%%CITATION = ARXIV:1411.5318;%%.

\bibitem{Roland:2015yoa}
S.~B. Roland, B.~Shakya and J.~D. Wells,
  \href{http://dx.doi.org/10.1103/PhysRevD.92.095018}{{\em Phys. Rev.}
  {\bfseries D92} no.~9, (2015) 095018},
\href{http://arxiv.org/abs/1506.08195}{{\ttfamily arXiv:1506.08195}}.
%%CITATION = ARXIV:1506.08195;%%.

\bibitem{Fiorentin:2016avj}
M.~Re~Fiorentin, V.~Niro and N.~Fornengo,
  \href{http://dx.doi.org/10.1007/JHEP11(2016)022}{{\em JHEP} {\bfseries 11}
  (2016) 022},
\href{http://arxiv.org/abs/1606.04445}{{\ttfamily arXiv:1606.04445}}.
%%CITATION = ARXIV:1606.04445;%%.

\bibitem{Chianese:2016smc}
M.~Chianese and A.~Merle,
  \href{http://dx.doi.org/10.1088/1475-7516/2017/04/017}{{\em JCAP} {\bfseries
  1704} no.~04, (2017) 017},
\href{http://arxiv.org/abs/1607.05283}{{\ttfamily arXiv:1607.05283}}.
%%CITATION = ARXIV:1607.05283;%%.

\bibitem{Albert:2017vtb}
A.~Albert {\em et~al.} (HAWC  Collaboration),
  \href{http://dx.doi.org/10.3847/1538-4357/aaa6d8}{{\em Astrophys. J.}
  {\bfseries 853} no.~2, (2018) 154},
  \href{http://arxiv.org/abs/1706.01277}{{\ttfamily arXiv:1706.01277}}.

\bibitem{Abeysekara:2017jxs}
A.~U. Abeysekara {\em et~al.} (HAWC  Collaboration),
  \href{http://dx.doi.org/10.1088/1475-7516/2018/02/049}{{\em JCAP} {\bfseries
  1802} no.~02, (2018) 049}, \href{http://arxiv.org/abs/1710.10288}{{\ttfamily
  arXiv:1710.10288}}.

\bibitem{Ackermann:2012rg}
M.~Ackermann {\em et~al.} (Fermi-LAT  Collaboration),
  \href{http://dx.doi.org/10.1088/0004-637X/761/2/91}{{\em Astrophys. J.}
  {\bfseries 761} (2012) 91}, \href{http://arxiv.org/abs/1205.6474}{{\ttfamily
  arXiv:1205.6474}}.

\bibitem{Aartsen:2018mxl}
M.~G. Aartsen {\em et~al.} (IceCube  Collaboration),
  \href{http://arxiv.org/abs/1804.03848}{{\ttfamily arXiv:1804.03848}}.

\bibitem{Bai:2013nga}
Y.~Bai, R.~Lu and J.~Salvado,
  \href{http://dx.doi.org/10.1007/JHEP01(2016)161}{{\em JHEP} {\bfseries 01}
  (2016) 161},
\href{http://arxiv.org/abs/1311.5864}{{\ttfamily arXiv:1311.5864}}.
%%CITATION = ARXIV:1311.5864;%%.

\bibitem{Esmaili:2014rma}
A.~Esmaili, S.~K. Kang and P.~D. Serpico,
  \href{http://dx.doi.org/10.1088/1475-7516/2014/12/054}{{\em JCAP} {\bfseries
  1412} no.~12, (2014) 054},
\href{http://arxiv.org/abs/1410.5979}{{\ttfamily arXiv:1410.5979}}.
%%CITATION = ARXIV:1410.5979;%%.

\bibitem{Ahlers:2015moa}
M.~Ahlers, Y.~Bai, V.~Barger and R.~Lu,
  \href{http://dx.doi.org/10.1103/PhysRevD.93.013009}{{\em Phys. Rev.}
  {\bfseries D93} no.~1, (2016) 013009},
\href{http://arxiv.org/abs/1505.03156}{{\ttfamily arXiv:1505.03156}}.
%%CITATION = ARXIV:1505.03156;%%.

\bibitem{Murase:2015gea}
K.~Murase, R.~Laha, S.~Ando and M.~Ahlers,
  \href{http://dx.doi.org/10.1103/PhysRevLett.115.071301}{{\em Phys. Rev.
  Lett.} {\bfseries 115} no.~7, (2015) 071301},
\href{http://arxiv.org/abs/1503.04663}{{\ttfamily arXiv:1503.04663}}.
%%CITATION = ARXIV:1503.04663;%%.

\bibitem{Esmaili:2015xpa}
A.~Esmaili and P.~D. Serpico,
  \href{http://dx.doi.org/10.1088/1475-7516/2015/10/014}{{\em JCAP} {\bfseries
  1510} no.~10, (2015) 014},
\href{http://arxiv.org/abs/1505.06486}{{\ttfamily arXiv:1505.06486}}.
%%CITATION = ARXIV:1505.06486;%%.

\bibitem{Cohen:2016uyg}
T.~Cohen, K.~Murase, N.~L. Rodd, B.~R. Safdi and Y.~Soreq,
  \href{http://dx.doi.org/10.1103/PhysRevLett.119.021102}{{\em Phys. Rev.
  Lett.} {\bfseries 119} no.~2, (2017) 021102},
\href{http://arxiv.org/abs/1612.05638}{{\ttfamily arXiv:1612.05638}}.
%%CITATION = ARXIV:1612.05638;%%.

\bibitem{Blanco:2017sbc}
C.~Blanco, J.~P. Harding and D.~Hooper,
  \href{http://dx.doi.org/10.1088/1475-7516/2018/04/060}{{\em JCAP} {\bfseries
  1804} no.~04, (2018) 060},
\href{http://arxiv.org/abs/1712.02805}{{\ttfamily arXiv:1712.02805}}.
%%CITATION = ARXIV:1712.02805;%%.

\bibitem{Dirac:1928hu}
P.~A.~M. Dirac,
\href{http://dx.doi.org/10.1098/rspa.1928.0023}{{\em Proc. Roy. Soc. Lond.}
  {\bfseries A117} (1928) 610--624}.
%%CITATION = PRSLA,A117,610;%%.

\bibitem{Dirac:1931kp}
P.~A.~M. Dirac,
\href{http://dx.doi.org/10.1098/rspa.1931.0130}{{\em Proc. Roy. Soc. Lond.}
  {\bfseries A133} (1931) 60--72}.
%%CITATION = PRSLA,A133,60;%%.

\bibitem{Preskill:1984gd}
J.~Preskill,
\href{http://dx.doi.org/10.1146/annurev.ns.34.120184.002333}{{\em Ann. Rev.
  Nucl. Part. Sci.} {\bfseries 34} (1984) 461--530}.
%%CITATION = ARNUA,34,461;%%.

\bibitem{tHooft:1974kcl}
G.~'t~Hooft, \href{http://dx.doi.org/10.1016/0550-3213(74)90486-6}{{\em Nucl.
  Phys.} {\bfseries B79} (1974) 276--284}.
[,291(1974)].
%%CITATION = NUPHA,B79,276;%%.

\bibitem{Polyakov:1974ek}
A.~M. Polyakov, {\em JETP Lett.} {\bfseries 20} (1974) 194--195.
[,300(1974)].
%%CITATION = JTPLA,20,194;%%.

\bibitem{Georgi:1974sy}
H.~Georgi and S.~L. Glashow,
\href{http://dx.doi.org/10.1103/PhysRevLett.32.438}{{\em Phys. Rev. Lett.}
  {\bfseries 32} (1974) 438--441}.
%%CITATION = PRLTA,32,438;%%.

\bibitem{Wick:2000yc}
S.~D. Wick, T.~W. Kephart, T.~J. Weiler and P.~L. Biermann,
  \href{http://dx.doi.org/10.1016/S0927-6505(02)00200-1}{{\em Astropart. Phys.}
  {\bfseries 18} (2003) 663--687},
\href{http://arxiv.org/abs/astro-ph/0001233}{{\ttfamily
  arXiv:astro-ph/0001233}}.
%%CITATION = ASTRO-PH/0001233;%%.

\bibitem{Acharya:2014nyr}
B.~Acharya {\em et~al.} (MoEDAL  Collaboration),
  \href{http://dx.doi.org/10.1142/S0217751X14300506}{{\em Int. J. Mod. Phys.}
  {\bfseries A29} (2014) 1430050},
\href{http://arxiv.org/abs/1405.7662}{{\ttfamily arXiv:1405.7662}}.
%%CITATION = ARXIV:1405.7662;%%.

\bibitem{Cho:2013vba}
Y.~M. Cho, K.~Kim and J.~H. Yoon,
  \href{http://dx.doi.org/10.1140/epjc/s10052-015-3290-3}{{\em Eur. Phys. J.}
  {\bfseries C75} no.~2, (2015) 67},
\href{http://arxiv.org/abs/1305.1699}{{\ttfamily arXiv:1305.1699}}.
%%CITATION = ARXIV:1305.1699;%%.

\bibitem{Ellis:2016glu}
J.~Ellis, N.~E. Mavromatos and T.~You,
  \href{http://dx.doi.org/10.1016/j.physletb.2016.02.048}{{\em Phys. Lett.}
  {\bfseries B756} (2016) 29--35},
\href{http://arxiv.org/abs/1602.01745}{{\ttfamily arXiv:1602.01745}}.
%%CITATION = ARXIV:1602.01745;%%.

\bibitem{Kibble:1976sj}
T.~W.~B. Kibble,
\href{http://dx.doi.org/10.1088/0305-4470/9/8/029}{{\em J. Phys.} {\bfseries
  A9} (1976) 1387--1398}.
%%CITATION = JPAGA,A9,1387;%%.

\bibitem{Kibble:1980mv}
T.~W.~B. Kibble,
\href{http://dx.doi.org/10.1016/0370-1573(80)90091-5}{{\em Phys. Rept.}
  {\bfseries 67} (1980) 183}.
%%CITATION = PRPLC,67,183;%%.

\bibitem{Parker:1970xv}
E.~N. Parker,
\href{http://dx.doi.org/10.1086/150442}{{\em Astrophys. J.} {\bfseries 160}
  (1970) 383}.
%%CITATION = ASJOA,160,383;%%.

\bibitem{Turner:1982ag}
M.~S. Turner, E.~N. Parker and T.~J. Bogdan,
\href{http://dx.doi.org/10.1103/PhysRevD.26.1296}{{\em Phys. Rev.} {\bfseries
  D26} (1982) 1296}.
%%CITATION = PHRVA,D26,1296;%%.

\bibitem{Rephaeli:1982nv}
Y.~Rephaeli and M.~S. Turner,
\href{http://dx.doi.org/10.1016/0370-2693(83)90897-3}{{\em Phys. Lett.}
  {\bfseries 121B} (1983) 115--118}.
%%CITATION = PHLTA,121B,115;%%.

\bibitem{Adams:1993fj}
F.~C. Adams, M.~Fatuzzo, K.~Freese, G.~Tarle, R.~Watkins and M.~S. Turner,
\href{http://dx.doi.org/10.1103/PhysRevLett.70.2511}{{\em Phys. Rev. Lett.}
  {\bfseries 70} (1993) 2511--2514}.
%%CITATION = PRLTA,70,2511;%%.

\bibitem{Patrizii:2015uea}
L.~Patrizii and M.~Spurio,
  \href{http://dx.doi.org/10.1146/annurev-nucl-102014-022137}{{\em Ann. Rev.
  Nucl. Part. Sci.} {\bfseries 65} (2015) 279--302},
\href{http://arxiv.org/abs/1510.07125}{{\ttfamily arXiv:1510.07125}}.
%%CITATION = ARXIV:1510.07125;%%.

\bibitem{ObertackePollmann:2016uvi}
A.~Obertacke~Pollmann (IceCube  Collaboration),
  \href{http://dx.doi.org/10.1051/epjconf/201716407019}{{\em EPJ Web Conf.}
  {\bfseries 164} (2017) 07019},
\href{http://arxiv.org/abs/1610.06397}{{\ttfamily arXiv:1610.06397}}.
%%CITATION = ARXIV:1610.06397;%%.

\bibitem{Rubakov:1982fp}
V.~A. Rubakov,
\href{http://dx.doi.org/10.1016/0550-3213(82)90034-7}{{\em Nucl. Phys.}
  {\bfseries B203} (1982) 311--348}.
%%CITATION = NUPHA,B203,311;%%.

\bibitem{Callan:1983nx}
C.~G. Callan, Jr. and E.~Witten,
\href{http://dx.doi.org/10.1016/0550-3213(84)90088-9}{{\em Nucl. Phys.}
  {\bfseries B239} (1984) 161--176}.
%%CITATION = NUPHA,B239,161;%%.

\bibitem{Arafune:1983uz}
J.~Arafune and M.~Fukugita,
\href{http://dx.doi.org/10.1103/PhysRevLett.50.1901}{{\em Phys. Rev. Lett.}
  {\bfseries 50} (1983) 1901}.
%%CITATION = PRLTA,50,1901;%%.

\bibitem{Nath:2006ut}
P.~Nath and P.~Fileviez~Perez,
  \href{http://dx.doi.org/10.1016/j.physrep.2007.02.010}{{\em Phys. Rept.}
  {\bfseries 441} (2007) 191--317},
\href{http://arxiv.org/abs/hep-ph/0601023}{{\ttfamily arXiv:hep-ph/0601023}}.
%%CITATION = HEP-PH/0601023;%%.

\bibitem{Lauber:2017}
F.~Lauber, in {\em {6th International Conference on New Frontiers in Physics
  (ICNFP 2017)}}.
\newblock 2017.

\bibitem{Aartsen:2014awd}
M.~G. Aartsen {\em et~al.} (IceCube  Collaboration),
  \href{http://dx.doi.org/10.1140/epjc/s10052-014-2938-8}{{\em Eur. Phys. J.}
  {\bfseries C74} no.~7, (2014) 2938},
\href{http://arxiv.org/abs/1402.3460}{{\ttfamily arXiv:1402.3460}}.
%%CITATION = ARXIV:1402.3460;%%.

\bibitem{Aartsen:2015exf}
M.~G. Aartsen {\em et~al.} (IceCube  Collaboration),
  \href{http://dx.doi.org/10.1140/epjc/s10052-016-3953-8}{{\em Eur. Phys. J.}
  {\bfseries C76} no.~3, (2016) 133},
\href{http://arxiv.org/abs/1511.01350}{{\ttfamily arXiv:1511.01350}}.
%%CITATION = ARXIV:1511.01350;%%.

\bibitem{Wu:1976ge}
T.~T. Wu and C.~N. Yang,
\href{http://dx.doi.org/10.1016/0550-3213(76)90143-7}{{\em Nucl. Phys.}
  {\bfseries B107} (1976) 365}.
%%CITATION = NUPHA,B107,365;%%.

\bibitem{Kazama:1976fm}
Y.~Kazama, C.~N. Yang and A.~S. Goldhaber,
\href{http://dx.doi.org/10.1103/PhysRevD.15.2287}{{\em Phys. Rev.} {\bfseries
  D15} (1977) 2287--2299}.
%%CITATION = PHRVA,D15,2287;%%.

\bibitem{ICRC-Lumi:2017aaa}
A.~Pollmann and S.~Pieper (IceCube  Collaboration), {\em Proceedings, 35th
  International Cosmic Ray Conference (ICRC 2017). Busan, Korea, 10-20 July,
  2017.} {\bfseries PoS(ICRC2017)} (2017) 1060.

\bibitem{Baikal08}
V.~Aynutdinov {\em et~al.} (BAIKAL  Collaboration),
  \href{http://dx.doi.org/10.1016/j.astropartphys.2008.03.006}{{\em Astrophys.
  J.} {\bfseries 29} (2008) 366--372}.

\bibitem{Quickenden82}
T.~Quickenden, S.~Trotman and D.~Sangster,
  \href{http://dx.doi.org/10.1063/1.444352}{{\em J. Chem. Phys.} {\bfseries 77}
  (1982) 3790}.

\bibitem{Abbasi:2012eda}
R.~Abbasi {\em et~al.} (IceCube  Collaboration),
  \href{http://dx.doi.org/10.1103/PhysRevD.87.022001}{{\em Phys. Rev.}
  {\bfseries D87} no.~2, (2013) 022001},
\href{http://arxiv.org/abs/1208.4861}{{\ttfamily arXiv:1208.4861}}.
%%CITATION = ARXIV:1208.4861;%%.

\bibitem{AdrianMartinez:2011xr}
S.~Adri\'an-Mart\'inez {\em et~al.} (ANTARES  Collaboration),
  \href{http://dx.doi.org/10.1016/j.astropartphys.2012.02.007}{{\em Astropart.
  Phys.} {\bfseries 35} (2012) 634--640},
\href{http://arxiv.org/abs/1110.2656}{{\ttfamily arXiv:1110.2656}}.
%%CITATION = ARXIV:1110.2656;%%.

\bibitem{Albert:2017fud}
A.~Albert {\em et~al.} (ANTARES  Collaboration),
  \href{http://dx.doi.org/10.1007/JHEP07(2017)054}{{\em JHEP} {\bfseries 07}
  (2017) 054},
\href{http://arxiv.org/abs/1703.00424}{{\ttfamily arXiv:1703.00424}}.
%%CITATION = ARXIV:1703.00424;%%.

\bibitem{Hogan:2008sx}
D.~P. Hogan, D.~Z. Besson, J.~P. Ralston, I.~Kravchenko and D.~Seckel,
  \href{http://dx.doi.org/10.1103/PhysRevD.78.075031}{{\em Phys. Rev.}
  {\bfseries D78} (2008) 075031},
\href{http://arxiv.org/abs/0806.2129}{{\ttfamily arXiv:0806.2129}}.
%%CITATION = ARXIV:0806.2129;%%.

\bibitem{Detrixhe:2010xi}
M.~Detrixhe {\em et~al.} (ANITA-II  Collaboration),
  \href{http://dx.doi.org/10.1103/PhysRevD.83.023513}{{\em Phys. Rev.}
  {\bfseries D83} (2011) 023513},
\href{http://arxiv.org/abs/1008.1282}{{\ttfamily arXiv:1008.1282}}.
%%CITATION = ARXIV:1008.1282;%%.

\bibitem{Aab:2016poe}
A.~Aab {\em et~al.} (Pierre Auger  Collaboration),
  \href{http://dx.doi.org/10.1103/PhysRevD.94.082002}{{\em Phys. Rev.}
  {\bfseries D94} no.~8, (2016) 082002},
\href{http://arxiv.org/abs/1609.04451}{{\ttfamily arXiv:1609.04451}}.
%%CITATION = ARXIV:1609.04451;%%.

\bibitem{Ackermann:2006gp}
M.~Ackermann {\em et~al.} (AMANDA  Collaboration),
  \href{http://dx.doi.org/10.1016/j.nima.2005.10.029}{{\em Nucl. Instrum.
  Meth.} {\bfseries A556} (2006) 169--181},
\href{http://arxiv.org/abs/astro-ph/0601397}{{\ttfamily
  arXiv:astro-ph/0601397}}.
%%CITATION = ASTRO-PH/0601397;%%.

\bibitem{Ambrosio:2002qu}
M.~Ambrosio {\em et~al.} (MACRO  Collaboration),
  \href{http://dx.doi.org/10.1140/epjc/s2002-01045-x}{{\em Eur. Phys. J.}
  {\bfseries C26} (2002) 163--172},
\href{http://arxiv.org/abs/hep-ex/0207024}{{\ttfamily arXiv:hep-ex/0207024}}.
%%CITATION = HEP-EX/0207024;%%.

\bibitem{Ambrosio:2002qq}
M.~Ambrosio {\em et~al.} (MACRO  Collaboration),
  \href{http://dx.doi.org/10.1140/epjc/s2002-01046-9}{{\em Eur. Phys. J.}
  {\bfseries C25} (2002) 511--522},
\href{http://arxiv.org/abs/hep-ex/0207020}{{\ttfamily arXiv:hep-ex/0207020}}.
%%CITATION = HEP-EX/0207020;%%.

\bibitem{Ueno:2012md}
K.~Ueno {\em et~al.} (Super-Kamiokande  Collaboration),
  \href{http://dx.doi.org/10.1016/j.astropartphys.2012.05.008}{{\em Astropart.
  Phys.} {\bfseries 36} (2012) 131--136},
\href{http://arxiv.org/abs/1203.0940}{{\ttfamily arXiv:1203.0940}}.
%%CITATION = ARXIV:1203.0940;%%.

\bibitem{Coleman:1985ki}
S.~R. Coleman, \href{http://dx.doi.org/10.1016/0550-3213(85)90286-X,
  10.1016/0550-3213(86)90520-1}{{\em Nucl. Phys.} {\bfseries B262} (1985) 263}.
[Erratum: Nucl. Phys.B269,744(1986)].
%%CITATION = NUPHA,B262,263;%%.

\bibitem{Lee:1991ax}
T.~D. Lee and Y.~Pang,
  \href{http://dx.doi.org/10.1016/0370-1573(92)90064-7}{{\em Phys. Rept.}
  {\bfseries 221} (1992) 251--350}.
[,169(1991)].
%%CITATION = PRPLC,221,251;%%.

\bibitem{Kusenko:1997zq}
A.~Kusenko, \href{http://dx.doi.org/10.1016/S0370-2693(97)00584-4}{{\em Phys.
  Lett.} {\bfseries B405} (1997) 108},
\href{http://arxiv.org/abs/hep-ph/9704273}{{\ttfamily arXiv:hep-ph/9704273}}.
%%CITATION = HEP-PH/9704273;%%.

\bibitem{Martin:1997ns}
S.~P. Martin, \href{http://arxiv.org/abs/hep-ph/9709356}{{\ttfamily
  arXiv:hep-ph/9709356}}.
[Adv. Ser. Direct. High Energy Phys.18,1(1998)].
%%CITATION = HEP-PH/9709356;%%.

\bibitem{Kusenko:1997vp}
A.~Kusenko, V.~Kuzmin, M.~E. Shaposhnikov and P.~G. Tinyakov,
  \href{http://dx.doi.org/10.1103/PhysRevLett.80.3185}{{\em Phys. Rev. Lett.}
  {\bfseries 80} (1998) 3185--3188},
\href{http://arxiv.org/abs/hep-ph/9712212}{{\ttfamily arXiv:hep-ph/9712212}}.
%%CITATION = HEP-PH/9712212;%%.

\bibitem{Affleck:1984fy}
I.~Affleck and M.~Dine,
\href{http://dx.doi.org/10.1016/0550-3213(85)90021-5}{{\em Nucl. Phys.}
  {\bfseries B249} (1985) 361--380}.
%%CITATION = NUPHA,B249,361;%%.

\bibitem{Kusenko:1997si}
A.~Kusenko and M.~E. Shaposhnikov,
  \href{http://dx.doi.org/10.1016/S0370-2693(97)01375-0}{{\em Phys. Lett.}
  {\bfseries B418} (1998) 46--54},
\href{http://arxiv.org/abs/hep-ph/9709492}{{\ttfamily arXiv:hep-ph/9709492}}.
%%CITATION = HEP-PH/9709492;%%.

\bibitem{Dvali:1997qv}
G.~R. Dvali, A.~Kusenko and M.~E. Shaposhnikov,
  \href{http://dx.doi.org/10.1016/S0370-2693(97)01378-6}{{\em Phys. Lett.}
  {\bfseries B417} (1998) 99--106},
\href{http://arxiv.org/abs/hep-ph/9707423}{{\ttfamily arXiv:hep-ph/9707423}}.
%%CITATION = HEP-PH/9707423;%%.

\bibitem{Bezrukov:1990mr}
L.~B. Bezrukov {\em et~al.}, {\em Sov. J. Nucl. Phys.} {\bfseries 52} (1990)
  54--59.
[Yad. Fiz.52,86(1990)].
%%CITATION = SJNCA,52,54;%%.

\bibitem{Belolaptikov:1998mn}
I.~A. Belolaptikov {\em et~al.},
\href{http://arxiv.org/abs/astro-ph/9802223}{{\ttfamily
  arXiv:astro-ph/9802223}}.
%%CITATION = ASTRO-PH/9802223;%%.

\bibitem{Bodmer:1971we}
A.~R. Bodmer,
\href{http://dx.doi.org/10.1103/PhysRevD.4.1601}{{\em Phys. Rev.} {\bfseries
  D4} (1971) 1601--1606}.
%%CITATION = PHRVA,D4,1601;%%.

\bibitem{Witten:1984rs}
E.~Witten,
\href{http://dx.doi.org/10.1103/PhysRevD.30.272}{{\em Phys. Rev.} {\bfseries
  D30} (1984) 272--285}.
%%CITATION = PHRVA,D30,272;%%.

\bibitem{Farhi:1984qu}
E.~Farhi and R.~L. Jaffe,
\href{http://dx.doi.org/10.1103/PhysRevD.30.2379}{{\em Phys. Rev.} {\bfseries
  D30} (1984) 2379}.
%%CITATION = PHRVA,D30,2379;%%.

\bibitem{Madsen:2001fu}
J.~Madsen, \href{http://dx.doi.org/10.1103/PhysRevLett.87.172003}{{\em Phys.
  Rev. Lett.} {\bfseries 87} (2001) 172003},
\href{http://arxiv.org/abs/hep-ph/0108036}{{\ttfamily arXiv:hep-ph/0108036}}.
%%CITATION = HEP-PH/0108036;%%.

\bibitem{Madsen:2000kb}
J.~Madsen, \href{http://dx.doi.org/10.1103/PhysRevLett.85.4687}{{\em Phys. Rev.
  Lett.} {\bfseries 85} (2000) 4687--4690},
\href{http://arxiv.org/abs/hep-ph/0008217}{{\ttfamily arXiv:hep-ph/0008217}}.
%%CITATION = HEP-PH/0008217;%%.

\bibitem{DeRujula:1984axn}
A.~De~R\'ujula and S.~L. Glashow,
\href{http://dx.doi.org/10.1038/312734a0}{{\em Nature} {\bfseries 312} (1984)
  734--737}.
%%CITATION = NATUA,312,734;%%.

\bibitem{Ambrosio:1999gj}
M.~Ambrosio {\em et~al.} (MACRO  Collaboration),
  \href{http://dx.doi.org/10.1007/s100520050708}{{\em Eur. Phys. J.} {\bfseries
  C13} (2000) 453--458},
\href{http://arxiv.org/abs/hep-ex/9904031}{{\ttfamily arXiv:hep-ex/9904031}}.
%%CITATION = HEP-EX/9904031;%%.

\bibitem{Pavalas:2015nab}
G.~E. Pavalas, (ANTARES  Collaboration),
{\em Proceedings, 34th International Cosmic Ray Conference (ICRC 2015). The
  Hague, The Netherlands, July 30-August 6, 2015.} {\bfseries PoS(ICRC2015)}
  (2016) 1060.
%%CITATION = POSCI,ICRC2015,1060;%%.

\bibitem{Albuquerque:2008zs}
I.~F.~M. Albuquerque, G.~Burdman, C.~A. Krenke and B.~Nosratpour,
  \href{http://dx.doi.org/10.1103/PhysRevD.78.015010}{{\em Phys. Rev.}
  {\bfseries D78} (2008) 015010},
\href{http://arxiv.org/abs/0803.3479}{{\ttfamily arXiv:0803.3479}}.
%%CITATION = ARXIV:0803.3479;%%.

\bibitem{Jittoh:2005pq}
T.~Jittoh, J.~Sato, T.~Shimomura and M.~Yamanaka,
  \href{http://dx.doi.org/10.1103/PhysRevD.73.055009,
  10.1103/PhysRevD.87.019901}{{\em Phys. Rev.} {\bfseries D73} (2006) 055009},
  \href{http://arxiv.org/abs/hep-ph/0512197}{{\ttfamily arXiv:hep-ph/0512197}}.
[Erratum: Phys. Rev.D87,no.1,019901(2013)].
%%CITATION = HEP-PH/0512197;%%.

\bibitem{Brandenburg:2005he}
A.~Brandenburg, L.~Covi, K.~Hamaguchi, L.~Roszkowski and F.~D. Steffen,
  \href{http://dx.doi.org/10.1016/j.physletb.2005.04.072}{{\em Phys. Lett.}
  {\bfseries B617} (2005) 99--111},
\href{http://arxiv.org/abs/hep-ph/0501287}{{\ttfamily arXiv:hep-ph/0501287}}.
%%CITATION = HEP-PH/0501287;%%.

\bibitem{Albuquerque:2003mi}
I.~F.~M. Albuquerque, G.~Burdman and Z.~Chacko,
  \href{http://dx.doi.org/10.1103/PhysRevLett.92.221802}{{\em Phys. Rev. Lett.}
  {\bfseries 92} (2004) 221802},
\href{http://arxiv.org/abs/hep-ph/0312197}{{\ttfamily arXiv:hep-ph/0312197}}.
%%CITATION = HEP-PH/0312197;%%.

\bibitem{Bi:2004ys}
X.-J. Bi, J.-X. Wang, C.~Zhang and X.-m. Zhang,
  \href{http://dx.doi.org/10.1103/PhysRevD.70.123512}{{\em Phys. Rev.}
  {\bfseries D70} (2004) 123512},
\href{http://arxiv.org/abs/hep-ph/0404263}{{\ttfamily arXiv:hep-ph/0404263}}.
%%CITATION = HEP-PH/0404263;%%.

\bibitem{Ahlers:2006pf}
M.~Ahlers, J.~Kersten and A.~Ringwald,
  \href{http://dx.doi.org/10.1088/1475-7516/2006/07/005}{{\em JCAP} {\bfseries
  0607} (2006) 005},
\href{http://arxiv.org/abs/hep-ph/0604188}{{\ttfamily arXiv:hep-ph/0604188}}.
%%CITATION = HEP-PH/0604188;%%.

\bibitem{Albuquerque:2006am}
I.~F.~M. Albuquerque, G.~Burdman and Z.~Chacko,
  \href{http://dx.doi.org/10.1103/PhysRevD.75.035006}{{\em Phys. Rev.}
  {\bfseries D75} (2007) 035006},
\href{http://arxiv.org/abs/hep-ph/0605120}{{\ttfamily arXiv:hep-ph/0605120}}.
%%CITATION = HEP-PH/0605120;%%.

\bibitem{Ando:2007ds}
S.~Ando, J.~F. Beacom, S.~Profumo and D.~Rainwater,
  \href{http://dx.doi.org/10.1088/1475-7516/2008/04/029}{{\em JCAP} {\bfseries
  0804} (2008) 029},
\href{http://arxiv.org/abs/0711.2908}{{\ttfamily arXiv:0711.2908}}.
%%CITATION = ARXIV:0711.2908;%%.

\bibitem{Canadas:2008ey}
B.~Ca\~nadas, D.~G. Cerde\~no, C.~Mu\~noz and S.~Panda,
  \href{http://dx.doi.org/10.1088/1475-7516/2009/04/028}{{\em JCAP} {\bfseries
  0904} (2009) 028},
\href{http://arxiv.org/abs/0812.1067}{{\ttfamily arXiv:0812.1067}}.
%%CITATION = ARXIV:0812.1067;%%.

\bibitem{Ahlers:2007js}
M.~Ahlers, J.~I. Illana, M.~Masip and D.~Meloni,
  \href{http://dx.doi.org/10.1088/1475-7516/2007/08/008}{{\em JCAP} {\bfseries
  0708} (2007) 008},
\href{http://arxiv.org/abs/0705.3782}{{\ttfamily arXiv:0705.3782}}.
%%CITATION = ARXIV:0705.3782;%%.

\bibitem{Kopper:2015rrp}
S.~Kopper, (IceCube  Collaboration),
{\em Proceedings, 34th International Cosmic Ray Conference (ICRC 2015). The
  Hague, The Netherlands, July 30-August 6, 2015.} {\bfseries PoS(ICRC2015)}
  (2016) 1104.
%%CITATION = POSCI,ICRC2015,1104;%%.

\bibitem{vanderDrift:2013zga}
D.~van~der Drift and S.~R. Klein,
  \href{http://dx.doi.org/10.1103/PhysRevD.88.033013}{{\em Phys. Rev.}
  {\bfseries D88} (2013) 033013},
\href{http://arxiv.org/abs/1305.5277}{{\ttfamily arXiv:1305.5277}}.
%%CITATION = ARXIV:1305.5277;%%.

\bibitem{Pati:1974yy}
J.~C. Pati and A.~Salam, \href{http://dx.doi.org/10.1103/PhysRevD.10.275,
  10.1103/PhysRevD.11.703.2}{{\em Phys. Rev.} {\bfseries D10} (1974) 275--289}.
[Erratum: Phys. Rev.D11,703(1975)].
%%CITATION = PHRVA,D10,275;%%.

\bibitem{Birkel:1998nx}
M.~Birkel and S.~Sarkar,
  \href{http://dx.doi.org/10.1016/S0927-6505(98)00028-0}{{\em Astropart. Phys.}
  {\bfseries 9} (1998) 297--309},
\href{http://arxiv.org/abs/hep-ph/9804285}{{\ttfamily arXiv:hep-ph/9804285}}.
%%CITATION = HEP-PH/9804285;%%.

\bibitem{Ambrosio:2004ub}
M.~Ambrosio {\em et~al.} (MACRO  Collaboration), {\em hep-ex/0402006} (2004) ,
\href{http://arxiv.org/abs/hep-ex/0402006}{{\ttfamily arXiv:hep-ex/0402006}}.
%%CITATION = HEP-EX/0402006;%%.

\bibitem{Aartsen:2014njl}
M.~G. Aartsen {\em et~al.} (IceCube  Collaboration),
\href{http://arxiv.org/abs/1412.5106}{{\ttfamily arXiv:1412.5106}}.
%%CITATION = ARXIV:1412.5106;%%.

\bibitem{TheIceCube-Gen2:2016cap}
M.~G. Aartsen {\em et~al.} (IceCube  Collaboration),
  \href{http://dx.doi.org/10.1088/1361-6471/44/5/054006}{{\em J. Phys.}
  {\bfseries G44} no.~5, (2017) 054006},
\href{http://arxiv.org/abs/1607.02671}{{\ttfamily arXiv:1607.02671}}.
%%CITATION = ARXIV:1607.02671;%%.

\end{thebibliography}
\providecommand{\href}[2]{#2}\begingroup\raggedright\endgroup

\end{document}